\documentclass[aps,prl,twocolumn,superscriptaddress,amsmath,amssymb]{revtex4}
\usepackage{graphicx} 
\usepackage{epstopdf} 
\usepackage{dcolumn} 
\usepackage{xcolor} 
\usepackage{amsmath,amssymb} 
\usepackage{hyperref} 
\usepackage[utf8]{inputenc} 
\usepackage[english]{babel} 
\usepackage{ifthen} 
\usepackage{orcidlink} 

\graphicspath{{figures/}} 

\newboolean{articletitles}
\setboolean{articletitles}{true} 

\input{belle2-symbols}

\def \BptoKS {\hbox{$B^+\to K^+S(\to x^+x^-)$}\xspace}
\def \BstarztoKS {\hbox{$B^0\to \Kstarz(\to K^+\pi^-) S(\to x^+x^-)$}\xspace}

\usepackage{subfigure}
\usepackage{placeins}

\newboolean{wordcount}
\setboolean{wordcount}{false} 

\ifthenelse{\boolean{wordcount}}%
{
\usepackage{bibentry} 
\usepackage{comment} 
\excludecomment{align*} 
\excludecomment{align} 
\excludecomment{equation*} 
\excludecomment{equation} 
\excludecomment{eqnarray*} 
\excludecomment{eqnarray} 
\excludecomment{acknowledgments} 
\def\maketitle{} 
}{}

\begin{document}

\title{
Search for a long-lived spin-0 mediator in \hbox{$b\to s$} transitions at the Belle~II experiment
}

\ifthenelse{\boolean{wordcount}}{}{
  \author{I.~Adachi\,\orcidlink{0000-0003-2287-0173}} 
  \author{K.~Adamczyk\,\orcidlink{0000-0001-6208-0876}} 
  \author{L.~Aggarwal\,\orcidlink{0000-0002-0909-7537}} 
  \author{H.~Aihara\,\orcidlink{0000-0002-1907-5964}} 
  \author{N.~Akopov\,\orcidlink{0000-0002-4425-2096}} 
  \author{A.~Aloisio\,\orcidlink{0000-0002-3883-6693}} 
  \author{N.~Anh~Ky\,\orcidlink{0000-0003-0471-197X}} 
  \author{D.~M.~Asner\,\orcidlink{0000-0002-1586-5790}} 
  \author{H.~Atmacan\,\orcidlink{0000-0003-2435-501X}} 
  \author{T.~Aushev\,\orcidlink{0000-0002-6347-7055}} 
% ...
\collaboration{The Belle II Collaboration}

}

\begin{abstract}
Additional spin-0 particles appear in many extensions of the standard model. 
We search for long-lived spin-0 particles $S$ in $B$-meson decays mediated by a \hbox{$b\to s$} quark transition in $e^+e^-$ collisions at the $\Upsilon(4S)$ resonance at the \belletwo experiment. 
Based on a sample corresponding to an integrated luminosity of $189 \invfb$, we observe no evidence for signal. 
We set model-independent upper limits on the product of branching fractions \hbox{$\mathcal{B}(B^0\to K^*(892)^0(\to K^+\pi^-)S)\times \mathcal{B}(S\to x^+x^-)$} and \hbox{$\mathcal{B}(B^+\to K^+S)\times \mathcal{B}(S\to x^+x^-)$}, where $x^+x^-$ indicates $e^+e^-, \mu^+\mu^-, \pi^+\pi^-$, or $K^+K^-$, as functions of $S$ mass and lifetime at the level of $10^{-7}$.
\end{abstract}

\maketitle

Minimal renormalizable extensions of the standard model (SM) allow for the existence of an additional light spin-0 (scalar) $S$ that may give mass to dark matter particles~\cite{Pospelov:2007mp}.
Such a new scalar would mix with the SM Higgs boson through a mixing angle $\theta$~\cite{OConnell:2006rsp,Beacham:2019nyx}.
However, for masses $m_S$ below the $B$-meson mass, decays of $S$ into dark matter particles must be kinematically forbidden to provide the correct relic density~\cite{Krnjaic:2015mbs}.
This motivates a search for $S$ decays into SM particles.
For couplings much weaker than the electroweak interaction, the scalar is long-lived.
Another possible extension of the SM introduces a spin-0 (pseudoscalar) axionlike particle (ALP) that couples to photons, fermions, or gluons~\cite{Jaeckel:2010ni}.
ALPs share the quantum numbers of axions, but differ in that their masses and couplings are independent.
The set of possible ALP couplings is large and includes  models with a predominant coupling $f_{a}$ to fermions that results in long-lived ALPs decaying into SM leptons~\cite{Freytsis:2009ct,Dolan:2014ska,Beacham:2019nyx}.

To distinguish between different models, lifetime-dependent results for different final states and different production modes are needed~\cite{Kachanovich:2020yhi, Dobrich:2018jyi, Ferber:2022rsf}.
To date, almost all direct searches or reinterpretations of previous analyses have focused on the minimal scalar model, with some reinterpretations in the context of ALPs~\cite{Beacham:2019nyx}.
The current best limits from colliders exclude mixing angles $\sin\theta$ larger than $10^{-3}$ to $10^{-4}$.
For $m_S\gtrsim0.3$\,\gevcc the best limits
 are reported in Ref.~\cite{Winkler:2018qyg}, which are based on searches by the LHCb Collaboration for displaced scalars in $B\to \Kstarz S(\to\mu^+\mu^-)$ and $B\to K^+S(\to\mu^+\mu^-)$ decays~\cite{LHCb:2015nkv, LHCb:2016awg}. 
In this letter, $\Kstarz$ indicates a $K^*(892)^0$ meson and charge conjugated processes are included implicitly.
The CMS experiment has reported limits competitive with LHCb~\cite{CMS:2021sch}.
For lighter $S$ masses, $m_S\lesssim 0.3$\,\gevcc, the best upper limits are provided by reinterpretations~\cite{Winkler:2018qyg} of searches for the decays $K^0_L\to\pi^0\mu^+\mu^-$~\cite{KTEV:2000ngj},
$K^+ \to \pi^+\nu\bar{\nu}$~\cite{BNL-E949:2009dza}, and searches for displaced lepton-pairs in beam-dump experiments~\cite{CHARM:1985anb,Gorbunov:2021ccu}, as well as by direct searches by the experiments NA62~\cite{NA62:2021zjw,NA62:2020pwi} and MicroBooNE~\cite{MicroBooNE:2021usw}.
An inclusive search for $B\to X_sS$ decays by the \babar experiment excludes a small parameter region around $m_S\approx 0.9\,\gevcc$ not covered by the other results~\cite{BaBar:2015jvu,Winkler:2018qyg}.
Model-dependent studies of supernova SN1987A and primordial nucleosynthesis constrain the mixing angle $\theta$ to values larger than $10^{-5}$ (for $m_S \lesssim 0.2$\,\gevcc) to $10^{-7}$ (for $0.2 < m_S \lesssim 4$\,\gevcc)~\cite{Winkler:2018qyg}.
To date only LHCb has studied scalar decays separated into exclusive production modes, whereas only the \babar experiment has studied scalar decays into hadronic final states but not separated into exclusive production modes.

In this Letter, we search for a long-lived particle (LLP) \hbox{$S\to x^+x^-$}, where $x^+x^-$ indicates $e^+e^-, \mu^+\mu^-, \pi^+\pi^-$, or $K^+K^-$, in \BptoKS and \BstarztoKS decays mediated by a flavor-changing neutral current \hbox{$b\to s$} transition~\cite{Winkler:2018qyg,Filimonova:2019tuy}.
We search for the signal as a narrow enhancement in the invariant $S$ mass distribution in events with $B$ decays at the $\Upsilon(4S)$ resonance.
The search is conducted for masses between 25\,\mevcc ($S\to e^+e^-$), 211\,\mevcc ($S\to \mu^+\mu^-$), 280\,\mevcc ($S\to \pi^+\pi^-$), or 988\,\mevcc ($S\to K^+K^-$), and 4.78\,\gevcc for \BptoKS or 4.38\,\gevcc for \BstarztoKS.
We present our results as model-independent limits on the products of branching fractions $\mathcal{B}(B^+\to K^+S)\times \mathcal{B}(S\to x^+x^-)$ and $\mathcal{B}(B^0\to \Kstarz(\to K^+\pi^-) S)\times \mathcal{B}(S\to x^+x^-)$ for various lifetimes $0.001 < c\tau < 100~\mathrm{cm}$.
In addition to the model-independent search, we report our results as limits on the mixing angle $\theta$ and on the ALP coupling for the aforementioned dark scalar and ALP models. 

We use a sample of $N_{B\bar{B}}= (198\pm3)\times 10^6$~$B$-meson pairs corresponding to an integrated luminosity of 189~\invfb.
The data is collected at a center-of-mass~(c.m.) energy of $\sqrt{s}=10.58\gev$ by the \belletwo experiment\,\cite{Abe:2010gxa} at the SuperKEKB $e^+e^-$ collider\,\cite{Akai:2018mbz}.
The beam energies are $7\gev$ for $e^-$ and $4\gev$ for $e^+$, resulting in a boost $\beta\gamma=0.28$ of the c.m.~frame relative to the laboratory frame.
The \belletwo detector is a hermetic magnetic spectrometer surrounded by particle-identification detectors, a electromagnetic calorimeter, and a $K^0_L$ and muon detector, arranged around the beam pipe in a cylindrical structure\,\cite{Kou:2018nap}. 
The longitudinal direction, the transverse plane, and the polar angle $\theta_{\rm{polar}}$ are defined with respect to the detector’s solenoidal axis in the direction of the electron beam. 
In the following, quantities are defined in the laboratory frame unless specified otherwise.

We use simulated events to determine efficiencies and signal-shape parameters.
Signal events are simulated using \textsc{EvtGen}\,\cite{Lange:2001uf} for various scalar masses $0.025 < m_S < 4.78$\,\gevcc in  about 90 steps of varying size, and various lifetimes $0.001 < c\tau < 400$\,\cm in variable steps.
We simulate the following background processes: $e^+e^-\to\FourS\to\BBbar$ where $B$ indicates a $B^0$ or a $B^+$ meson with \textsc{EvtGen}\,\cite{Lange:2001uf}; $e^+e^-\to q\bar{q} (\gamma)$ where  $q\bar{q}$ indicates $u\bar{u}, d\bar{d}, s\bar{s}$, or $c\bar{c}$ quark pairs with \textsc{KKMC}\,\cite{Jadach:1999vf} interfaced with  \textsc{Pythia8}\,\cite{Sjostrand:2014zea} and  \textsc{EvtGen}; $e^+e^-\to\tau\tau(\gamma)$ with \textsc{KKMC} interfaced with \textsc{Tauola}\,\cite{Jadach:1990mz}.
Electromagnetic final-state radiation is simulated using \textsc{Photos}~\cite{Barberio:1990ms} for all charged particles generated by \textsc{EvtGen}.
The detector geometry and interactions of final-state particles with detector material are simulated using \textsc{Geant4}\,\cite{Agostinelli:2002hh}. 
Both experimental and simulated events are reconstructed and analyzed using the \belletwo software~\cite{Kuhr:2018lps, basf2-zenodo}.
To avoid experimenter's bias, we examine the experimental data only after finalizing the analysis selection.
However, we observed negative background yields in fits to data that led to a modification of the fit strategy restricting the background probability density function~(pdf) to non-negative values. 
Simulations show that this does not introduce any significant bias in the signal yield or expected limits.
All selection criteria are chosen by optimizing the figure-of-merit for a discovery with a significance of five standard deviations\,\cite{Punzi:2003bu}.

We reconstruct $B$-meson candidates from charged-particle trajectories (tracks) originating either from the $e^+e^-$ interaction point (prompt), or from a vertex separated from it by a macroscopic distance (displaced).
We require each prompt track to correspond to a transverse momentum of more than 0.15\,\gevc. 
In addition, it must have a distance of closest approach to the interaction point~(IP) of less than 0.5\,\cm in the plane transverse to the beam and 2.0\,\cm in the direction orthogonal to it in order to remove charged particles not associated with the $e^+e^-$ interaction.
Displaced tracks are required to correspond to a transverse momentum of greater than 0.25\,\gevc, but have no restrictions on their distance of closest approach to the IP.
We reconstruct $S$ candidates by combining pairs of oppositely charged displaced particles, both identified as electrons, muons, pions, or kaons.
A fit constrains the pair of displaced tracks to come from a common vertex.
The displaced vertex must have a transverse distance $d_{\text{v}}$ to the IP of at least $0.05$\,\cm.
All displaced tracks must have an extrapolated polar angle $32^{\circ} < \theta_{\rm{polar}} < 150^{\circ}$, calculated by extrapolating the track from the displaced vertex to the calorimeter surface.
The distance of the displaced vertex from the IP should exceed three times its resolution.
For signal \BstarztoKS decays, we combine two oppositely charged prompt particles, identified as a kaon and a pion, in a vertex fit to form a $\Kstarz$ candidate with mass $0.796 < M({K^+\pi^-}) < 0.996$\,\gevcc that is then combined with the $S$ candidate.
For signal \BptoKS candidates, we combine the $S$ candidate with a prompt track identified as a kaon.

Particle identification (PID) information from all relevant sub-detectors is combined to separate final states into exclusive samples and to further reduce backgrounds~\cite{Kou:2018nap}. 
We exclude the time-of-propagation detector from the PID determination when separating $e^+e^-$, $\mu^+\mu^-$, and $\pi^+\pi^-$ pairs because it tends to misidentify the $S$ decay products as heavier particles due to the long $S$ time-of-flight.
The prompt pion PID efficiency is about 84\% for all $S$ masses; the prompt kaon PID efficiency is about 80\% and decreases to 40\% for the highest $S$ masses.
For displaced tracks we give the PID efficiency for pairs, corresponding to the identification of $S\to x^+x^-$. It ranges in 96\%--99\% for $e^+e^-$, 89\%--96\% for $\mu^+\mu^-$, 75\%--90\% for $\pi^+\pi^-$, and 50\%--80\% for $K^+K^-$.
The differences in displaced-pair efficiencies between $\Kstarz$ and $K^+$ final states for the same $S$ mass do not exceed 2\%.
The dominant backgrounds are from light quark pair production, followed by $c\bar{c}$ pair production and then $e^+e^-\to\FourS\to\BBbar$.
The above selections reduce the backgrounds by factors between 7 (for $K^+\pi^+\pi^-$) and about 370 (for $\Kstarz\mu^+\mu^-$).
The probability to misidentify signal is generally less than 1\% in our simulated samples.
To suppress prompt peaking SM backgrounds, the transverse-distance requirement of the displaced vertex is increased to $d_{\text{v}}>0.2$\,\cm in $S$-mass regions close to known two-body decays of SM particles like $J/\psi\to\mu^+\mu^-$ or $\phi\to K^+K^-$~\cite{aux:2023}.
To suppress background from $\gamma\to e^+e^-$ conversions, we veto events in the $e^+e^-$ final state for \hbox{$m_S<0.05$\,\gevcc} if the vertex is close to inner tracking layers.

The cosine of the angle between the vector connecting the IP with the decay vertex and the momentum vector of the scalar candidate in the transverse plane must exceed 0.95 for $e^+e^-$, $\mu^+\mu^-$, and $K^+K^-$ candidates to reject background from events with missing particles and random track combinations; it must exceed 0.99 for $\pi^+\pi^-$ to further reduce the higher backgrounds in this final state.
To suppress $e^+e^-\to q\bar{q}(\gamma)$ and $e^+e^-\to\tau^+\tau^-(\gamma)$ backgrounds, we require each $B$-meson candidate to have a beam-constrained mass value $M_{\text{bc}} = \sqrt{s/4-\vert\vec{p}^{\,*}_B\vert^2} > 5.27$\,\gevcc, where $\vec{p}^{\,*}_B$ is the three-momentum of the $B$-meson candidate in the c.m. system. 
We further require that the $B$-meson candidate has an energy difference $\vert \Delta E \vert= \vert E^{\,*}_B-\sqrt{s}/2\vert < 0.05$\,\gev, where $E^{\,*}_B$ is the energy of the $B$-meson candidate in the c.m.\,system; for $\pi^+\pi^-$ candidates, the requirement is tightened to $\vert \Delta E \vert < 0.035$\,\gev.
Displaced $K^+K^-$ pairs are selected with high purity by the $\vert\Delta E\vert$ requirement alone due to the larger $K$ mass. 
To reduce continuum background, events must have $R_2 < 0.45$, where $R_2$ is the ratio of the second and zeroth Fox-Wolfram moments. The ratio tends to small values for more spherical distributions of final-state particle momenta as expected from $B$-mesons, which are lightly boosted, while larger values are expected for the collimated momentum distribution of light-quarks, which are boosted\,\cite{PhysRevLett.41.1581}. The requirement is restricted to $R_2 < 0.35$ for $\pi^+\pi^-$ candidates.
We reject events with displaced track-pairs consistent with {$0.498 < M(\pi^+\pi^-) < 0.507$\,\gevcc} to reduce background from $K_S^0$ decays.
If multiple signal candidates pass the selections, which happens in less than 0.5\% of the events, we choose the candidate with the smallest value of $\vert\Delta E\vert$.

\begin{figure}[ht]
\centering
\includegraphics[width=0.48\textwidth]{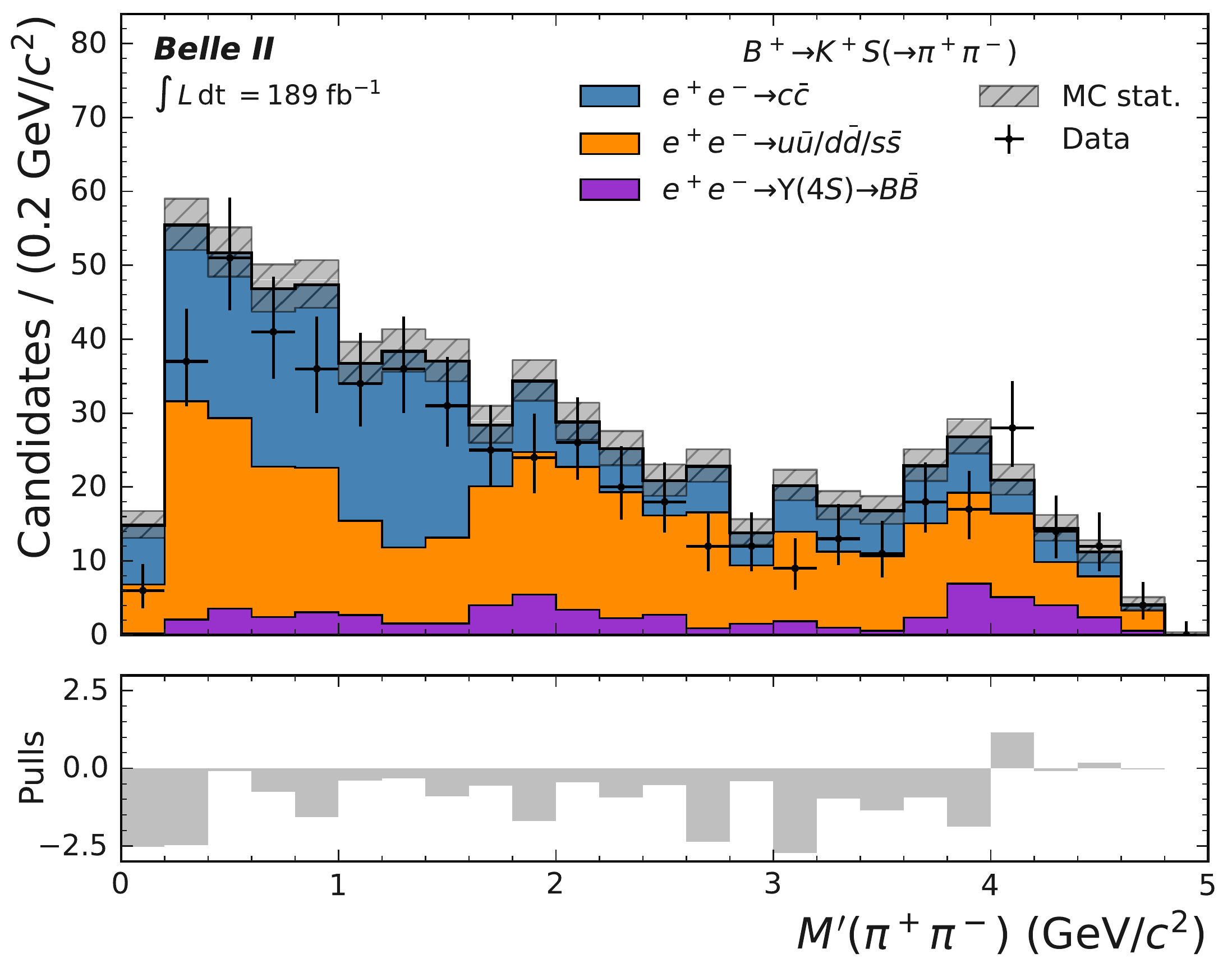}%
\caption{\label{fig:1} Distribution of $M^{\prime}(\pi^+\pi^-)$ together with  the  stacked  contributions  from  the  various simulated SM background samples for \hbox{$B^+\to K^+S(\to \pi^+\pi^-)$} candidates. 
Simulation  is  normalized  to a luminosity of 189~\invfb. The hatched area represents the statistical uncertainty of the SM background prediction.
The background from $e^+e^-\to\tau\tau(\gamma)$ is negligible.
The bottom panel shows the pulls per bin, defined as the difference between data and simulation, normalized to the statistical uncertainties added in quadrature.}
\end{figure}

For the signal extraction, we use a modified mass $M^{\prime} ({x^+x^-})= \sqrt{M^2({x^+x^-})-4m^2_{x}}$ with $m_x=m_{e^+}, m_{\mu^+}, m_{\pi^+}$, or $m_{K^+}$, to simplify the modeling of the signal width close to kinematic thresholds where the scalar mass approaches twice the rest mass of the final-state particles.
$M^{\prime}$ equals twice the $x$ momentum in the $x^+x^-$ rest frame.
An example of a modified-mass distribution for $B^+\to K^+S(\to\pi^+\pi^-)$ is shown in Fig.\,\ref{fig:1}.
Normalization discrepancies are not a concern since backgrounds are floating in all fits.

To validate the selection we compare simulation and data in the $\KS$ mass region rejected in the signal selection, in the displacement regions close to promptly decaying SM resonances rejected in the signal selection, and in sidebands formed by inverting the $M_{\text{bc}}$ and $\Delta E$ selections.
We find agreement for all selection variables.

We extract the \BptoKS and \BstarztoKS signal yields by performing extended  maximum likelihood fits to the unbinned modified $S$-mass distribution.
We fit for a narrow nonnegative-yield signal peak, at various values of $S$ mass and assuming various lifetimes, over a smooth background.
We perform independent fits~\cite{Eschle:2019jmu} for each of the eight final states and for each lifetime with a $S$-mass scan step-size equal to half the signal mass resolution $\sigma_{\mathrm{sig}}$.
For the model-dependent searches, we perform a combined fit in all relevant and kinematically accessible analysis channels, again separately for various lifetimes.
For the dark scalar model we fix the $B$-meson and scalar branching fractions to the theoretical values from Refs.\,\cite{Winkler:2018qyg,Filimonova:2019tuy,Gubernari:2018wyi}; for the ALP model the $B$-meson and ALP branching fractions are taken from  Refs.\,\cite{ Beacham:2019nyx, Batell:2009jf, Dolan:2014ska, Domingo:2016yih} using a cut-off scale of $\Lambda=1\,\tev$ and assuming identical coupling $f_a=f_q=f_{\ell}$ to quarks and leptons.
For $m_S$ greater than 2\,\gevcc, only $S\to\mu^+\mu^-$ is included in the combined scalar fit due to large uncertainties in the predicted branching fractions.

The signal is described by a double-sided Crystal Ball function\,\cite{Gaiser:Phd, Skwarnicki:1986xj} with all parameters determined from fits to the simulated signal samples.
Mass hypotheses that lack a simulation sample are interpolated from adjacent simulated samples.
The resolution $\sigma_{\mathrm{sig}}$ increases smoothly from about 2\,\mevcc for a light $S$ to about 10\,\mevcc for a heavy $S$ and does not depend significantly on lifetime or final state.
However, the tails of the signal distribution, especially for larger $m_S$, increase for larger lifetimes.
This is reflected in a variation of the corresponding parameter values.

The background is modeled by a straight line, with normalization and slope determined from the fit to data.
This model describes the background beneath any potential signal across the range of $S$ masses.
We restrict the linear function to non-negative values in the full fit range by limiting the slope parameter accordingly.
To account for a possible remaining conversion background, an exponential function is added to the background model when signal mass hypotheses below $m_S < 40$\,\mevcc are tested in the $e^+e^-$ final state.
Each likelihood fit is performed over an $M^{\prime}(x^+x^-)$ range with a width of $\pm 20\sigma_{\mathrm{sig}}$.
To improve the fit stability, we iteratively increase the fit range symmetrically in 10\% steps until it contains at least ten events.
We verify that small variations of the fit-interval extension have negligible effects on the results.

We include mass- and lifetime-dependent systematic uncertainties associated with the signal efficiency and with our signal model pdf as Gaussian nuisance parameters with widths equal to the systematic uncertainty.
The systematic uncertainties associated with the signal efficiency are typically around 4\% for most of the scalar masses and lifetimes, but can reach 10\% for the lightest scalar masses accessible in the $e^+e^-$ final state.
For large displacements, the dominant systematic uncertainty on the signal efficiency is due to the difference in track finding efficiency for displaced tracks between data and simulation. 
This uncertainty varies between zero (prompt) and 45\% per event depending linearly on the vertex position.
We correct for this efficiency difference based on a large $\KS$ control sample and assign the full efficiency difference as a systematic uncertainty, which is relevant mostly for small $m_S$.
For larger $m_S$ values, the 2.9\% contribution from the combination of the uncertainty on the $B\bar{B}$ yield and the uncertainty on the charged-to-neutral $B$-meson ratio from $\Upsilon(4S)$ decays\,\cite{Belle:2022hka}, along with the PID efficiency of low-momentum prompt kaons in the $\Kstarz$ channel (3\%) are the largest systematic uncertainties.
We verify the modeling and fitting procedure using pseudo-experiments and add an uncertainty of 3\% to the signal efficiency to account for a small bias in the independent fits; the uncertainty is 4\%  for the combined fit.
We also include systematic uncertainties due to differences between simulation and data that affect the signal model. 
For this we correct the difference between simulation and data of the signal pdf parameters using a large $\KS$ control sample and assign the full difference between simulation and data as a systematic uncertainty.
The typical total uncertainty is around 15\% for the signal width and around 10\% for the tail parameters.

The local significance $\mathcal{S}$ of the signal for a given mass and lifetime hypothesis is given by $\mathcal{S}=\sqrt{2(\log\mathcal{L}-\log\mathcal{L}_{\text{bkg}})}$, where $\mathcal{L}$ is the maximum likelihood for the full fit and $\mathcal{L}_{\text{bkg}}$ is the maximum likelihood for a fit to the background-only hypothesis.

\begin{figure}[ht]
\centering
\includegraphics[width=0.45\textwidth]{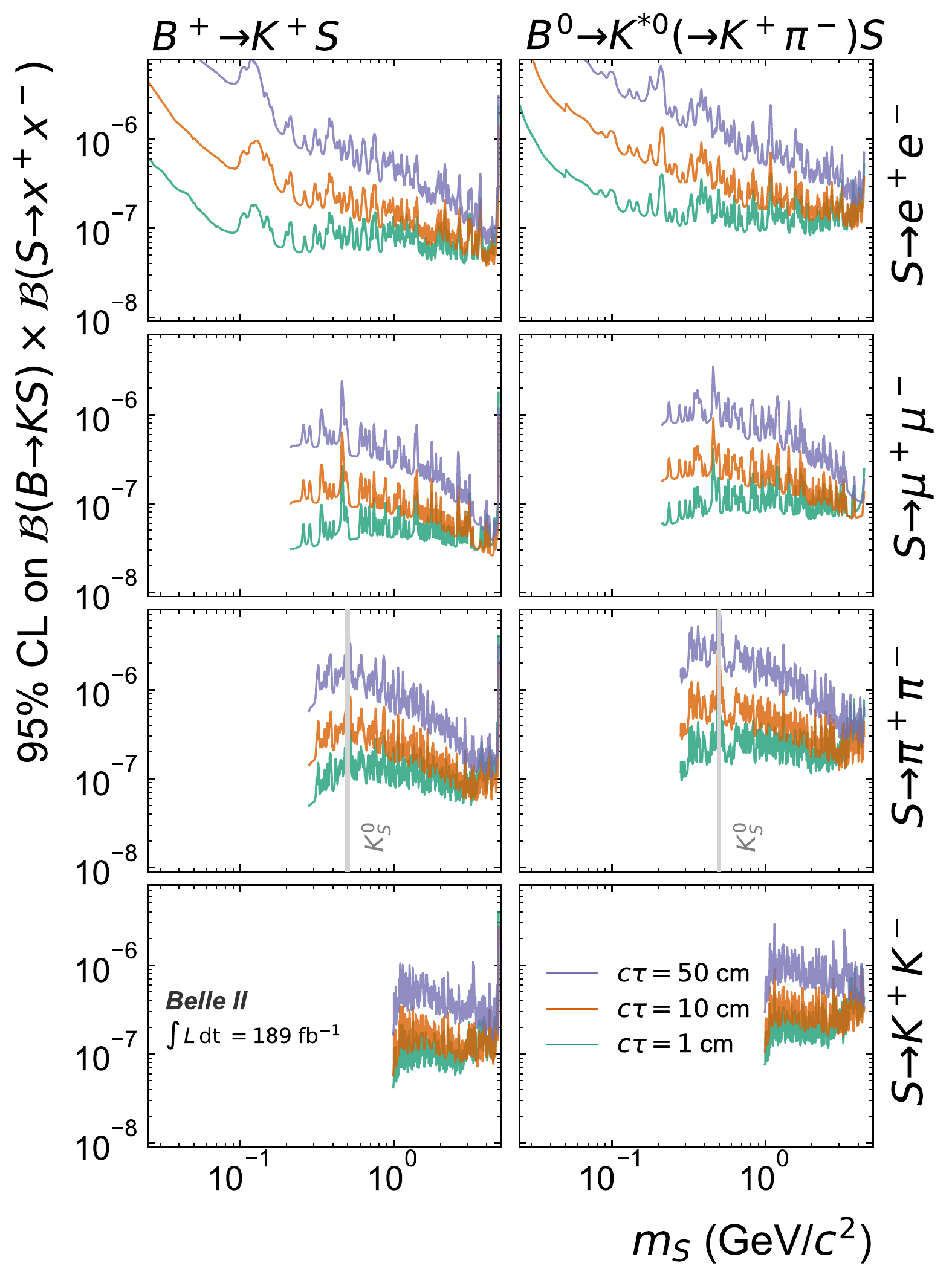}%
\caption{\label{fig:2} Upper limits (95\% CL) on the product of branching fractions $\mathcal{B}(B^+\to K^+S) \times \mathcal{B}(S\to x^+x^-)$ (left) and $\mathcal{B}(B^0\to \Kstarz(\to K^+\pi^-) S) \times \mathcal{B}(S\to x^+x^-)$  (right) as functions of scalar mass $m_S$ for \hbox{$c\tau =1\,\text{cm}$}~(green), \hbox{$c\tau =10\,\text{cm}$}~(orange), and \hbox{$c\tau =50\,\text{cm}$}~(lavender). 
The region corresponding to the fully-vetoed $\KS$ for $S\to\pi^+\pi^-$ is marked in gray.
}
\end{figure}

The largest local significance for the model-independent search is $3.6\sigma$, including systematic uncertainties, found near $m_S=1.061$\,\gevcc for  $K^+\pi^+\pi^-$ for a lifetime of $c\tau=0.05\cm$.
Taking into account the look-elsewhere effect\,\cite{Gross:2010qma}, this excess has a global significance of $1.0\sigma$.
By dividing the signal yield by the signal efficiency and $N_{B\bar{B}}$, we obtain the products of branching fractions $\mathcal{B}(B^+\to K^+S)\times \mathcal{B}(S\to x^+x^-)$ and $\mathcal{B}(B^0\to \Kstarz(\to K^+\pi^-) S)\times \mathcal{B}(S\to x^+x^-)$.
To convert the latter to upper limits on the product of branching fractions $\mathcal{B}(B^0\to \Kstarz S) \times \mathcal{B}(S\to x^+x^-)$, the limits are multiplied by 3/2~\cite{Workman:2022ynf}.
We  compute  the  95\% confidence level~(CL) upper limits~\cite{Rodrigues:2020syo} as functions of scalar mass $m_S$ using a one-sided modified frequentist $CL_S$ method\,\cite{Read:2002hq} with asymptotic approximation\,\cite{Cowan:2010js}.
The observed upper limits are shown in Fig.\,\ref{fig:2}.
Systematic uncertainties weaken the limits by about 2\% for light $S$ and large lifetime; for heavier $S$ or short lifetimes, the reduction is less than 1\%.
A direct comparison of our model-independent limits with the inclusive \babar~\cite{BaBar:2015jvu} limits are possible whenever the \babar limits are stronger than ours and the knowledge of the production mode is not important.

\begin{figure}[ht]
\centering
\includegraphics[width=0.48\textwidth]{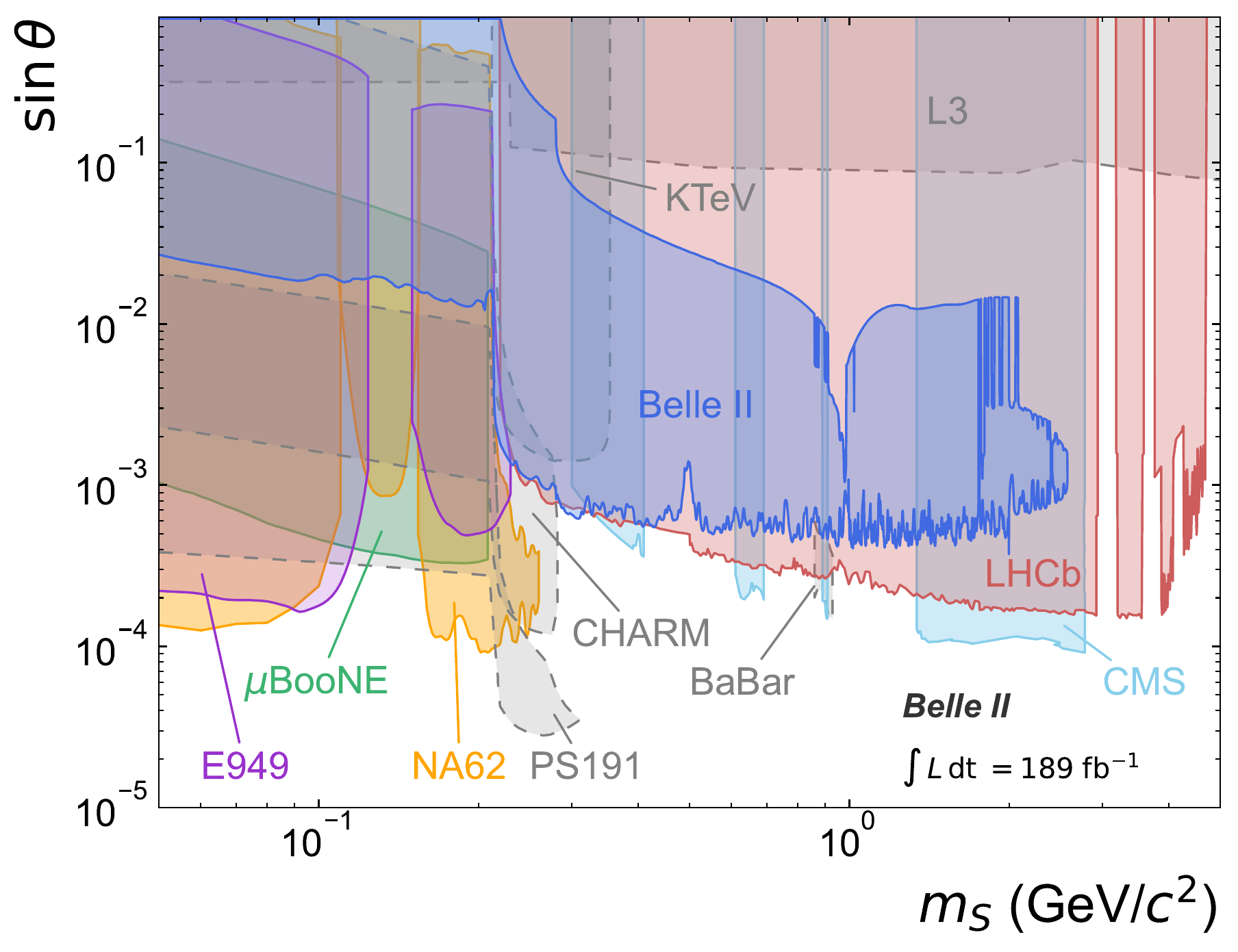}%
\caption{\label{fig:4} Exclusion regions in the plane of the sine of the mixing angle $\theta$ and scalar mass $m_S$ from this work (blue) together with existing constraints from LHCb~\cite{LHCb:2015nkv,LHCb:2016awg}, 
CMS~\cite{CMS:2021sch},
KTeV~\cite{KTEV:2000ngj}, 
E949~\cite{BNL-E949:2009dza},
CHARM~\cite{CHARM:1985anb}, 
PS191~\cite{Gorbunov:2021ccu},
NA62~\cite{NA62:2021zjw,NA62:2020pwi}
\babar~\cite{BaBar:2015jvu},
MicroBooNE~\cite{MicroBooNE:2021usw}, and L3~\cite{L3:1996ome}.
The exclusion regions from \belletwo, CMS, LHCb, CHARM, and MicroBooNE correspond to 95\%\,CL, while PS191, KTeV, E949, NA62, and \babar correspond to 90\%~CL.
The CMS constraint should be interpreted with caution since it is based on different $B$-meson and scalar branching fractions.
Constraints colored in gray with dashed outline are reinterpretations not performed by the experimental collaborations.}
\end{figure}

The largest local significance for the combined scalar and ALP fit is $3.3\sigma$, including systematic uncertainties, found near $m_S=2.619\gevcc$ for a lifetime of $c\tau=100\cm$; the global significance is $0.3\sigma$.
For each scalar or ALP mass hypothesis, we determine the value of $\sin\theta$ or $f_a$ such that the resulting predicted product of branching-fraction ratios equals the 95\% excluded branching fraction.
The observed upper limit on $\sin\theta$ is shown in Fig.\,\ref{fig:4}.
Our limit is competitive with that set by LHCb for $m_S\approx 0.3\,\gevcc$.
The observed upper limit on $f_a$, as well as additional plots and detailed numerical results can be found in the supplemental material~\cite{aux:2023}.

In conclusion, we report the first \belletwo search for long-lived particles.
We search for a long-lived spin-0 mediator $S$ in $B$-meson decays mediated by a \hbox{$b\to s$} transition using 189~\invfb of \belletwo data.
We  do  not  observe  any  significant  excess  of events consistent with a signal process.
We set 95\%~CL upper limits on the product of branching fractions $\mathcal{B}(B^+\to K^+S)\times \mathcal{B}(S\to x^+x^-)$ and $\mathcal{B}(B^0\to \Kstarz S)\times \mathcal{B}(S\to x^+x^-)$ that are the first for exclusive hadronic and for $e^+e^-$ final states.\\

\ifthenelse{\boolean{wordcount}}%
{ {}}
{ \begin{acknowledgments}
We thank Felix Kahlhoefer for useful discussions during the preparation of this manuscript.
This work, based on data collected using the Belle II detector, which was built and commissioned prior to March 2019, was supported by
Science Committee of the Republic of Armenia Grant No.~20TTCG-1C010;
Australian Research Council and research Grants
No.~DP200101792, 
No.~DP210101900, 
No.~DP210102831, 
No.~DE220100462, 
No.~LE210100098, 
and
No.~LE230100085; 
Austrian Federal Ministry of Education, Science and Research,
Austrian Science Fund
No.~P~31361-N36
and
No.~J4625-N,
and
Horizon 2020 ERC Starting Grant No.~947006 ``InterLeptons'';
Natural Sciences and Engineering Research Council of Canada, Compute Canada and CANARIE;
National Key R\&D Program of China under Contract No.~2022YFA1601903,
National Natural Science Foundation of China and research Grants
No.~11575017,
No.~11761141009,
No.~11705209,
No.~11975076,
No.~12135005,
No.~12150004,
No.~12161141008,
and
No.~12175041,
and Shandong Provincial Natural Science Foundation Project~ZR2022JQ02;
the Ministry of Education, Youth, and Sports of the Czech Republic under Contract No.~LTT17020 and
Charles University Grant No.~SVV 260448 and
the Czech Science Foundation Grant No.~22-18469S;
European Research Council, Seventh Framework PIEF-GA-2013-622527,
Horizon 2020 ERC-Advanced Grants No.~267104 and No.~884719,
Horizon 2020 ERC-Consolidator Grant No.~819127,
Horizon 2020 Marie Sklodowska-Curie Grant Agreement No.~700525 "NIOBE"
and
No.~101026516,
and
Horizon 2020 Marie Sklodowska-Curie RISE project JENNIFER2 Grant Agreement No.~822070 (European grants);
L'Institut National de Physique Nucl\'{e}aire et de Physique des Particules (IN2P3) du CNRS (France);
BMBF, DFG, HGF, MPG, and AvH Foundation (Germany);
Department of Atomic Energy under Project Identification No.~RTI 4002 and Department of Science and Technology (India);
Israel Science Foundation Grant No.~2476/17,
U.S.-Israel Binational Science Foundation Grant No.~2016113, and
Israel Ministry of Science Grant No.~3-16543;
Istituto Nazionale di Fisica Nucleare and the research grants BELLE2;
Japan Society for the Promotion of Science, Grant-in-Aid for Scientific Research Grants
No.~16H03968,
No.~16H03993,
No.~16H06492,
No.~16K05323,
No.~17H01133,
No.~17H05405,
No.~18K03621,
No.~18H03710,
No.~18H05226,
No.~19H00682, 
No.~22H00144,
No.~26220706,
and
No.~26400255,
the National Institute of Informatics, and Science Information NETwork 5 (SINET5), 
and
the Ministry of Education, Culture, Sports, Science, and Technology (MEXT) of Japan;  
National Research Foundation (NRF) of Korea Grants
No.~2016R1\-D1A1B\-02012900,
No.~2018R1\-A2B\-3003643,
No.~2018R1\-A6A1A\-06024970,
No.~2018R1\-D1A1B\-07047294,
No.~2019R1\-I1A3A\-01058933,
No.~2022R1\-A2C\-1003993,
and
No.~RS-2022-00197659,
Radiation Science Research Institute,
Foreign Large-size Research Facility Application Supporting project,
the Global Science Experimental Data Hub Center of the Korea Institute of Science and Technology Information
and
KREONET/GLORIAD;
Universiti Malaya RU grant, Akademi Sains Malaysia, and Ministry of Education Malaysia;
Frontiers of Science Program Contracts
No.~FOINS-296,
No.~CB-221329,
No.~CB-236394,
No.~CB-254409,
and
No.~CB-180023, and No.~SEP-CINVESTAV research Grant No.~237 (Mexico);
the Polish Ministry of Science and Higher Education and the National Science Center;
the Ministry of Science and Higher Education of the Russian Federation,
Agreement No.~14.W03.31.0026, and
the HSE University Basic Research Program, Moscow;
University of Tabuk research Grants
No.~S-0256-1438 and No.~S-0280-1439 (Saudi Arabia);
Slovenian Research Agency and research Grants
No.~J1-9124
and
No.~P1-0135;
Agencia Estatal de Investigacion, Spain
Grant No.~RYC2020-029875-I
and
Generalitat Valenciana, Spain
Grant No.~CIDEGENT/2018/020
Ministry of Science and Technology and research Grants
No.~MOST106-2112-M-002-005-MY3
and
No.~MOST107-2119-M-002-035-MY3,
and the Ministry of Education (Taiwan);
Thailand Center of Excellence in Physics;
TUBITAK ULAKBIM (Turkey);
National Research Foundation of Ukraine, project No.~2020.02/0257,
and
Ministry of Education and Science of Ukraine;
the U.S. National Science Foundation and research Grants
No.~PHY-1913789 
and
No.~PHY-2111604, 
and the U.S. Department of Energy and research Awards
No.~DE-AC06-76RLO1830, 
No.~DE-SC0007983, 
No.~DE-SC0009824, 
No.~DE-SC0009973, 
No.~DE-SC0010007, 
No.~DE-SC0010073, 
No.~DE-SC0010118, 
No.~DE-SC0010504, 
No.~DE-SC0011784, 
No.~DE-SC0012704, 
No.~DE-SC0019230, 
No.~DE-SC0021274, 
No.~DE-SC0022350, 
No.~DE-SC0023470; 
and
the Vietnam Academy of Science and Technology (VAST) under Grant No.~DL0000.05/21-23.

These acknowledgements are not to be interpreted as an endorsement of any statement made
by any of our institutes, funding agencies, governments, or their representatives.

We thank the SuperKEKB team for delivering high-luminosity collisions;
the KEK cryogenics group for the efficient operation of the detector solenoid magnet;
the KEK computer group and the NII for on-site computing support and SINET6 network support;
and the raw-data centers at BNL, DESY, GridKa, IN2P3, INFN, and the University of Victoria for offsite computing support.
\end{acknowledgments}
 }

\ifthenelse{\boolean{wordcount}}%
{ \nobibliography{references} }
{ \bibliography{references} }

\ifthenelse{\boolean{wordcount}}%
{ {} }
{ \clearpage
\renewcommand\thefigure{\thesection X \arabic{figure}}    
\renewcommand\theHfigure{\thesection X \arabic{figure}}    
\setcounter{figure}{0}    

\section*{Supplementary information}

This material is submitted as supplementary information for the Electronic Physics Auxiliary Publication Service.

\subsection*{Peaking background selection}
The transverse distance requirement of the displaced vertex is increased to $d_{\text{v}}>0.2$\,\cm in $S$ mass regions in the vicinity of various known narrow SM states with two-body decays $D^0\to K^+K^-$ and $D^0\to K^-\pi^+$, $J/\psi\to e^+e^-$ and $J/\psi\to \mu^+\mu^-$, $\psi(2S)\to e^+e^-$ and $\psi(2S)\to \mu^+\mu^-$, hadronic decays of $\eta_c$, $\chi_{c1}$, and $\eta_c(2S)$, and $\phi\to K^+K^-$.
These backgrounds include missing or misidentified particles leading to wrong mass hypotheses and hence invariant masses shift with respect to the known particle masses.
The resulting final-state dependent mass regions are summarized in Table\,\ref{sup:tab:selection}.

\begin{table}[b]
\caption{\label{sup:tab:selection}
Selection requirements on two-body masses (in \gevcc) to reject peaking backgrounds for the different $S$ final states.}
\begin{ruledtabular}
\begin{tabular}{lcccc}
 &\textrm{$e^+e^-$}&
\textrm{$\mu^+\mu^-$}&
\textrm{$\pi^+\pi^-$}&
\textrm{$K^+K^-$}\\
\hline\\[-3mm]
$D^0$ & [1.0, 1.3] & [1.7, 1.8] & [1.65, 1.75] & [1.75, 1.85]\\
$J/\psi$ & \multicolumn{2}{c}{[3.00, 3.15]} &  - & - \\
$\psi(2S)$ & \multicolumn{2}{c}{[3.65, 3.75]} & - & -\\
$\eta_c$ & - & - & [2.85, 3.15] & [2.80, 3.20]\\
$\chi_{c1},\eta_c(2S) $ &  - & -  & \multicolumn{2}{c}{[3.4, 3.8]}\\
$\phi$ &  - & - & - & [1.00, 1.04] 
\end{tabular}
\end{ruledtabular}
\end{table}

\subsection*{Axionlike particles with coupling to fermions }
For the model-dependent search for axionlike particles with coupling to fermions, we perform a combined fit in all relevant and kinematically accessible analysis channels, again separately for various lifetimes.
We follow the conventions of Refs.~\cite{Beacham:2019nyx, Batell:2009jf}.
For masses $m_a\leq 2m_{\mu}$ we use the pseudoscalar branching fractions calculation from Ref.~\cite{Dolan:2014ska}, neglecting interference with the $\pi^0$ meson.
For masses $2m_{\mu}<m_a\leq 1.2\,\gevcc$ we use a chiral model and for $m_a>1.2\,\gevcc$, we employ the spectator approach~\cite{Domingo:2016yih}.
The interference with the $\eta_c$ is not included in this calculation.
Our observed upper limits are shown in Fig.\,\ref{supl_fig:alps}.

\begin{figure}[ht]
\centering
\includegraphics[width=0.48\textwidth]{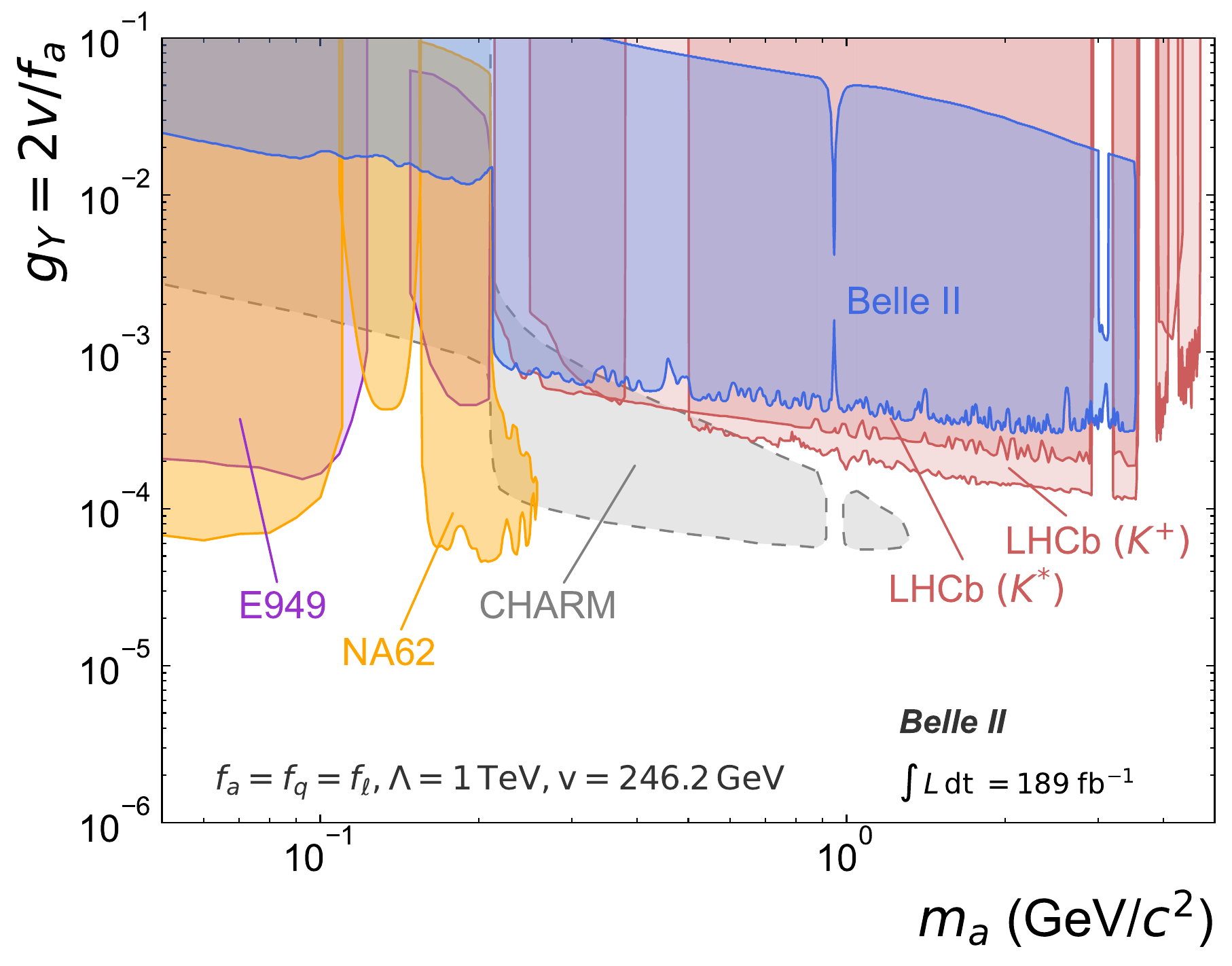}%
\caption{\label{supl_fig:alps}
Exclusion regions in the plane of the coupling $g_Y=2v/f_a$ with the vacuum expectation value $v$ and the ALP mass $m_a$ from this work (blue) together with existing constraints from LHCb~\cite{LHCb:2015nkv,LHCb:2016awg}, 
E949~\cite{BNL-E949:2009dza}, and 
CHARM~\cite{CHARM:1985anb}. 
The exclusion regions from \belletwo, LHCb, and CHARM correspond to 95\%\,CL, while E949 correspond to 90\%~CL.
All existing limits are taken from Ref.~\cite{Beacham:2019nyx}.
The constraint colored in gray with dashed outline is a reinterpretation not performed by the experimental collaboration and without access to raw data.}
\end{figure}

\subsection*{Example fits}
Figures~\ref{supl_fig:1a} and \ref{supl_fig:1b} show the fits that resulted in the largest local significance observed in the single channel fits.
Figure~\ref{supl_fig:2} shows the fit for $m_S=2.619\gevcc$ for a lifetime of $c\tau=100\cm$ that resulted in the largest local significance observed in the combined fits.

\begin{figure*}%
\hspace*{\fill}
\subfigure[$S\to e^+e^-$]{%
  \label{fig:subfig:mu:A}%
  \includegraphics[width=0.45\textwidth]{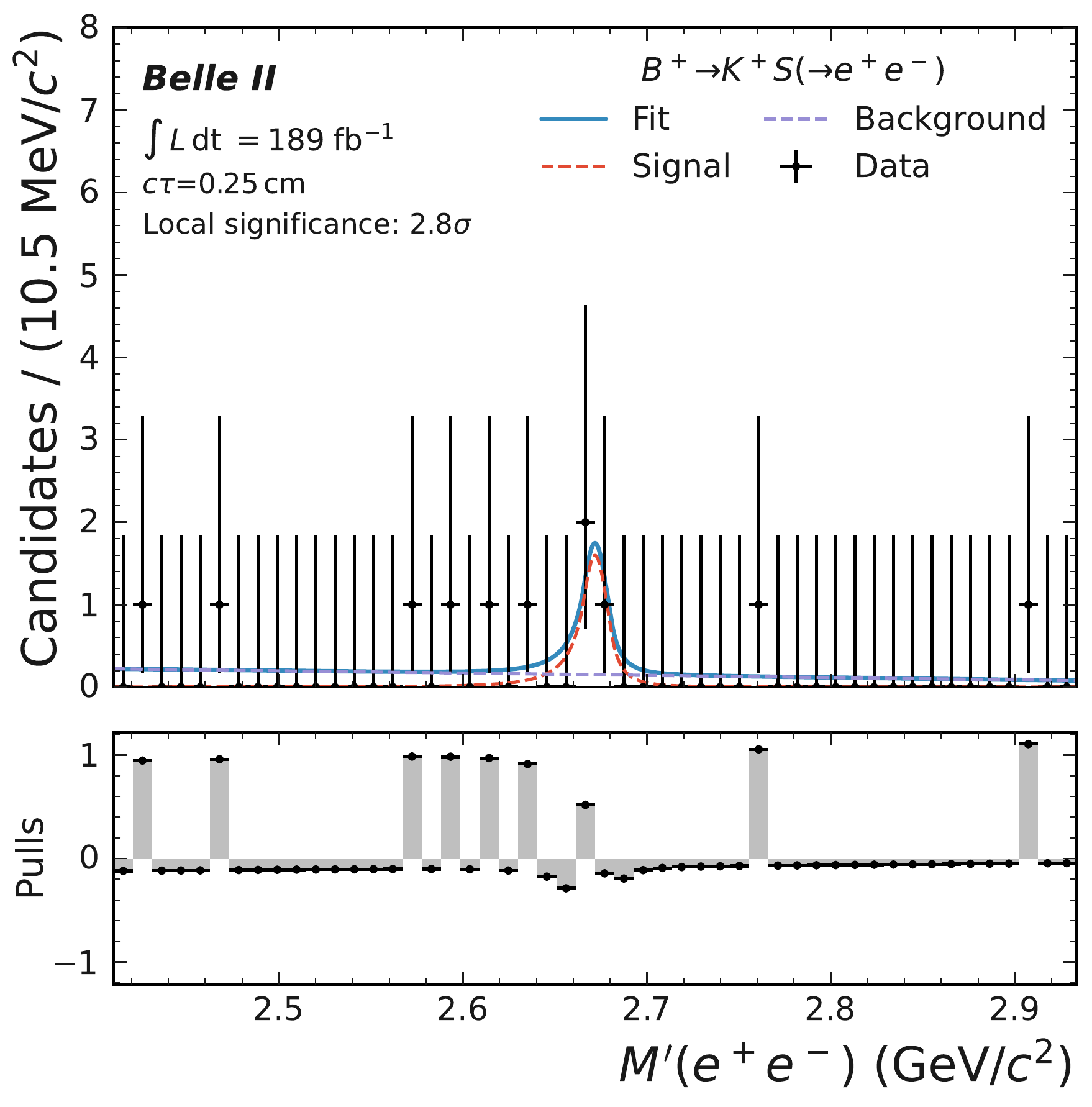}%
}%
\hspace*{\fill}
\subfigure[$S\to \mu^+\mu^-$]{
  \label{fig:subfig:mu:B}%
  \includegraphics[width=0.45\textwidth]{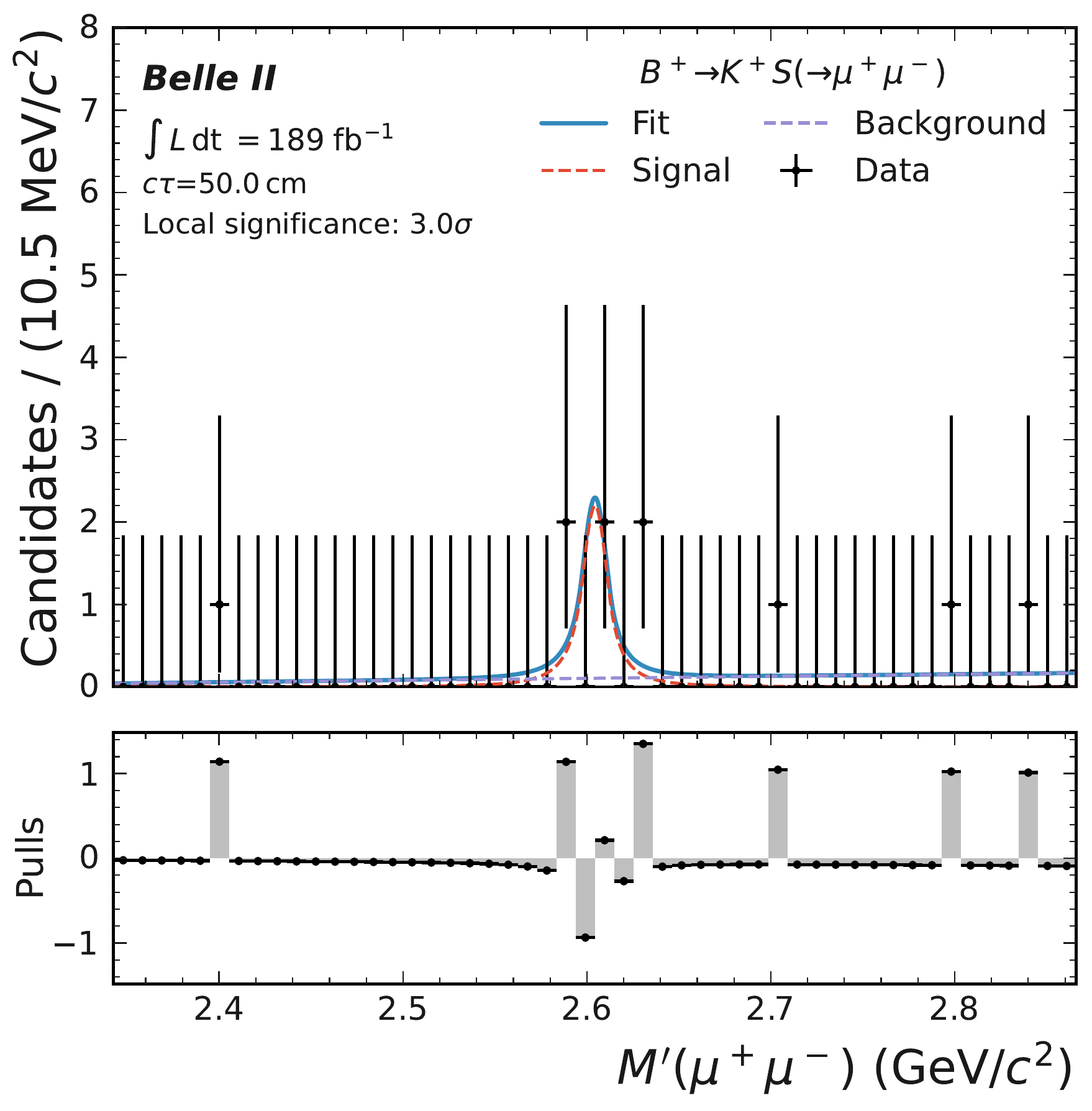}%
}

\hspace*{\fill}
\subfigure[$S\to \pi^+\pi^-$]{
  \label{fig:subfig:mu:C}%
  \includegraphics[width=0.45\textwidth]{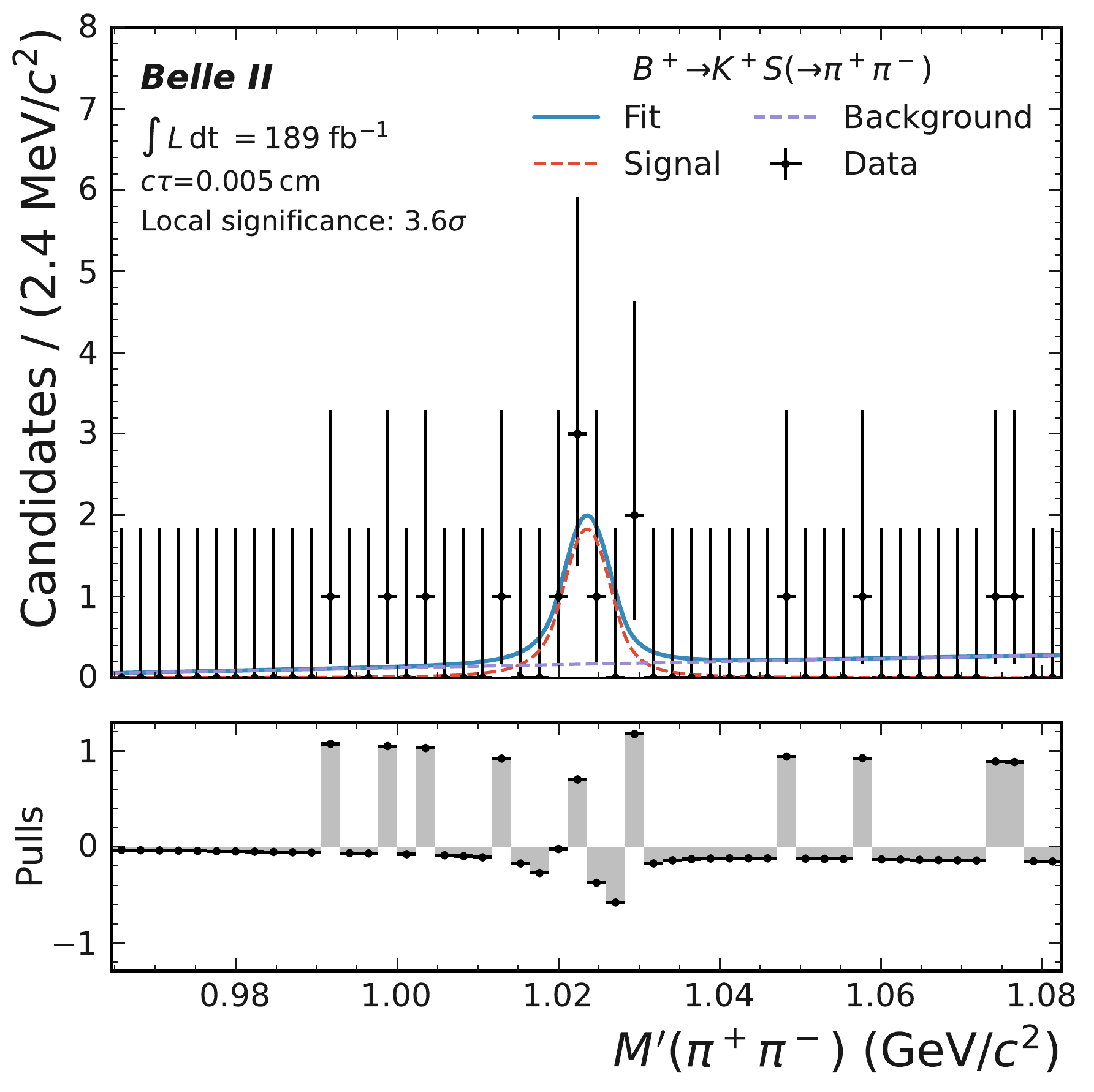}%
}
\hspace*{\fill}
\subfigure[$S\to K^+K^-$]{%
  \label{fig:subfig:mu:D}%
  \includegraphics[width=0.45\textwidth]{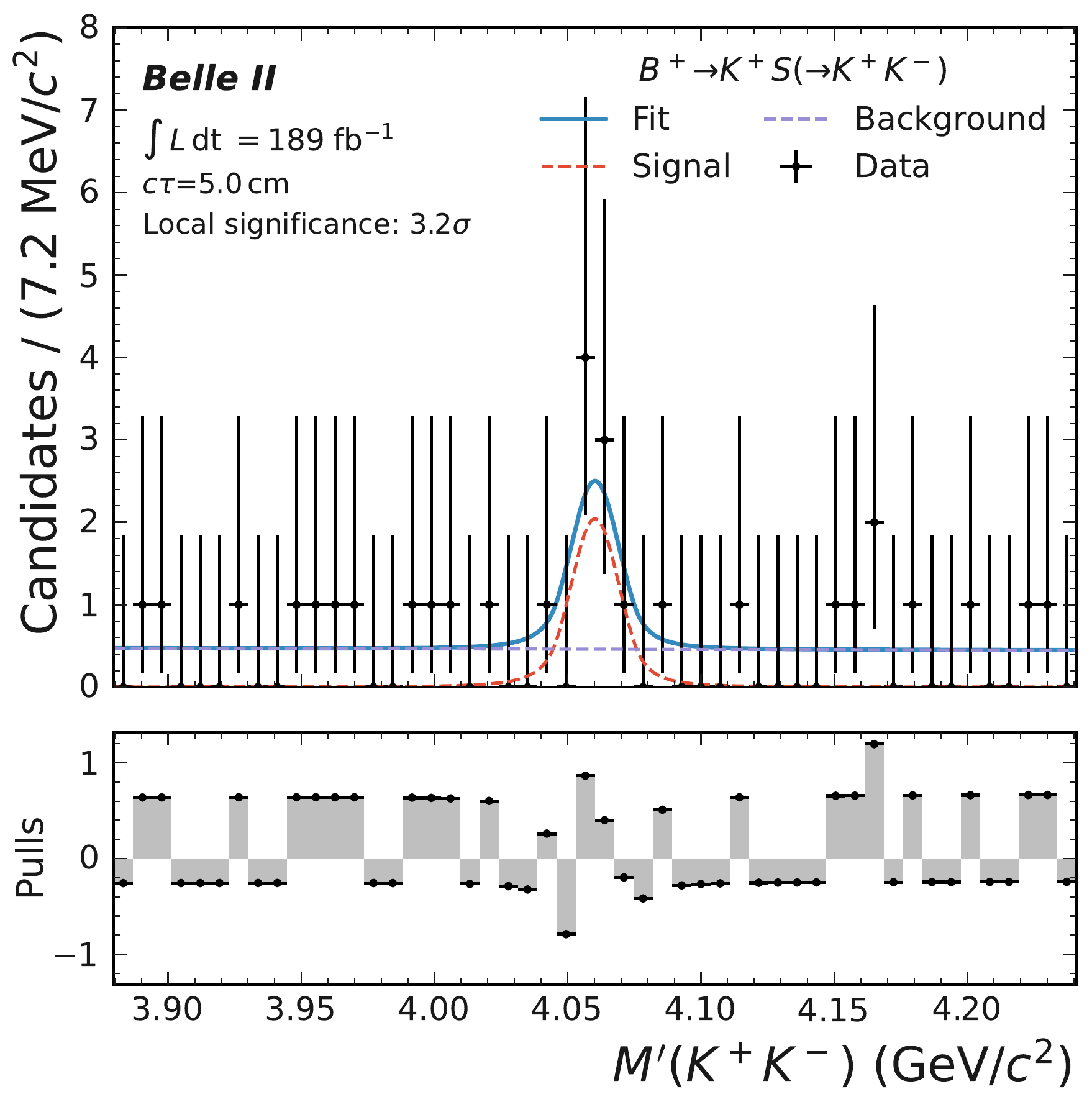}%
}%

\caption{\label{supl_fig:1a} Single channel fits with highest local significances for \BptoKS. 
The bottom panels show the pulls per bin, defined as the difference between data and simulation, divided by the statistical uncertainty of the data.}
\end{figure*}

\begin{figure*}%
\centering
\hspace*{\fill}
\subfigure[E]{
  \label{fig:subfig:mu:E}%
  \includegraphics[width=0.45\textwidth]{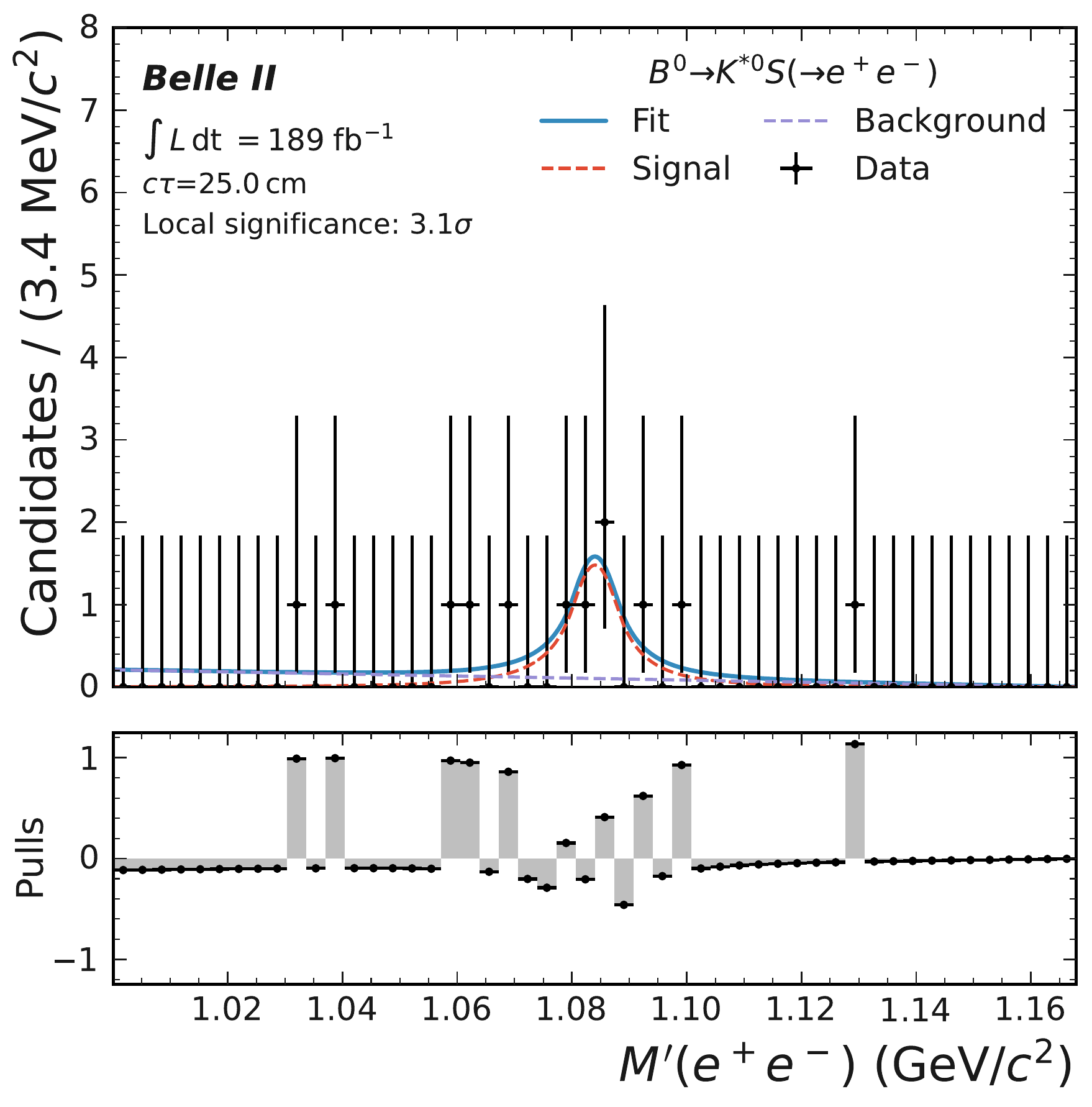}%
}%
\hspace*{\fill}
\subfigure[F]{
  \label{fig:subfig:mu:F}%
  \includegraphics[width=0.45\textwidth]{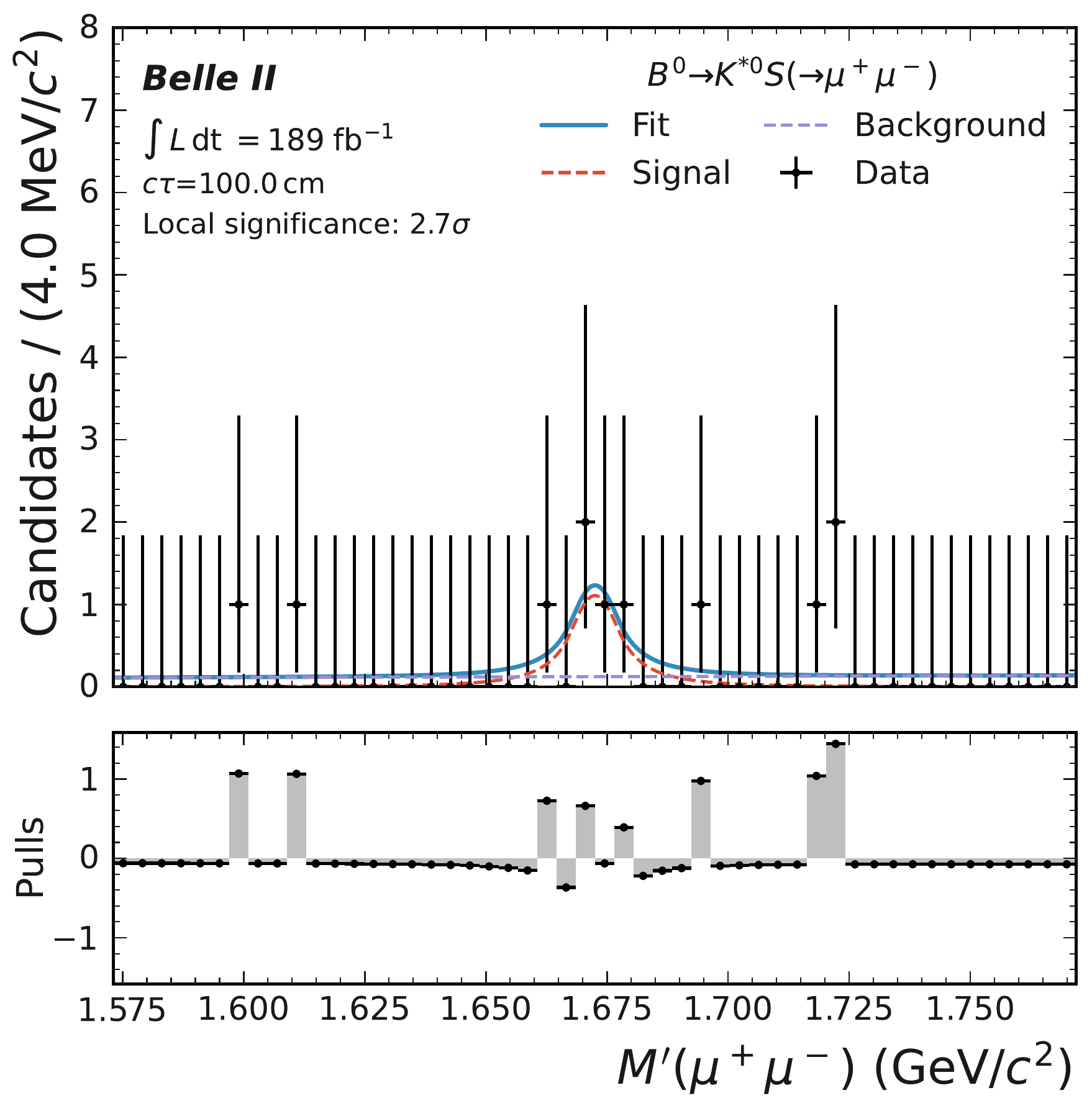}%
}

    \hspace*{\fill}
    \subfigure[G]{
      \label{fig:subfig:mu:G}%
      \includegraphics[width=0.45\textwidth]{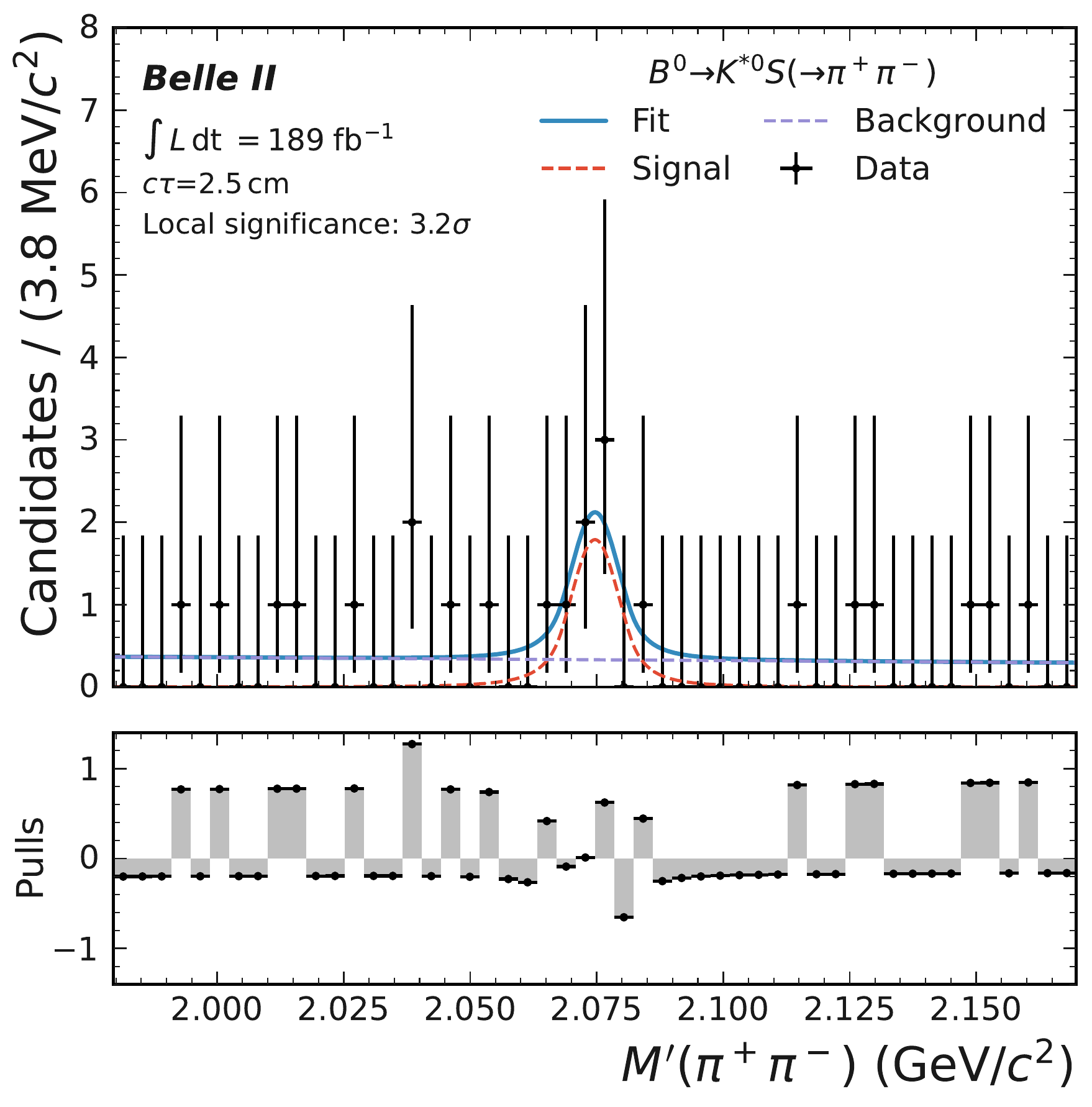}%
    }
    \hspace*{\fill}
    \subfigure[H]{
      \label{fig:subfig:mu:H}%
      \includegraphics[width=0.45\textwidth]{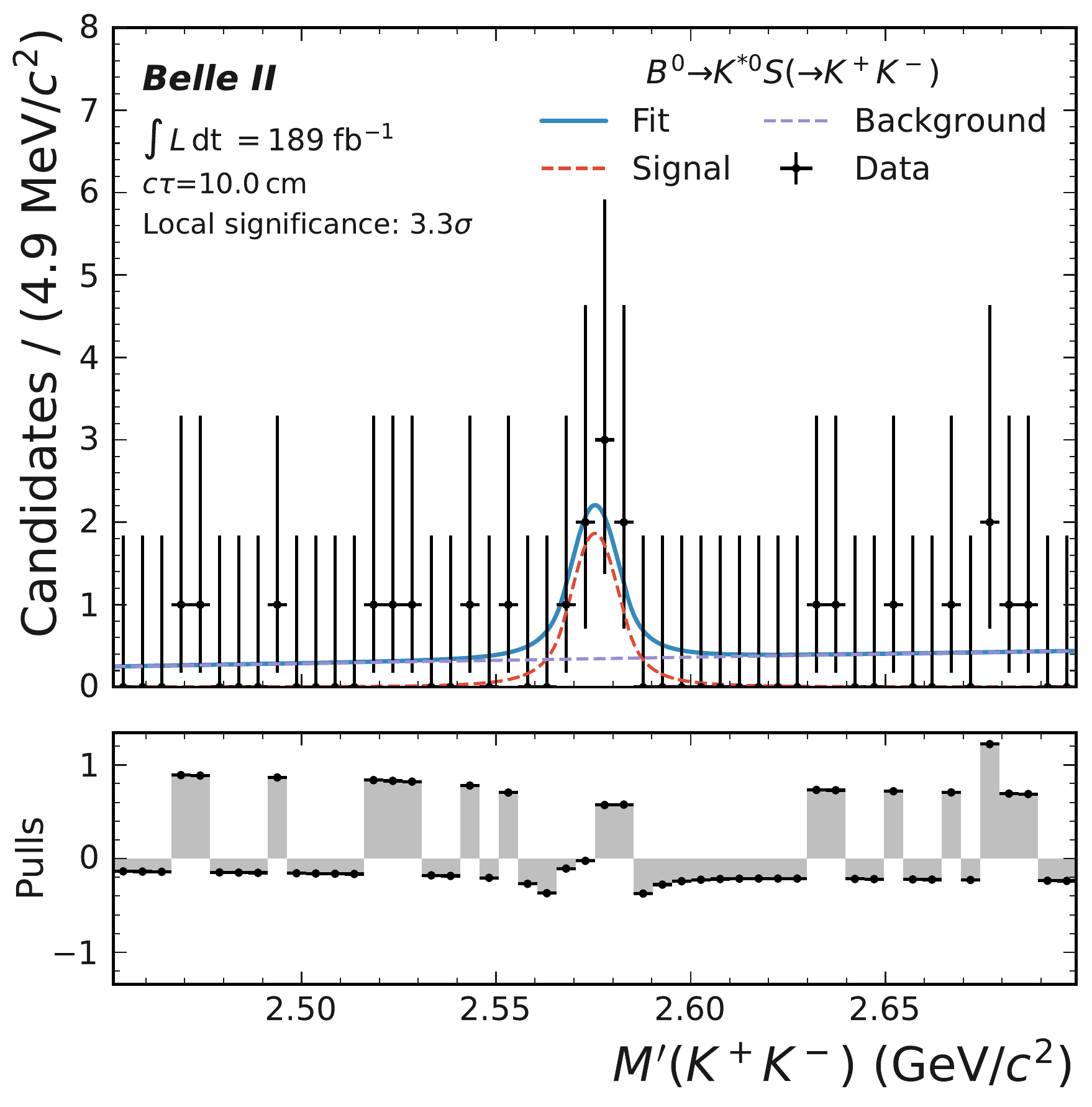}%
    }
\caption{Single channel fits with highest local significances for \BstarztoKS. 
The bottom panels show the pulls per bin, defined as the difference between data and simulation, divided by the statistical uncertainty of the data.}\label{supl_fig:1b}
\end{figure*}

\begin{figure*}
\centering
\includegraphics[width=0.9\textwidth]{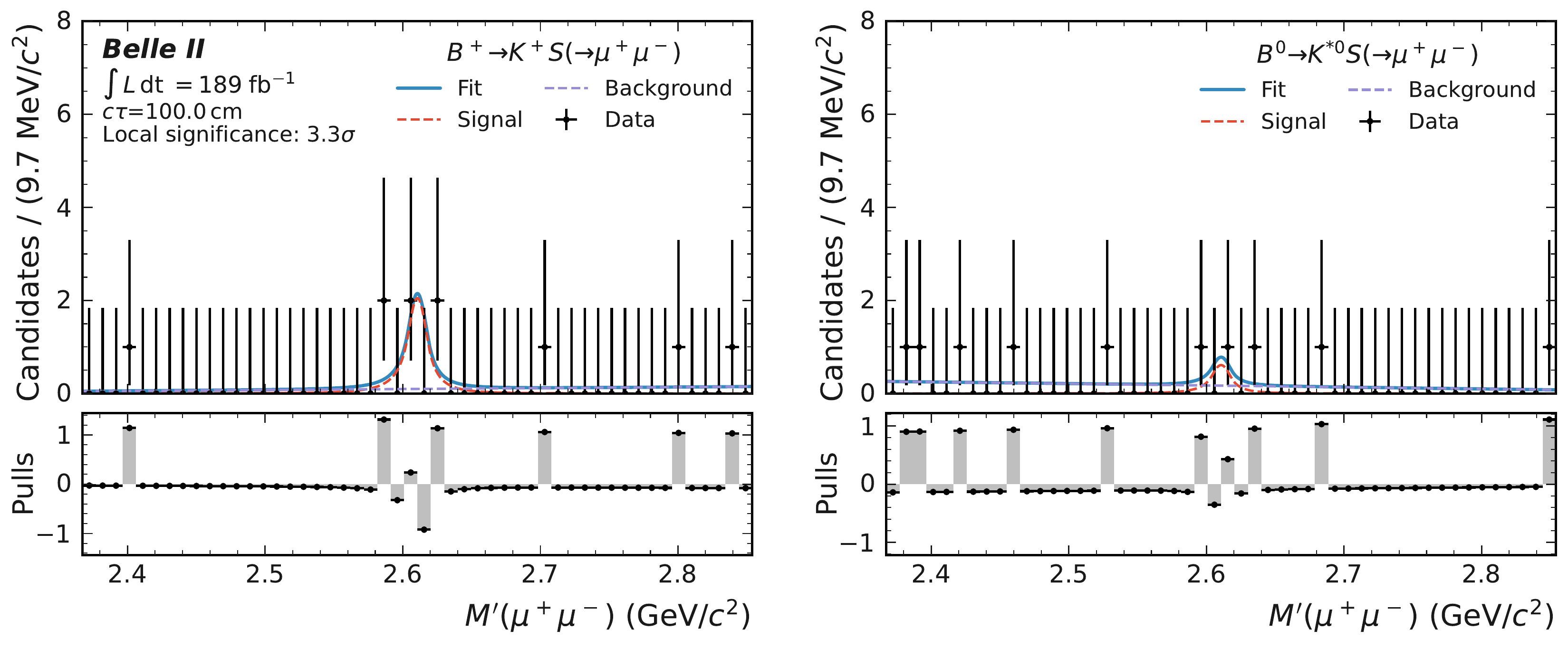}%
\caption{\label{supl_fig:2} Combined fit for $m_S=2.619\,\gevcc$ in the (left) $\Bp\to \Kp S, S\to\mu^+\mu^-$ and (right) $\Bz\to \Kstarz(\to K^+\pi^-) S, S\to\mu^+\mu^-$ states with background pdfs (purple dashed), signal pdfs (red dashed) and total pdfs (blue). The combined signal correspond to a local significance including systematic uncertainties of $S=3.3\sigma$.
The bottom panels show the pulls per bin, defined as the difference between data and simulation, divided by the statistical uncertainty of the data.}
\end{figure*}

\FloatBarrier
\subsection*{Distributions}
Figures~\ref{subfig:distributions_1} and \ref{subfig:distributions_2} show the $M^{\prime}(x^+x^-)$ distributions together  with  the  stacked  contributions  from  the  different simulated SM background samples.
Normalization discrepancies observed in some of the following plots are not a concern since backgrounds are floating in all fits.

\begin{figure*}[ht]%
\subfigure[$B^+\to K^+S, S\to e^+e^-$.]{%
  \label{subfit:distributions:Kp_e}%
  \includegraphics[width=0.45\textwidth]{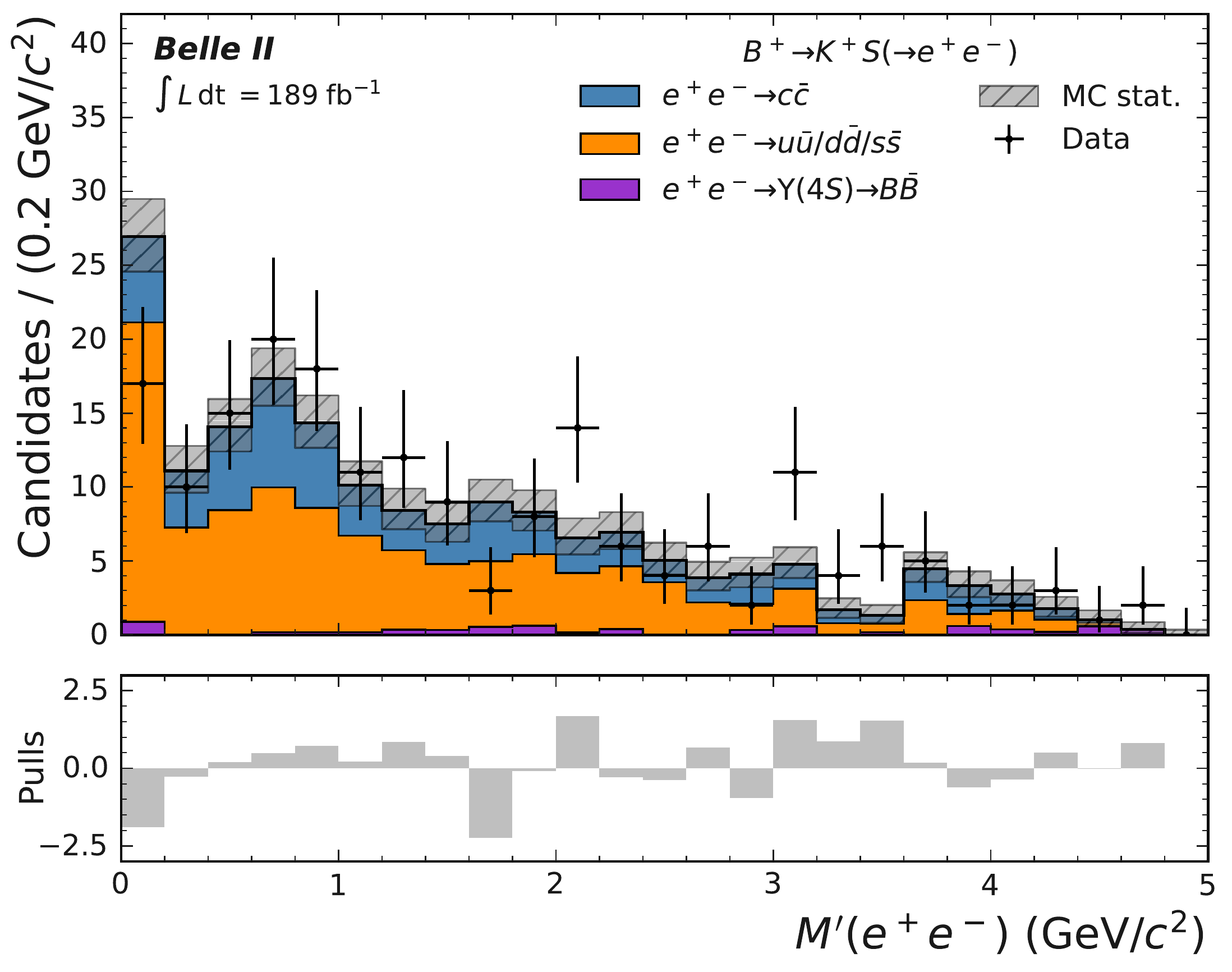}%
}%
\hspace*{\fill}
\subfigure[$B^0\to \Kstarz(\to K^+\pi^-) S, S\to e^+e^-$.]{
  \label{subfit:distributions:Kstar_e}%
  \includegraphics[width=0.45\textwidth]{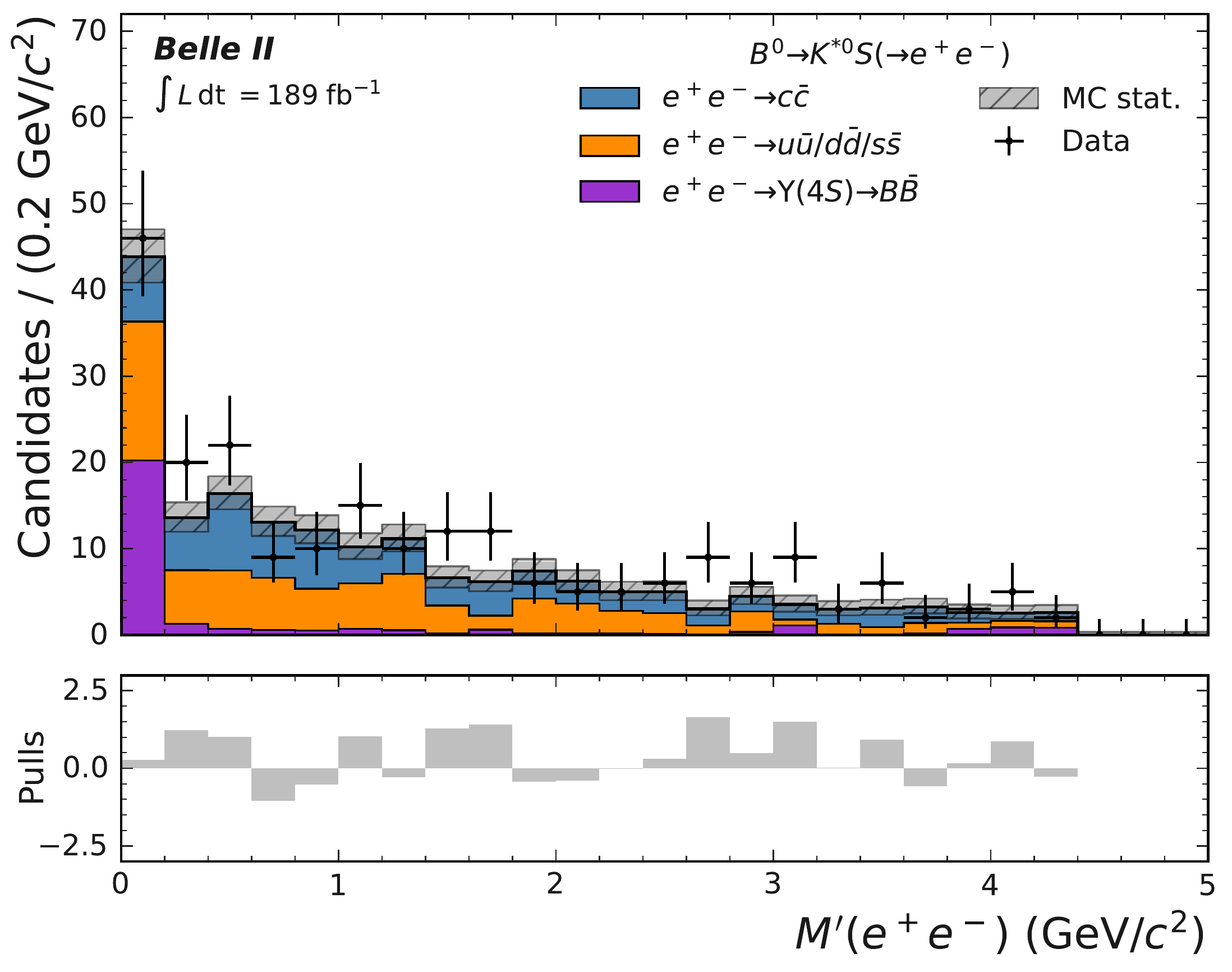}%
}%

\subfigure[$B^+\to K^+S, S\to \mu^+\mu^-$.]{%
  \label{subfit:distributions:Kp_mu}%
  \includegraphics[width=0.45\textwidth]{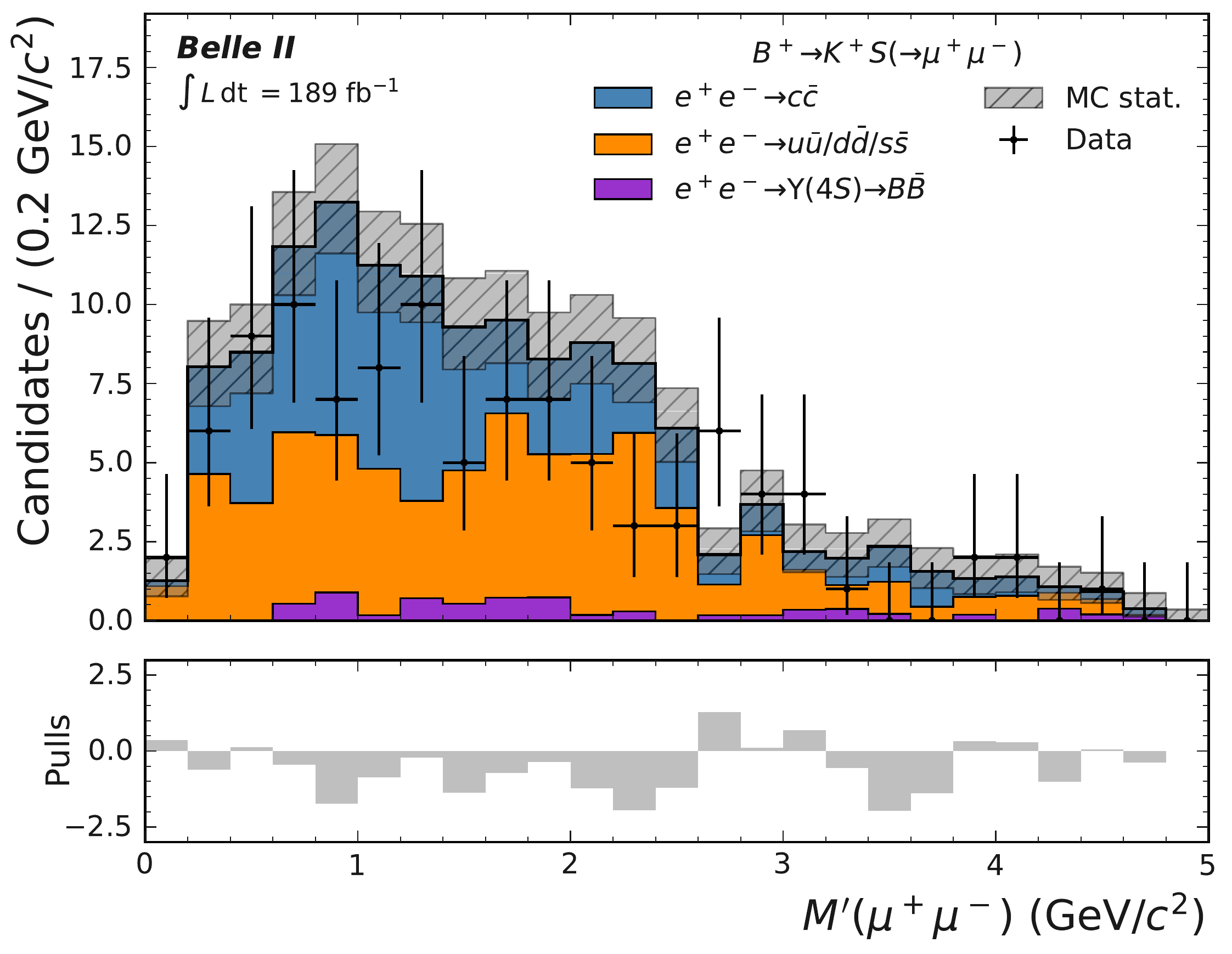}%
}%
\hspace*{\fill}
\subfigure[$B^0\to \Kstarz(\to K^+\pi^-) S, S\to \mu^+\mu^-$.]{
  \label{subfit:distributions:Kstar_mu}%
  \includegraphics[width=0.45\textwidth]{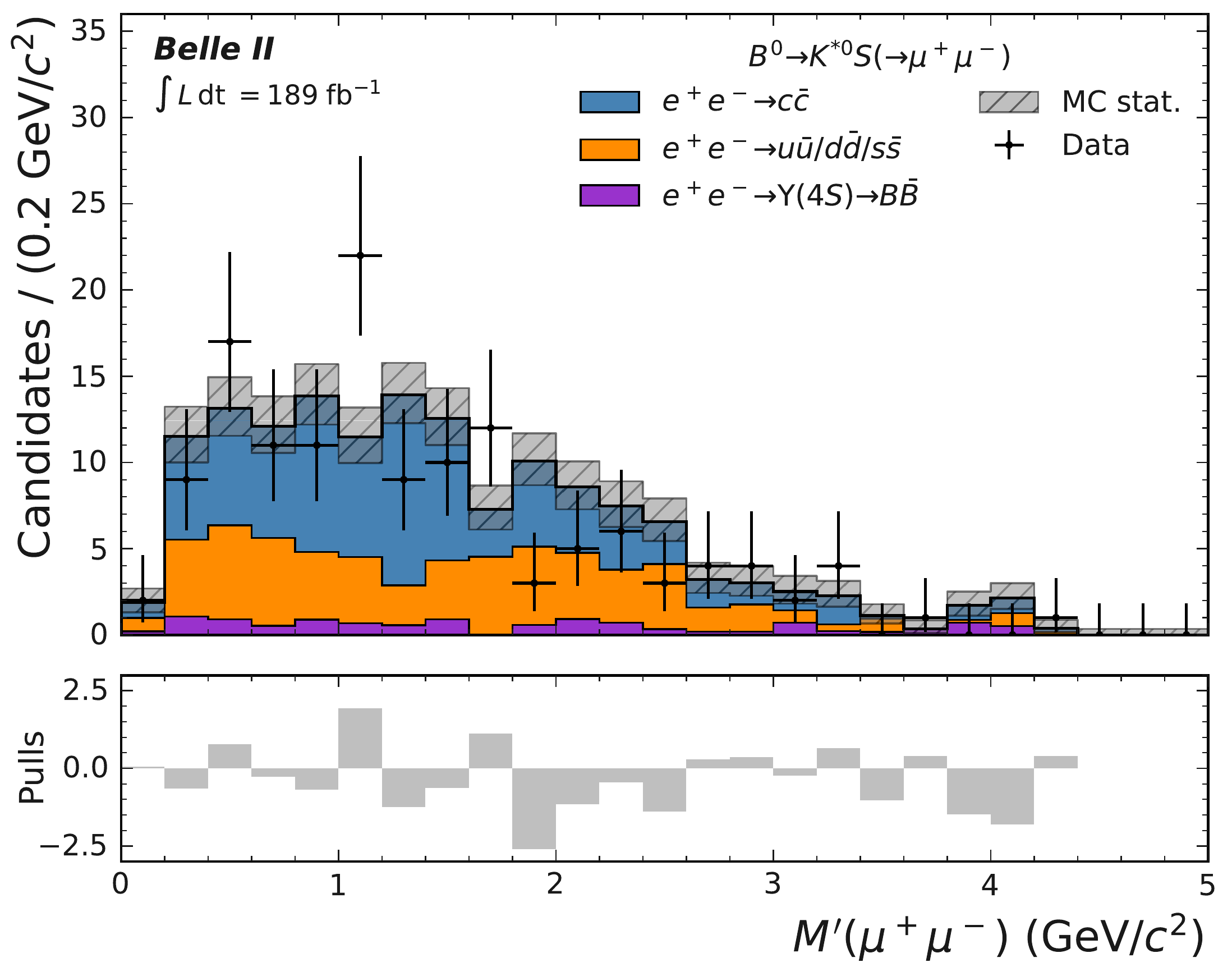}%
}%

\caption{Distribution of $M^{\prime}(x^+x^-)$ together  with  the  stacked  contributions  from  the  various simulated SM background samples. 
Simulation  is  normalized  to a luminosity of 189~\invfb. The hatched area represents the statistical uncertainty of the SM background prediction.
The bottom panels show the pulls per bin, defined as the difference between data and simulation, normalized to the statistical uncertainties added in quadrature.}\label{subfig:distributions_1}
\end{figure*}

\begin{figure*}[ht]%
\subfigure[$B^+\to K^+S, S\to \pi^+\pi^-$.]{%
  \label{subfit:distributions:Kp_pi}%
  \includegraphics[width=0.45\textwidth]{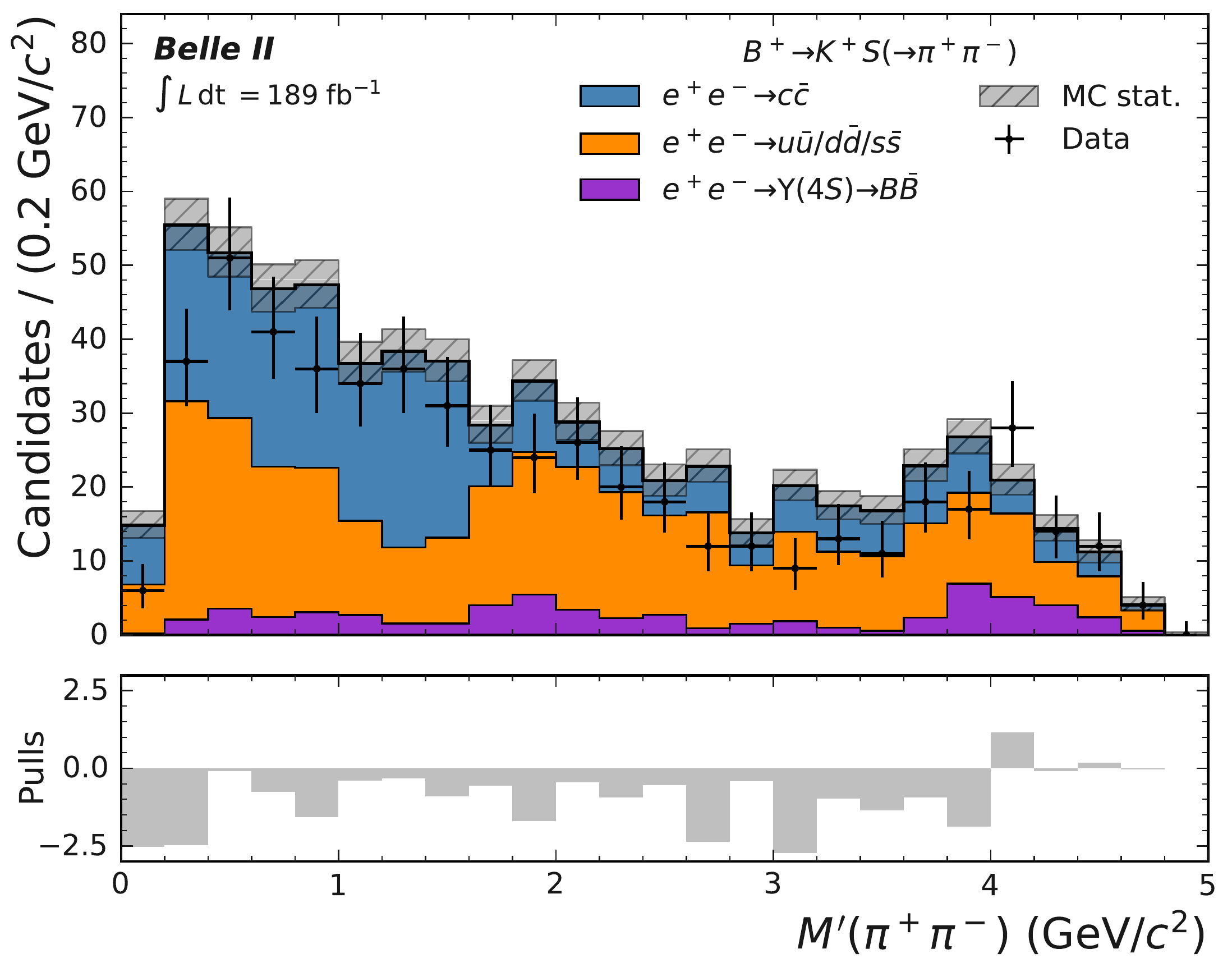}%
}%
\hspace*{\fill}
\subfigure[$B^0\to \Kstarz(\to K^+\pi^-) S, S\to \pi^+\pi^-$.]{
  \label{subfit:distributions:Kstar_pi}%
  \includegraphics[width=0.45\textwidth]{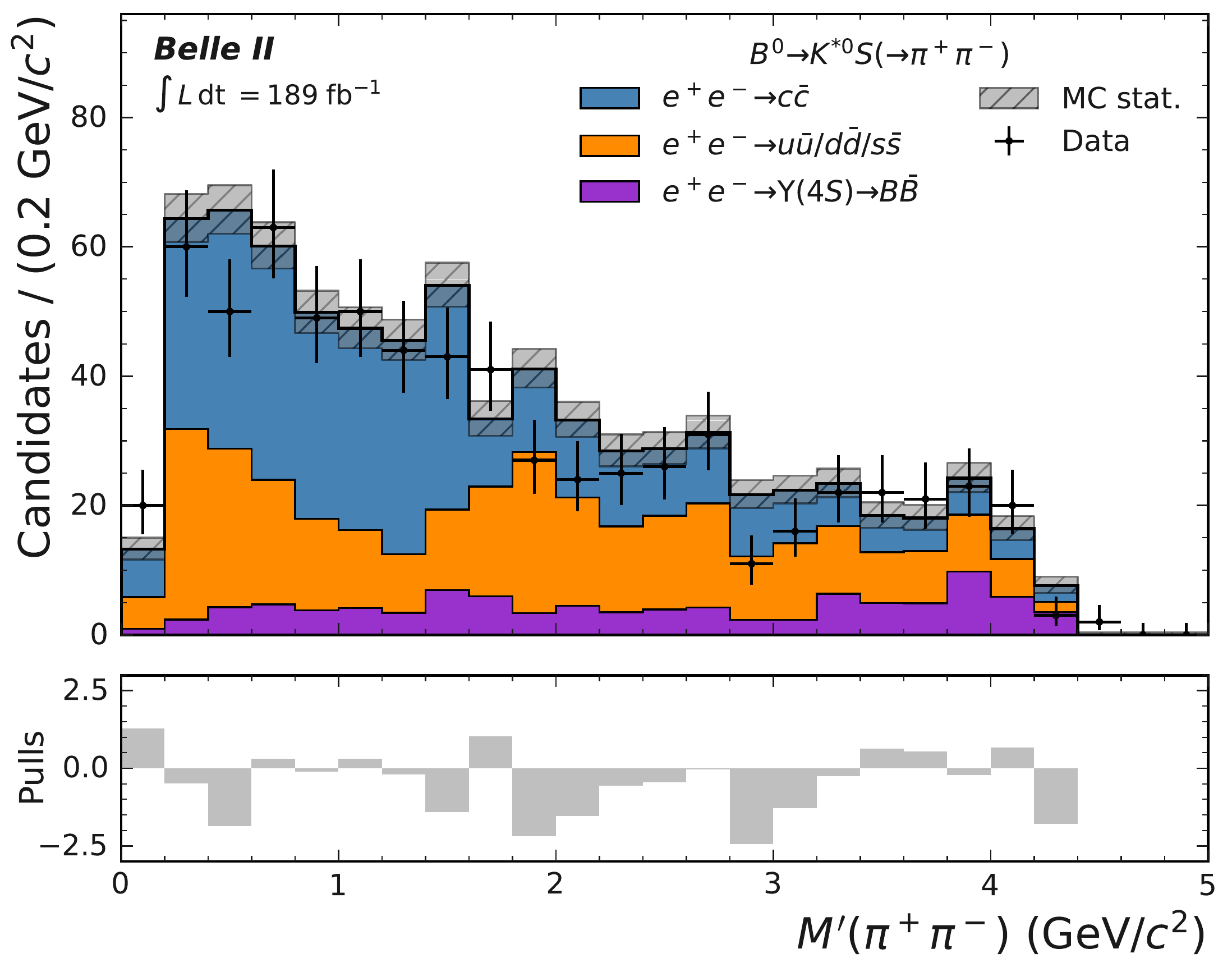}%
}%

\subfigure[$B^+\to K^+S, S\to K^+K^-$.]{%
  \label{subfit:distributions:Kp_K}%
  \includegraphics[width=0.45\textwidth]{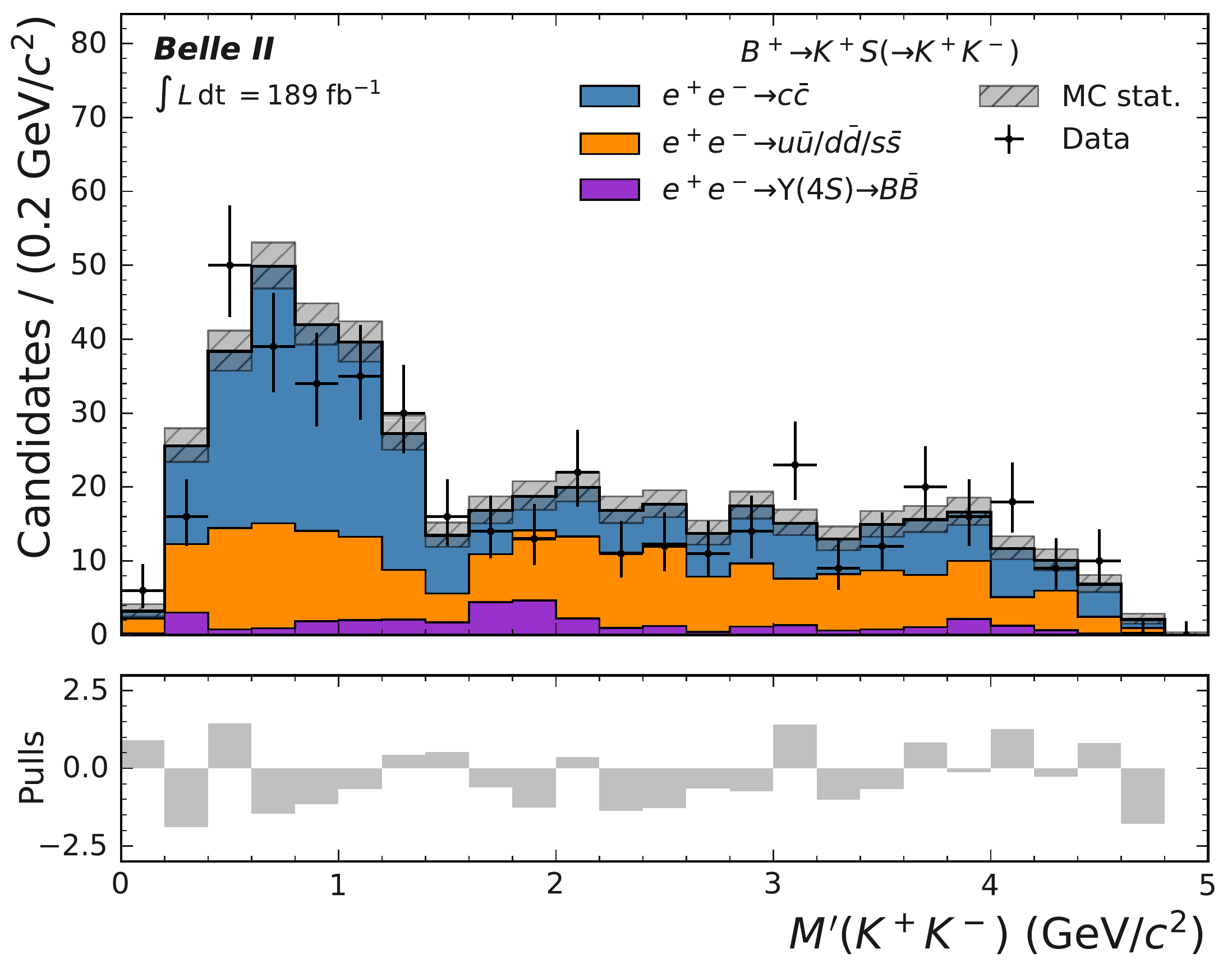}%
}%
\hspace*{\fill}
\subfigure[$B^0\to \Kstarz(\to K^+\pi^-) S, S\to K^+K^-$.]{
  \label{subfit:distributions:Kstar_K}%
  \includegraphics[width=0.45\textwidth]{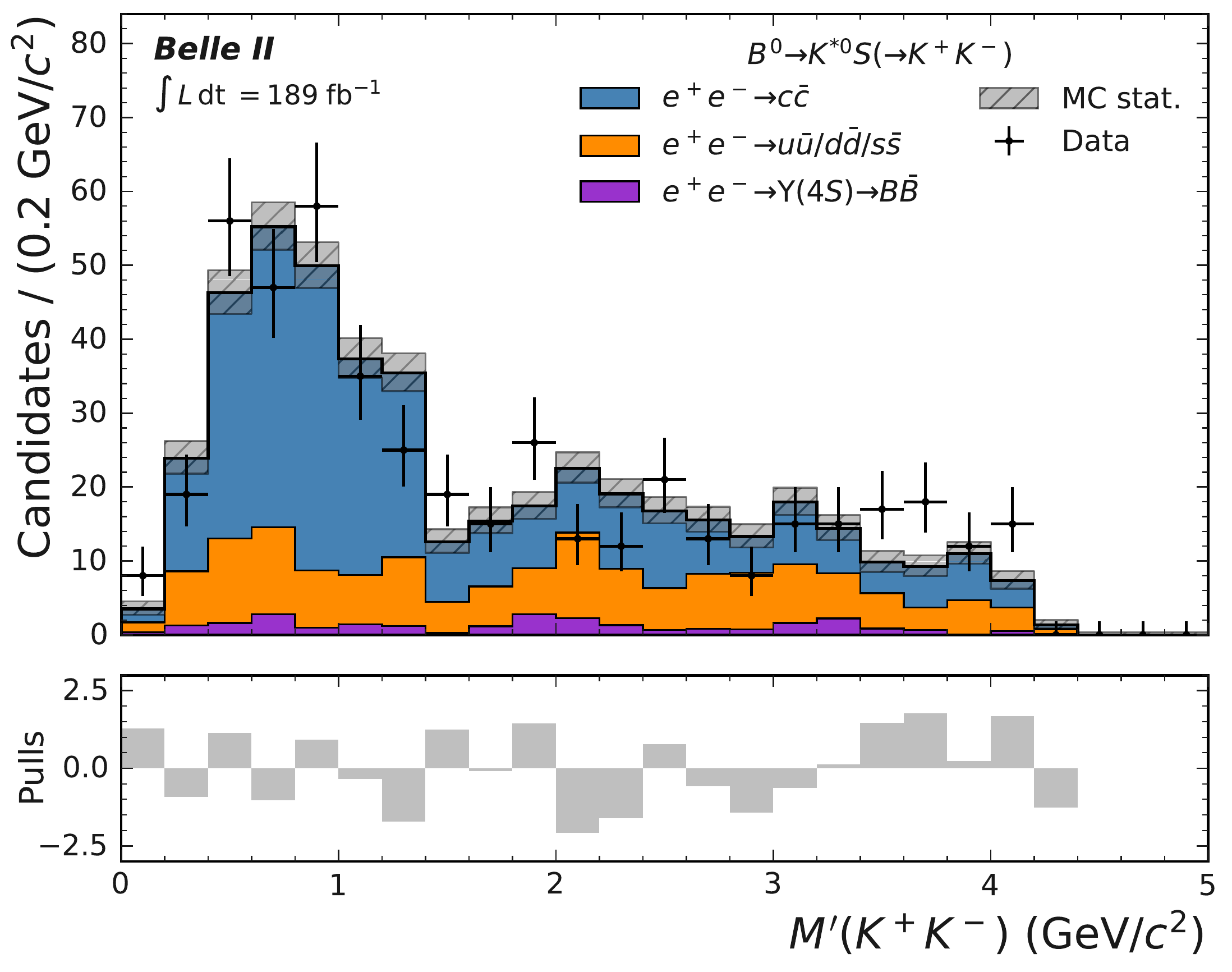}%
}%
\caption{Distribution of $M^{\prime}(x^+x^-)$ together  with  the  stacked  contributions  from  the  various simulated SM background samples. 
Simulation  is  normalized  to a luminosity of 189~\invfb. The hatched area represents the statistical uncertainty of the SM background prediction.
The bottom panels show the pulls per bin, defined as the difference between data and simulation, normalized to the statistical uncertainties added in quadrature.}\label{subfig:distributions_2}
\end{figure*}

\FloatBarrier
\subsection*{Expected and observed limits}
Expected and observed 95\% upper limits for various scalar masses, final states, and lifetimes are shown in Figures\,\ref{subfit:brazil:Kp_e_1}, \ref{subfit:brazil:Kp_e_2}, \ref{subfit:brazil:Kstar_e_1}, \ref{subfit:brazil:Kstar_e_2} ($e^+e^-$), \,\ref{subfit:brazil:Kp_mu_1}, \ref{subfit:brazil:Kp_mu_2}, \ref{subfit:brazil:Kstar_mu_1}, \ref{subfit:brazil:Kstar_mu_2} ($\mu^+\mu^-$), \,\ref{subfit:brazil:Kp_pi_1}, \ref{subfit:brazil:Kp_pi_2}, \ref{subfit:brazil:Kstar_pi_1}, \ref{subfit:brazil:Kstar_pi_2} ($\pi^+\pi^-$), \,\ref{subfit:brazil:Kp_K_1}, \ref{subfit:brazil:Kp_K_2}, \ref{subfit:brazil:Kstar_K_1}, \ref{subfit:brazil:Kstar_K_2} ($K^+K^-$). 

\begin{figure*}[ht]%
\subfigure[$B^+\to K^+S, S\to e^+e^-$, \newline lifetime of $c\tau=0.001\cm$.]{%
  \label{subfit:brazil:Kp_e_1:A}%
  \includegraphics[width=0.31\textwidth]{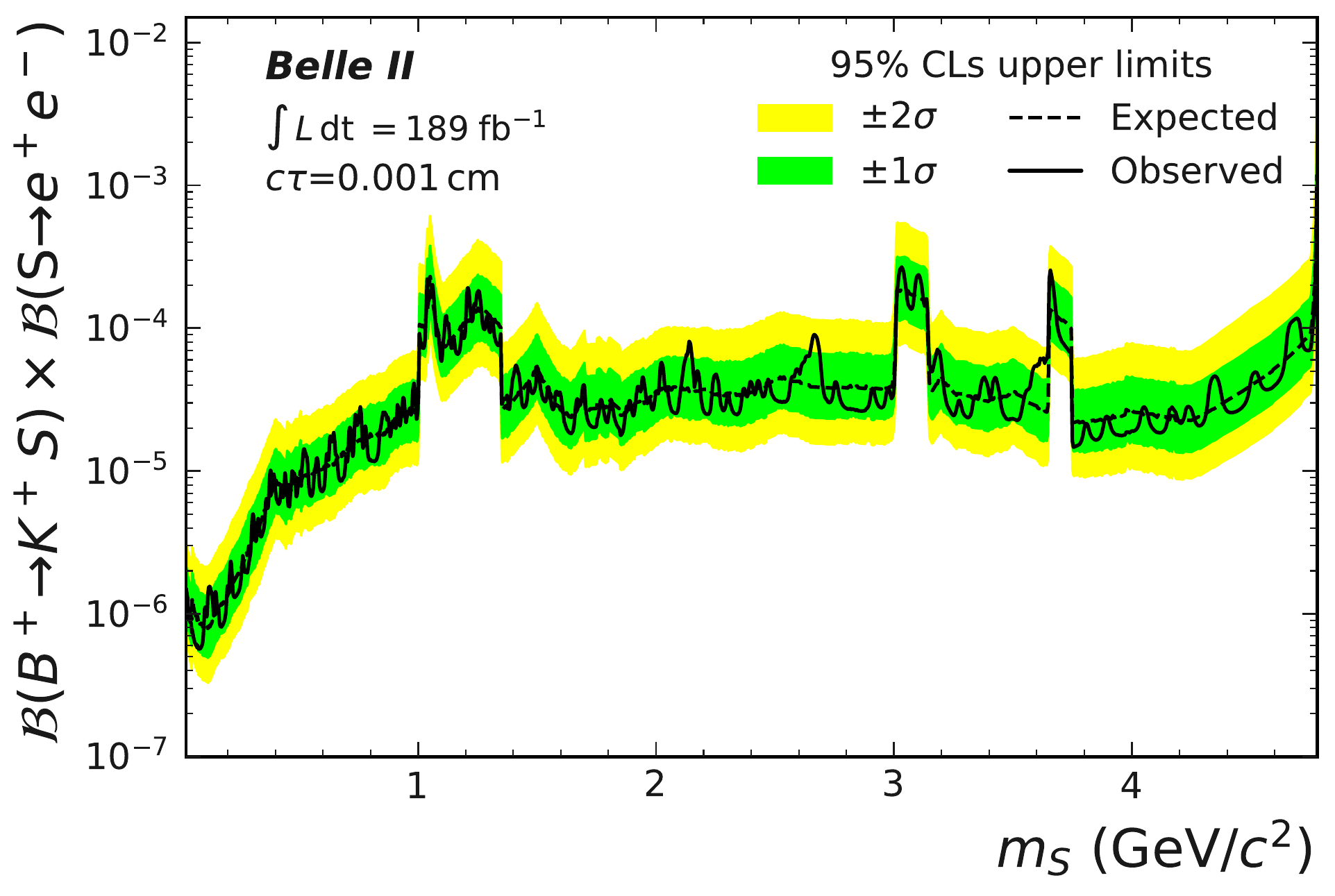}%
}%
\hspace*{\fill}
\subfigure[$B^+\to K^+S, S\to e^+e^-$, \newline lifetime of $c\tau=0.003\cm$.]{
  \label{subfit:brazil:Kp_e_1:B}%
  \includegraphics[width=0.31\textwidth]{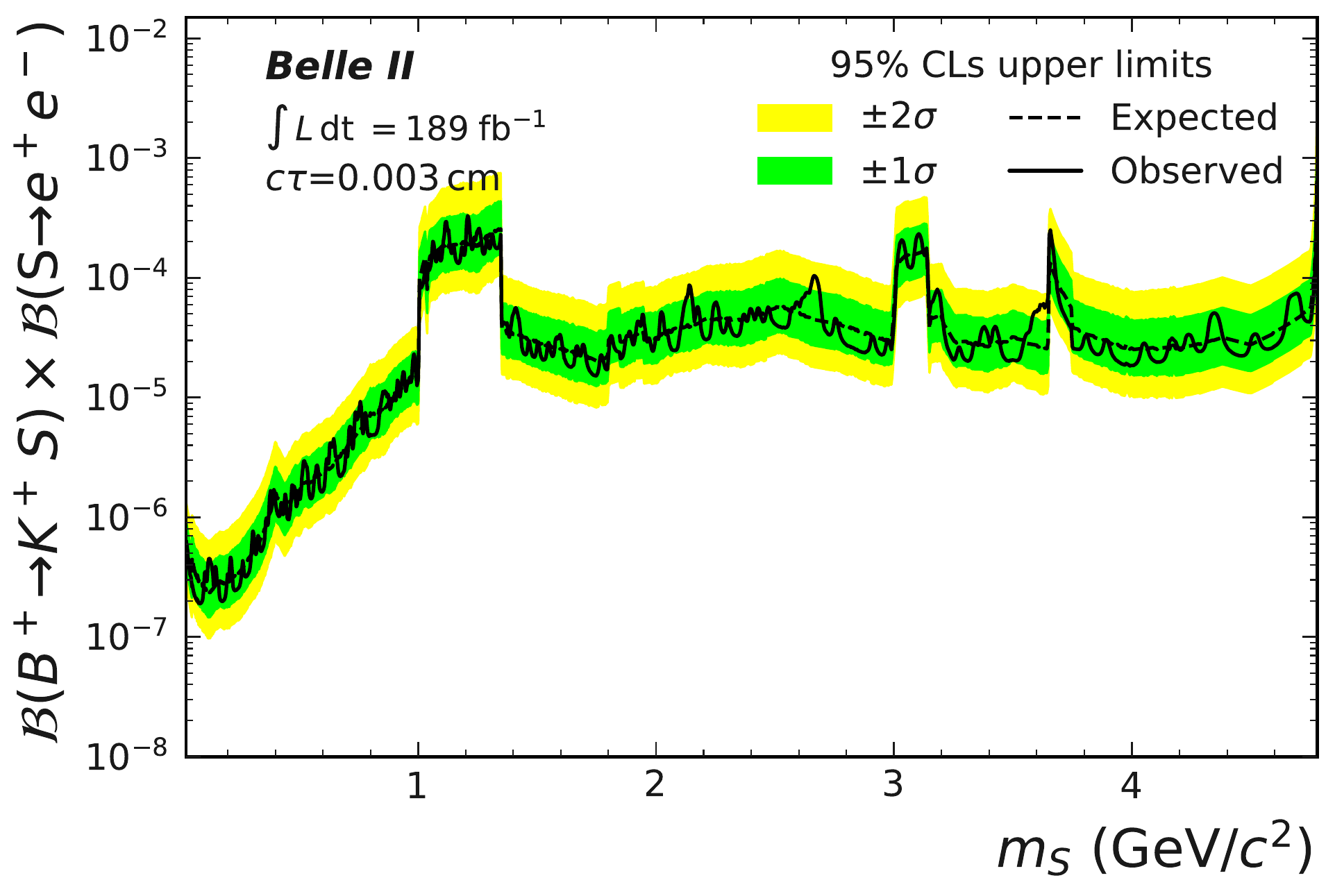}%
}%
\hspace*{\fill}
\subfigure[$B^+\to K^+S, S\to e^+e^-$, \newline lifetime of $c\tau=0.005\cm$.]{
  \label{subfit:brazil:Kp_e_1:C}%
  \includegraphics[width=0.31\textwidth]{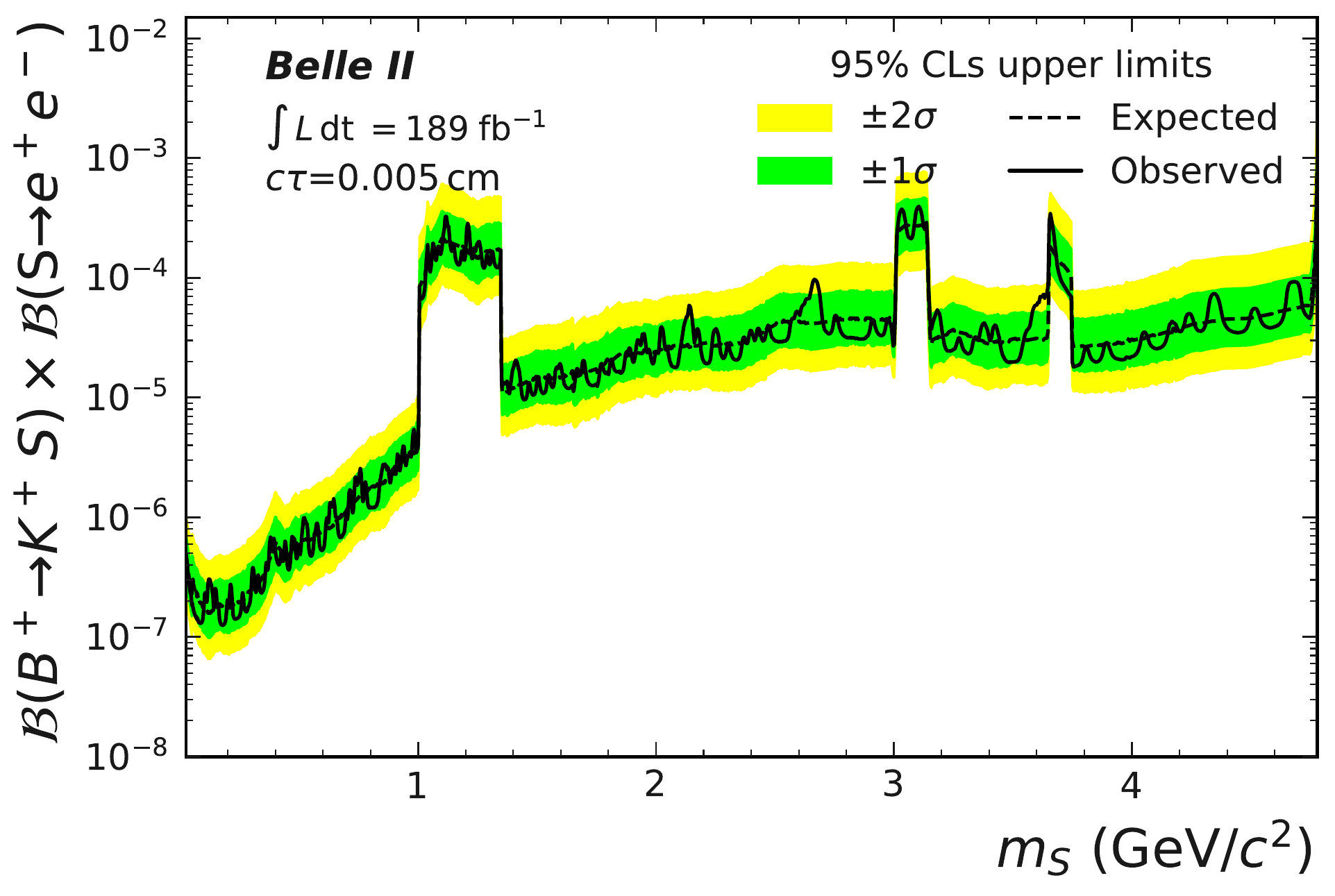}%
}
\subfigure[$B^+\to K^+S, S\to e^+e^-$, \newline lifetime of $c\tau=0.007\cm$.]{%
  \label{subfit:brazil:Kp_e_1:D}%
  \includegraphics[width=0.31\textwidth]{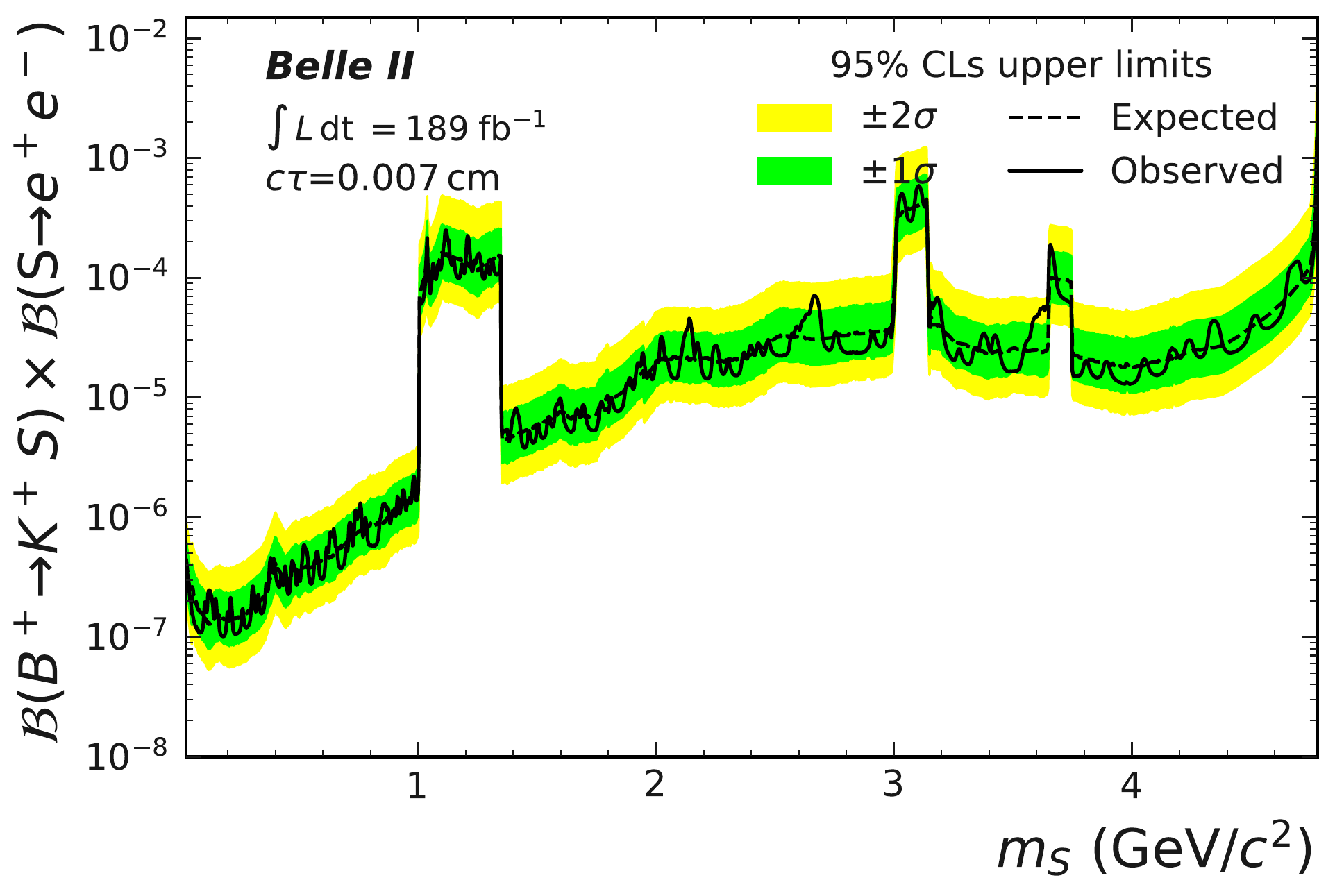}%
}%
\hspace*{\fill}
\subfigure[$B^+\to K^+S, S\to e^+e^-$, \newline lifetime of $c\tau=0.01\cm$.]{
  \label{subfit:brazil:Kp_e_1:E}%
  \includegraphics[width=0.31\textwidth]{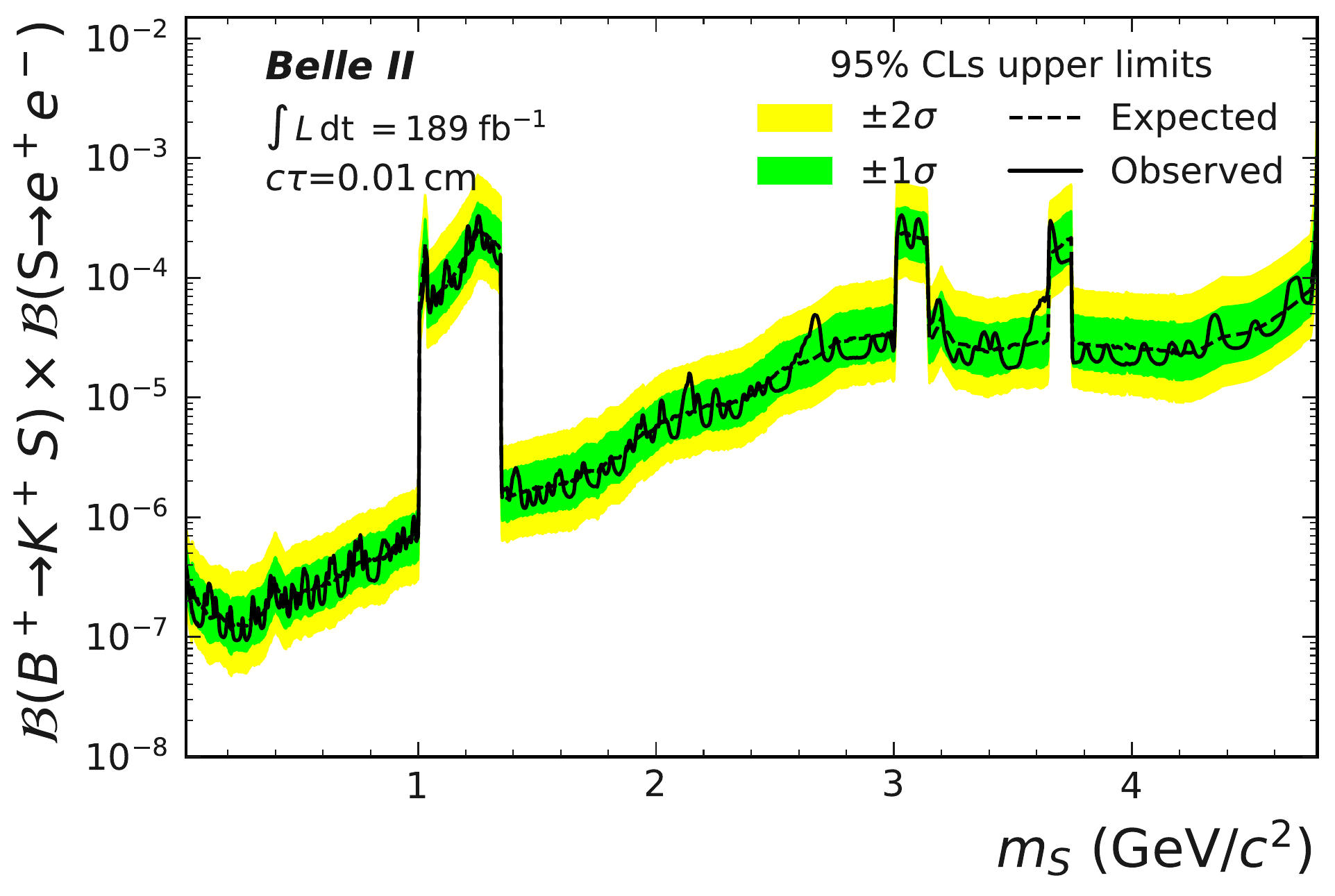}%
}%
\hspace*{\fill}
\subfigure[$B^+\to K^+S, S\to e^+e^-$, \newline lifetime of $c\tau=0.025\cm$.]{
  \label{subfit:brazil:Kp_e_1:F}%
  \includegraphics[width=0.31\textwidth]{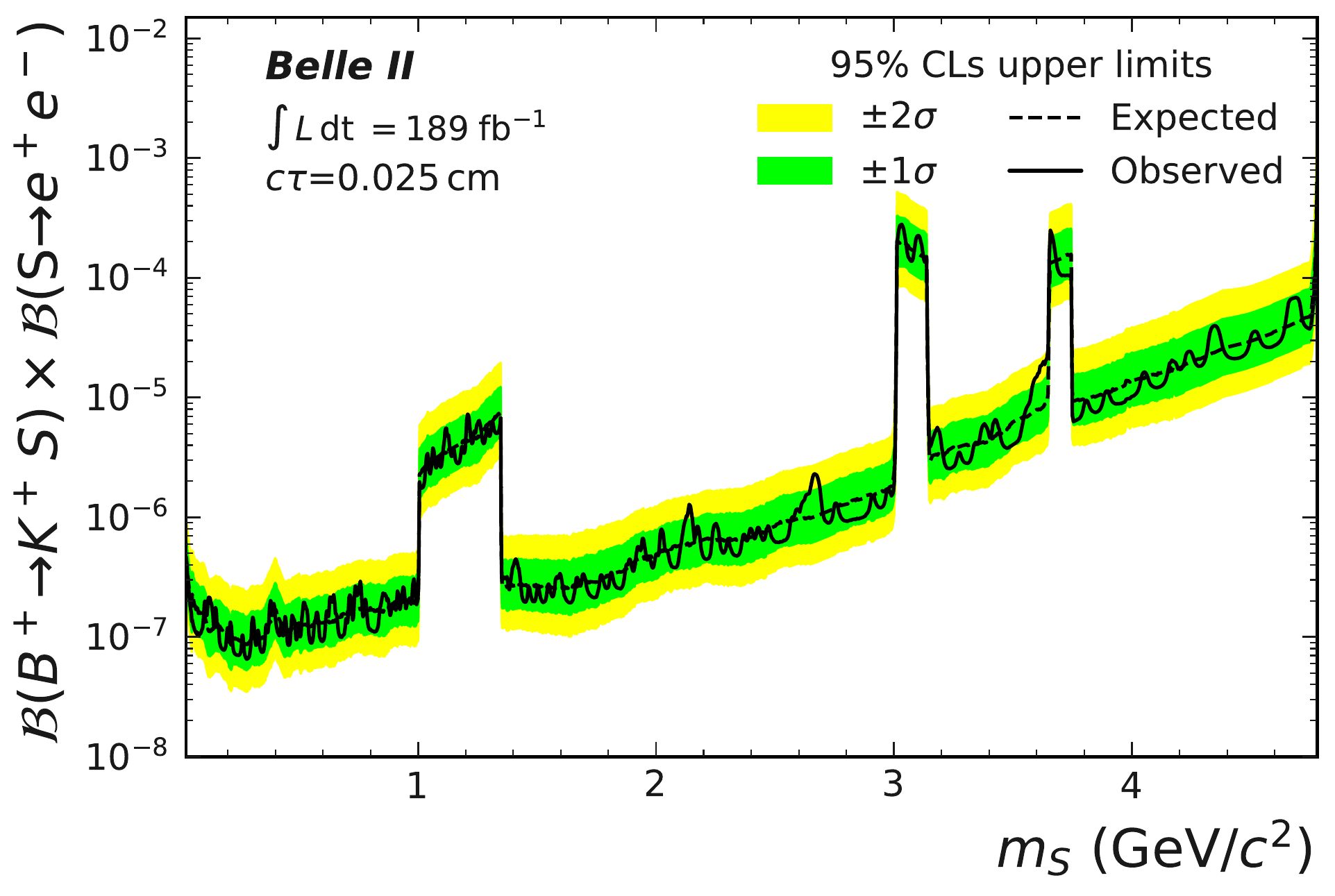}%
}
\subfigure[$B^+\to K^+S, S\to e^+e^-$, \newline lifetime of $c\tau=0.05\cm$.]{%
  \label{subfit:brazil:Kp_e_1:G}%
  \includegraphics[width=0.31\textwidth]{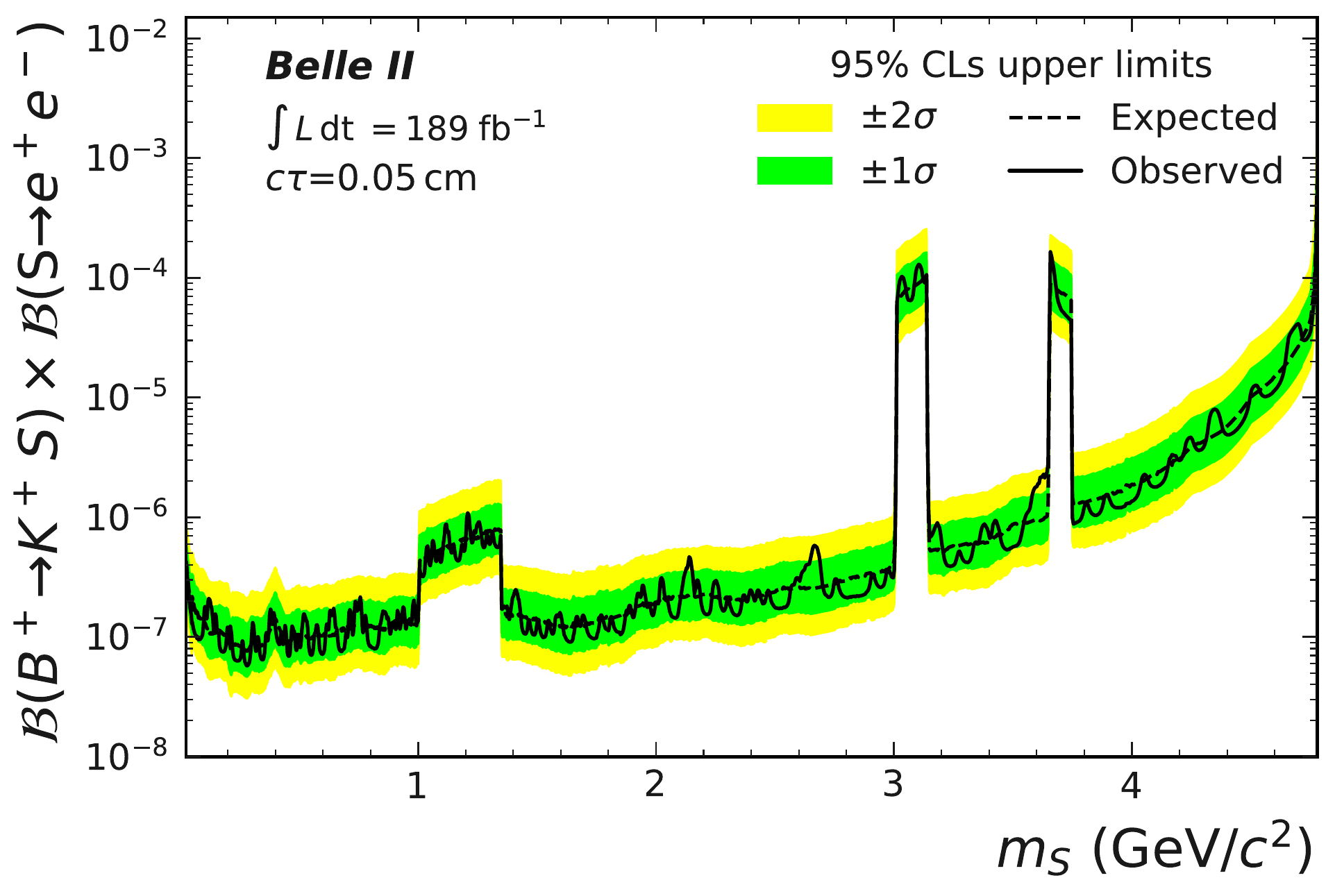}%
}%
\hspace*{\fill}
\subfigure[$B^+\to K^+S, S\to e^+e^-$, \newline lifetime of $c\tau=0.100\cm$.]{
  \label{subfit:brazil:Kp_e_1:H}%
  \includegraphics[width=0.31\textwidth]{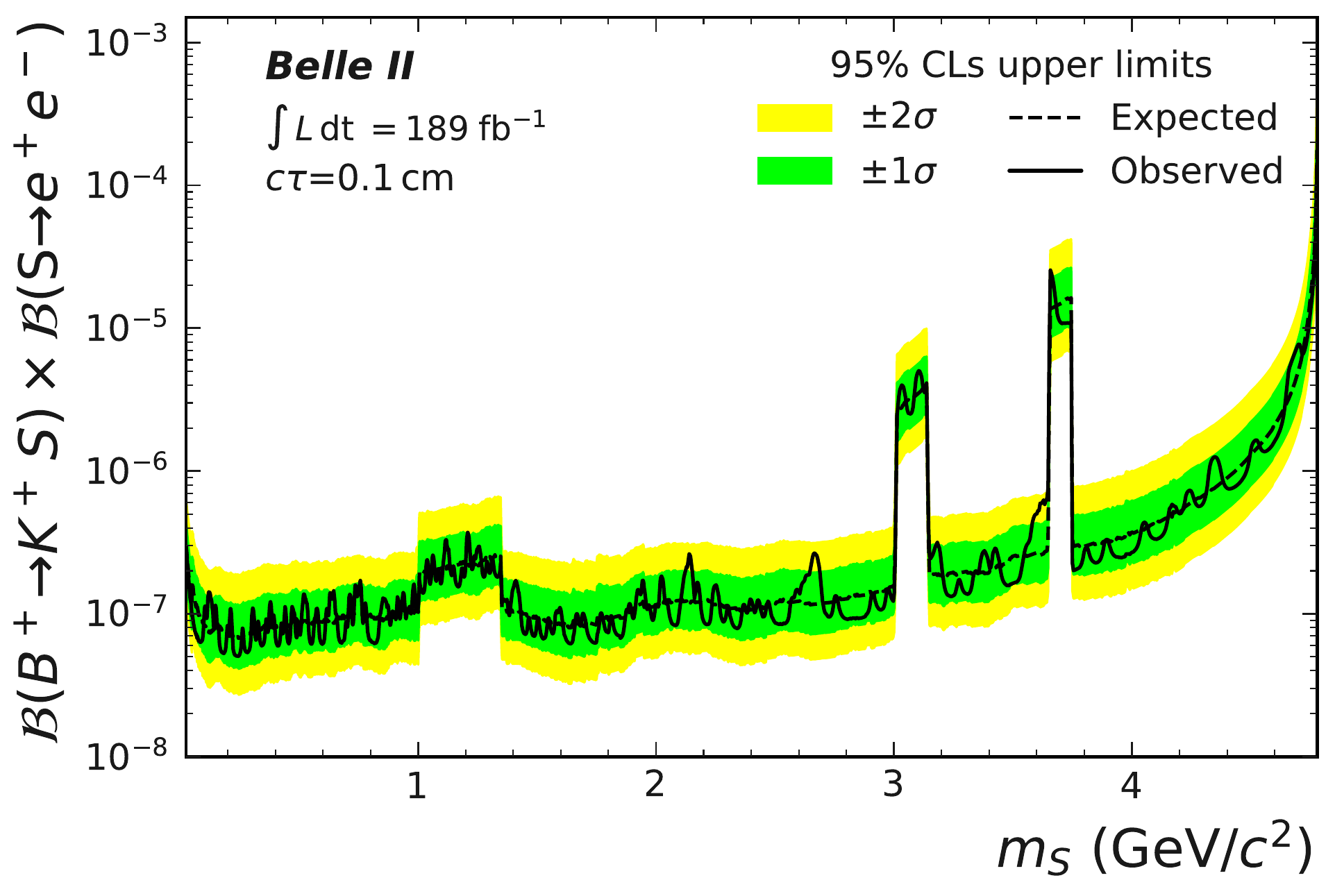}%
}%
\hspace*{\fill}
\subfigure[$B^+\to K^+S, S\to e^+e^-$, \newline lifetime of $c\tau=0.25\cm$.]{
  \label{subfit:brazil:Kp_e_1:I}%
  \includegraphics[width=0.31\textwidth]{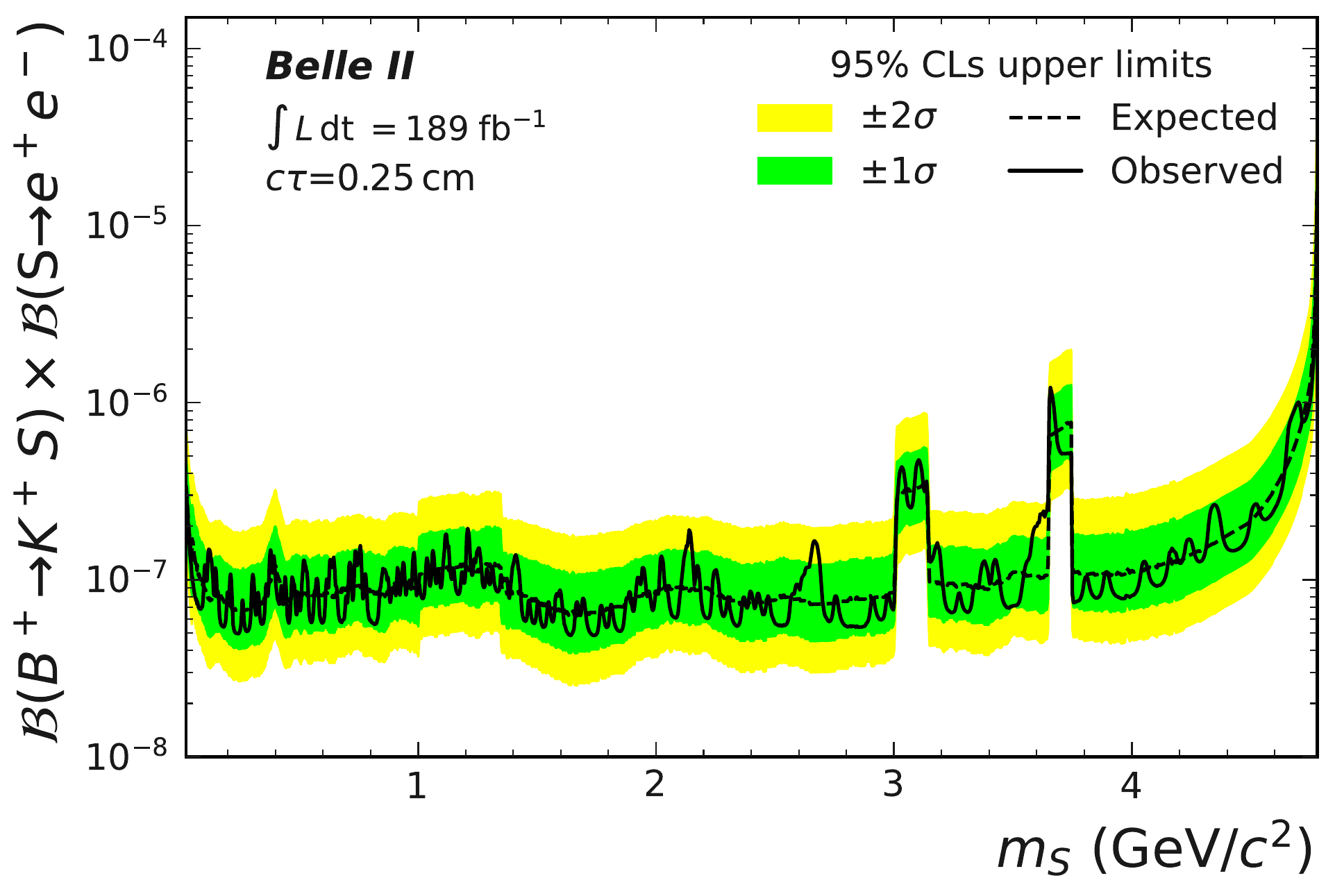}%
}
\subfigure[$B^+\to K^+S, S\to e^+e^-$, \newline lifetime of $c\tau=0.5\cm$.]{%
  \label{subfit:brazil:Kp_e_1:J}%
  \includegraphics[width=0.31\textwidth]{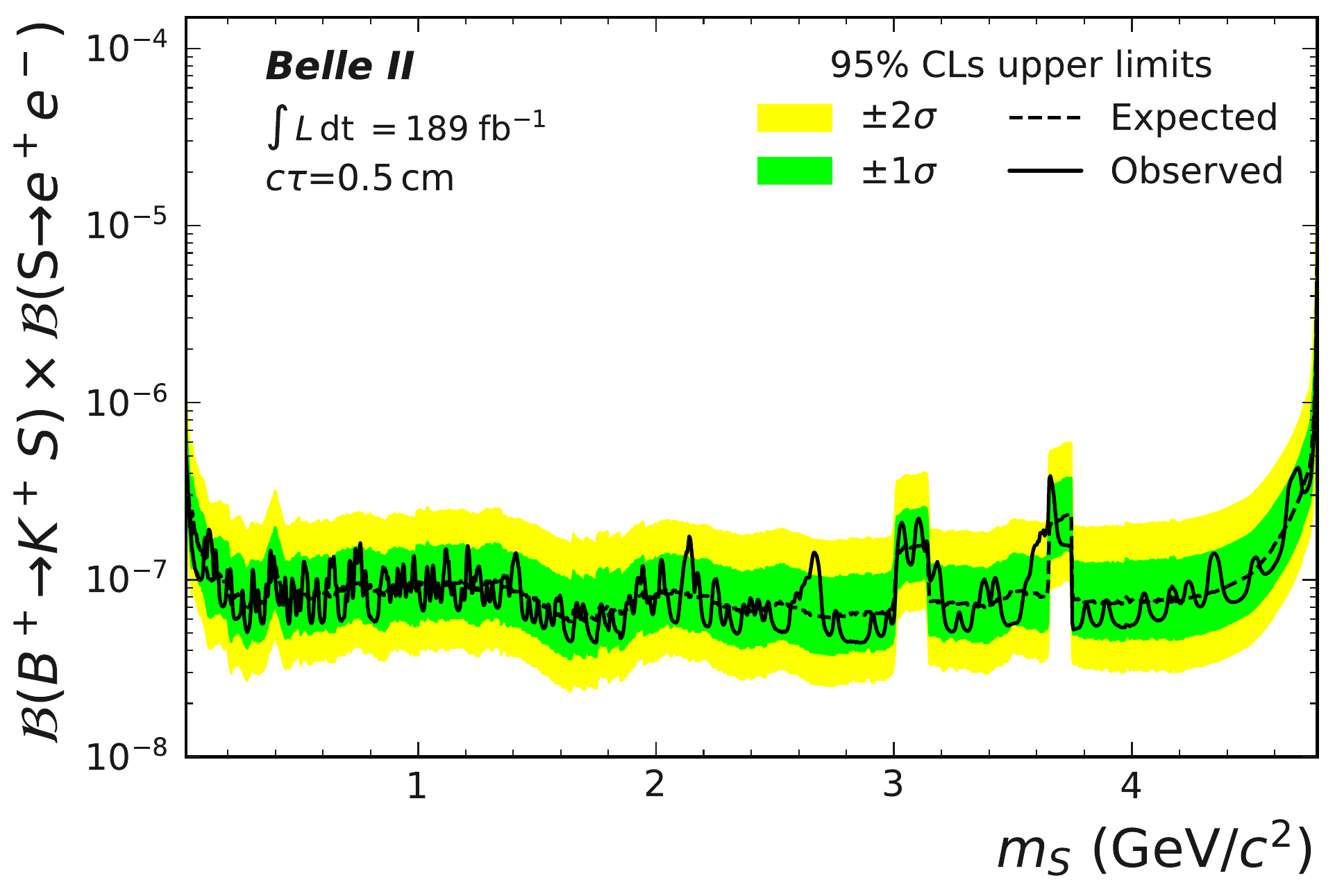}%
}%
\hspace*{\fill}
\subfigure[$B^+\to K^+S, S\to e^+e^-$, \newline lifetime of $c\tau=1\cm$.]{
  \label{subfit:brazil:Kp_e_1:K}%
  \includegraphics[width=0.31\textwidth]{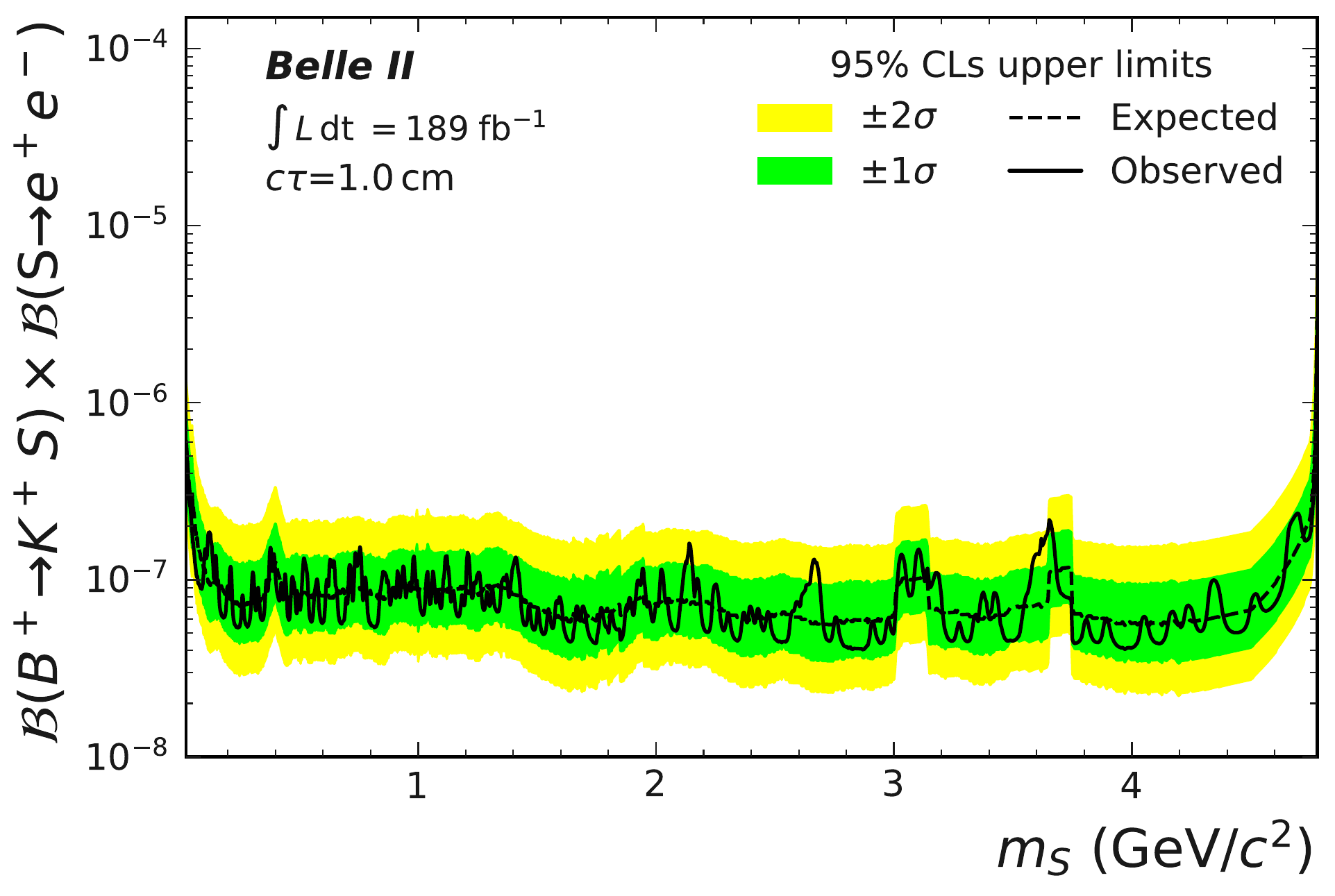}%
}%
\hspace*{\fill}
\subfigure[$B^+\to K^+S, S\to e^+e^-$, \newline lifetime of $c\tau=2.5\cm$.]{
  \label{subfit:brazil:Kp_e_1:L}%
  \includegraphics[width=0.31\textwidth]{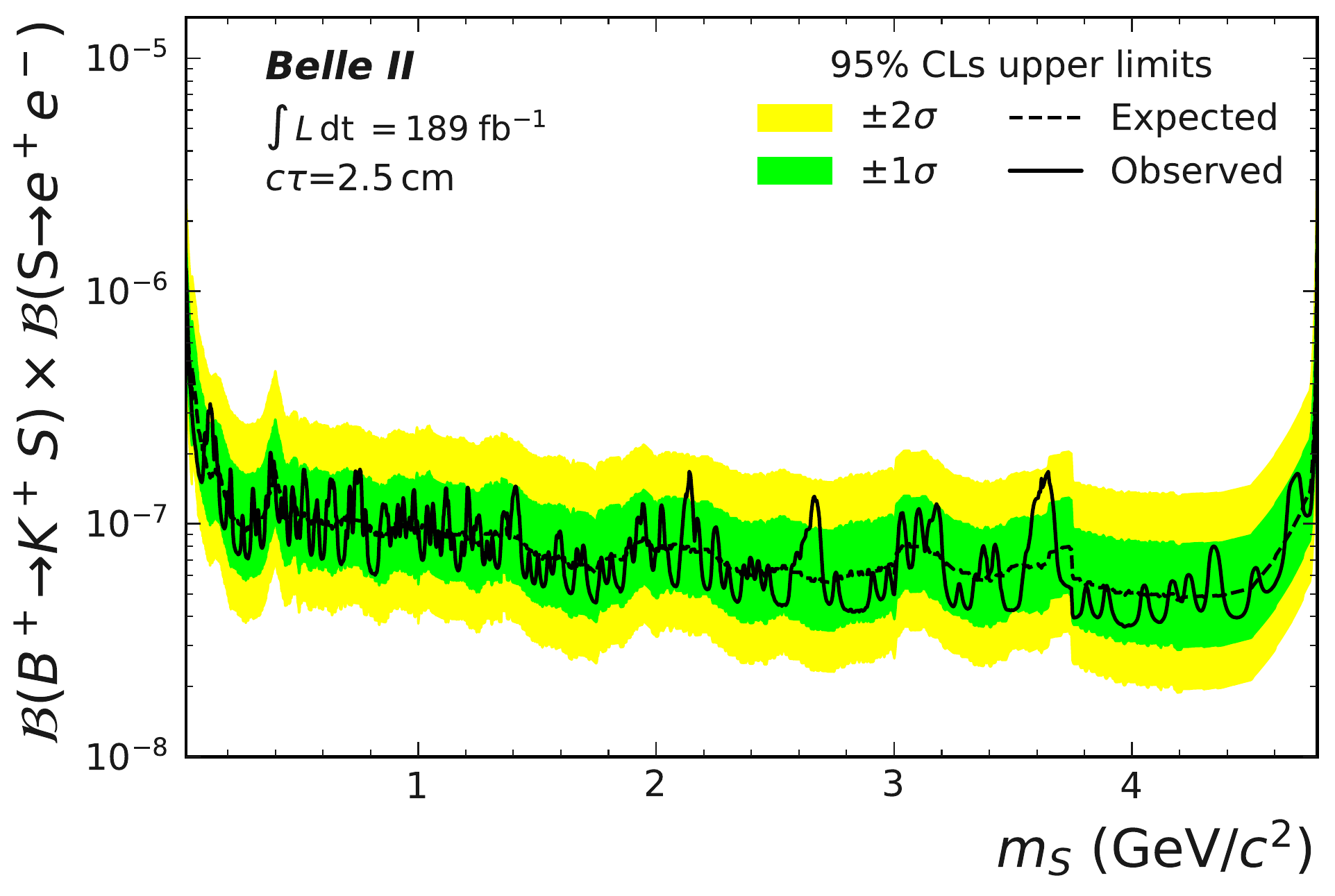}%
}
\caption{Expected and observed limits on the product branching fractions $\mathcal{B}(B^+\to K^+S) \times \mathcal{B}(S\to e^+e^-)$ for lifetimes \hbox{$0.001 < c\tau < 2.5\,\cm$}.}\label{subfit:brazil:Kp_e_1}
\end{figure*}

\begin{figure*}[ht]%
\subfigure[$B^+\to K^+S, S\to e^+e^-$, \newline lifetime of $c\tau=5\cm$.]{%
  \label{subfit:brazil:Kp_e_2:A}%
  \includegraphics[width=0.31\textwidth]{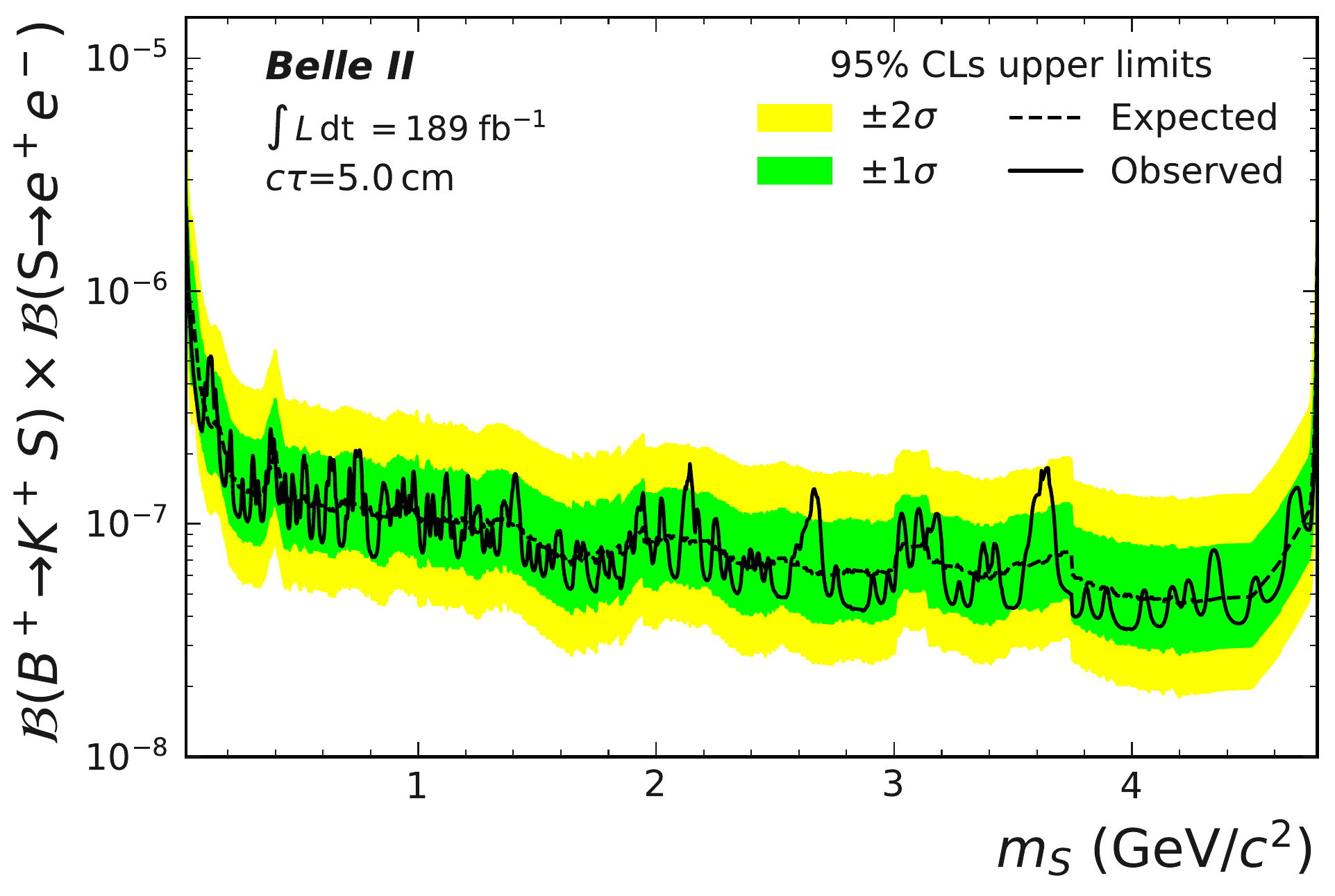}%
}%
\hspace*{\fill}
\subfigure[$B^+\to K^+S, S\to e^+e^-$, \newline lifetime of $c\tau=10\cm$.]{
  \label{subfit:brazil:Kp_e_2:B}%
  \includegraphics[width=0.31\textwidth]{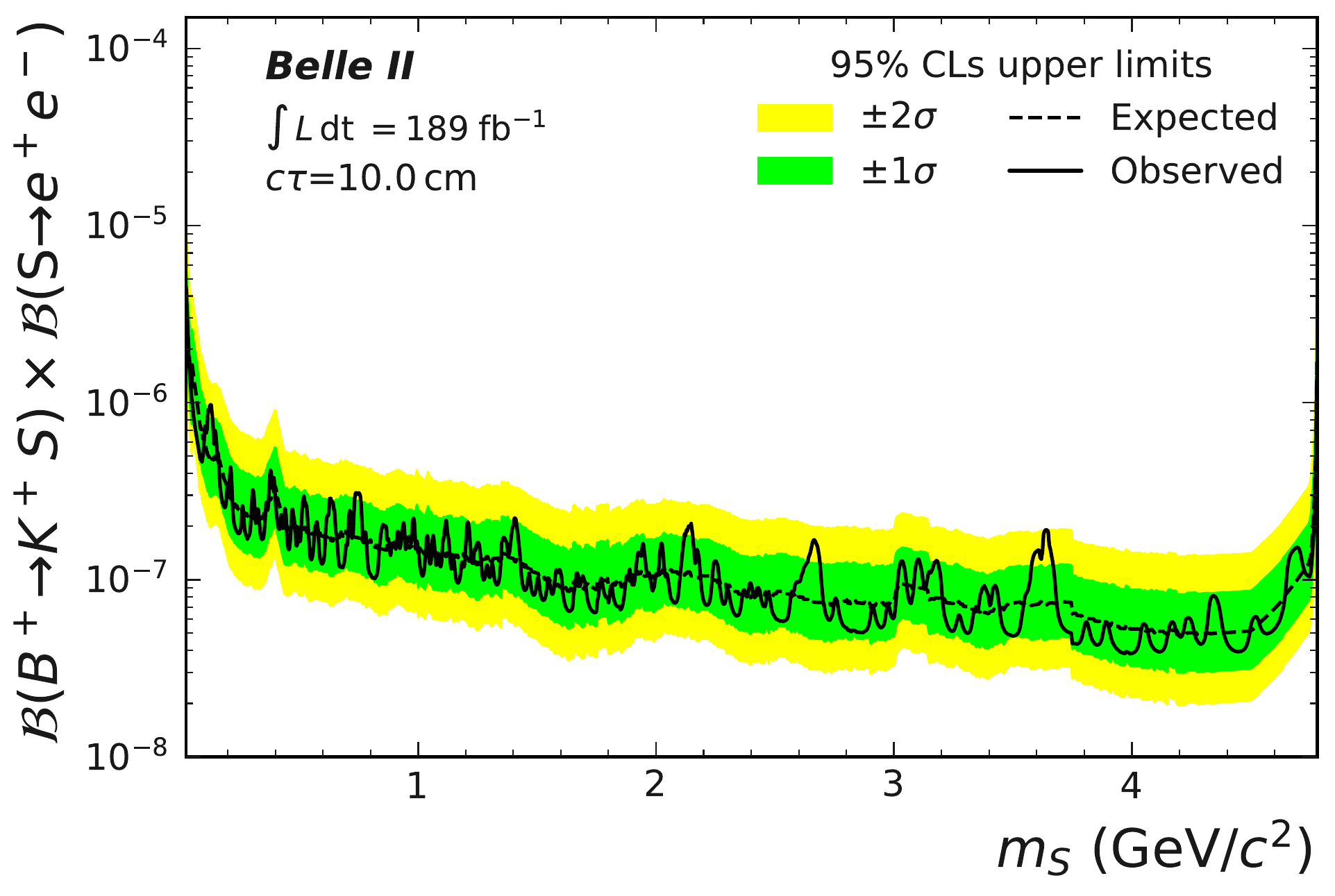}%
}%
\hspace*{\fill}
\subfigure[$B^+\to K^+S, S\to e^+e^-$, \newline lifetime of $c\tau=25\cm$.]{
  \label{subfit:brazil:Kp_e_2:C}%
  \includegraphics[width=0.31\textwidth]{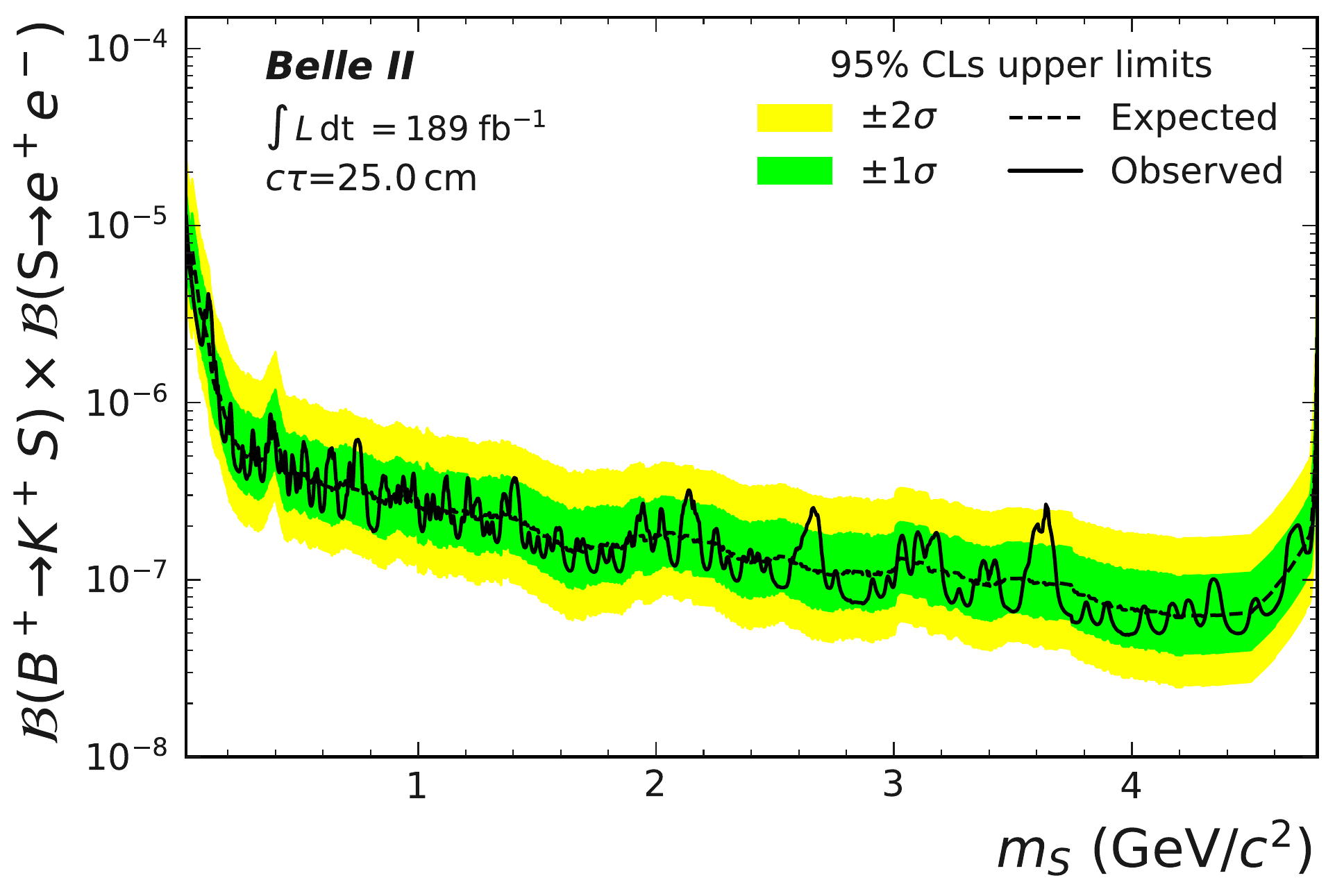}%
}
\subfigure[$B^+\to K^+S, S\to e^+e^-$, \newline lifetime of $c\tau=50\cm$.]{%
  \label{subfit:brazil:Kp_e_2:D}%
  \includegraphics[width=0.31\textwidth]{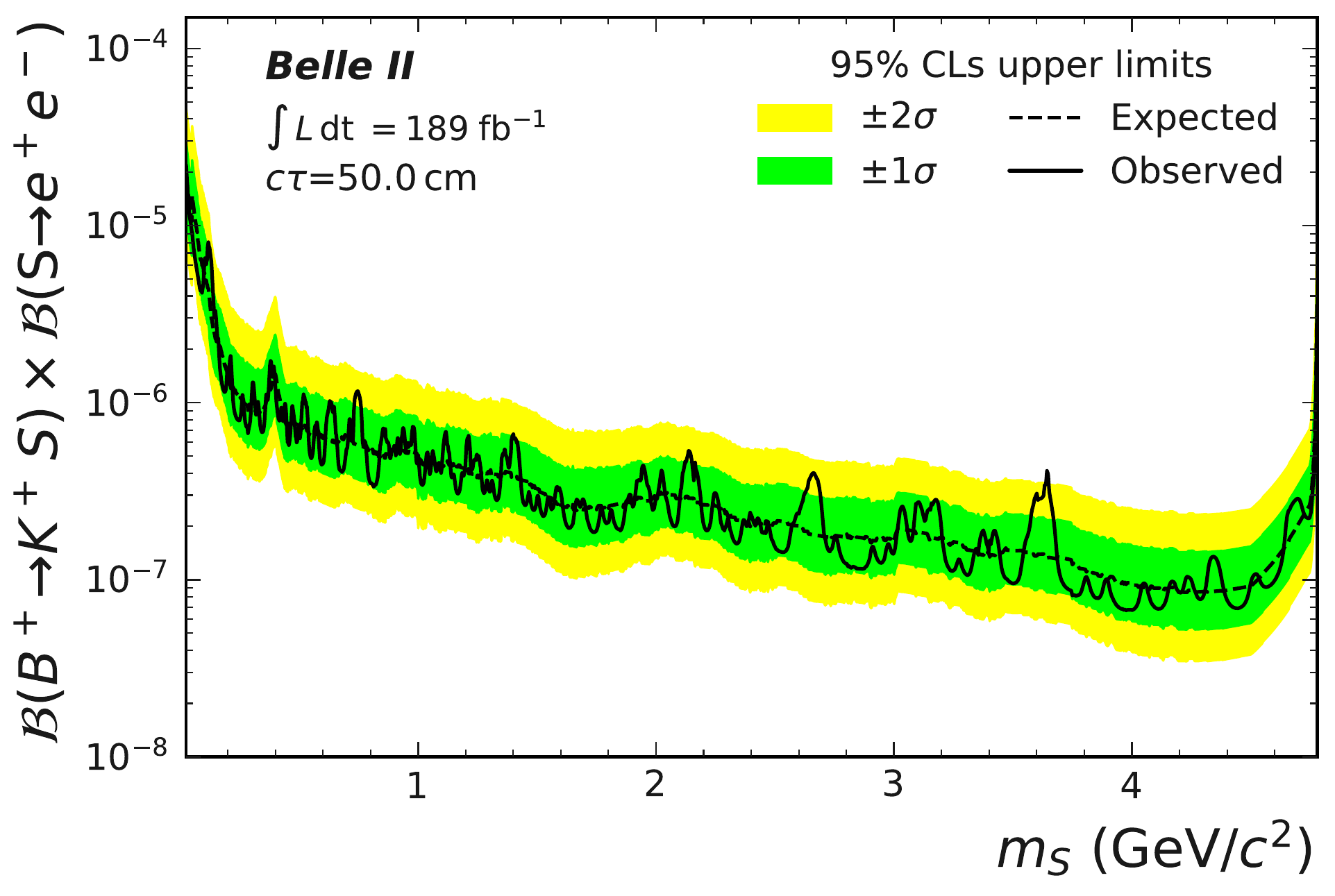}%
}%
\hspace*{\fill}
\subfigure[$B^+\to K^+S, S\to e^+e^-$, \newline lifetime of $c\tau=100\cm$.]{
  \label{subfit:brazil:Kp_e_2:E}%
  \includegraphics[width=0.31\textwidth]{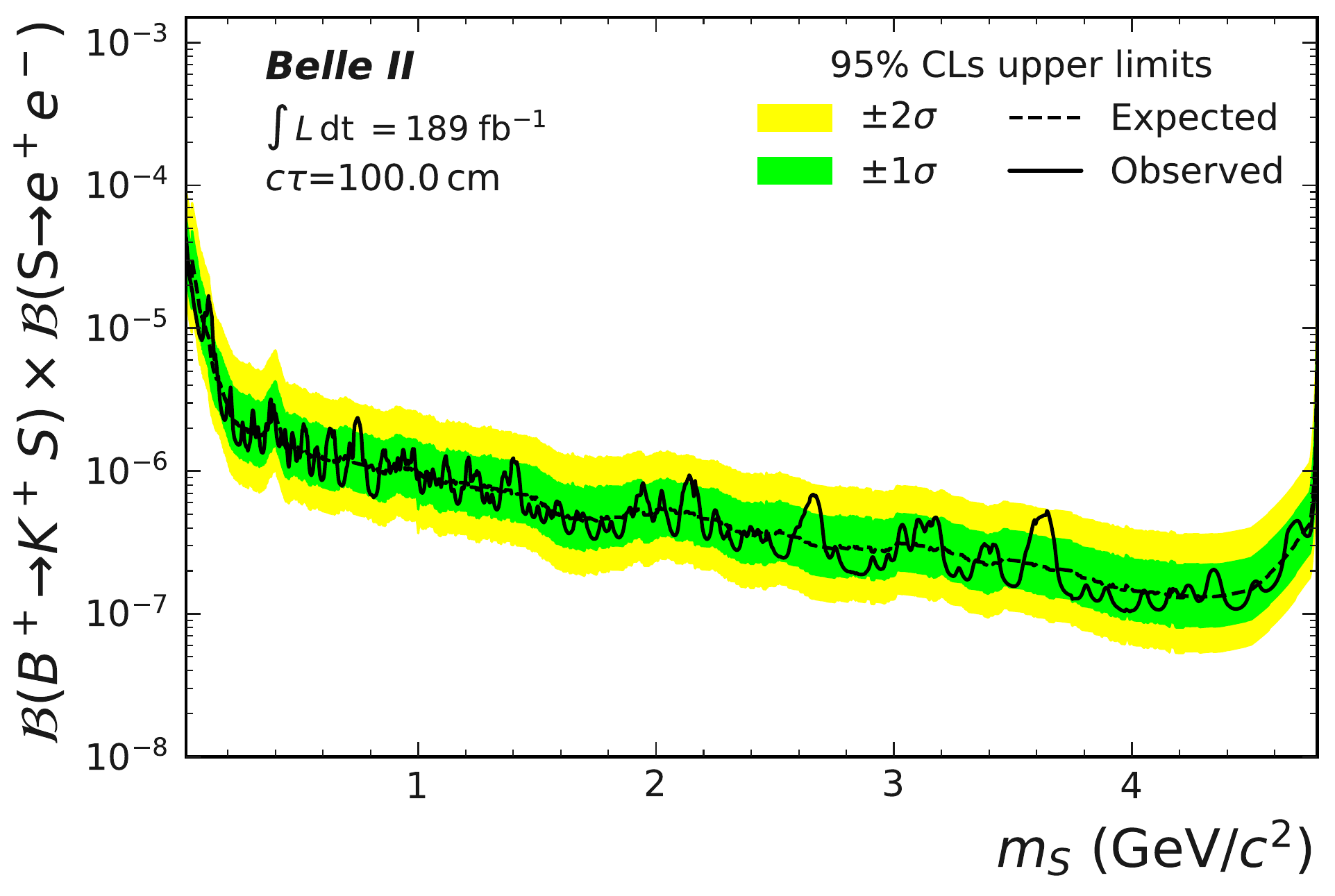}%
}%
\hspace*{\fill}
\subfigure[$B^+\to K^+S, S\to e^+e^-$, \newline lifetime of $c\tau=200\cm$.]{
  \label{subfit:brazil:Kp_e_2:F}%
  \includegraphics[width=0.31\textwidth]{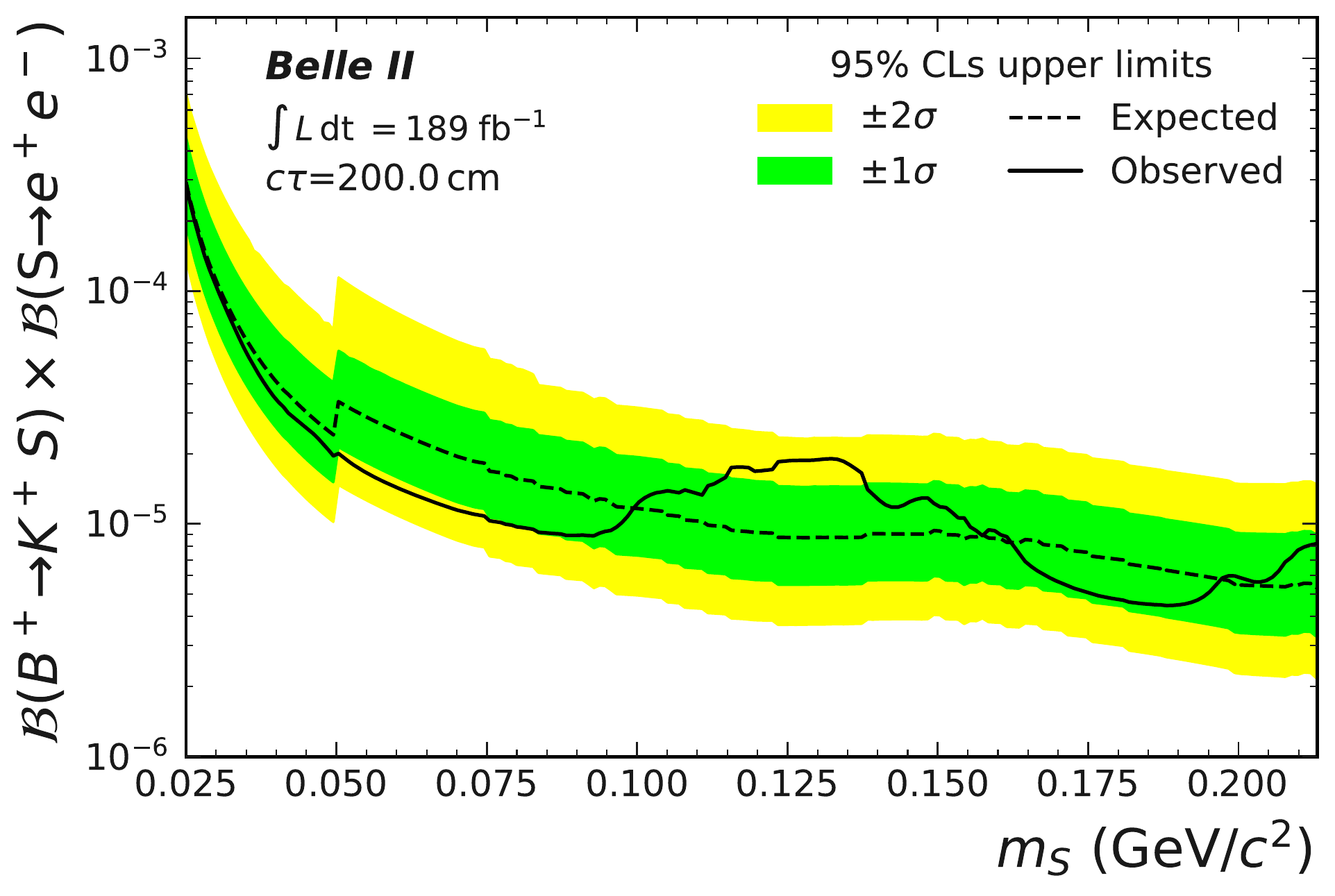}%
}

\subfigure[$B^+\to K^+S, S\to e^+e^-$, \newline lifetime of $c\tau=400\cm$.]{
  \label{subfit:brazil:Kp_e_2:G}%
  \includegraphics[width=0.31\textwidth]{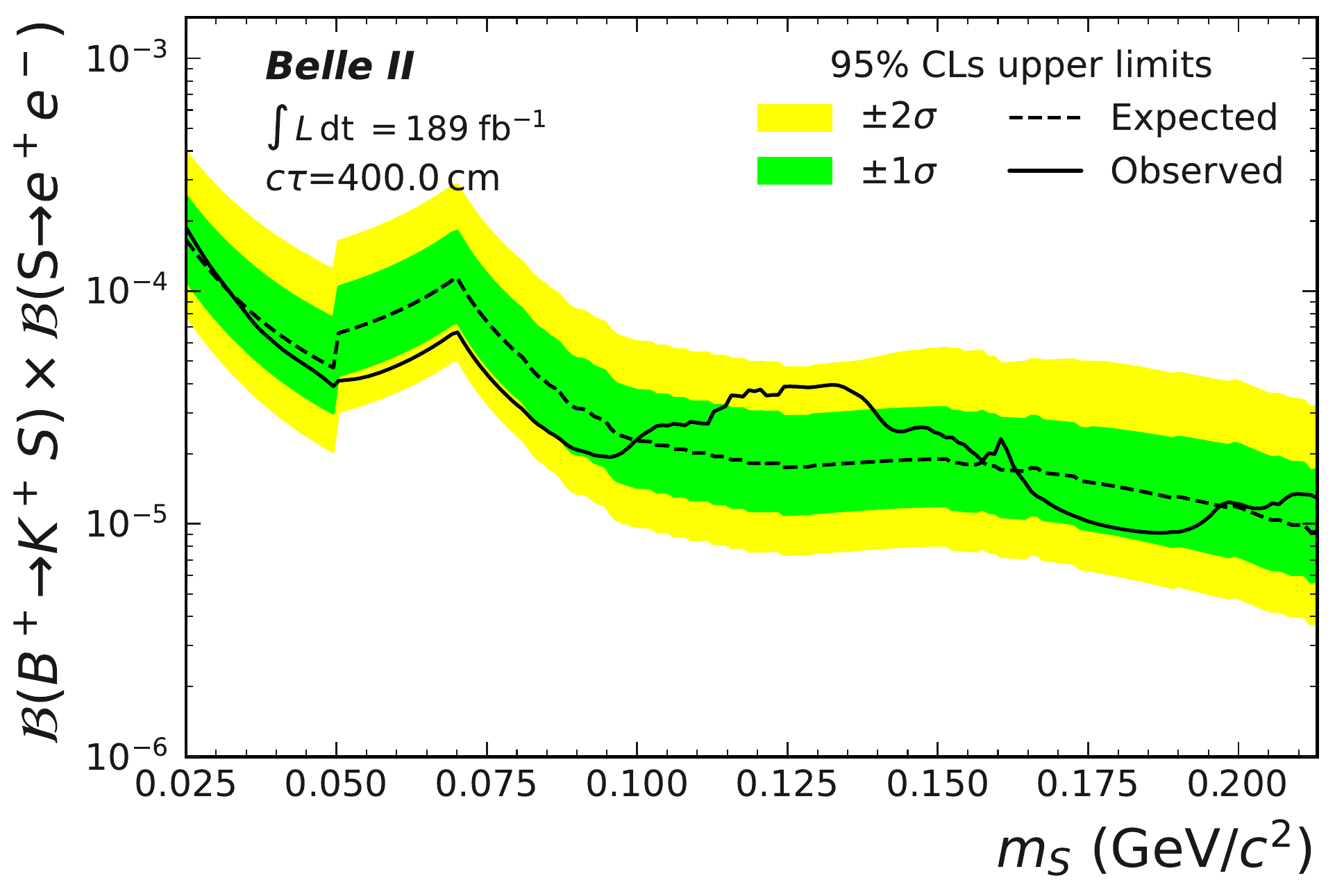}%
}
\hspace*{\fill}
\subfigure[$B^+\to K^+S, S\to e^+e^-$, \newline lifetime of $c\tau=700\cm$.]{%
  \label{subfit:brazil:Kp_e_2:H}%
  \includegraphics[width=0.31\textwidth]{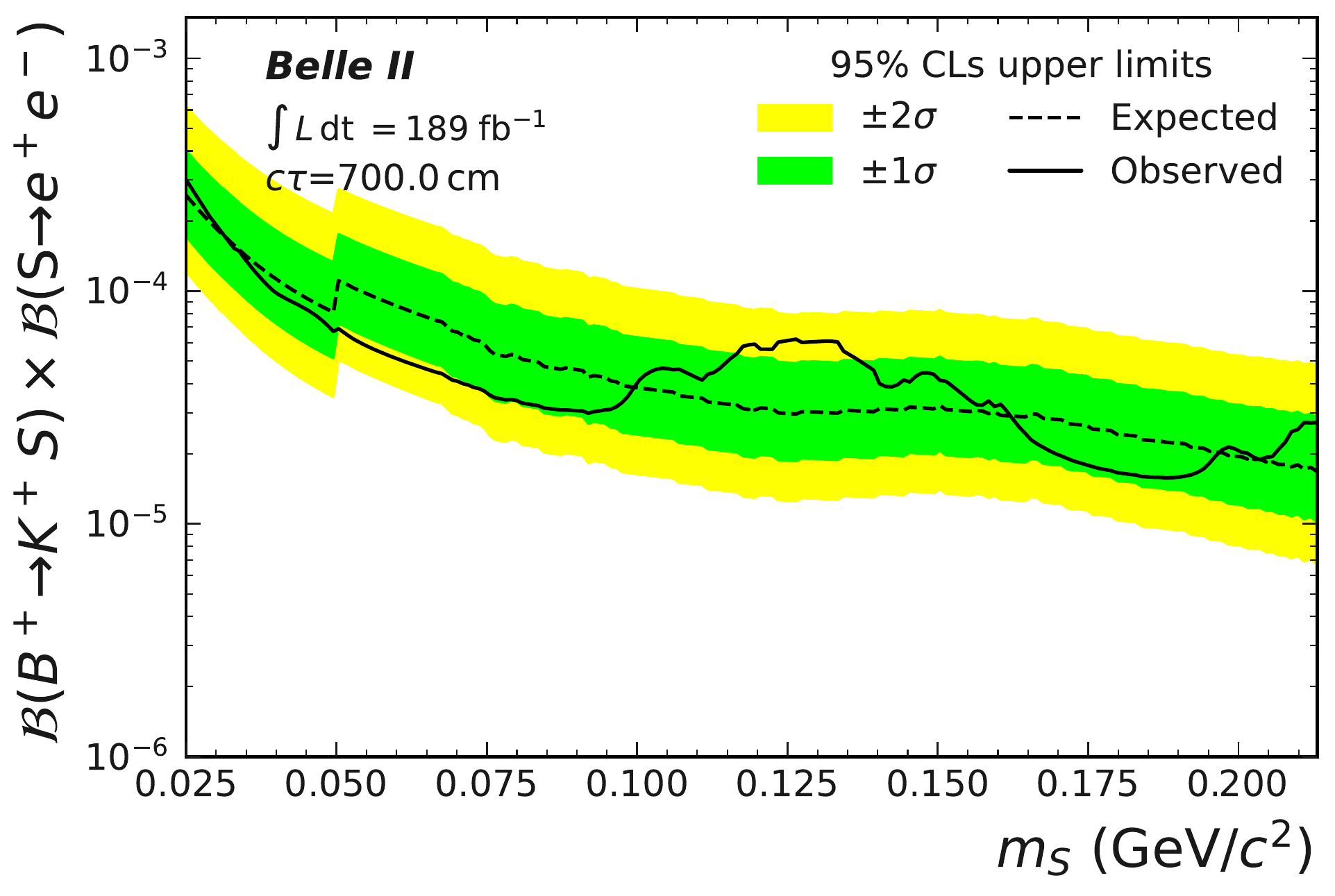}%
}%
\hspace*{\fill}
\subfigure[$B^+\to K^+S, S\to e^+e^-$, \newline lifetime of $c\tau=1000\cm$.]{
  \label{subfit:brazil:Kp_e_2:I}%
  \includegraphics[width=0.31\textwidth]{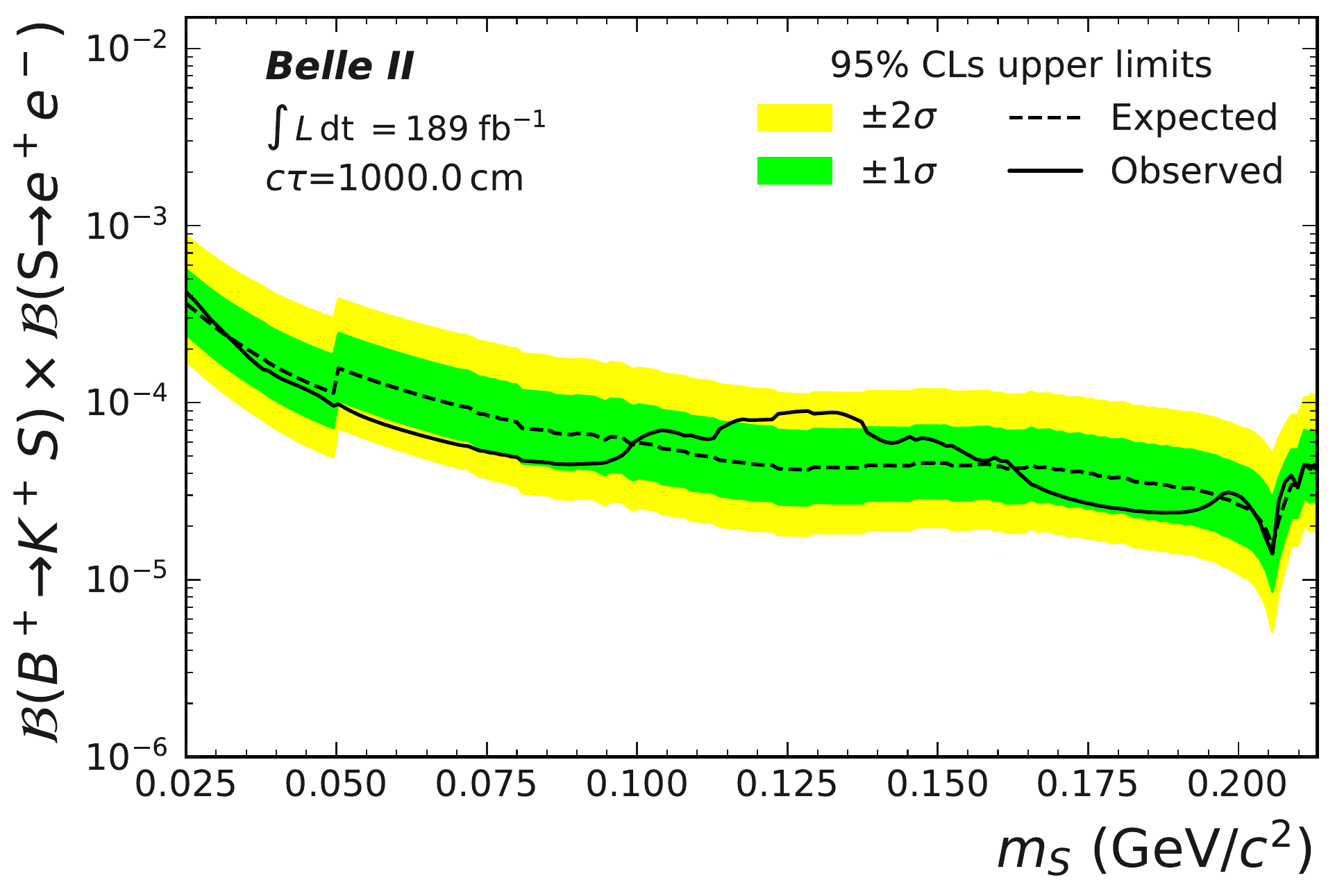}%
}%

\subfigure[$B^+\to K^+S, S\to e^+e^-$, \newline lifetime of $c\tau=2500\cm$.]{
  \label{subfit:brazil:Kp_e_2:J}%
  \includegraphics[width=0.31\textwidth]{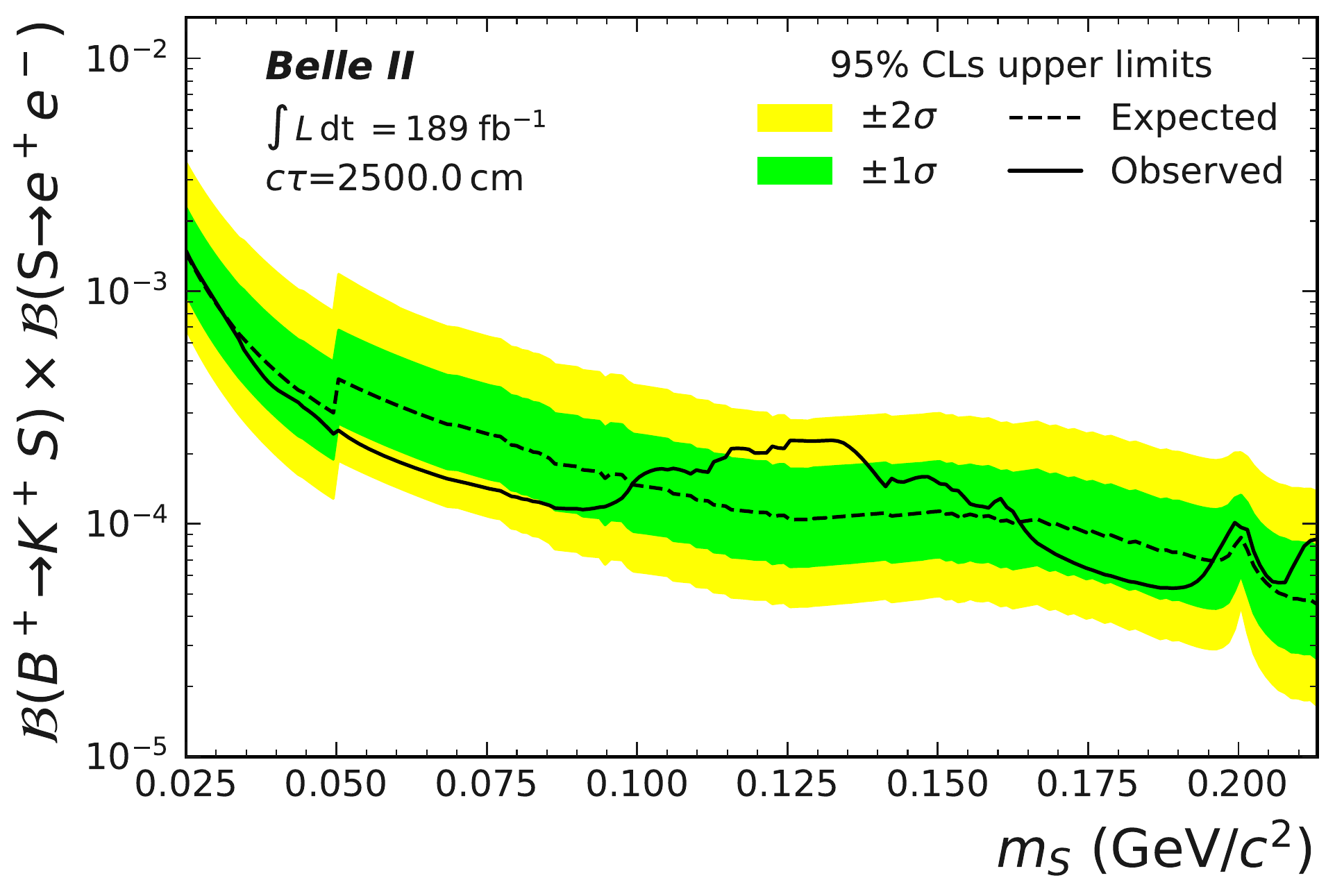}%
}
\hspace*{\fill}
\subfigure[$B^+\to K^+S, S\to e^+e^-$, \newline lifetime of $c\tau=5000\cm$.]{%
  \label{subfit:brazil:Kp_e_2:K}%
  \includegraphics[width=0.31\textwidth]{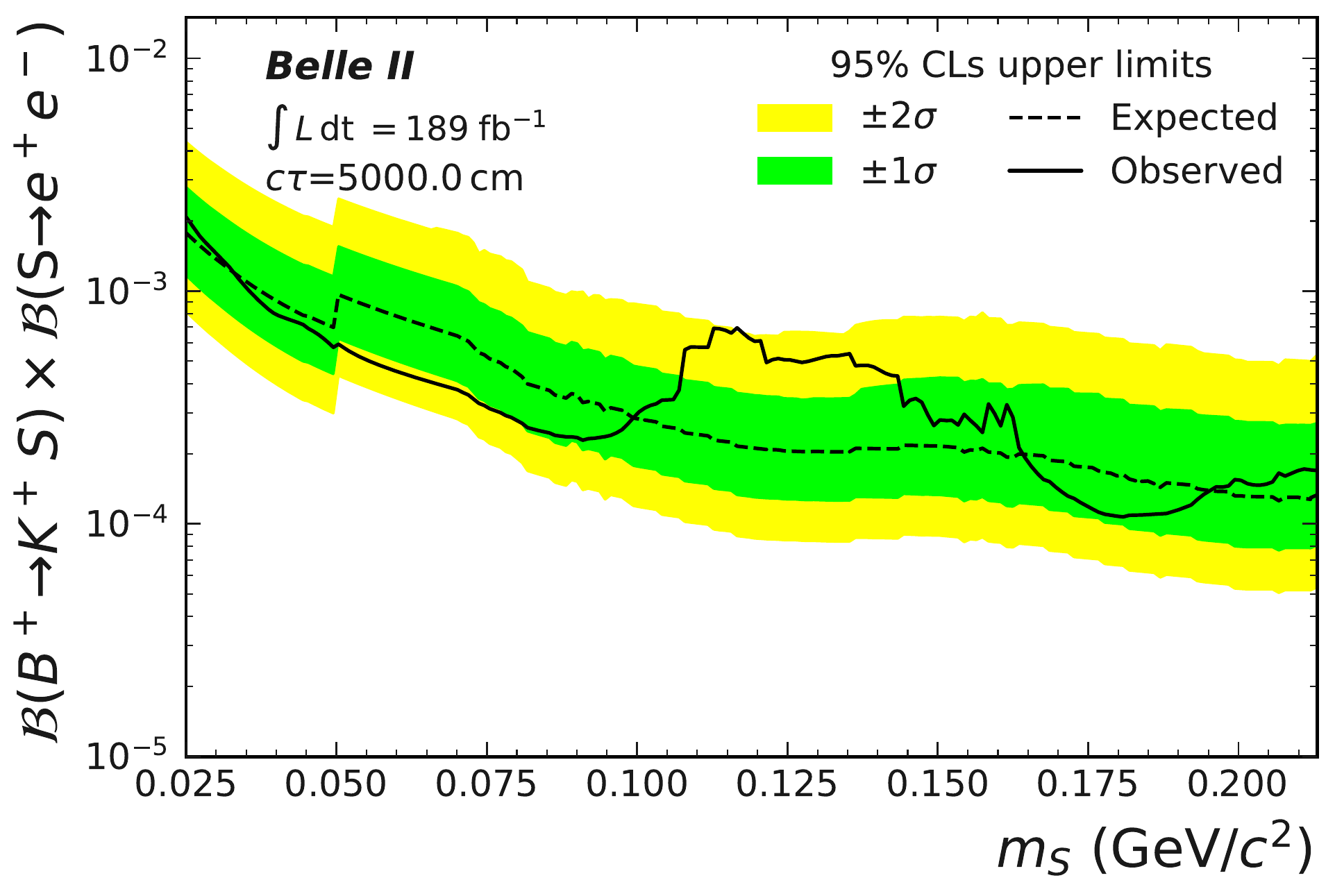}%
}%
\hspace*{\fill}
\subfigure[$B^+\to K^+S, S\to e^+e^-$, \newline lifetime of $c\tau=10000\cm$.]{
  \label{subfit:brazil:Kp_e_2:L}%
  \includegraphics[width=0.31\textwidth]{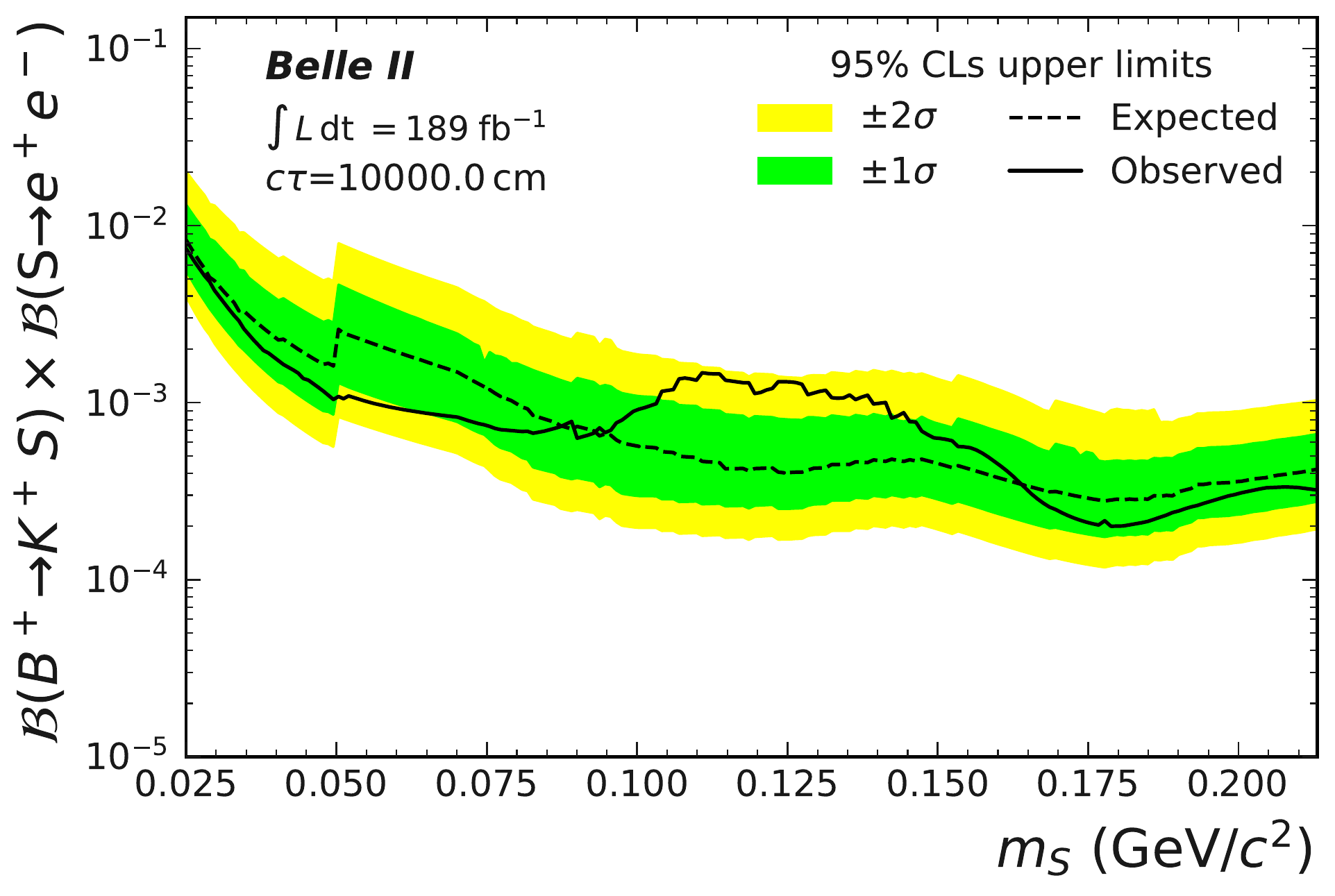}%
}%
\caption{Expected and observed limits on the product of branching fractions $\mathcal{B}(B^+\to K^+S) \times \mathcal{B}(S\to e^+e^-)$ for lifetimes \hbox{$5 < c\tau < 10000\,\cm$}.}\label{subfit:brazil:Kp_e_2}
\end{figure*}

\begin{figure*}[ht]%
\subfigure[$\Bz\to \Kstarz(\to K^+\pi^-) S, S\to e^+e^-$, \newline lifetime of $c\tau=0.001\cm$.]{%
  \label{subfit:brazil:Kstar_e_1:A}%
  \includegraphics[width=0.31\textwidth]{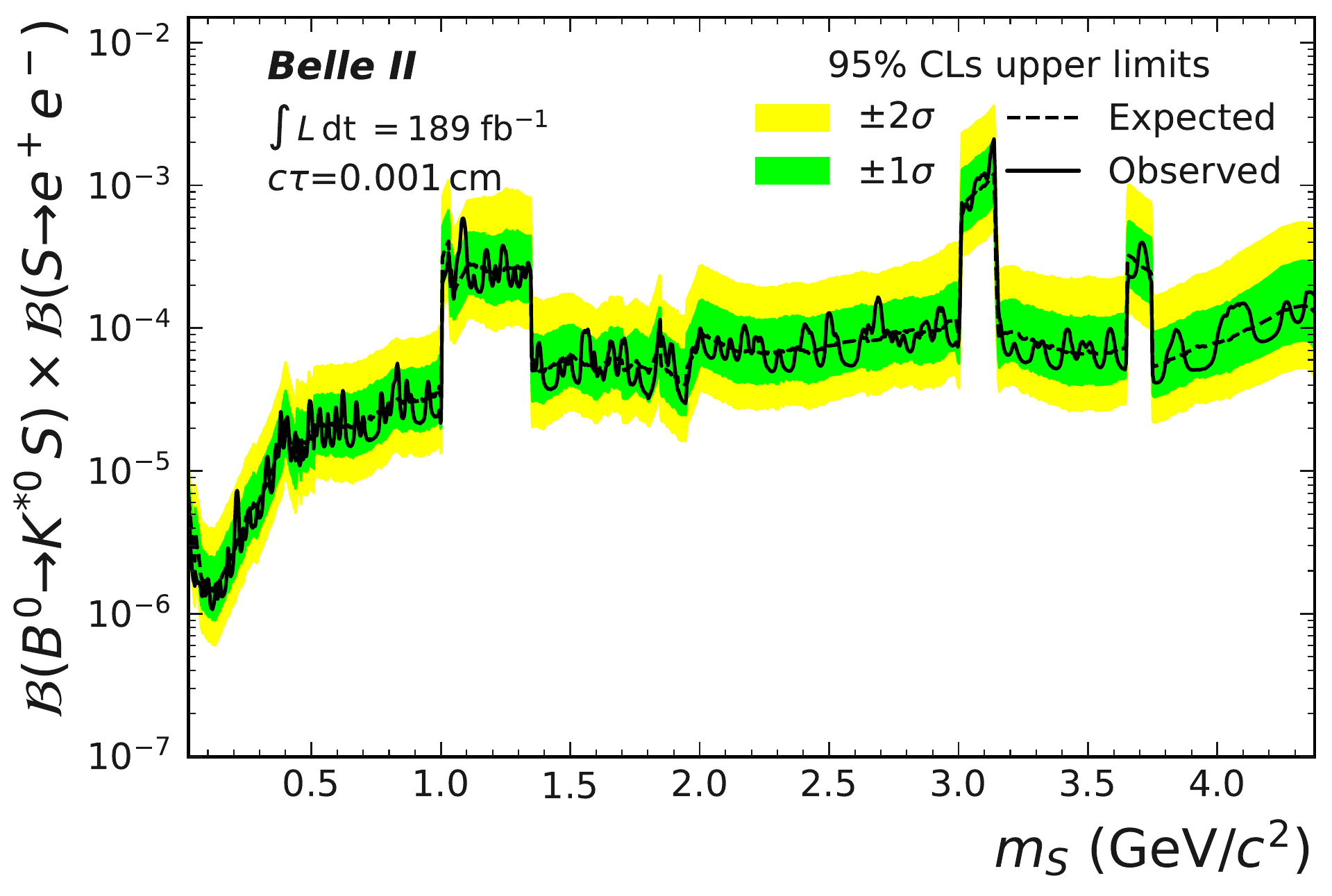}%
}%
\hspace*{\fill}
\subfigure[$\Bz\to \Kstarz(\to K^+\pi^-) S, S\to e^+e^-$, \newline lifetime of $c\tau=0.003\cm$.]{
  \label{subfit:brazil:Kstar_e_1:B}%
  \includegraphics[width=0.31\textwidth]{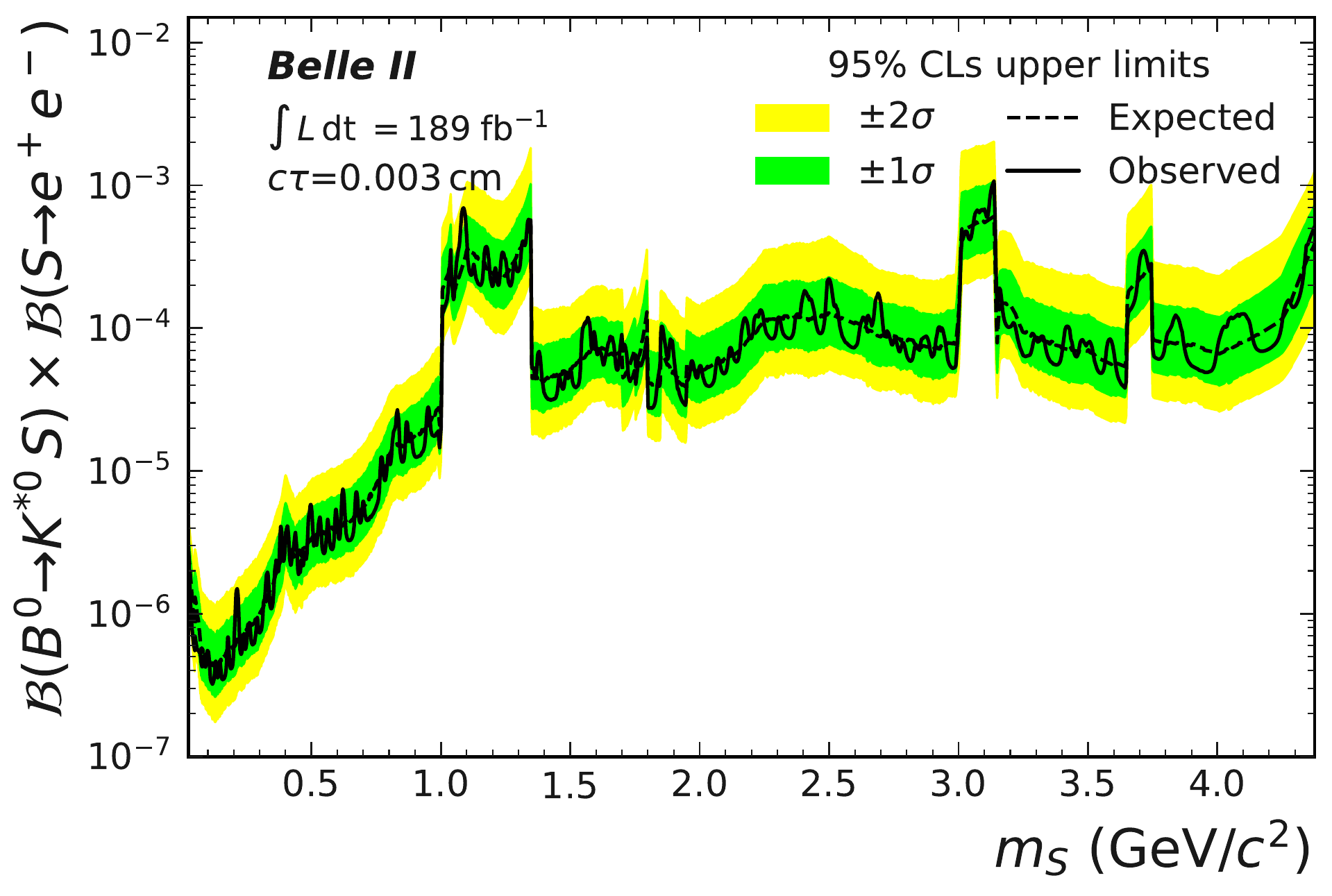}%
}%
\hspace*{\fill}
\subfigure[$\Bz\to \Kstarz(\to K^+\pi^-) S, S\to e^+e^-$, \newline lifetime of $c\tau=0.005\cm$.]{
  \label{subfit:brazil:Kstar_e_1:C}%
  \includegraphics[width=0.31\textwidth]{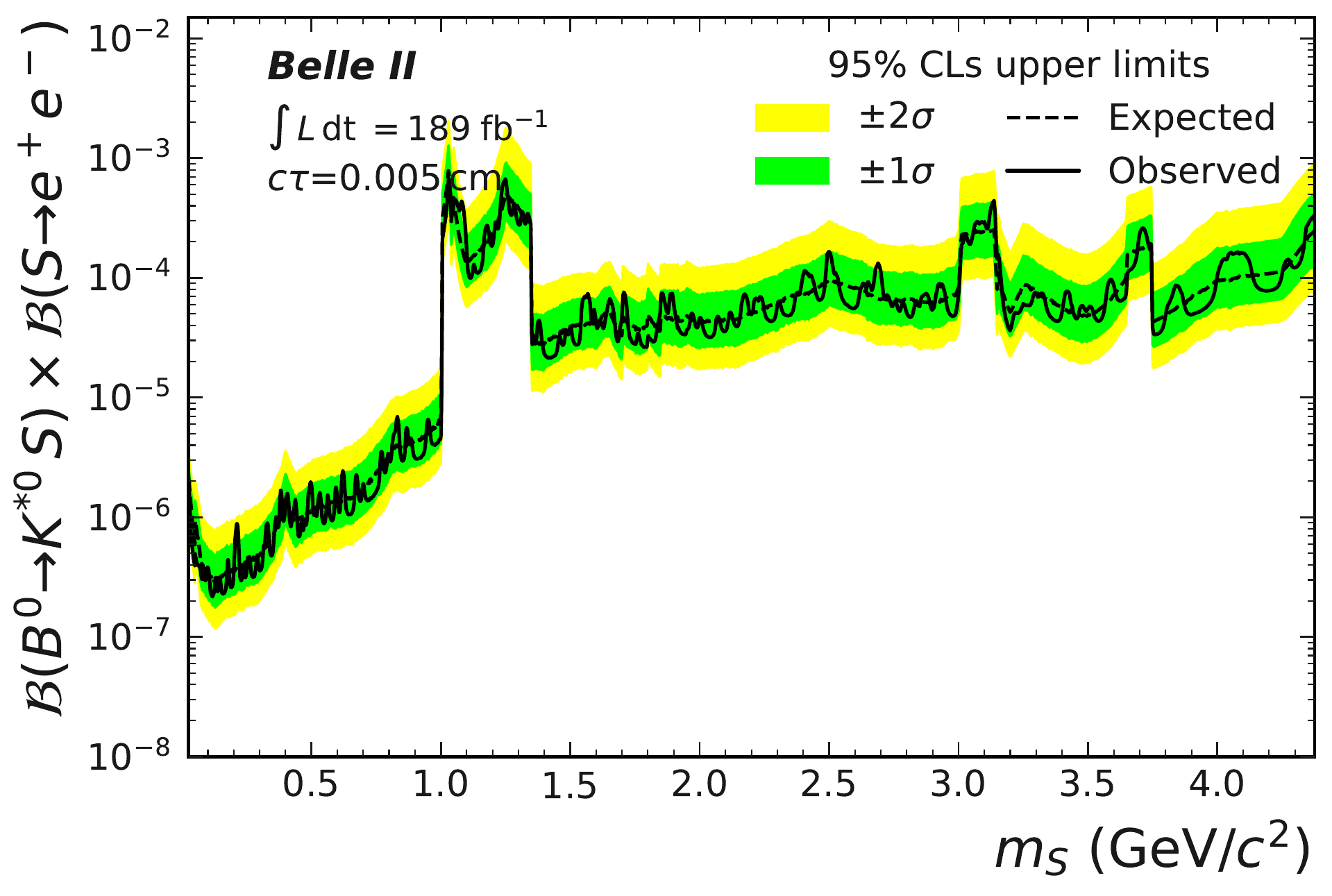}%
}
\subfigure[$\Bz\to \Kstarz(\to K^+\pi^-) S, S\to e^+e^-$, \newline lifetime of $c\tau=0.007\cm$.]{%
  \label{subfit:brazil:Kstar_e_1:D}%
  \includegraphics[width=0.31\textwidth]{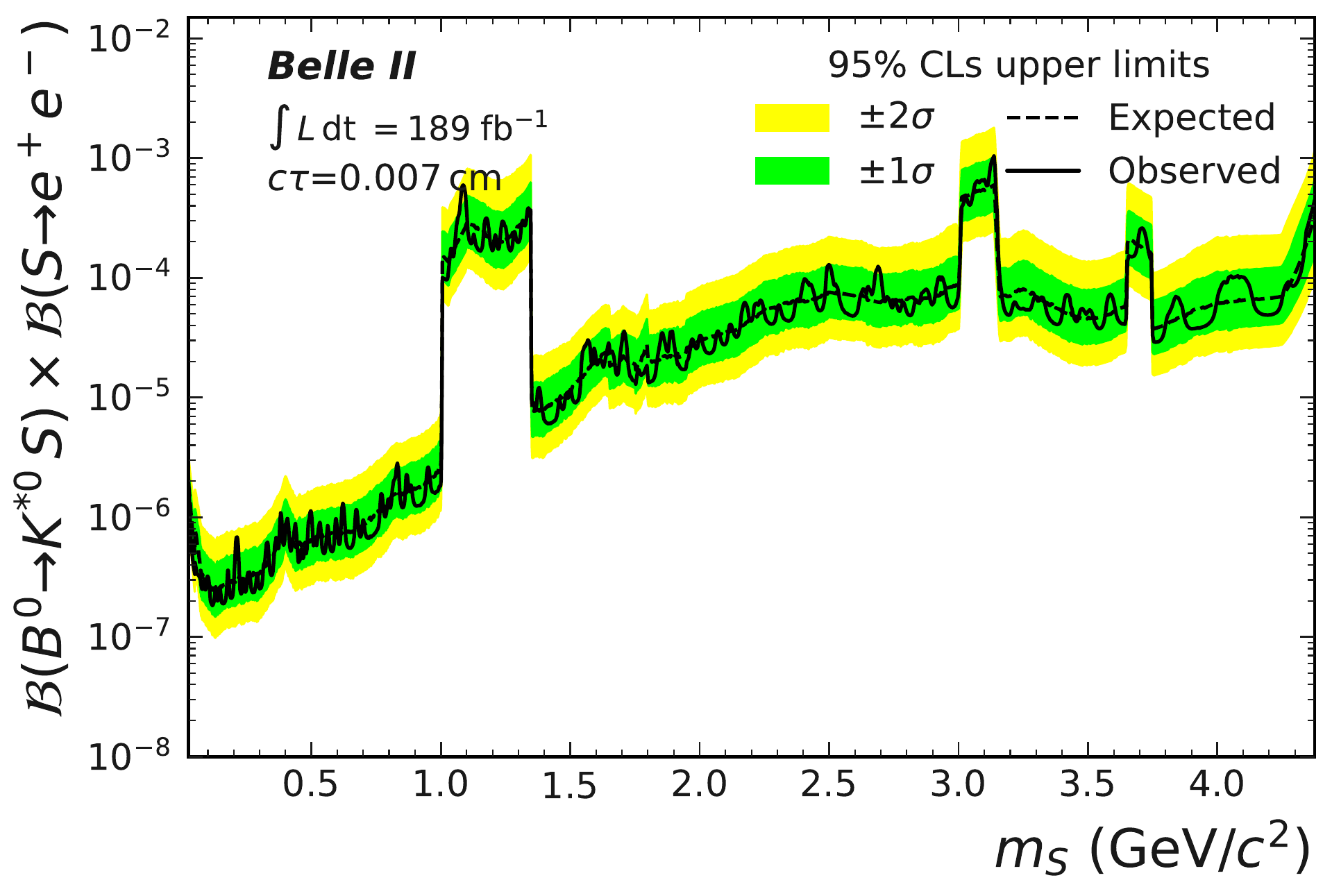}%
}%
\hspace*{\fill}
\subfigure[$\Bz\to \Kstarz(\to K^+\pi^-) S, S\to e^+e^-$, \newline lifetime of $c\tau=0.01\cm$.]{
  \label{subfit:brazil:Kstar_e_1:E}%
  \includegraphics[width=0.31\textwidth]{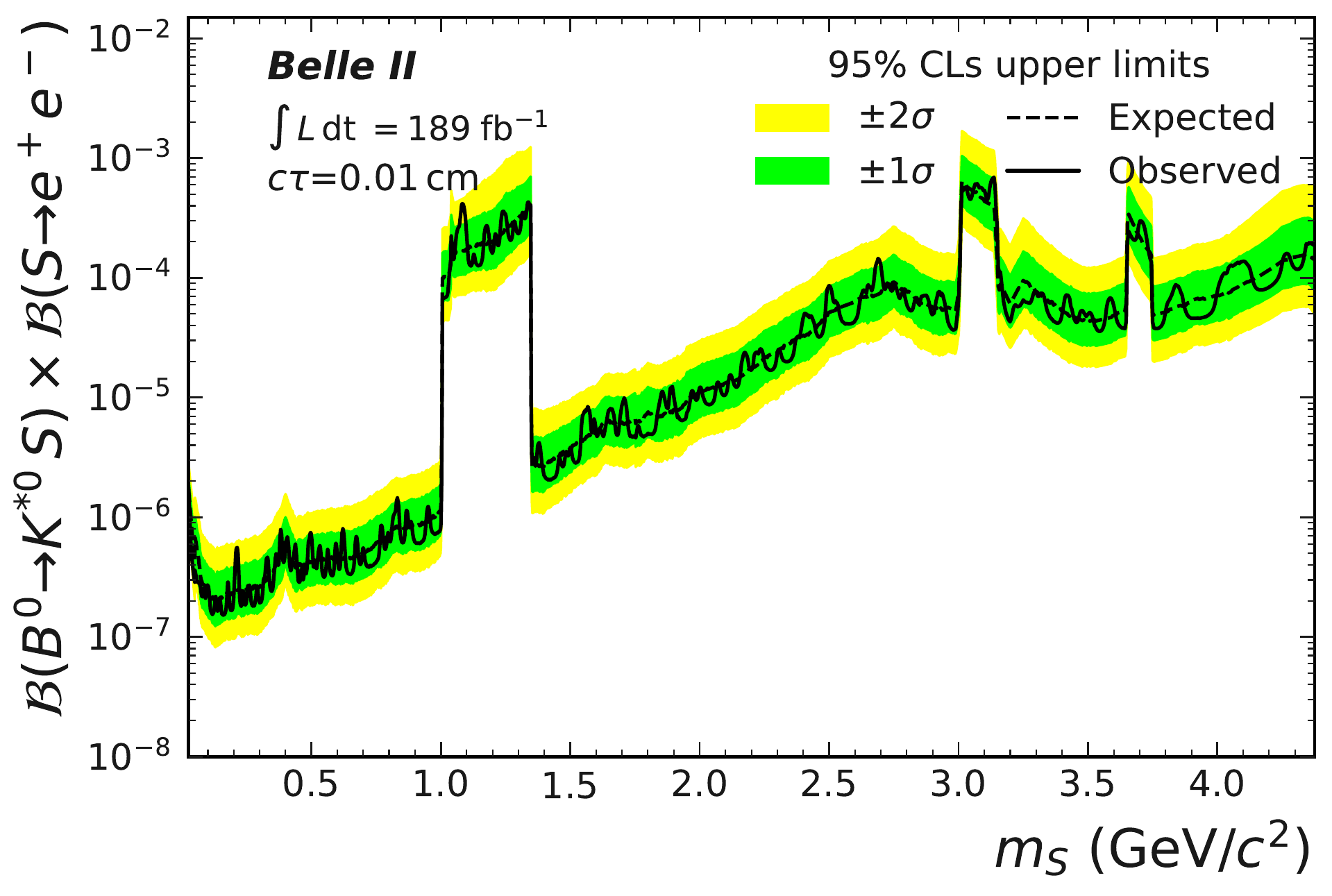}%
}%
\hspace*{\fill}
\subfigure[$\Bz\to \Kstarz(\to K^+\pi^-) S, S\to e^+e^-$, \newline lifetime of $c\tau=0.025\cm$.]{
  \label{subfit:brazil:Kstar_e_1:F}%
  \includegraphics[width=0.31\textwidth]{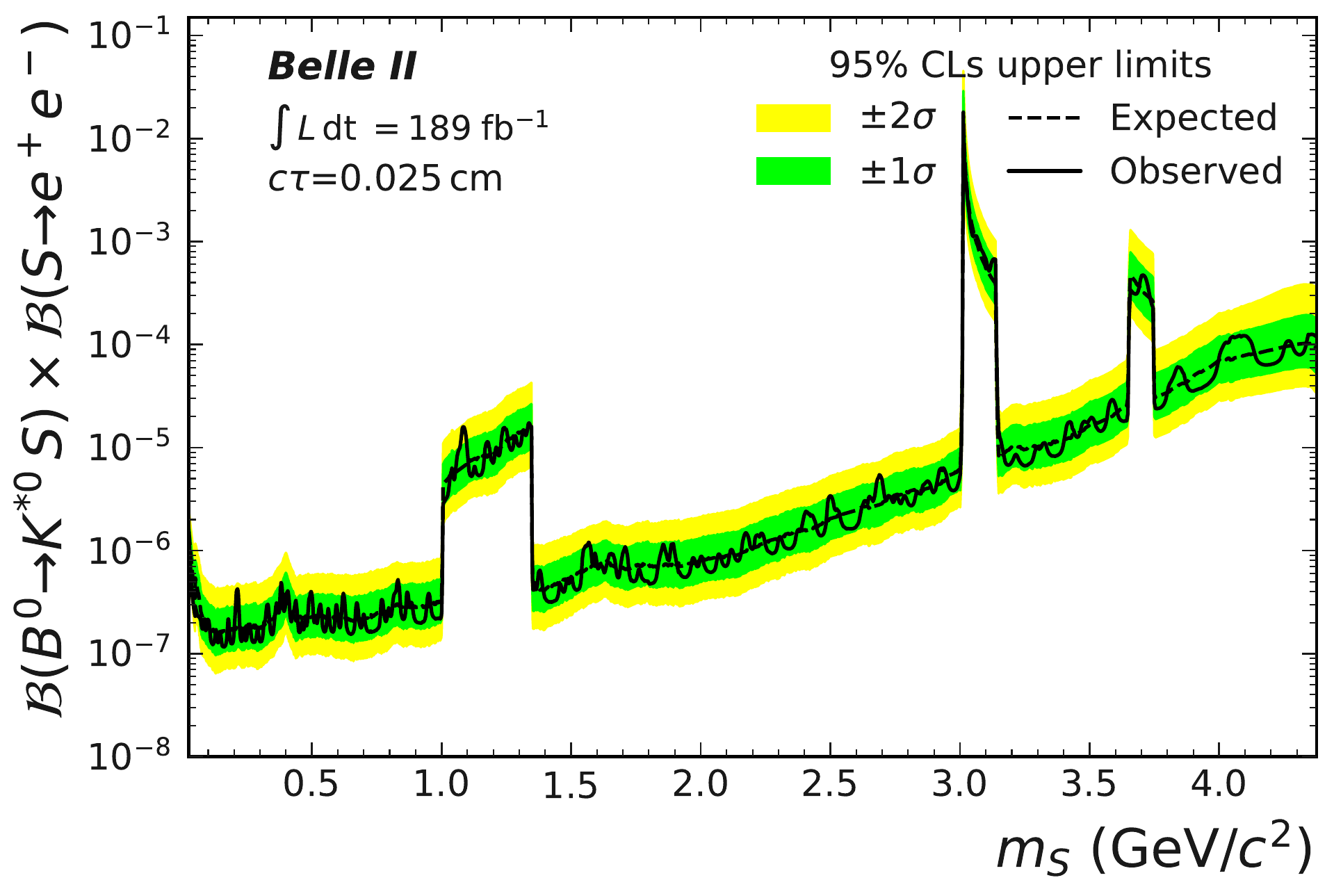}%
}
\subfigure[$\Bz\to \Kstarz(\to K^+\pi^-) S, S\to e^+e^-$, \newline lifetime of $c\tau=0.05\cm$.]{%
  \label{subfit:brazil:Kstar_e_1:G}%
  \includegraphics[width=0.31\textwidth]{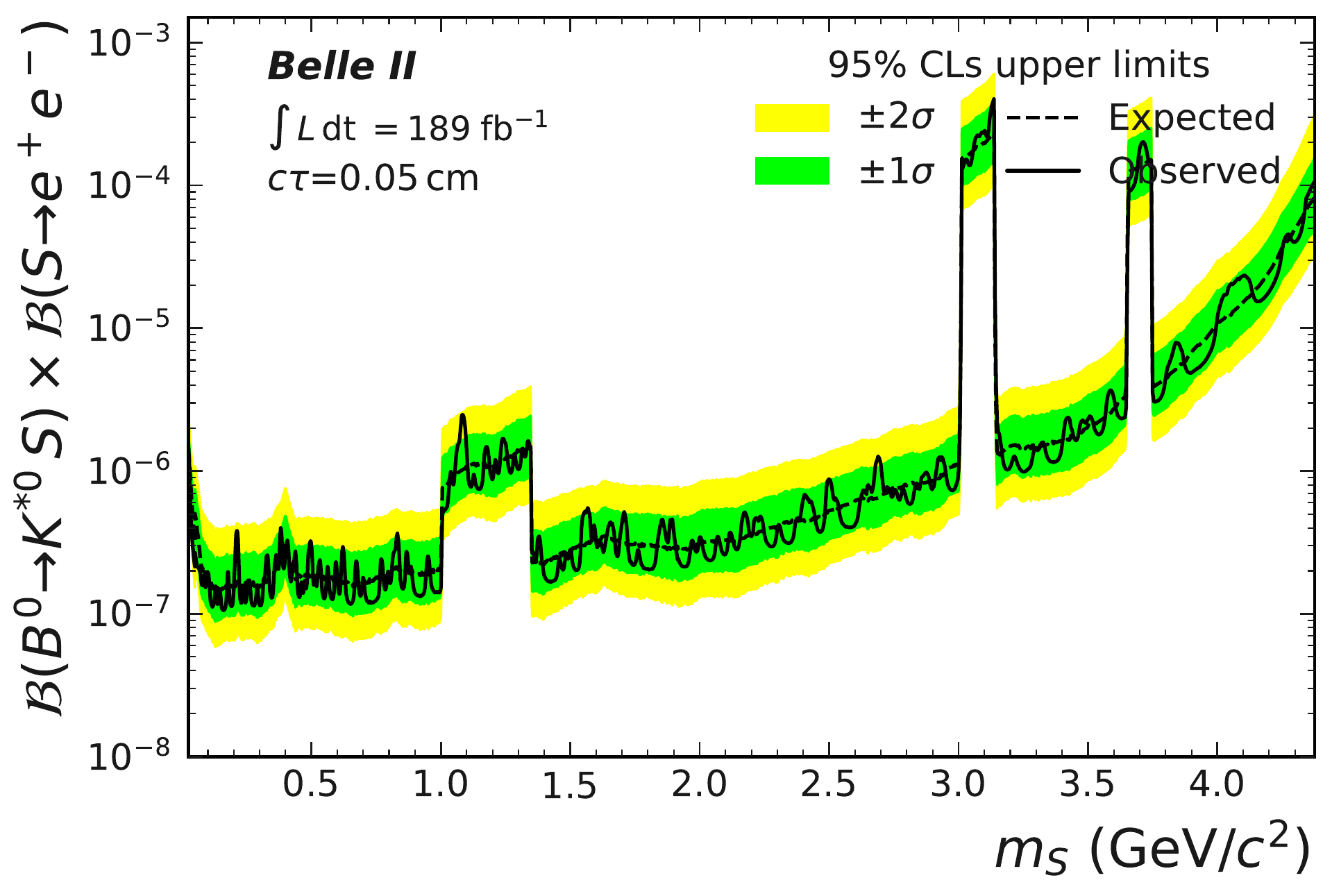}%
}%
\hspace*{\fill}
\subfigure[$\Bz\to \Kstarz(\to K^+\pi^-) S, S\to e^+e^-$, \newline lifetime of $c\tau=0.100\cm$.]{
  \label{subfit:brazil:Kstar_e_1:H}%
  \includegraphics[width=0.31\textwidth]{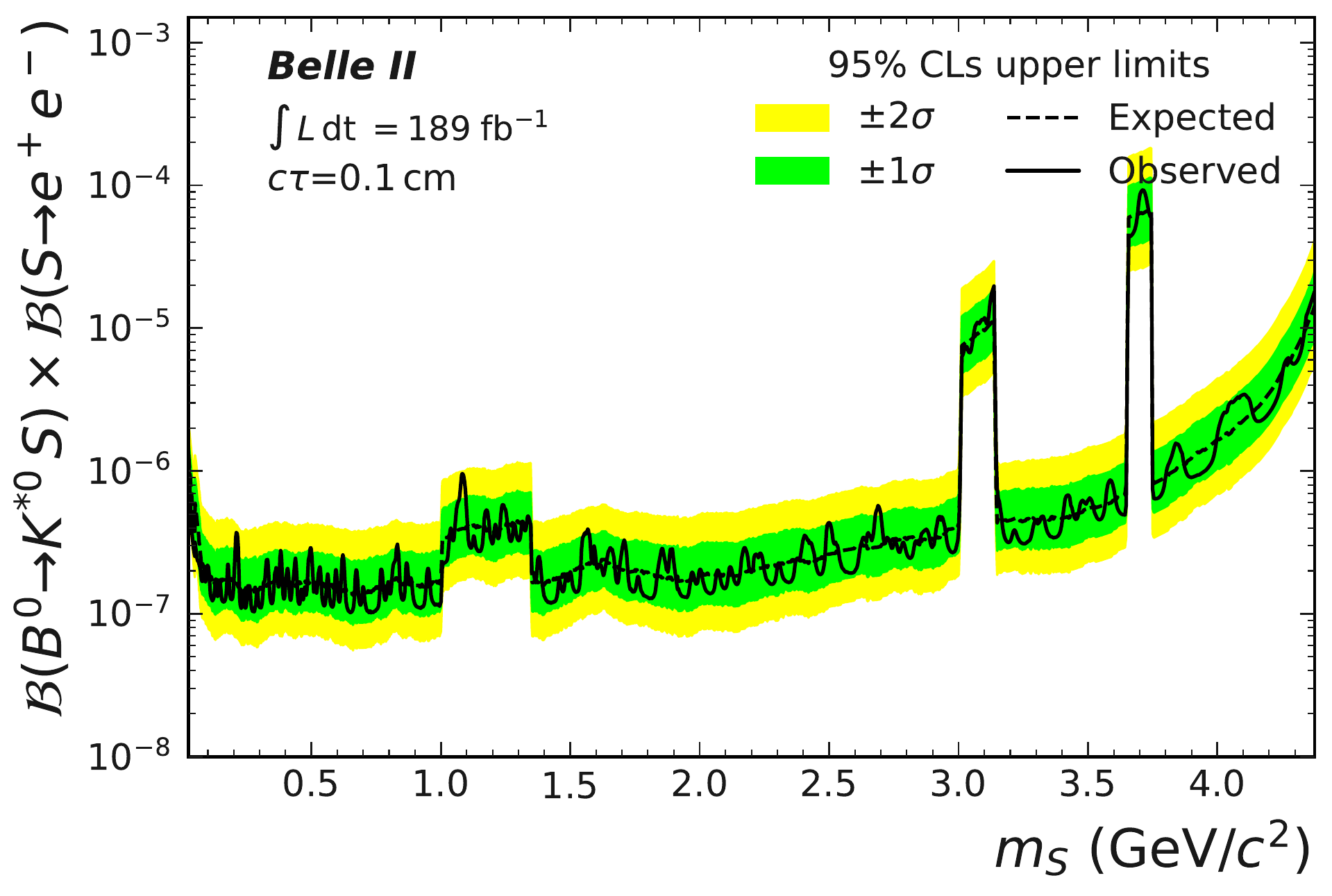}%
}%
\hspace*{\fill}
\subfigure[$\Bz\to \Kstarz(\to K^+\pi^-) S, S\to e^+e^-$, \newline lifetime of $c\tau=0.25\cm$.]{
  \label{subfit:brazil:Kstar_e_1:I}%
  \includegraphics[width=0.31\textwidth]{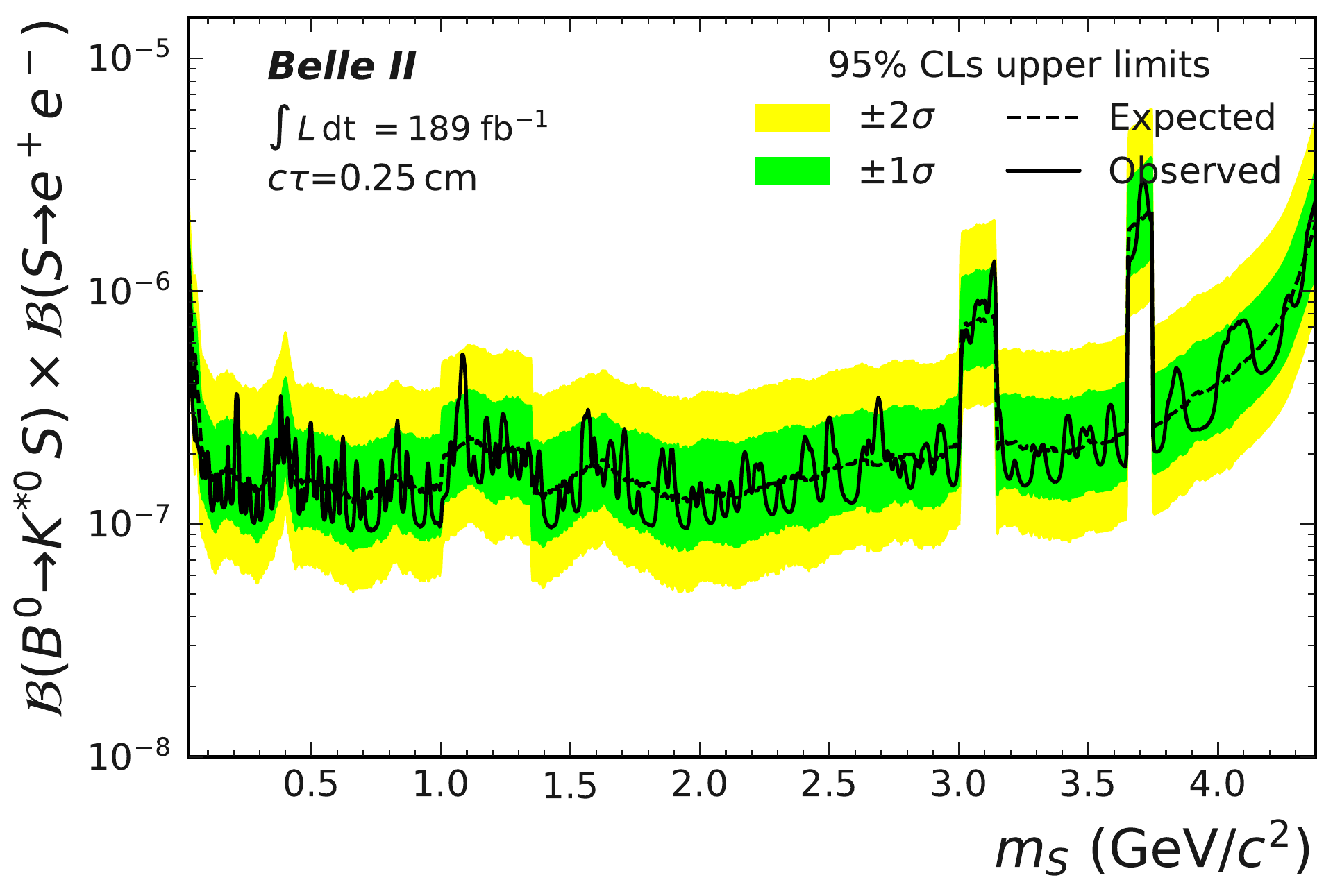}%
}
\subfigure[$\Bz\to \Kstarz(\to K^+\pi^-) S, S\to e^+e^-$, \newline lifetime of $c\tau=0.5\cm$.]{%
  \label{subfit:brazil:Kstar_e_1:J}%
  \includegraphics[width=0.31\textwidth]{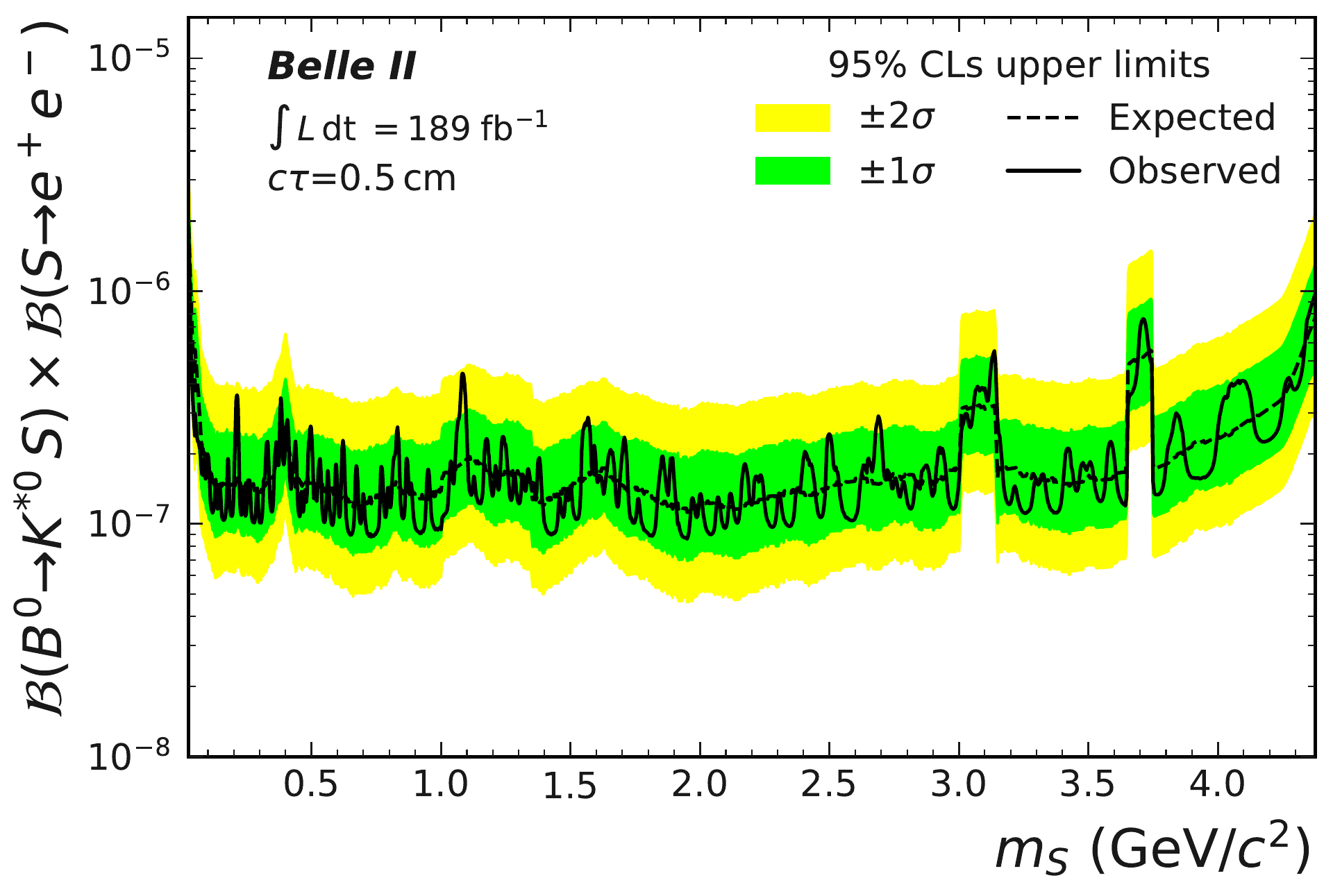}%
}%
\hspace*{\fill}
\subfigure[$\Bz\to \Kstarz(\to K^+\pi^-) S, S\to e^+e^-$, \newline lifetime of $c\tau=1\cm$.]{
  \label{subfit:brazil:Kstar_e_1:K}%
  \includegraphics[width=0.31\textwidth]{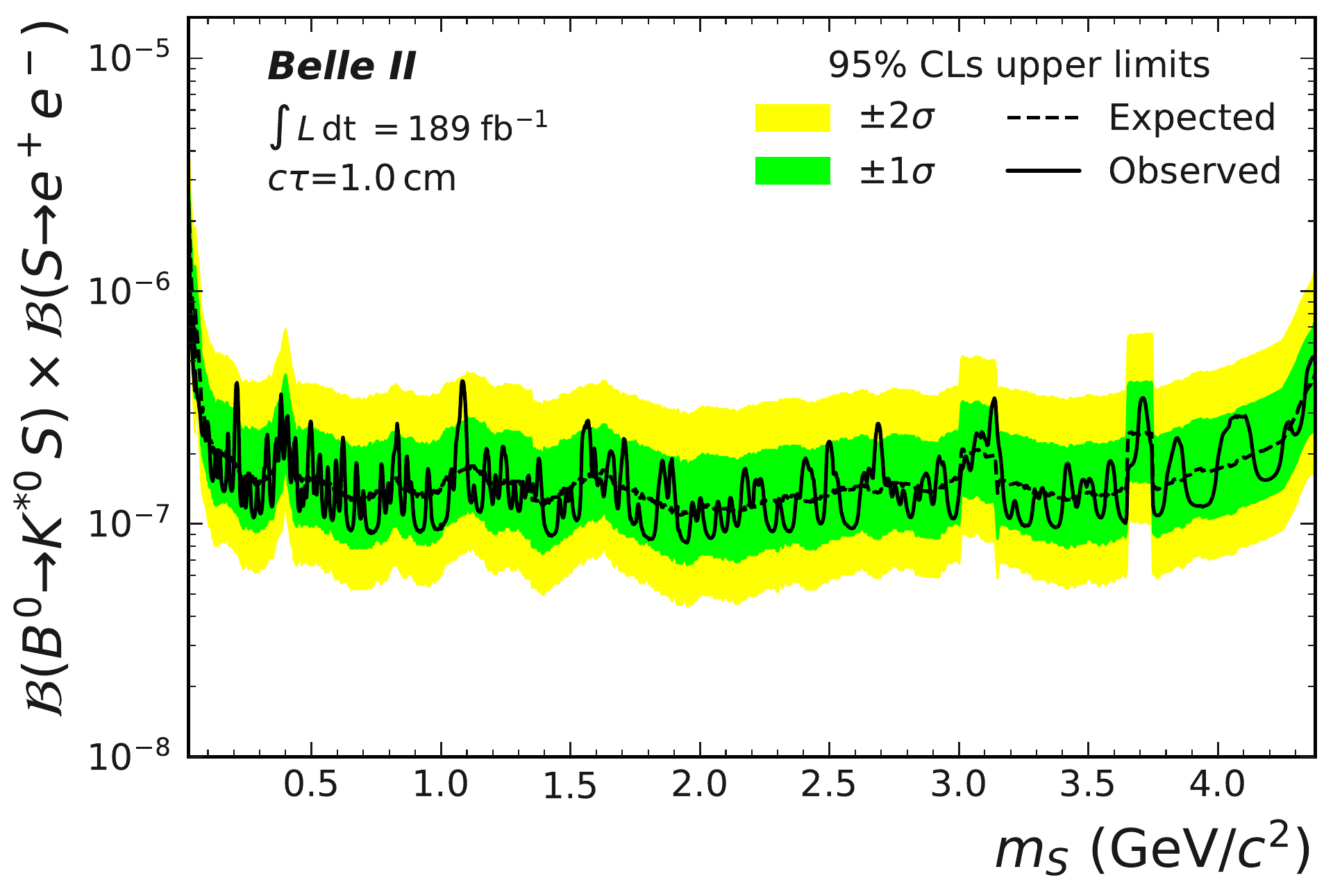}%
}%
\hspace*{\fill}
\subfigure[$\Bz\to \Kstarz(\to K^+\pi^-) S, S\to e^+e^-$, \newline lifetime of $c\tau=2.5\cm$.]{
  \label{subfit:brazil:Kstar_e_1:L}%
  \includegraphics[width=0.31\textwidth]{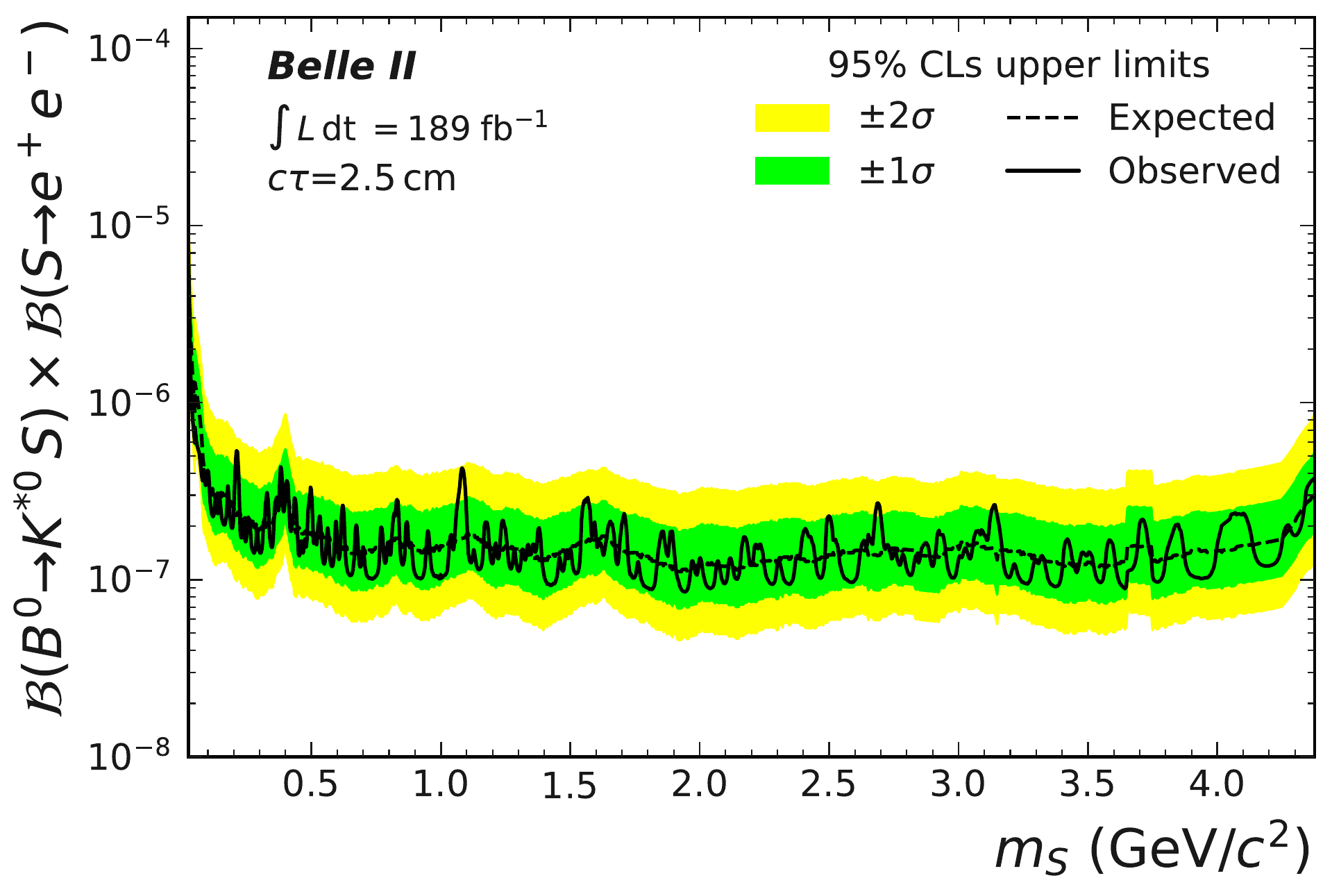}%
}
\caption{Expected and observed limits on the  product of branching fractions $\mathcal{B}(B^0\to \Kstarz(\to K^+\pi^-) S) \times \mathcal{B}(S\to e^+e^-)$ for \\lifetimes \hbox{$0.001 < c\tau < 2.5\,\cm$}.}\label{subfit:brazil:Kstar_e_1}
\end{figure*}

\begin{figure*}[ht]%
\subfigure[$\Bz\to \Kstarz(\to K^+\pi^-) S, S\to e^+e^-$, \newline lifetime of $c\tau=5\cm$.]{%
  \label{subfit:brazil:Kstar_e_2:A}%
  \includegraphics[width=0.31\textwidth]{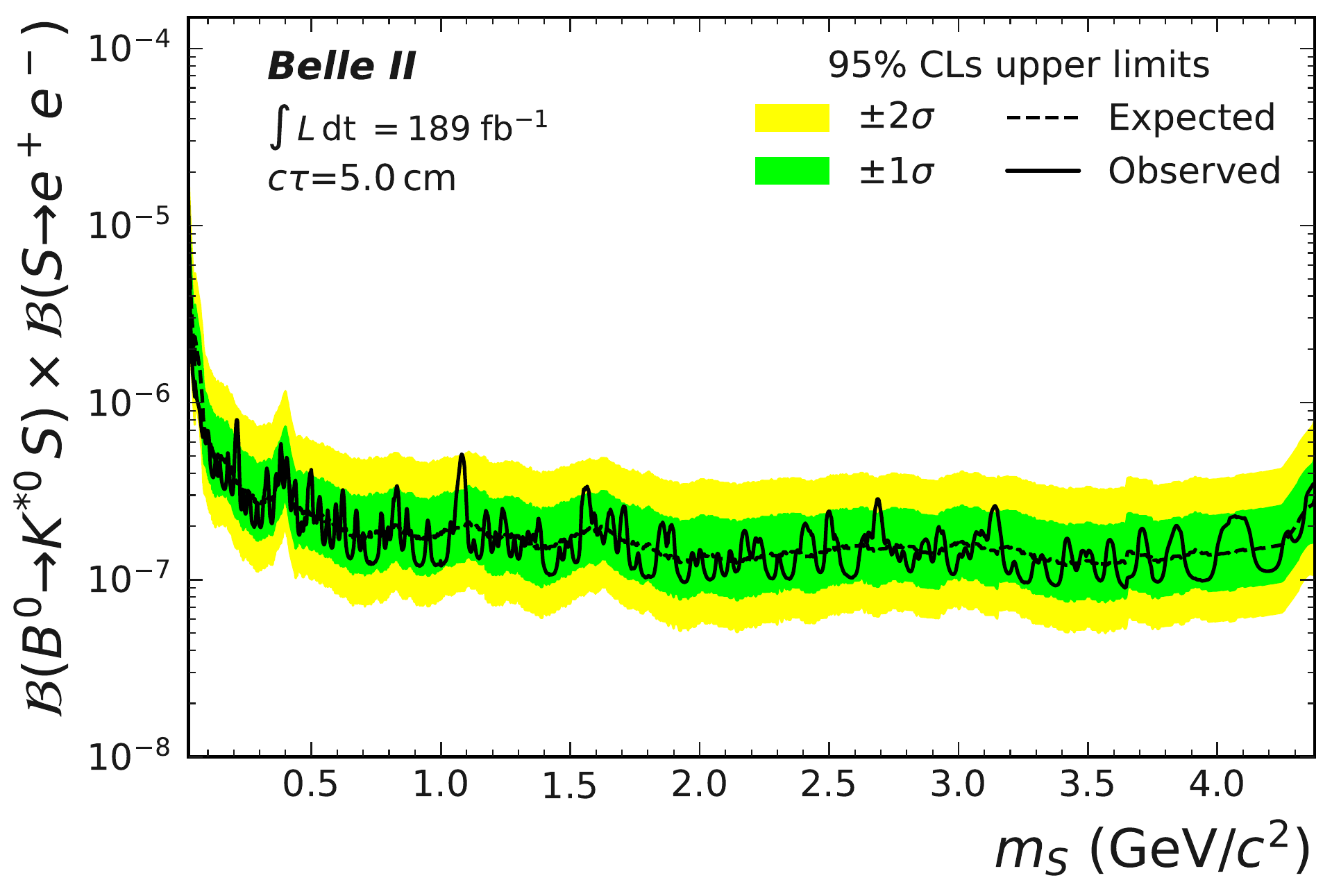}%
}%
\hspace*{\fill}
\subfigure[$\Bz\to \Kstarz(\to K^+\pi^-) S, S\to e^+e^-$, \newline lifetime of $c\tau=10\cm$.]{
  \label{subfit:brazil:Kstar_e_2:B}%
  \includegraphics[width=0.31\textwidth]{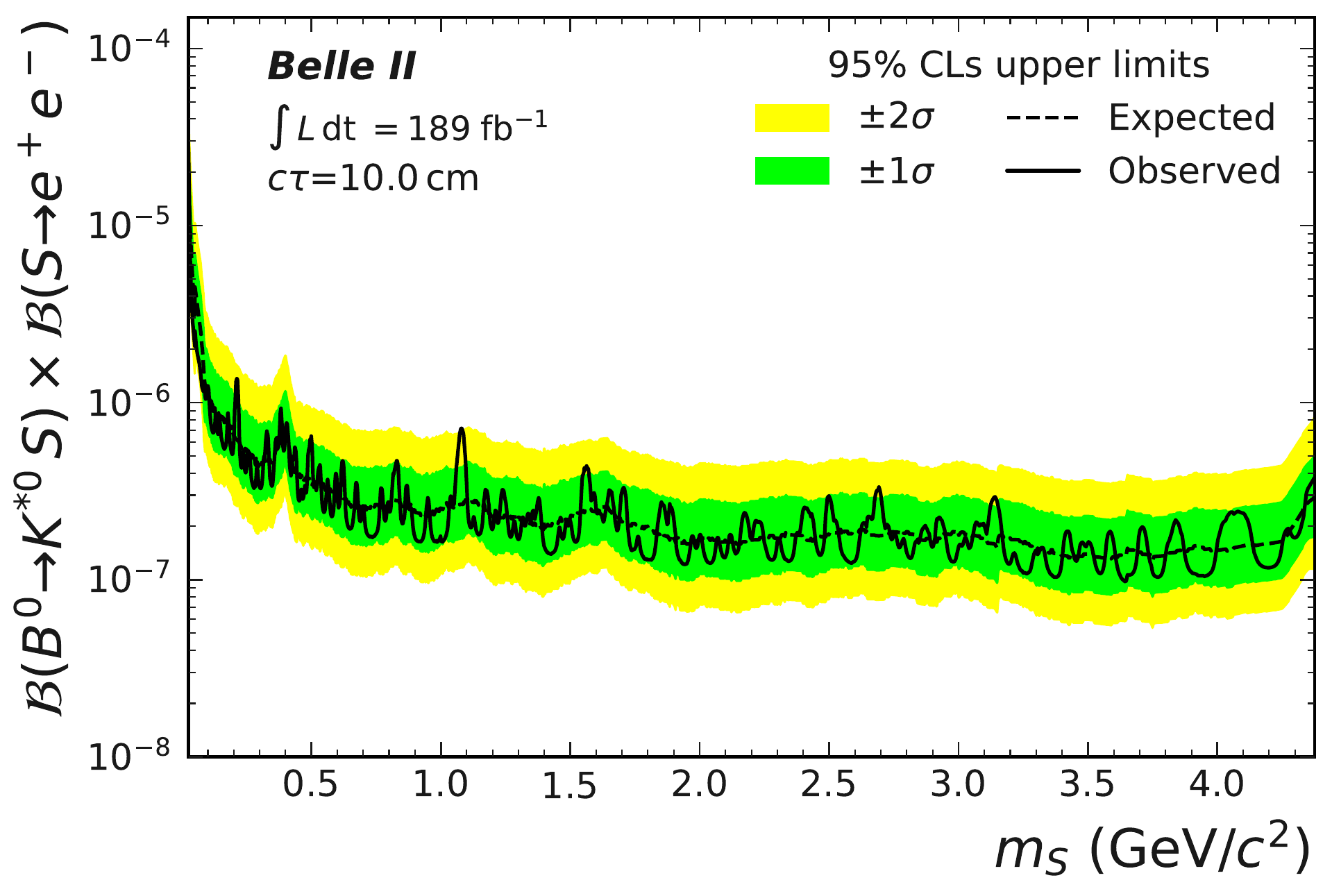}%
}%
\hspace*{\fill}
\subfigure[$\Bz\to \Kstarz(\to K^+\pi^-) S, S\to e^+e^-$, \newline lifetime of $c\tau=25\cm$.]{
  \label{subfit:brazil:Kstar_e_2:C}%
  \includegraphics[width=0.31\textwidth]{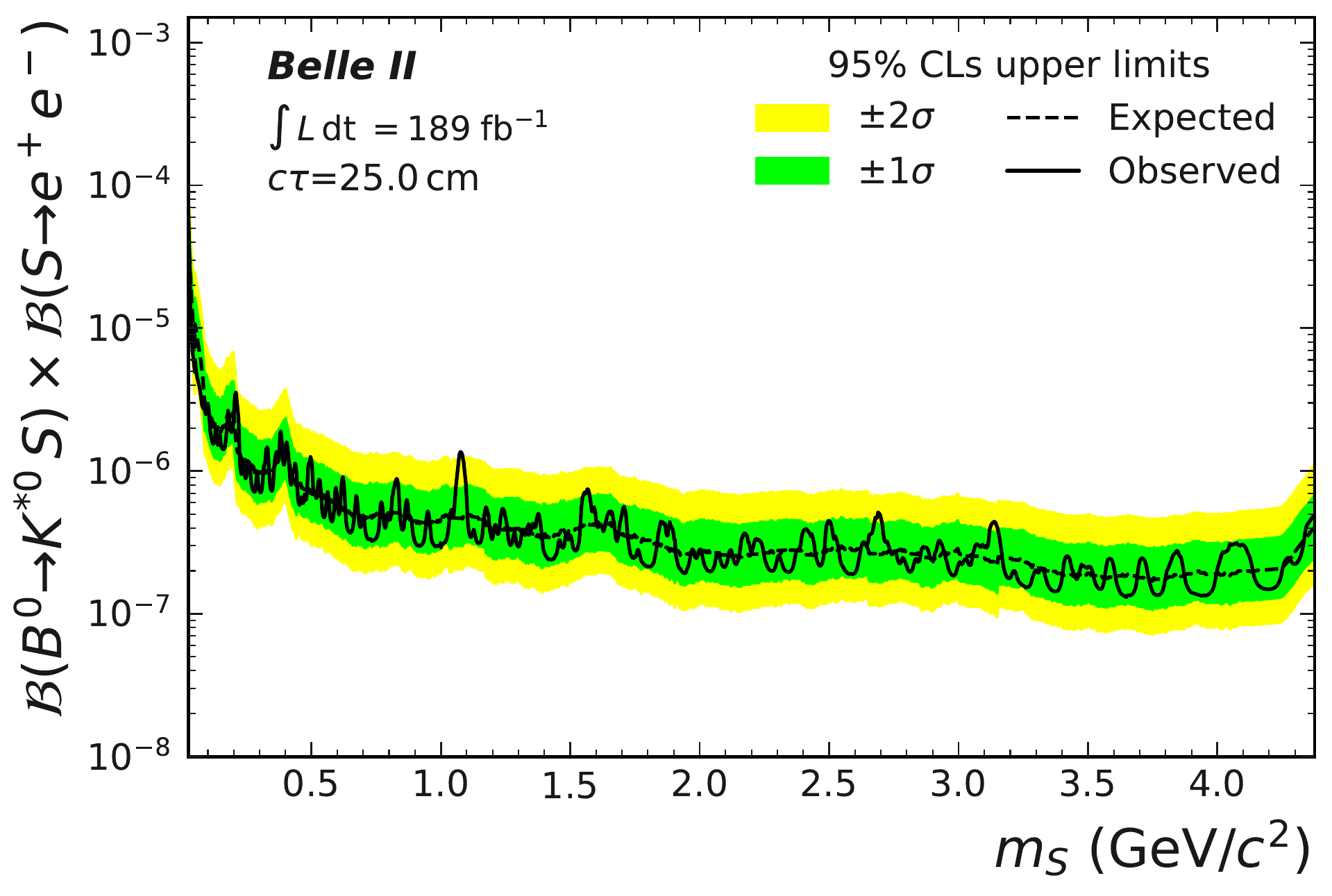}%
}
\subfigure[$\Bz\to \Kstarz(\to K^+\pi^-) S, S\to e^+e^-$, \newline lifetime of $c\tau=50\cm$.]{%
  \label{subfit:brazil:Kstar_e_2:D}%
  \includegraphics[width=0.31\textwidth]{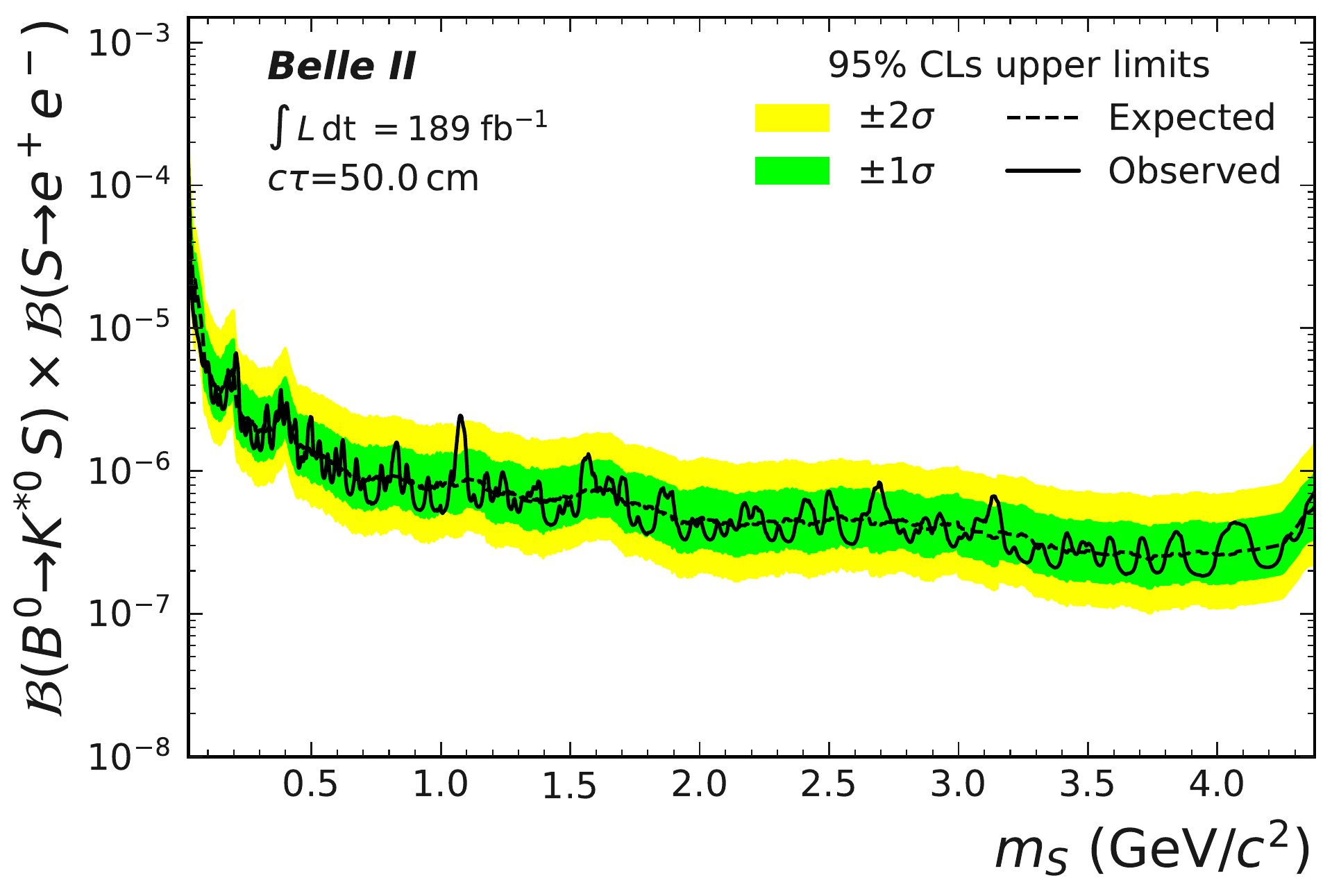}%
}%
\hspace*{\fill}
\subfigure[$\Bz\to \Kstarz(\to K^+\pi^-) S, S\to e^+e^-$, \newline lifetime of $c\tau=100\cm$.]{
  \label{subfit:brazil:Kstar_e_2:E}%
  \includegraphics[width=0.31\textwidth]{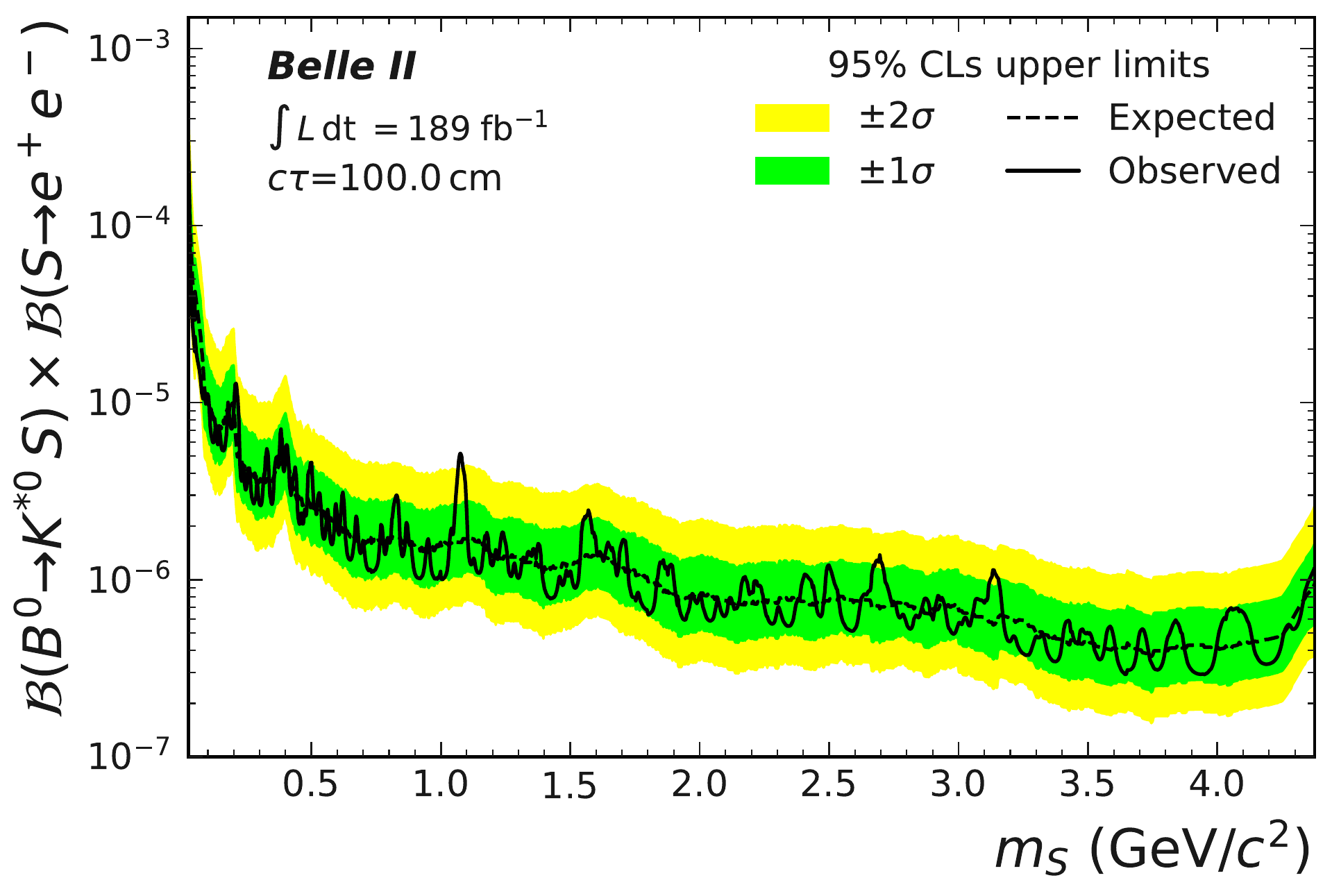}%
}%
\hspace*{\fill}
\subfigure[$\Bz\to \Kstarz(\to K^+\pi^-) S, S\to e^+e^-$, \newline lifetime of $c\tau=200\cm$.]{
  \label{subfit:brazil:Kstar_e_2:F}%
  \includegraphics[width=0.31\textwidth]{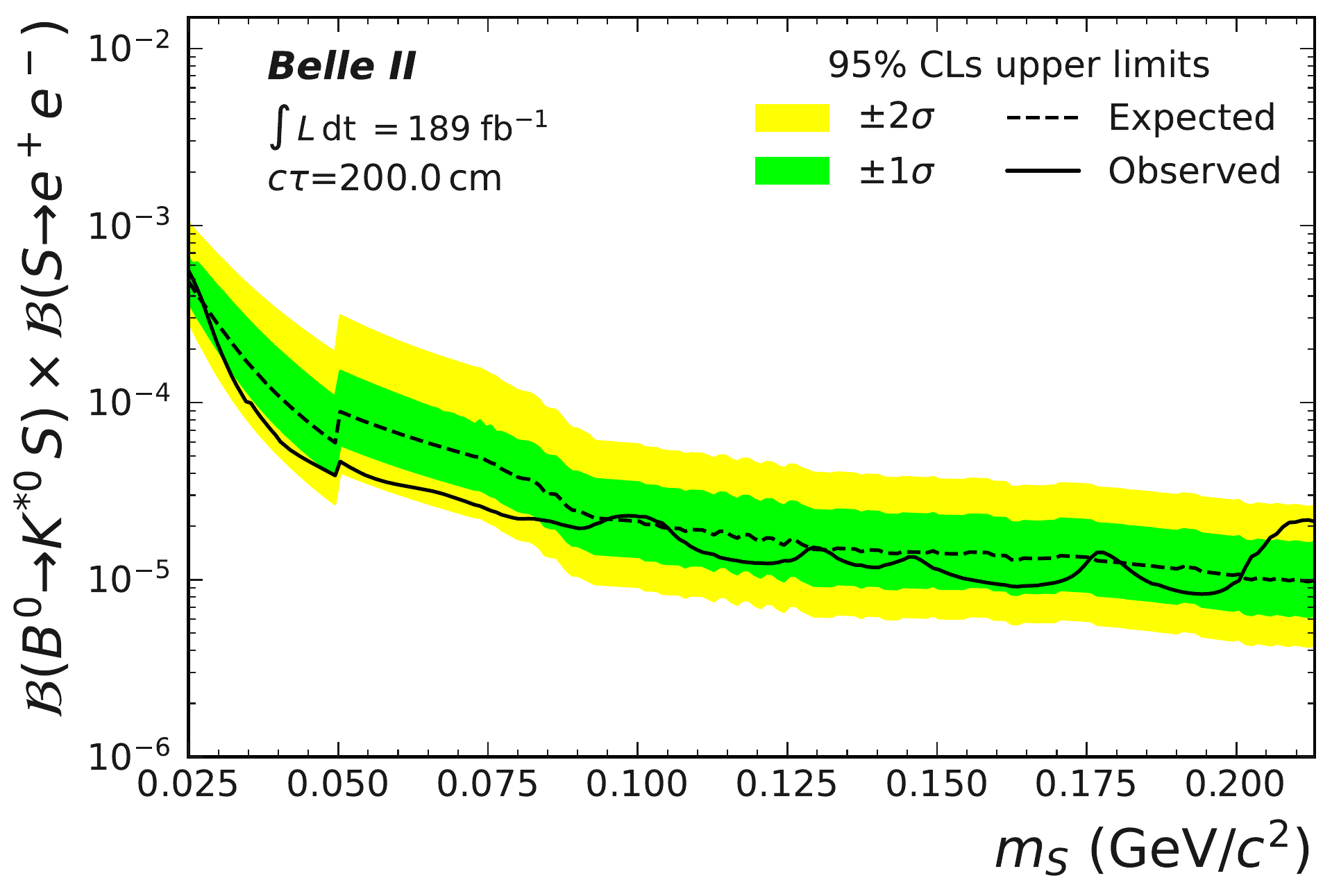}%
}

\subfigure[$\Bz\to \Kstarz(\to K^+\pi^-) S, S\to e^+e^-$, \newline lifetime of $c\tau=400\cm$.]{
  \label{subfit:brazil:Kstar_e_2:G}%
  \includegraphics[width=0.31\textwidth]{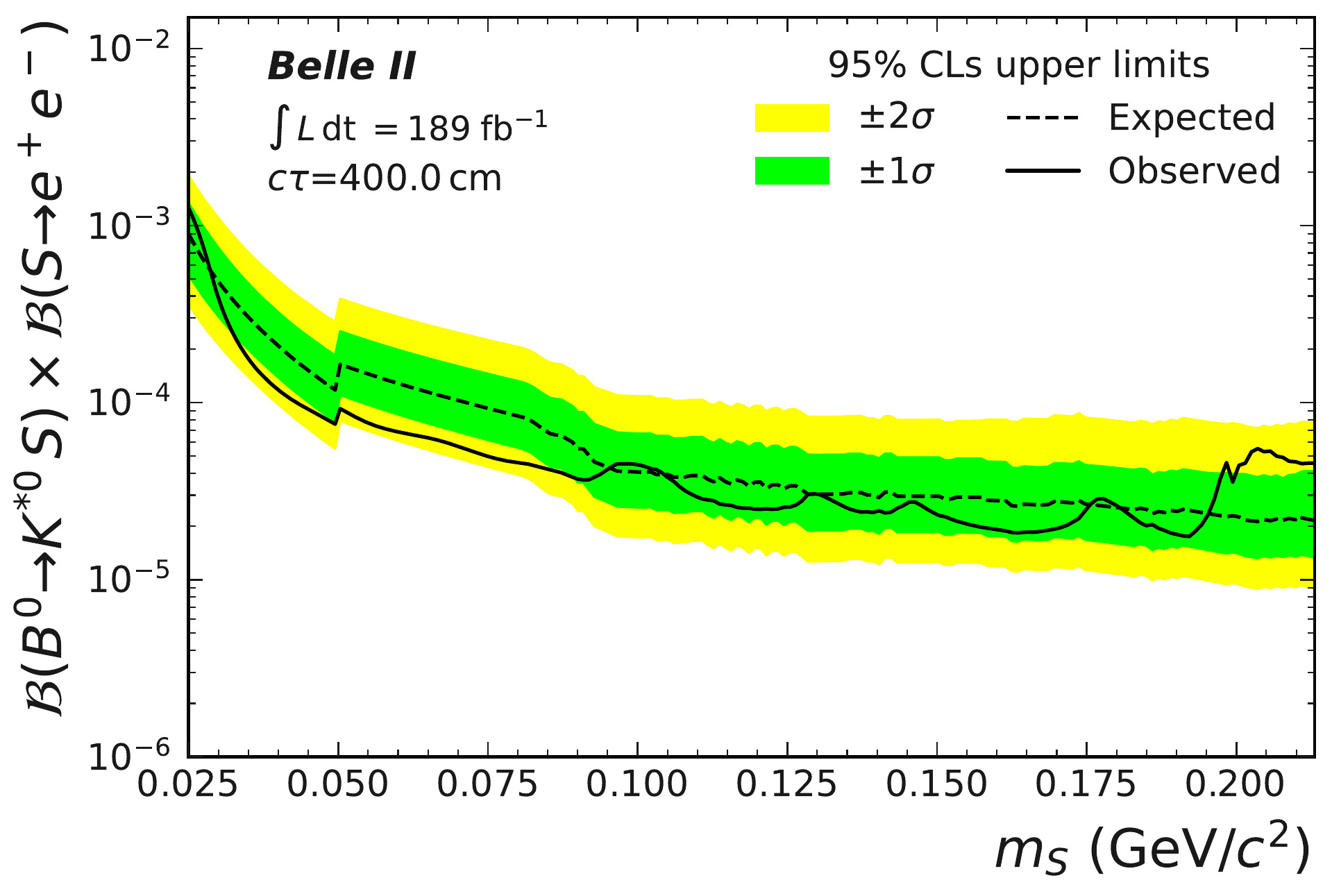}%
}
\hspace*{\fill}
\subfigure[$\Bz\to \Kstarz(\to K^+\pi^-) S, S\to e^+e^-$, \newline lifetime of $c\tau=700\cm$.]{%
  \label{subfit:brazil:Kstar_e_2:H}%
  \includegraphics[width=0.31\textwidth]{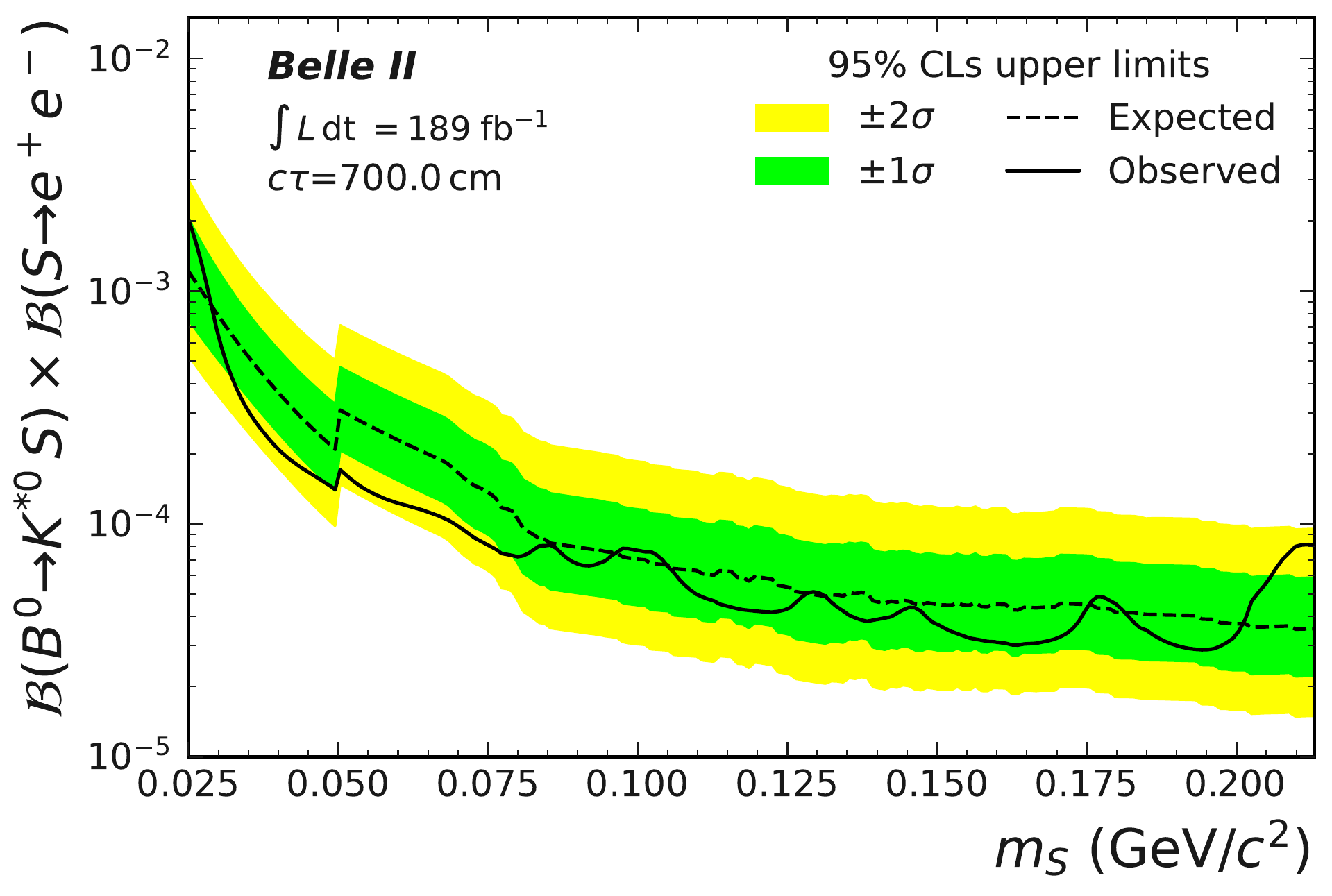}%
}%
\hspace*{\fill}
\subfigure[$\Bz\to \Kstarz(\to K^+\pi^-) S, S\to e^+e^-$, \newline lifetime of $c\tau=1000\cm$.]{
  \label{subfit:brazil:Kstar_e_2:I}%
  \includegraphics[width=0.31\textwidth]{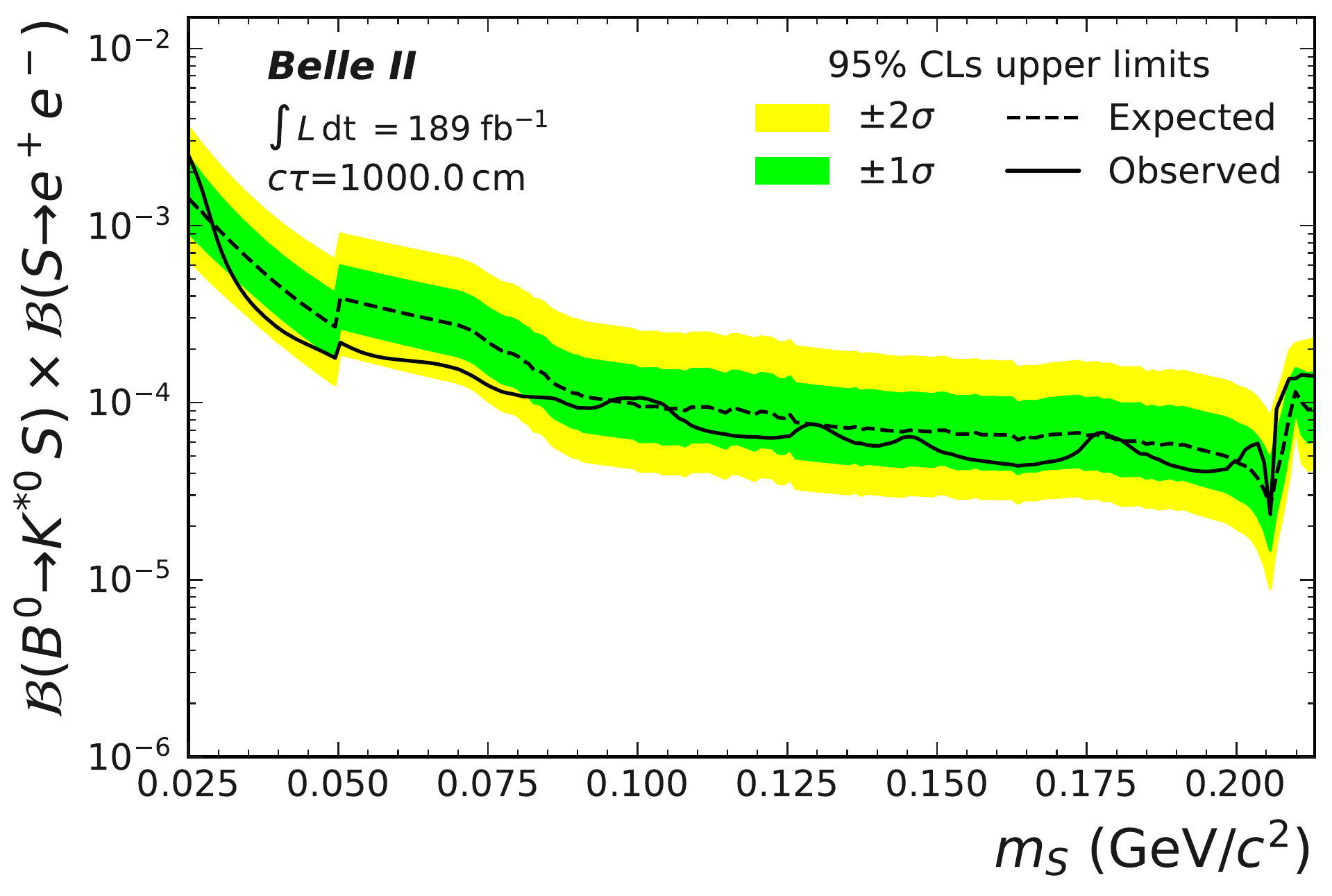}%
}%

\subfigure[$\Bz\to \Kstarz(\to K^+\pi^-) S, S\to e^+e^-$, \newline lifetime of $c\tau=2500\cm$.]{
  \label{subfit:brazil:Kstar_e_2:J}%
  \includegraphics[width=0.31\textwidth]{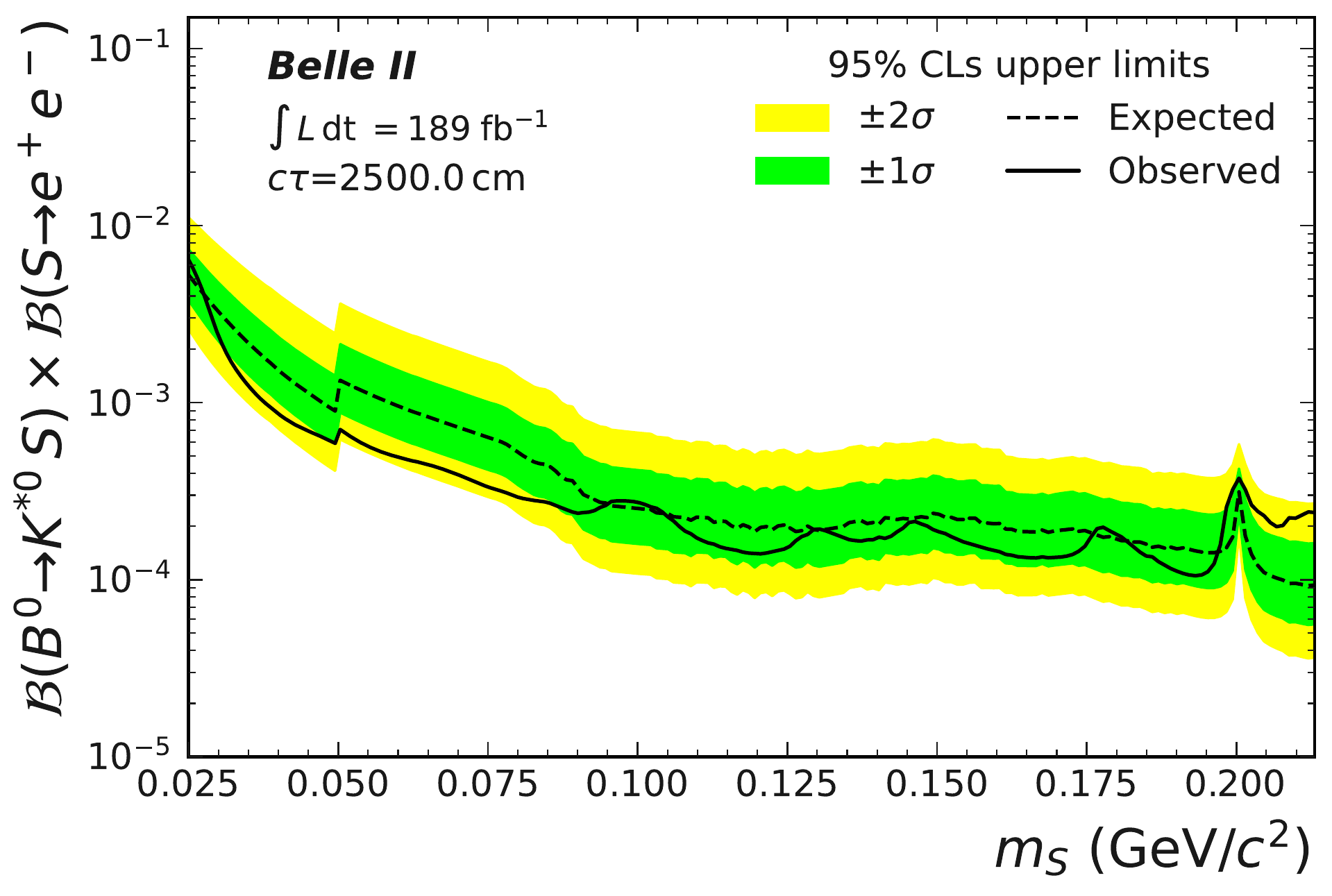}%
}
\hspace*{\fill}
\subfigure[$\Bz\to \Kstarz(\to K^+\pi^-) S, S\to e^+e^-$, \newline lifetime of $c\tau=5000\cm$.]{%
  \label{subfit:brazil:Kstar_e_2:K}%
  \includegraphics[width=0.31\textwidth]{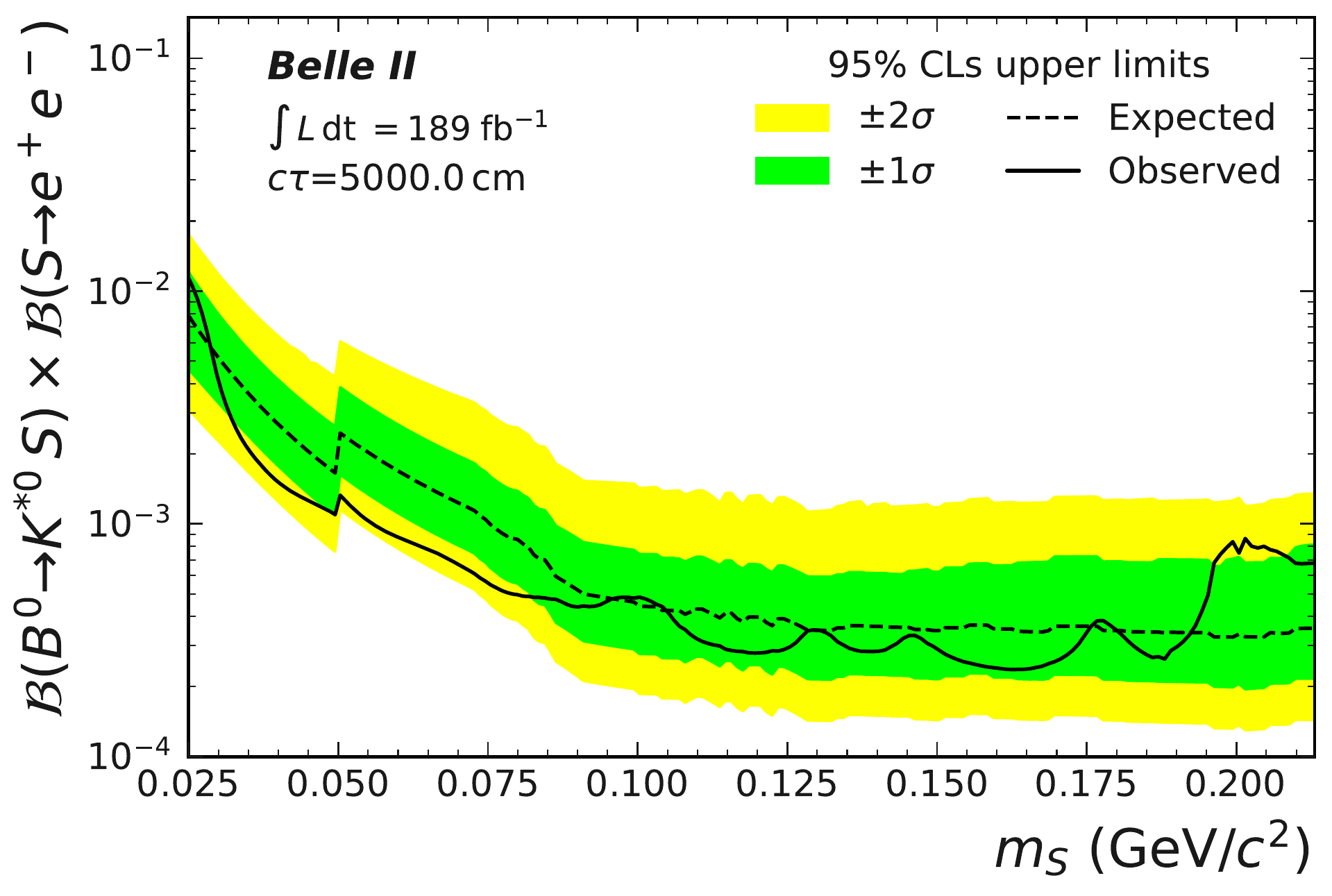}%
}%
\hspace*{\fill}
\subfigure[$\Bz\to \Kstarz(\to K^+\pi^-) S, S\to e^+e^-$, \newline lifetime of $c\tau=10000\cm$.]{
  \label{subfit:brazil:Kstar_e_2:L}%
  \includegraphics[width=0.31\textwidth]{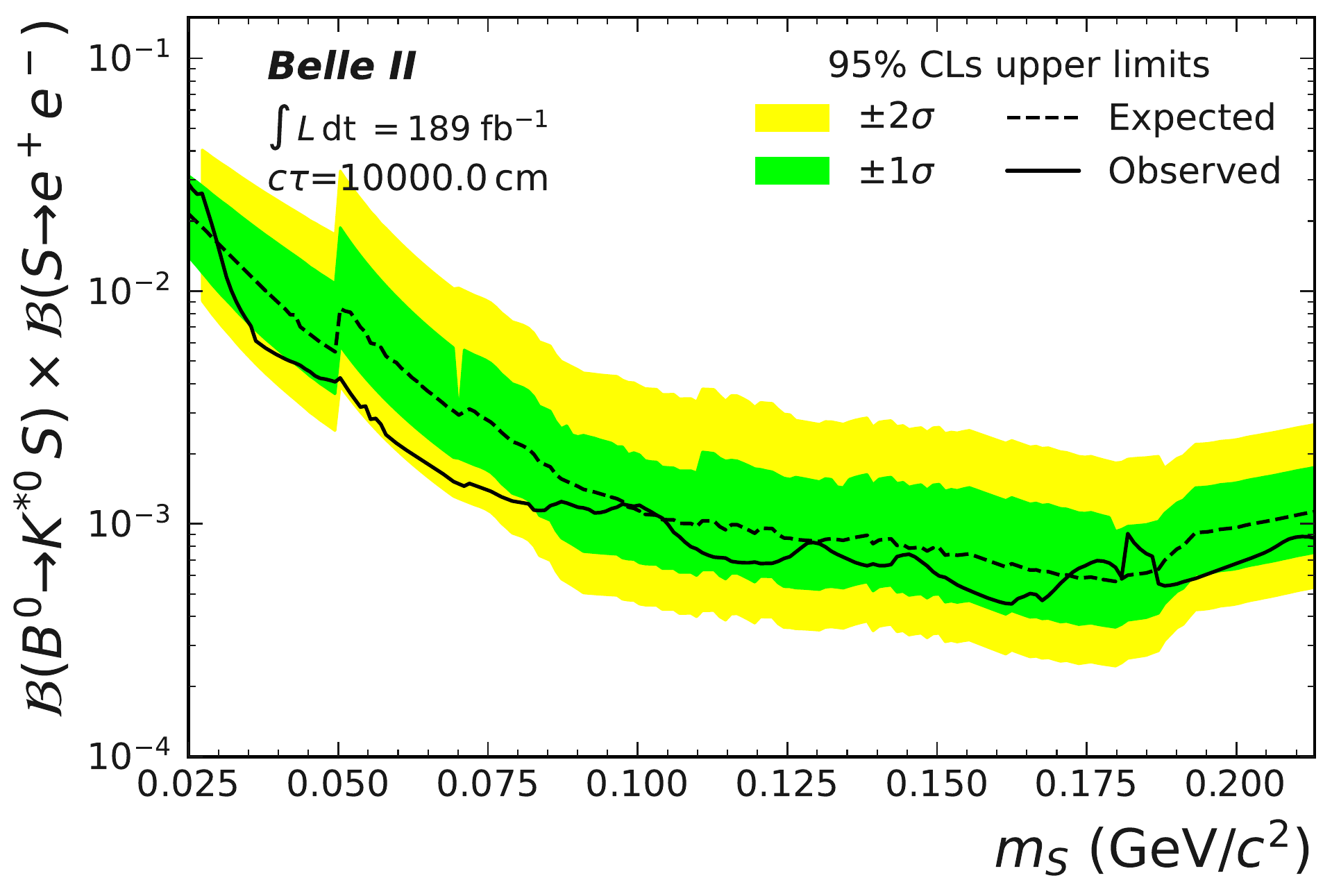}%
}%
\caption{Expected and observed limits on the product of branching fractions $\mathcal{B}(B^0\to \Kstarz(\to K^+\pi^-) S) \times \mathcal{B}(S\to e^+e^-)$ for lifetimes \hbox{$5 < c\tau < 10000\,\cm$}.}\label{subfit:brazil:Kstar_e_2}
\end{figure*}


\begin{figure*}[ht]%
\subfigure[$B^+\to K^+S, S\to \mu^+\mu^-$, \newline lifetime of $c\tau=0.001\cm$.]{%
  \label{subfit:brazil:Kp_mu_1:A}%
  \includegraphics[width=0.31\textwidth]{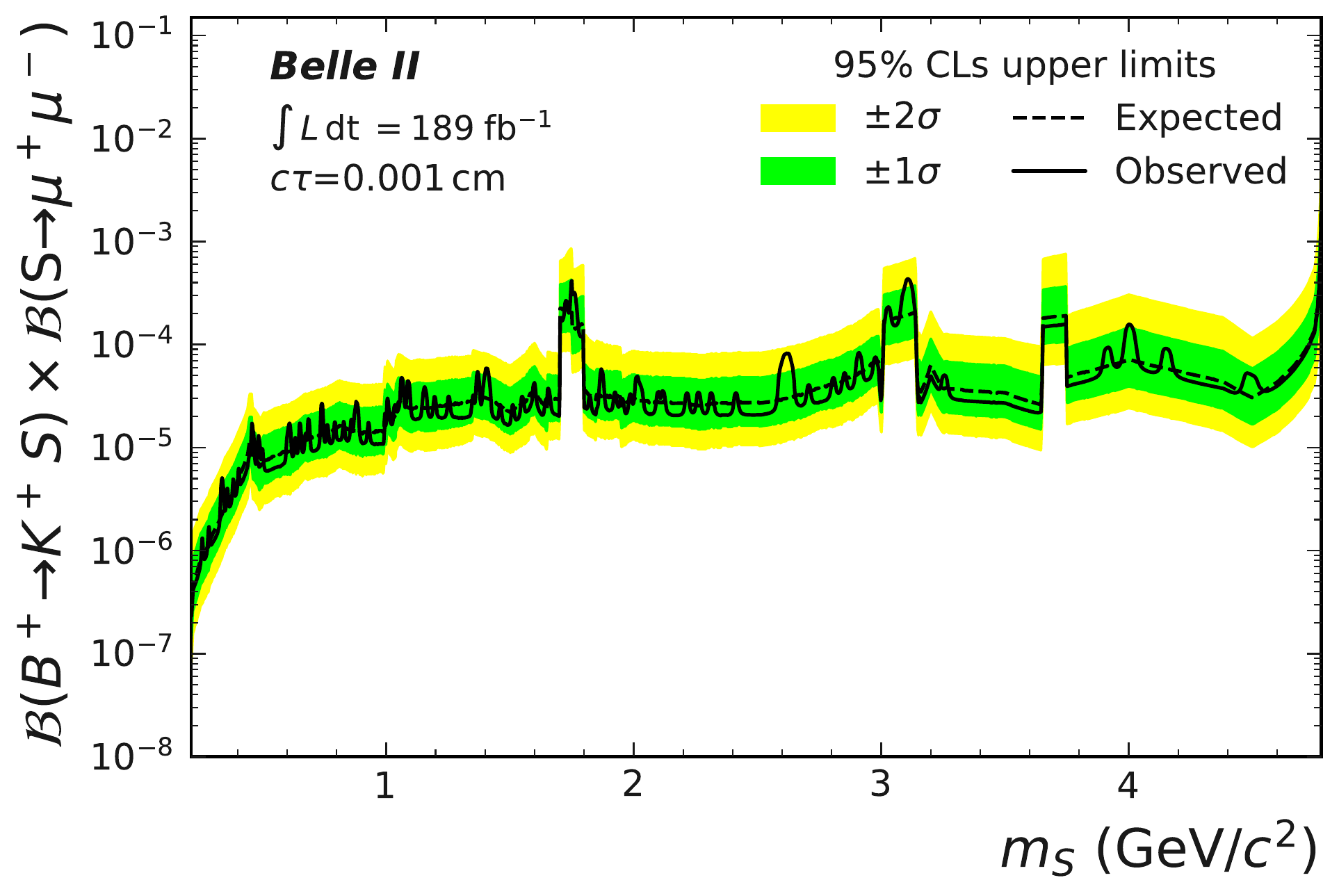}%
}%
\hspace*{\fill}
\subfigure[$B^+\to K^+S, S\to \mu^+\mu^-$, \newline lifetime of $c\tau=0.003\cm$.]{
  \label{subfit:brazil:Kp_mu_1:B}%
  \includegraphics[width=0.31\textwidth]{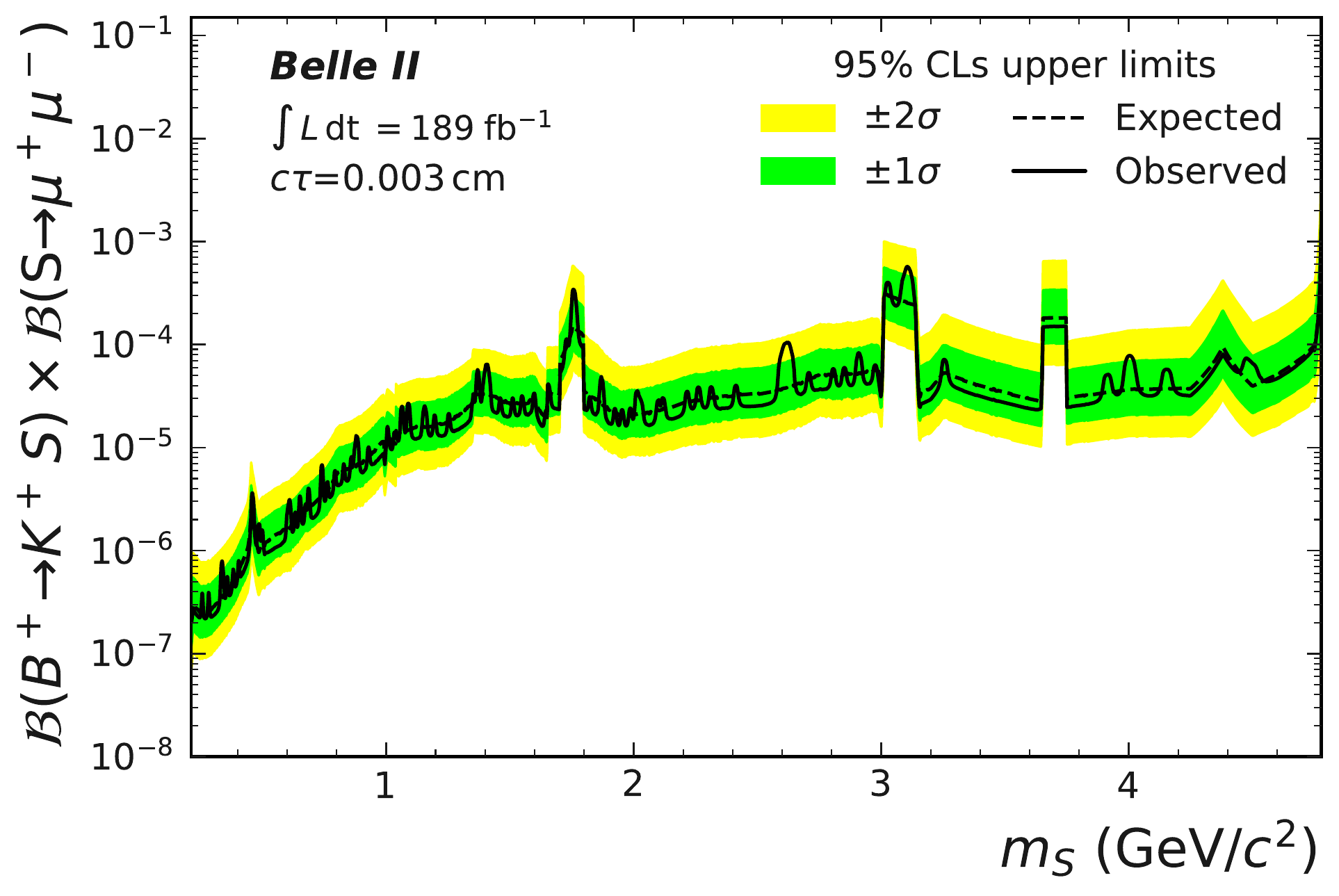}%
}%
\hspace*{\fill}
\subfigure[$B^+\to K^+S, S\to \mu^+\mu^-$, \newline lifetime of $c\tau=0.005\cm$.]{
  \label{subfit:brazil:Kp_mu_1:C}%
  \includegraphics[width=0.31\textwidth]{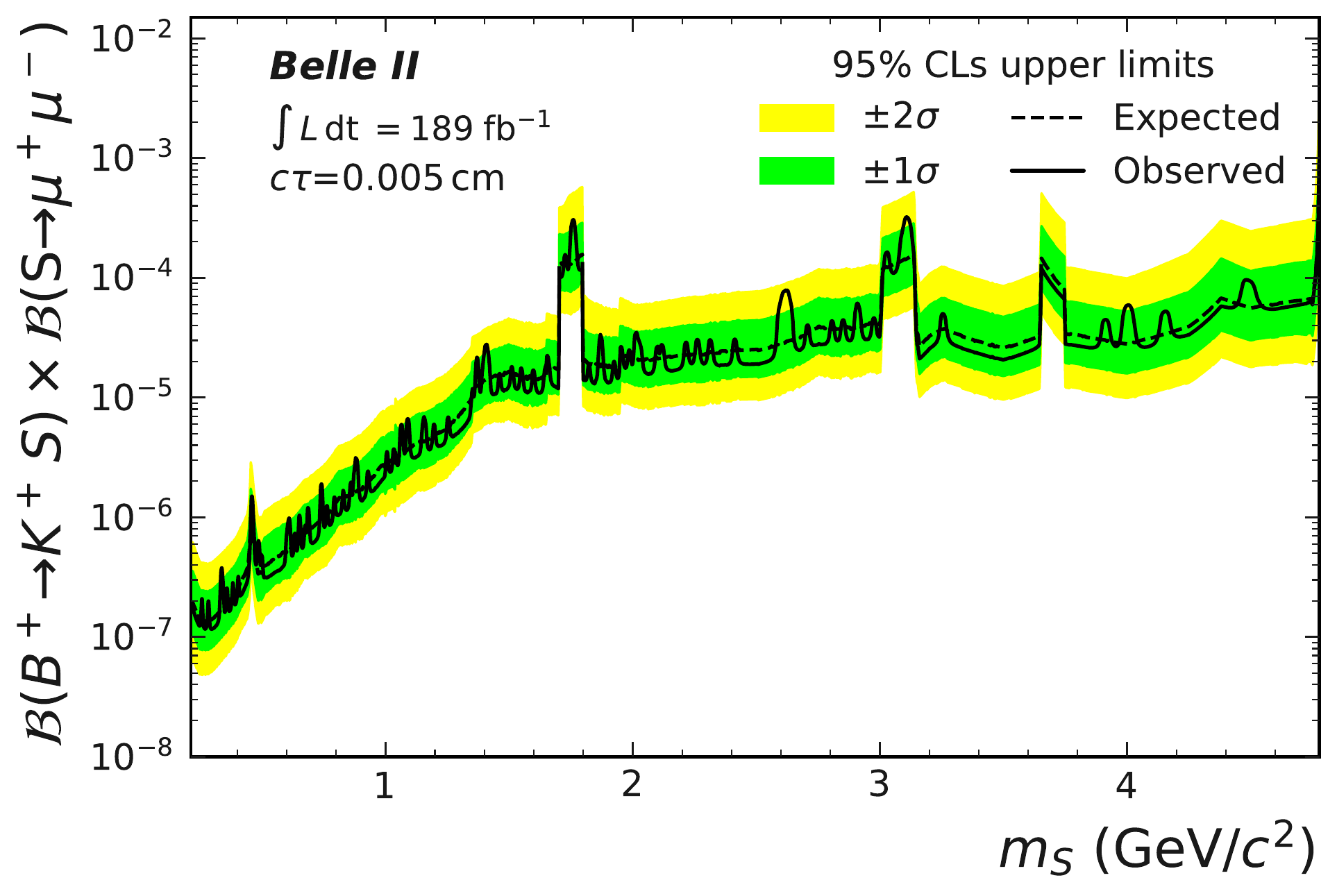}%
}
\subfigure[$B^+\to K^+S, S\to \mu^+\mu^-$, \newline lifetime of $c\tau=0.007\cm$.]{%
  \label{subfit:brazil:Kp_mu_1:D}%
  \includegraphics[width=0.31\textwidth]{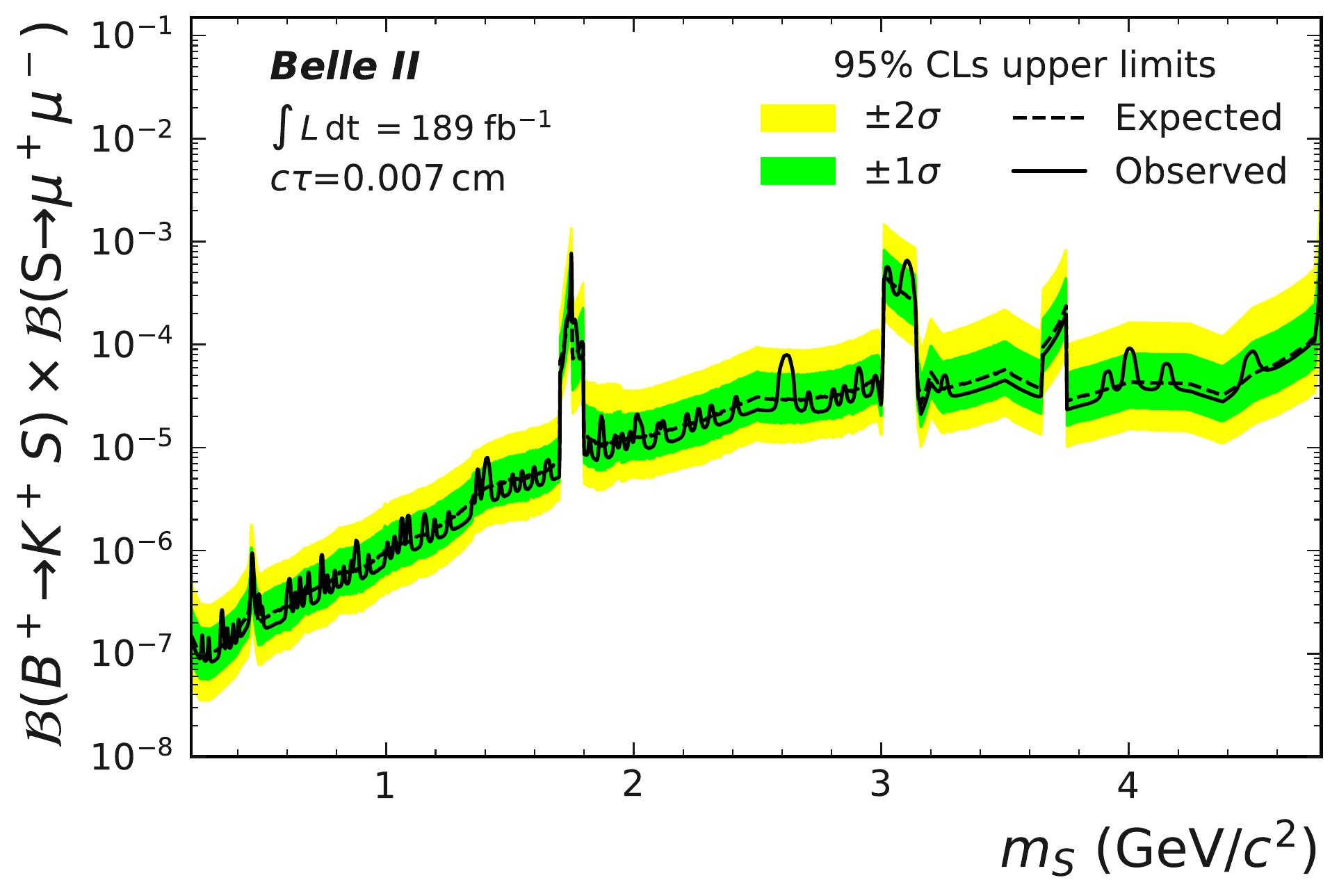}%
}%
\hspace*{\fill}
\subfigure[$B^+\to K^+S, S\to \mu^+\mu^-$, \newline lifetime of $c\tau=0.01\cm$.]{
  \label{subfit:brazil:Kp_mu_1:E}%
  \includegraphics[width=0.31\textwidth]{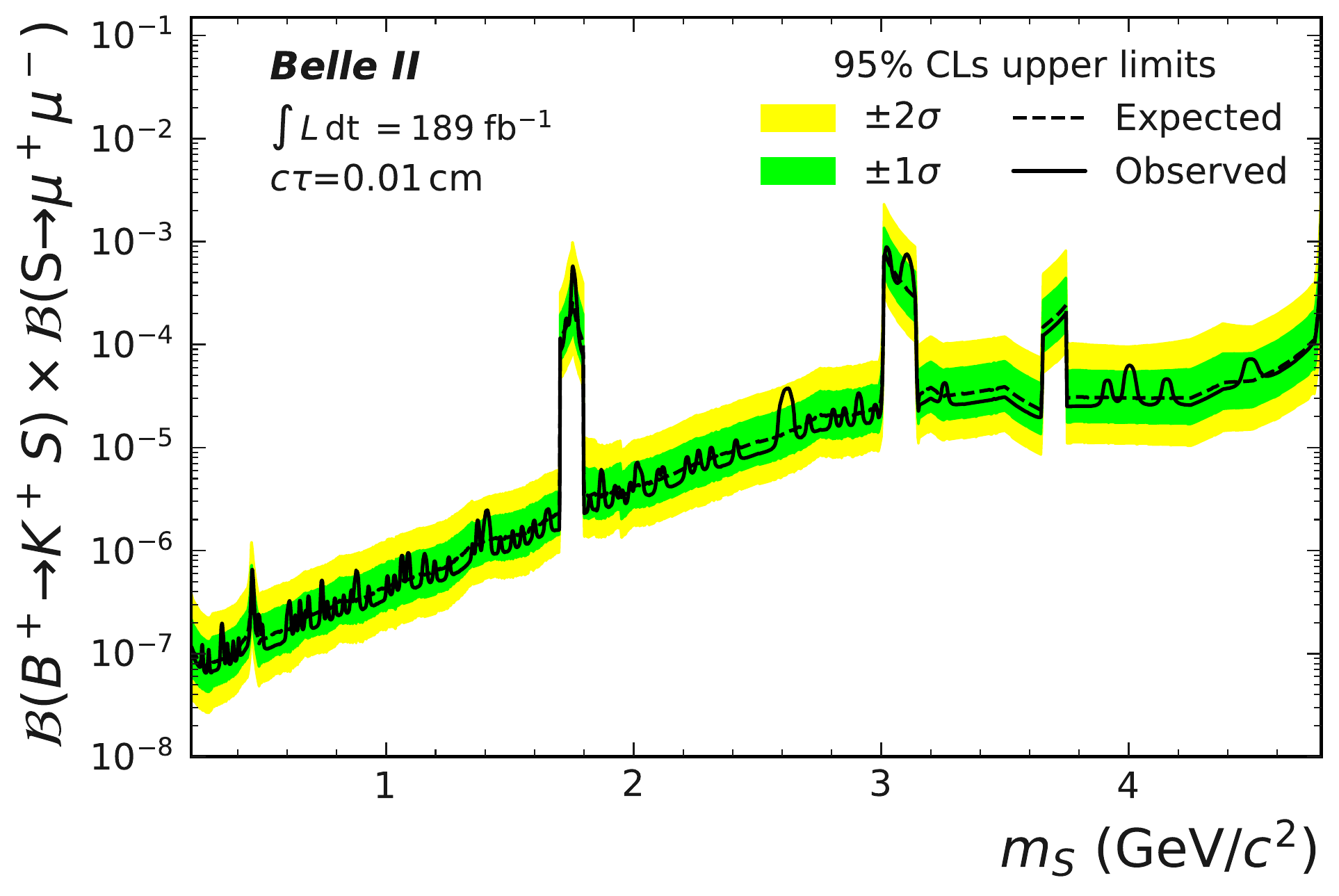}%
}%
\hspace*{\fill}
\subfigure[$B^+\to K^+S, S\to \mu^+\mu^-$, \newline lifetime of $c\tau=0.025\cm$.]{
  \label{subfit:brazil:Kp_mu_1:F}%
  \includegraphics[width=0.31\textwidth]{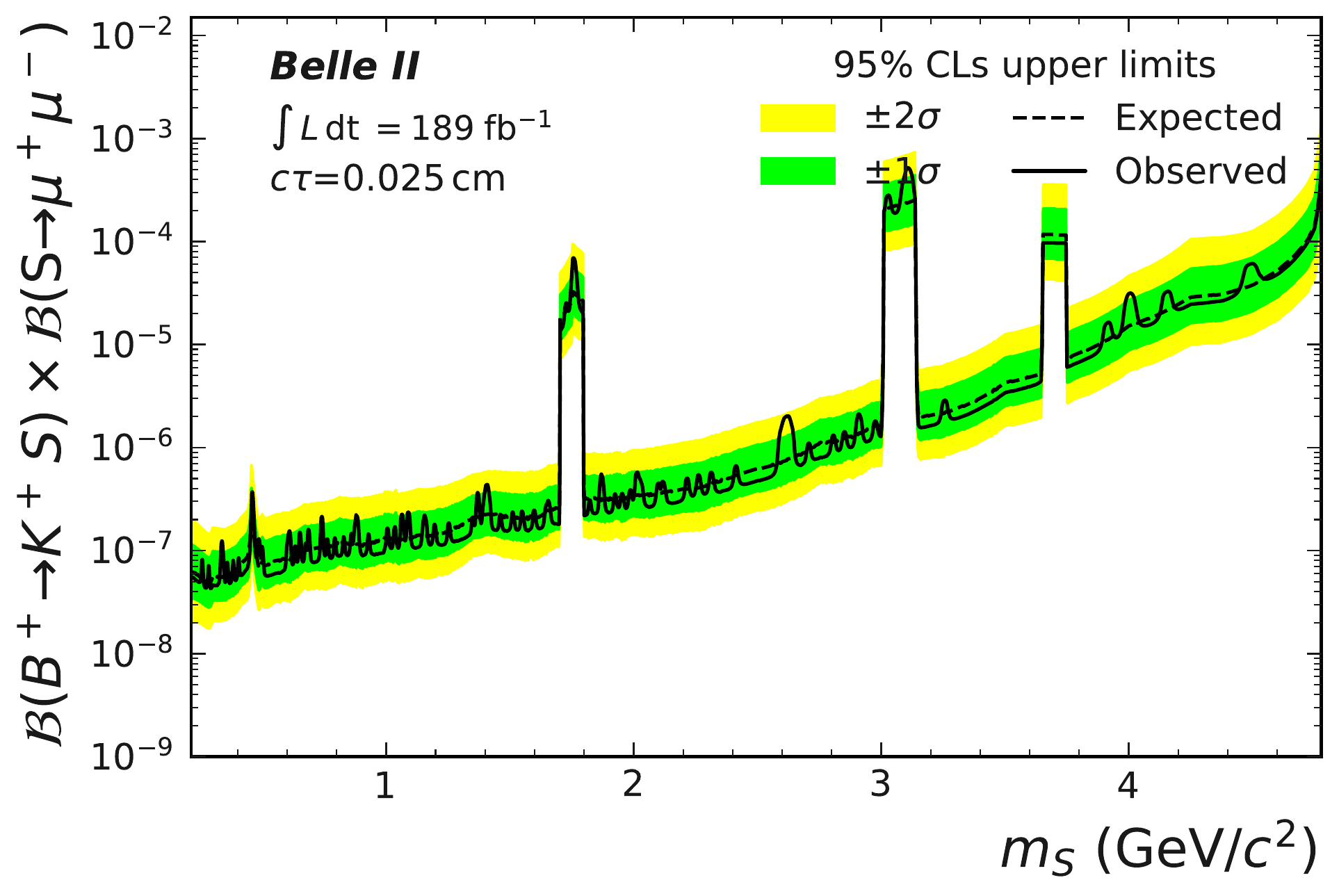}%
}
\subfigure[$B^+\to K^+S, S\to \mu^+\mu^-$, \newline lifetime of $c\tau=0.05\cm$.]{%
  \label{subfit:brazil:Kp_mu_1:G}%
  \includegraphics[width=0.31\textwidth]{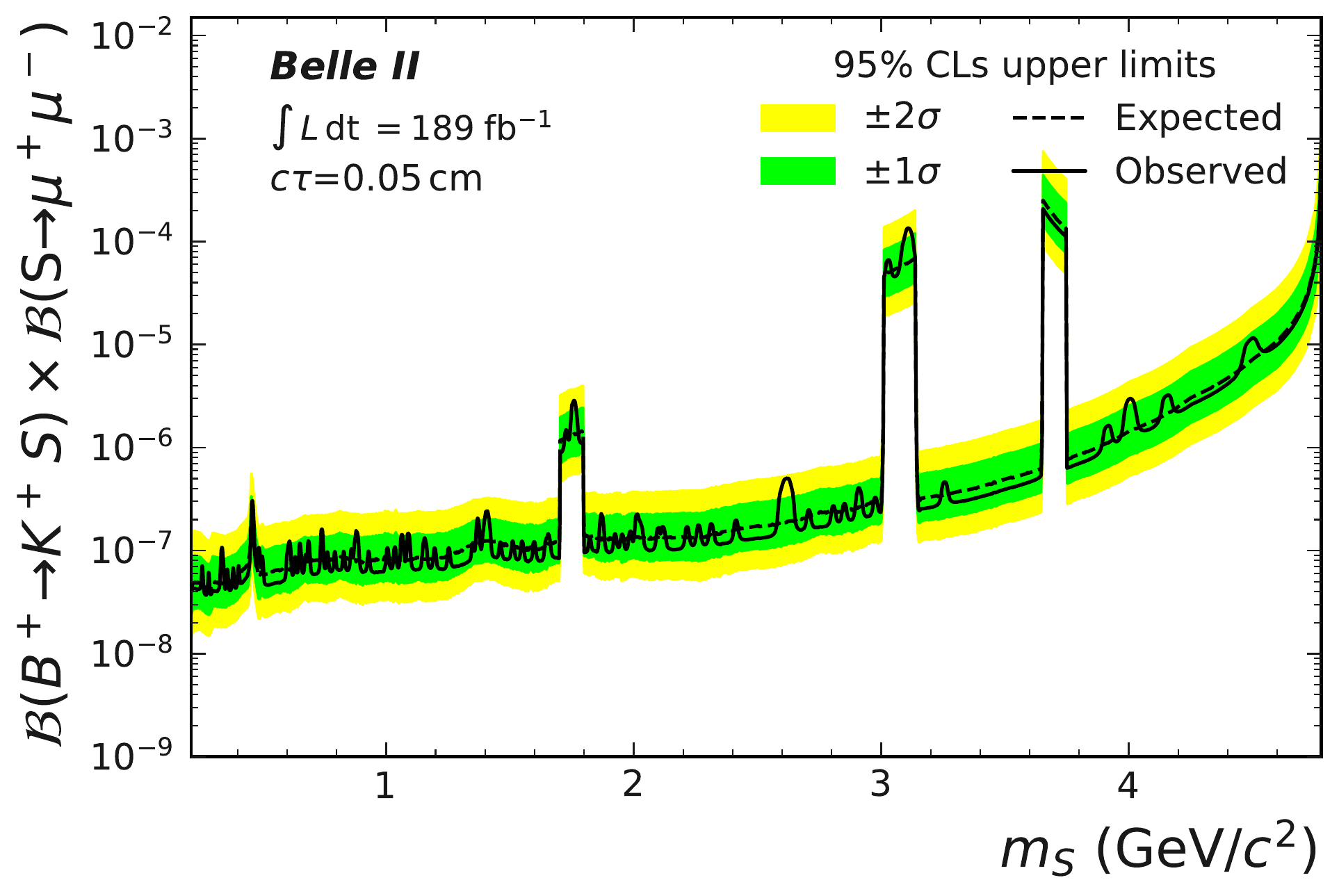}%
}%
\hspace*{\fill}
\subfigure[$B^+\to K^+S, S\to \mu^+\mu^-$, \newline lifetime of $c\tau=0.100\cm$.]{
  \label{subfit:brazil:Kp_mu_1:H}%
  \includegraphics[width=0.31\textwidth]{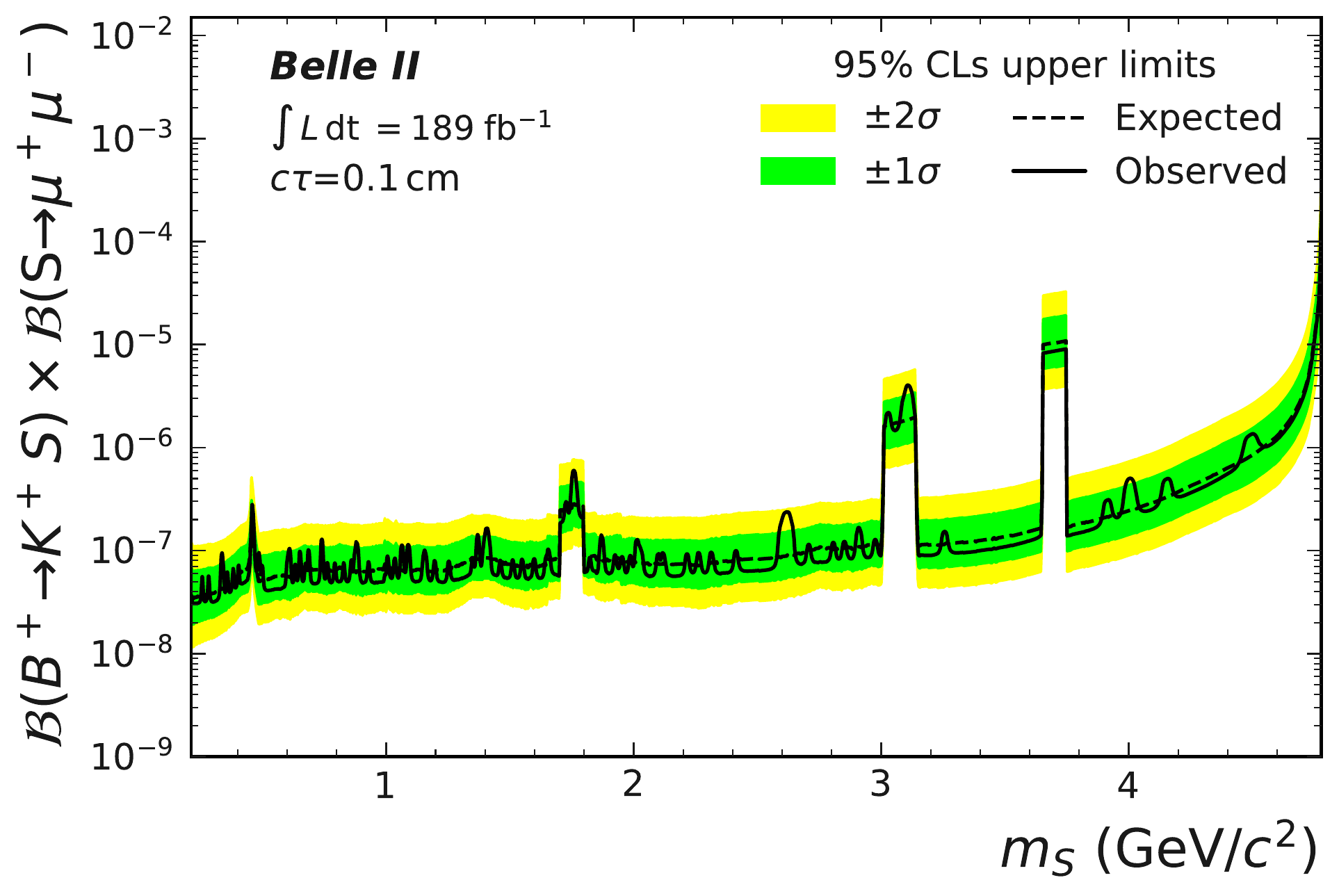}%
}%
\hspace*{\fill}
\subfigure[$B^+\to K^+S, S\to \mu^+\mu^-$, \newline lifetime of $c\tau=0.25\cm$.]{
  \label{subfit:brazil:Kp_mu_1:I}%
  \includegraphics[width=0.31\textwidth]{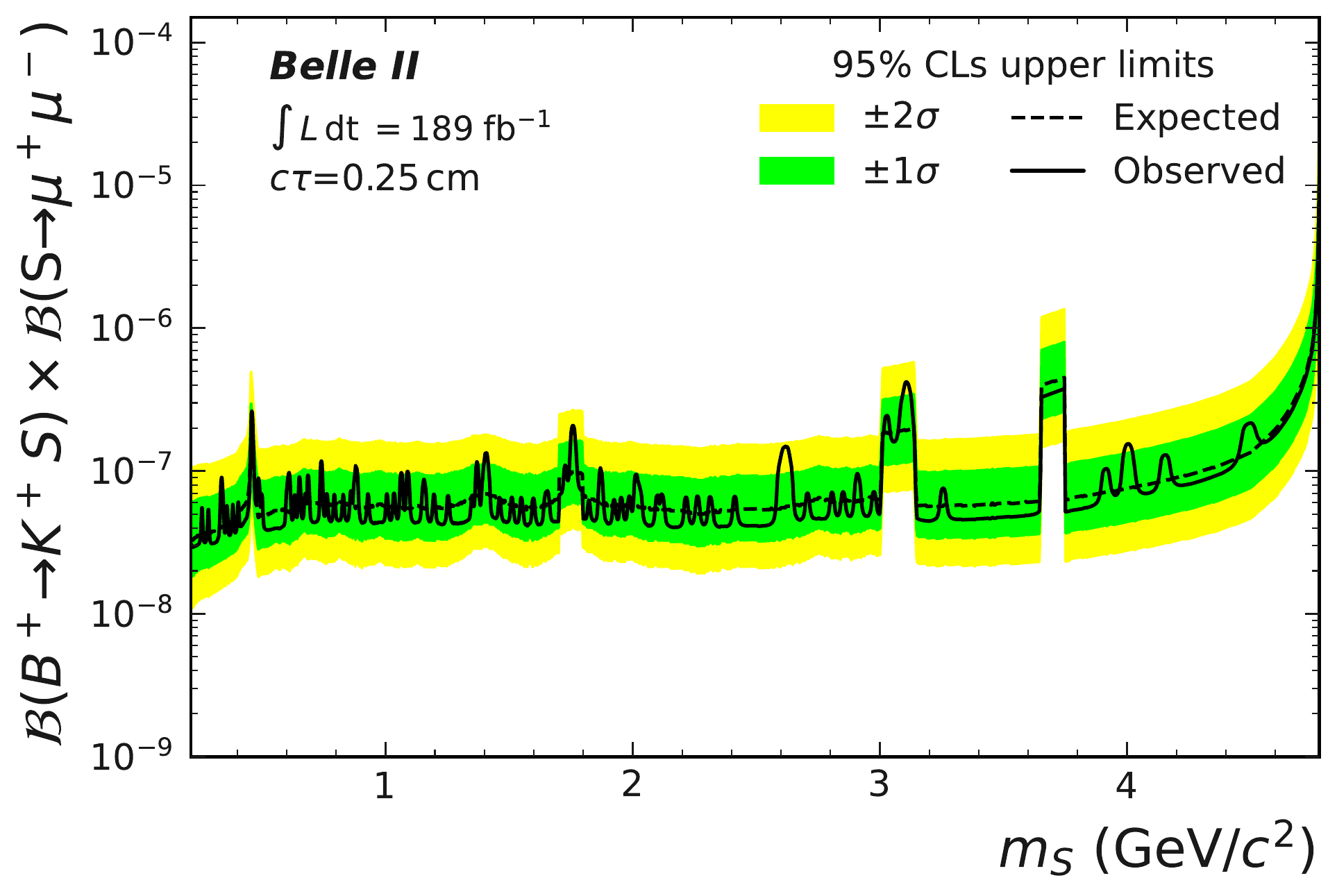}%
}
\subfigure[$B^+\to K^+S, S\to \mu^+\mu^-$, \newline lifetime of $c\tau=0.5\cm$.]{%
  \label{subfit:brazil:Kp_mu_1:J}%
  \includegraphics[width=0.31\textwidth]{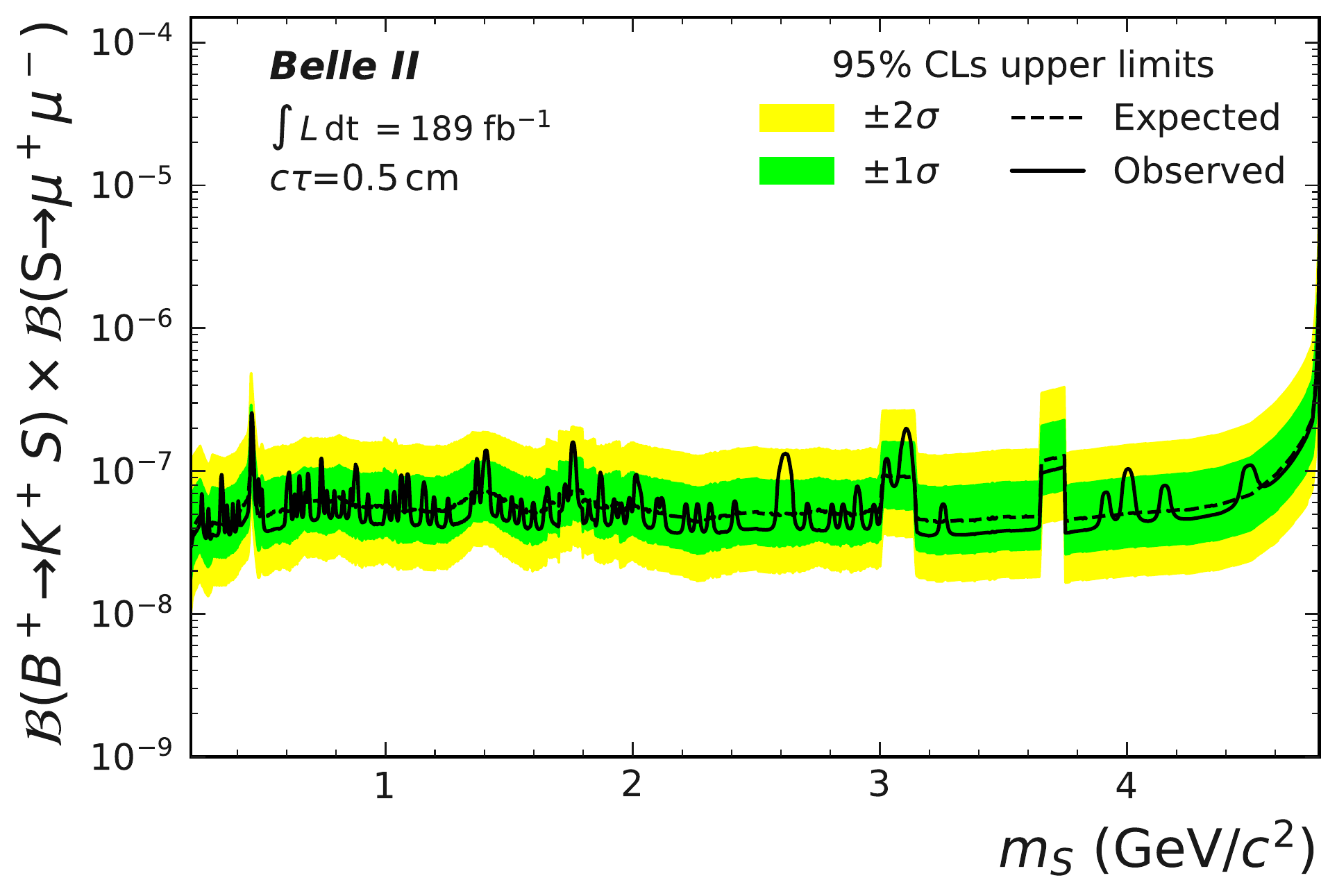}%
}%
\hspace*{\fill}
\subfigure[$B^+\to K^+S, S\to \mu^+\mu^-$, \newline lifetime of $c\tau=1\cm$.]{
  \label{subfit:brazil:Kp_mu_1:K}%
  \includegraphics[width=0.31\textwidth]{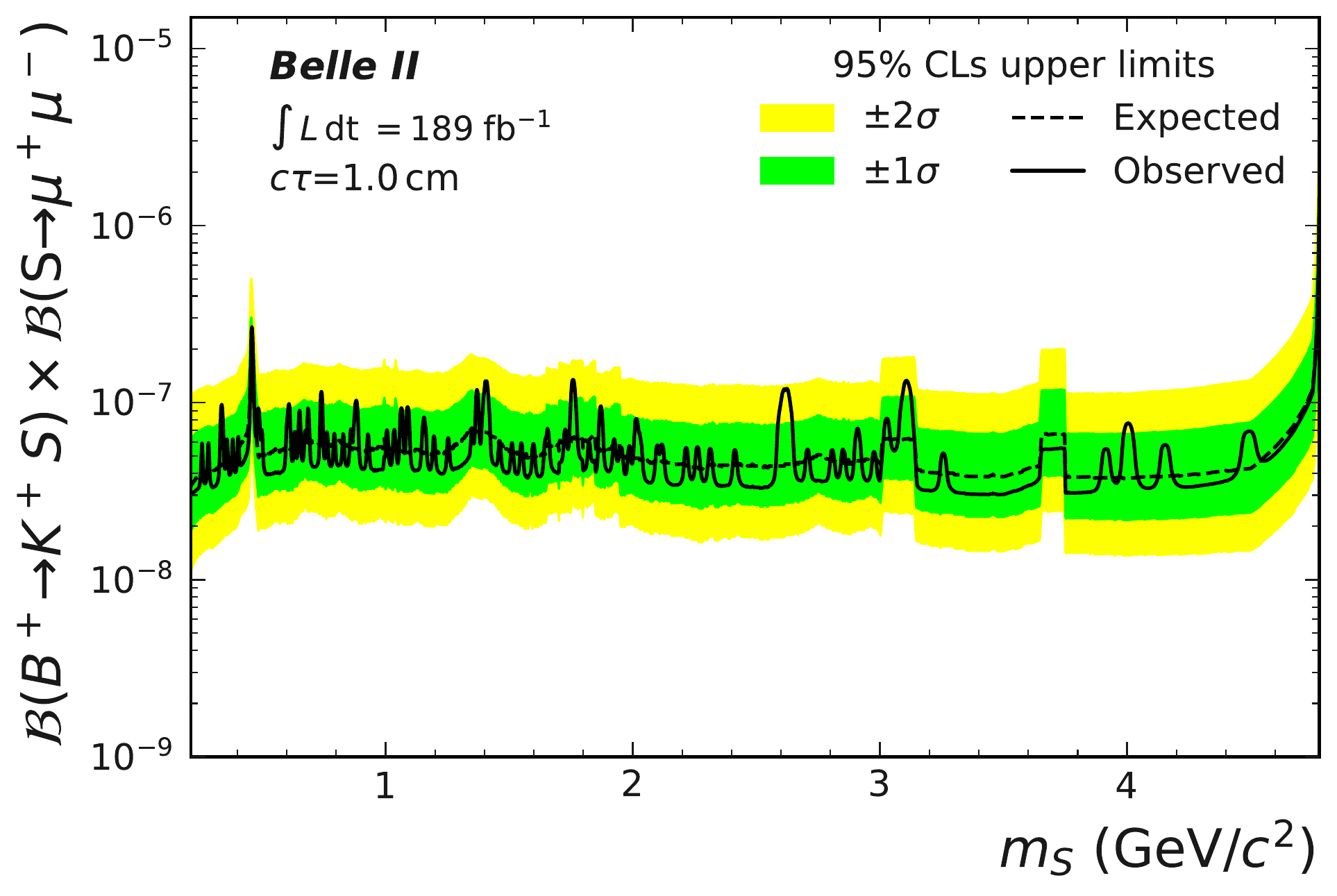}%
}%
\hspace*{\fill}
\subfigure[$B^+\to K^+S, S\to \mu^+\mu^-$, \newline lifetime of $c\tau=2.5\cm$.]{
  \label{subfit:brazil:Kp_mu_1:L}%
  \includegraphics[width=0.31\textwidth]{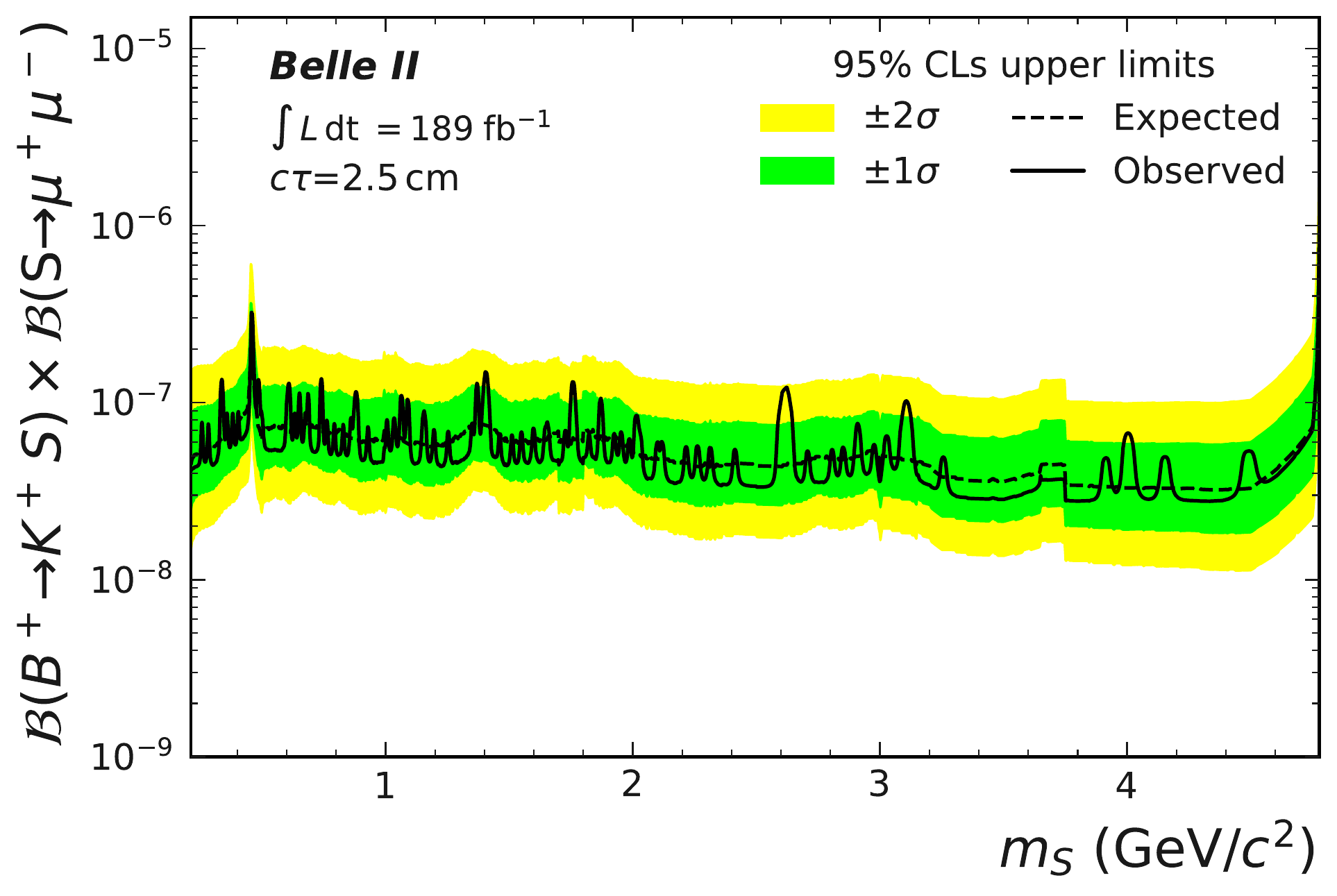}%
}
\caption{Expected and observed limits on the product of branching fractions $\mathcal{B}(B^+\to K^+S) \times \mathcal{B}(S\to \mu^+\mu^-)$ for lifetimes \hbox{$0.001 < c\tau < 2.5\,\cm$}.}\label{subfit:brazil:Kp_mu_1}
\end{figure*}

\begin{figure*}[ht]%
\subfigure[$B^+\to K^+S, S\to \mu^+\mu^-$, \newline lifetime of $c\tau=5\cm$.]{%
  \label{subfit:brazil:Kp_mu_2:A}%
  \includegraphics[width=0.31\textwidth]{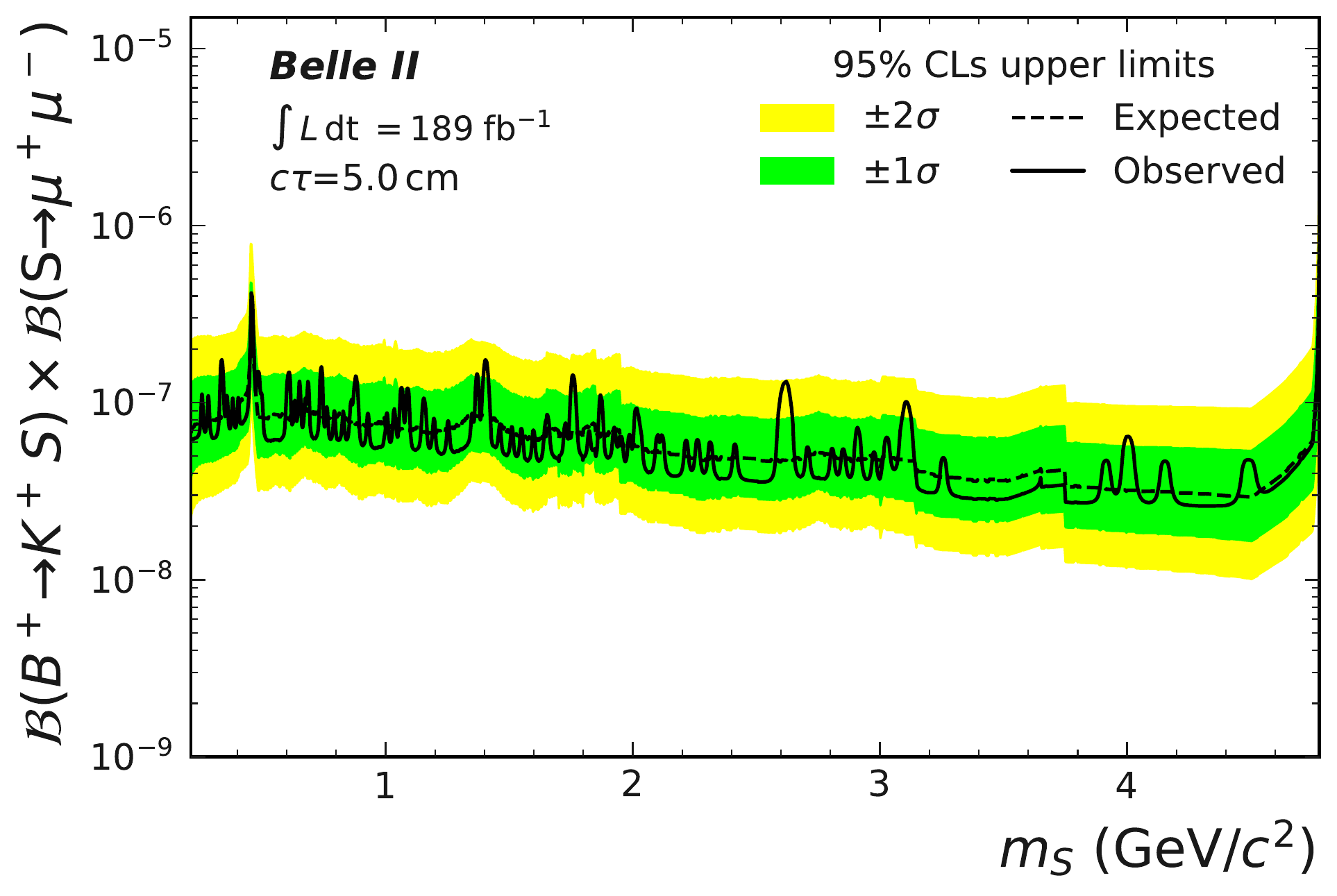}%
}%
\hspace*{\fill}
\subfigure[$B^+\to K^+S, S\to \mu^+\mu^-$, \newline lifetime of $c\tau=10\cm$.]{
  \label{subfit:brazil:Kp_mu_2:B}%
  \includegraphics[width=0.31\textwidth]{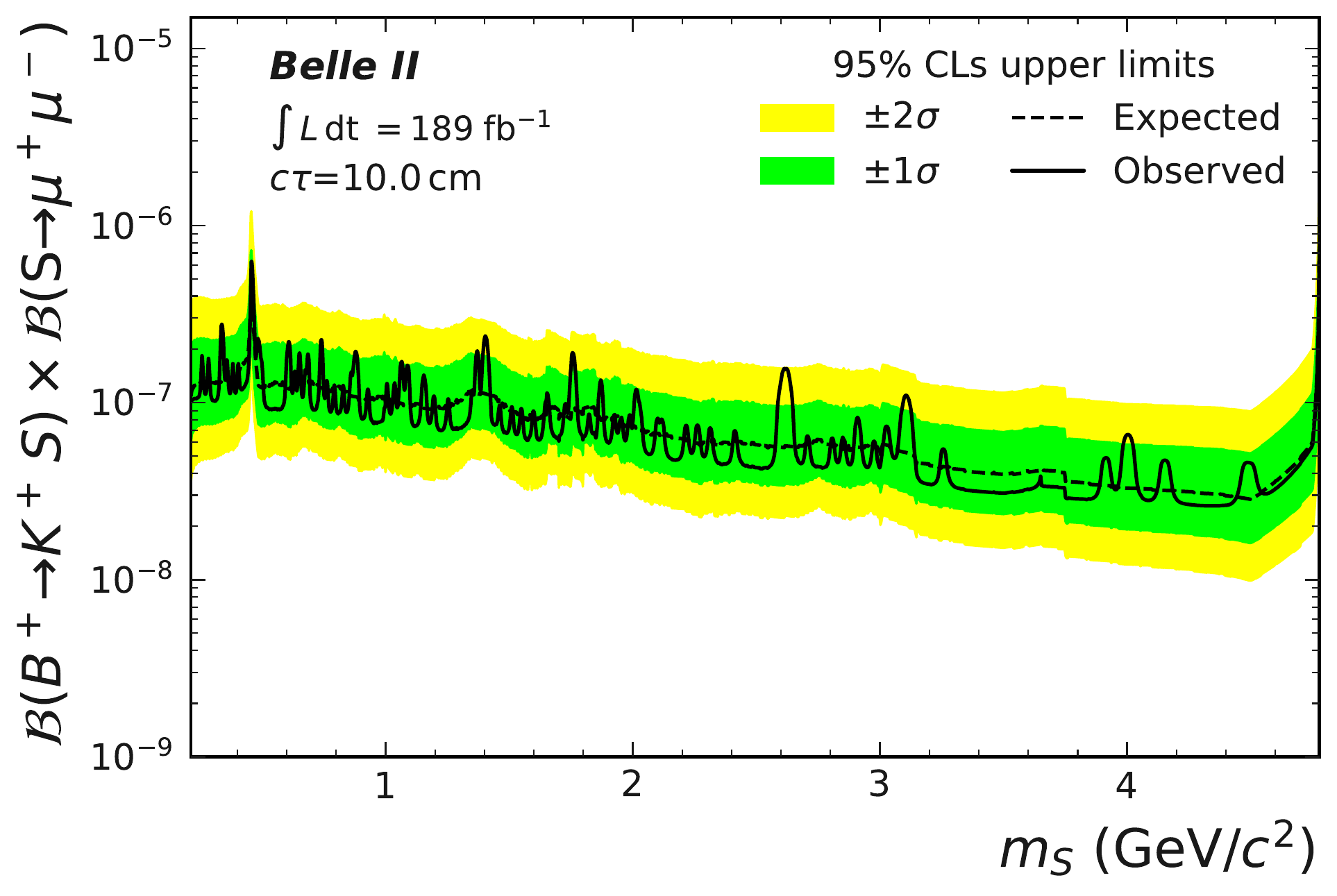}%
}%
\hspace*{\fill}
\subfigure[$B^+\to K^+S, S\to \mu^+\mu^-$, \newline lifetime of $c\tau=25\cm$.]{
  \label{subfit:brazil:Kp_mu_2:C}%
  \includegraphics[width=0.31\textwidth]{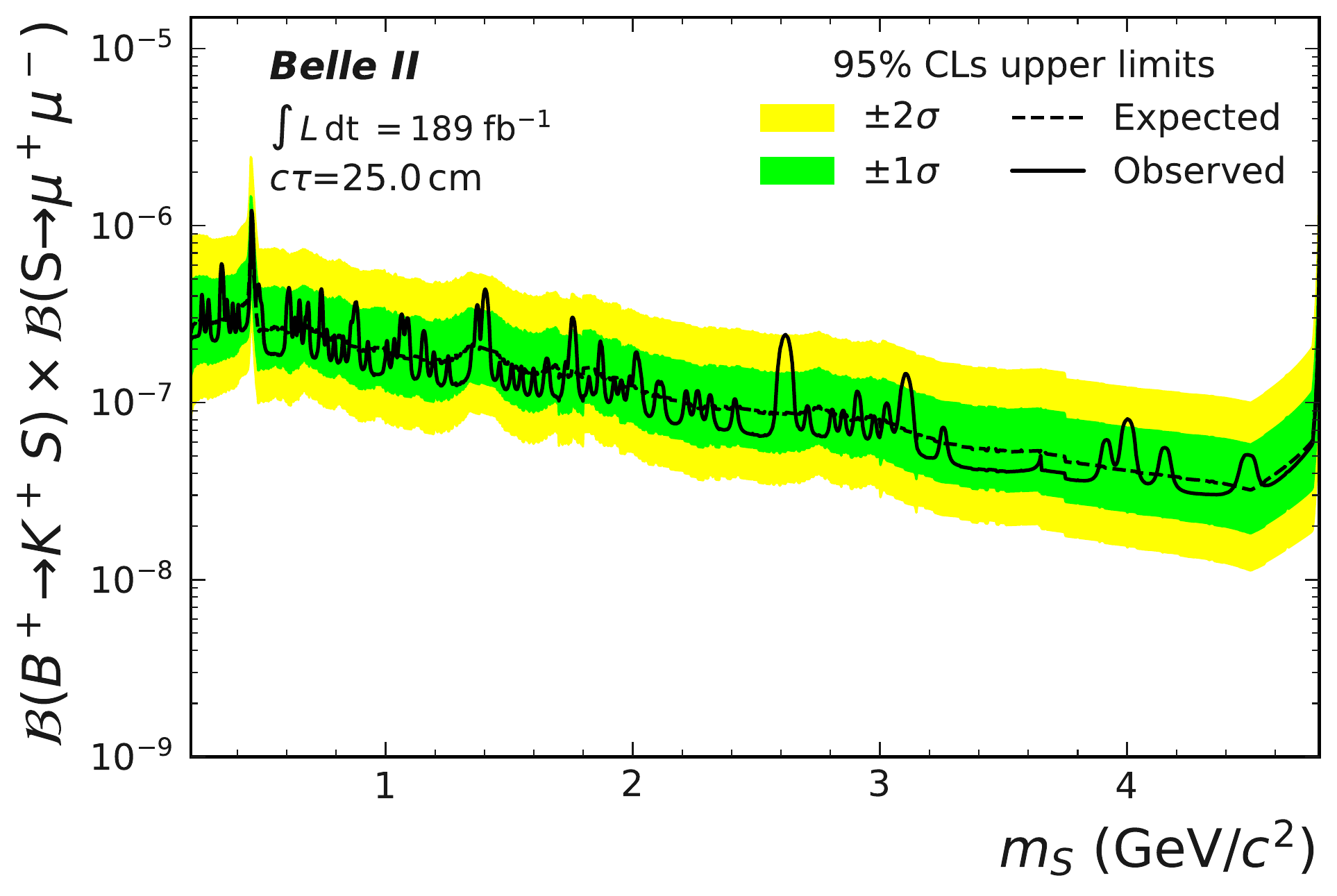}%
}
\subfigure[$B^+\to K^+S, S\to \mu^+\mu^-$, \newline lifetime of $c\tau=50\cm$.]{%
  \label{subfit:brazil:Kp_mu_2:D}%
  \includegraphics[width=0.31\textwidth]{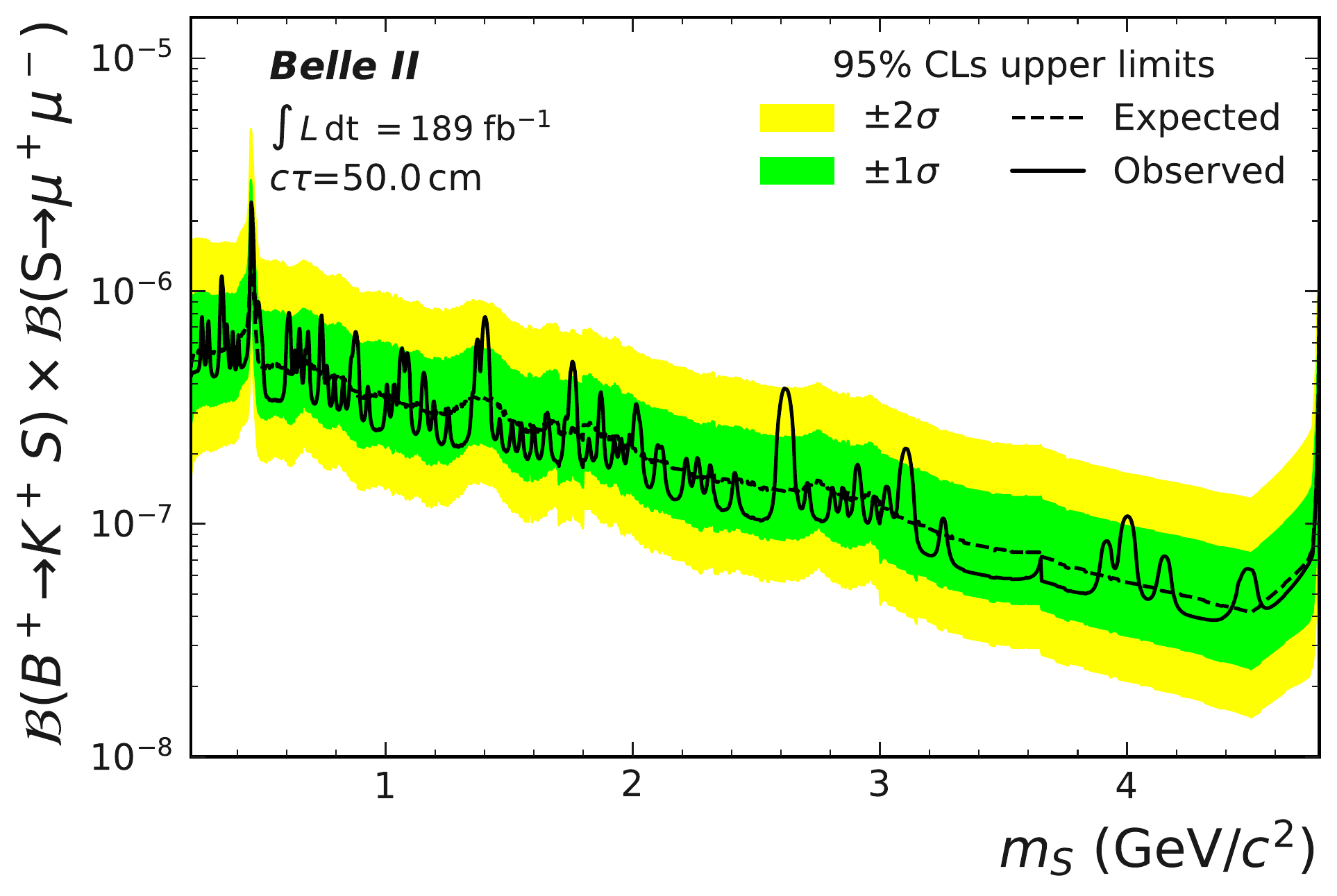}%
}%
\hspace*{\fill}
\subfigure[$B^+\to K^+S, S\to \mu^+\mu^-$, \newline lifetime of $c\tau=100\cm$.]{
  \label{subfit:brazil:Kp_mu_2:E}%
  \includegraphics[width=0.31\textwidth]{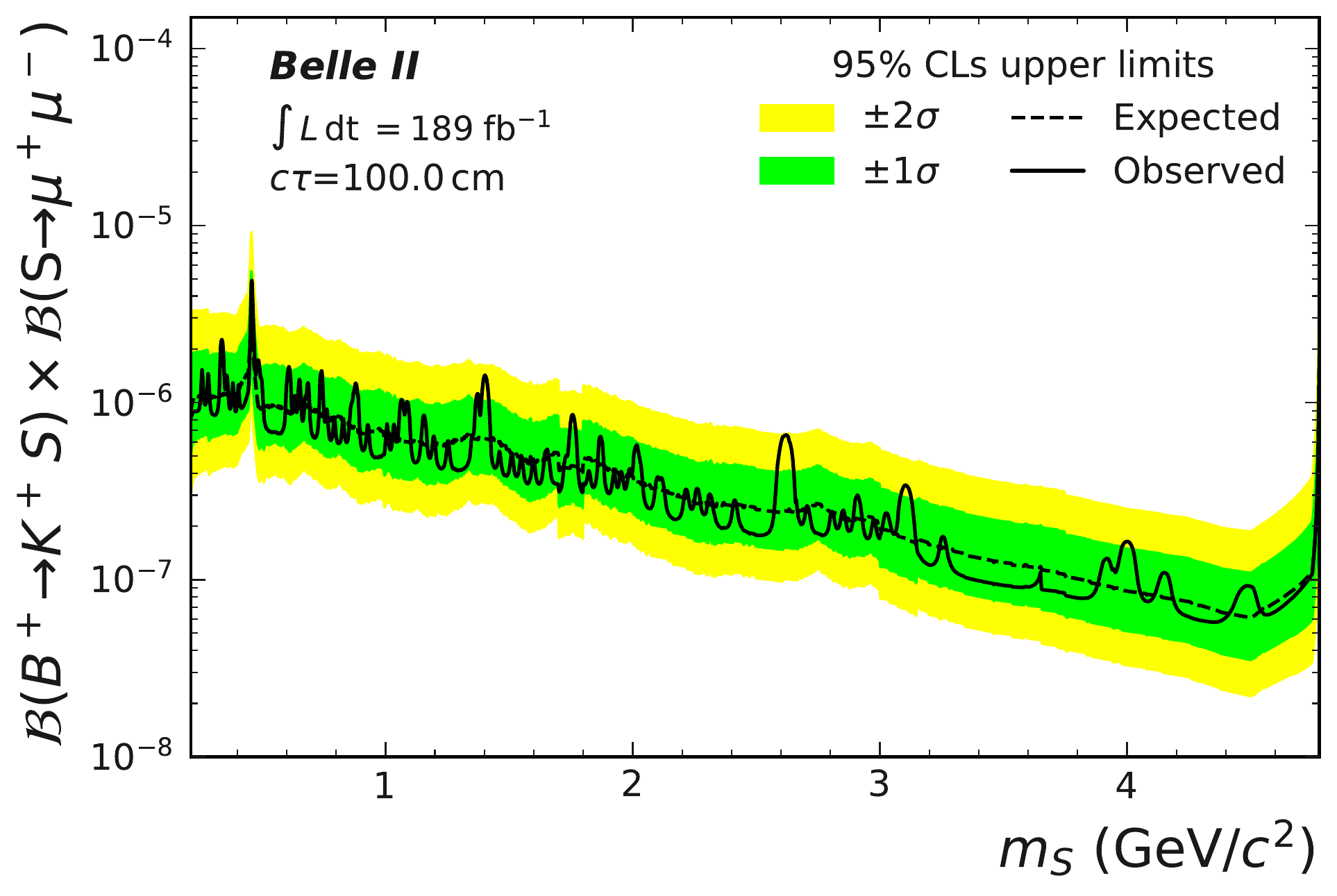}%
}%
\hspace*{\fill}
\subfigure[$B^+\to K^+S, S\to \mu^+\mu^-$, \newline lifetime of $c\tau=200\cm$.]{
  \label{subfit:brazil:Kp_mu_2:F}%
  \includegraphics[width=0.31\textwidth]{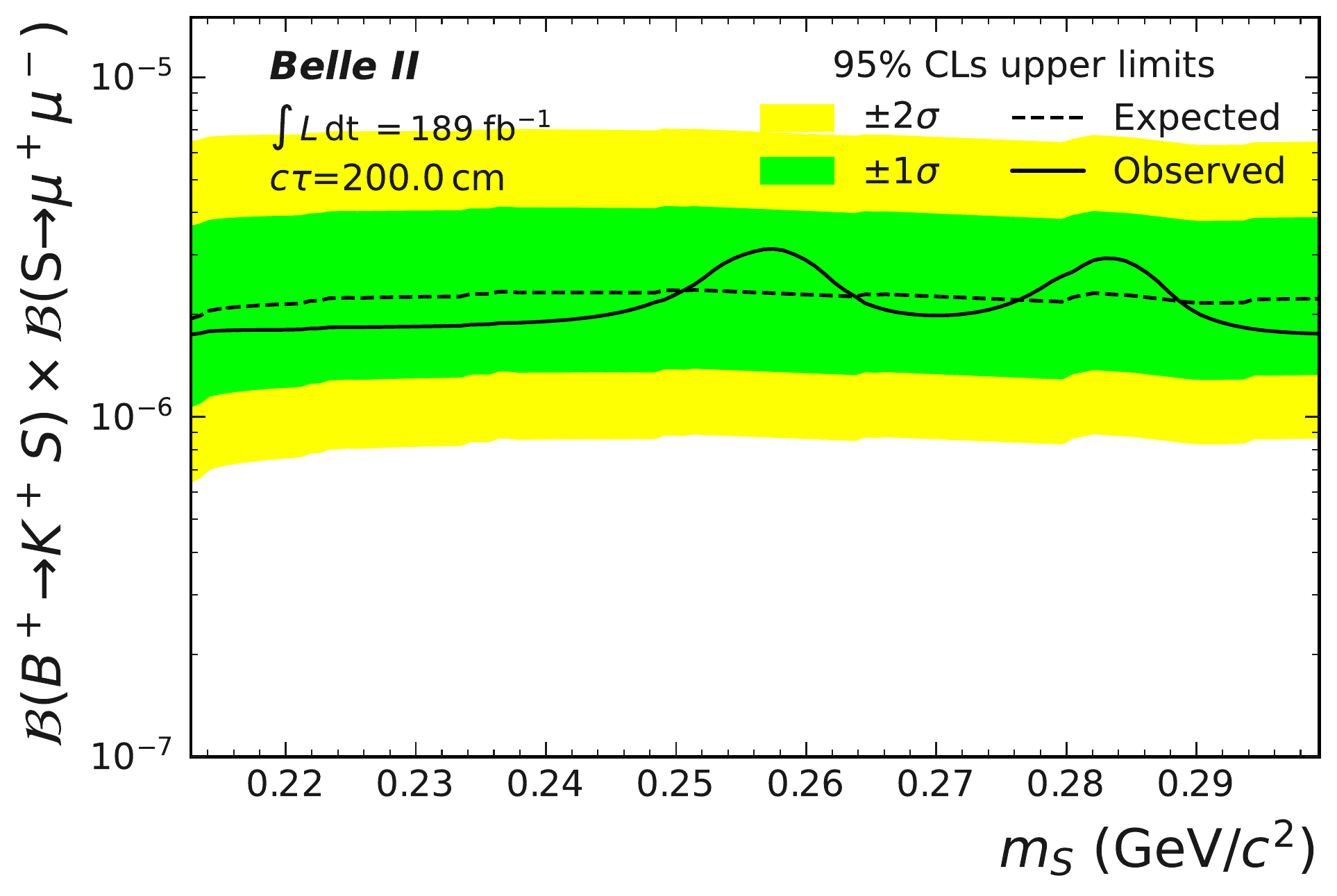}%
}
\subfigure[$B^+\to K^+S, S\to \mu^+\mu^-$, \newline lifetime of $c\tau=400\cm$.]{
  \label{subfit:brazil:Kp_mu_2:G}%
  \includegraphics[width=0.31\textwidth]{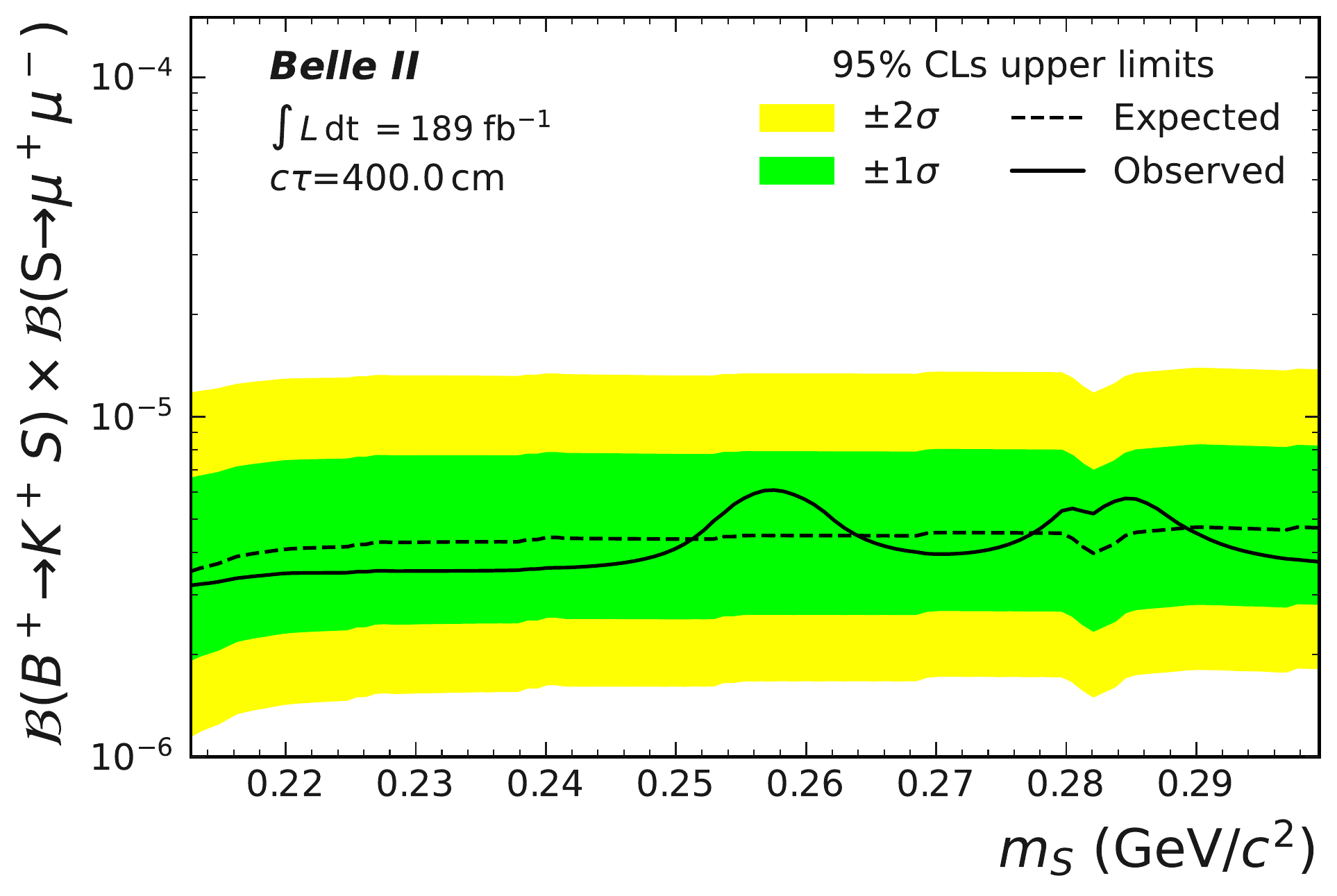}%
}
\caption{Expected and observed limits on the product of branching fractions $\mathcal{B}(B^+\to K^+S) \times \mathcal{B}(S\to \mu^+\mu^-)$ for lifetimes \hbox{$5 < c\tau < 400\,\cm$}.}\label{subfit:brazil:Kp_mu_2}
\end{figure*}

\begin{figure*}[ht]%
\subfigure[$\Bz\to \Kstarz(\to K^+\pi^-) S, S\to \mu^+\mu^-$, \newline lifetime of $c\tau=0.001\cm$.]{%
  \label{subfit:brazil:Kstar_mu_1:A}%
  \includegraphics[width=0.31\textwidth]{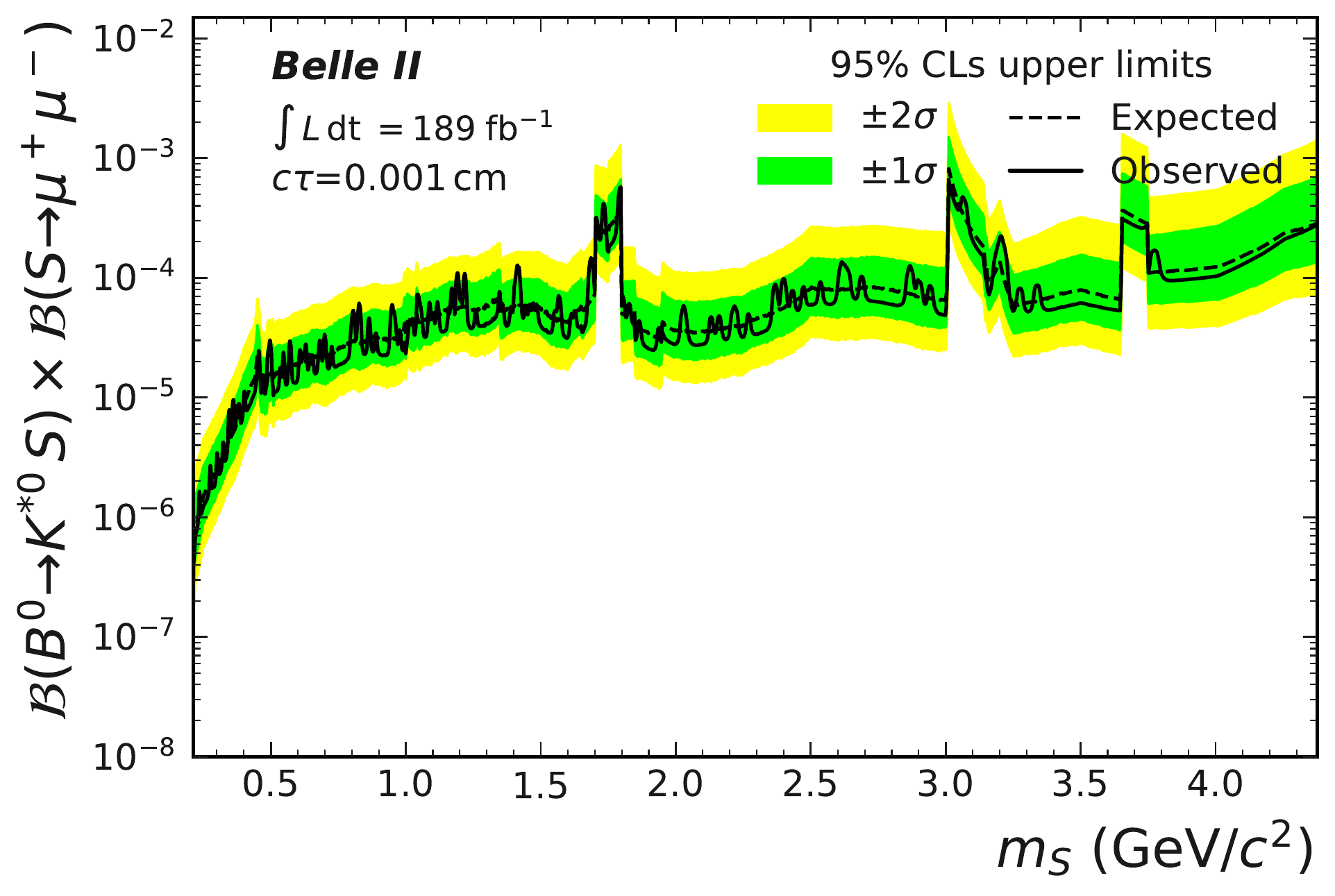}%
}%
\hspace*{\fill}
\subfigure[$\Bz\to \Kstarz(\to K^+\pi^-) S, S\to \mu^+\mu^-$, \newline lifetime of $c\tau=0.003\cm$.]{
  \label{subfit:brazil:Kstar_mu_1:B}%
  \includegraphics[width=0.31\textwidth]{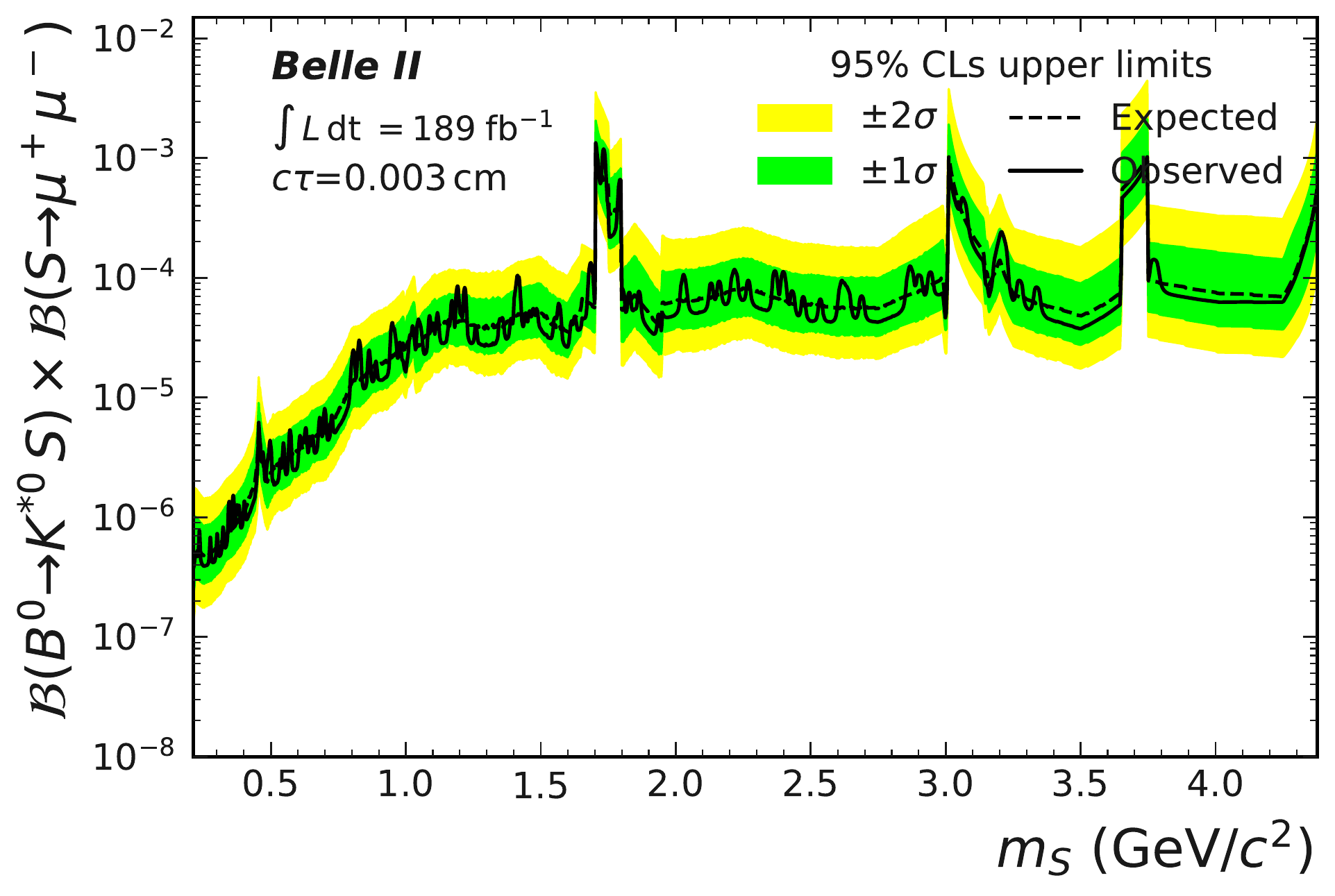}%
}%
\hspace*{\fill}
\subfigure[$\Bz\to \Kstarz(\to K^+\pi^-) S, S\to \mu^+\mu^-$, \newline lifetime of $c\tau=0.005\cm$.]{
  \label{subfit:brazil:Kstar_mu_1:C}%
  \includegraphics[width=0.31\textwidth]{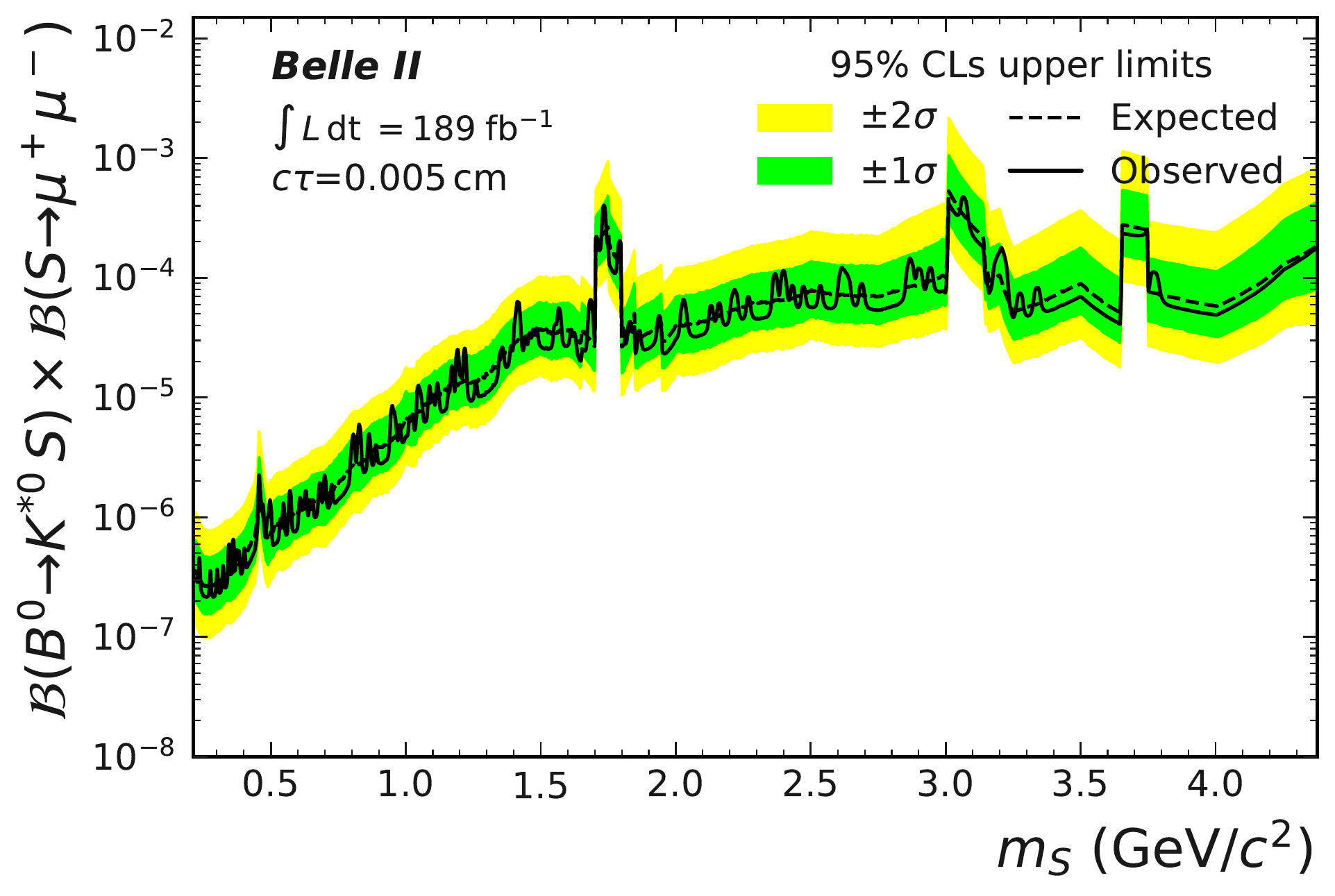}%
}
\subfigure[$\Bz\to \Kstarz(\to K^+\pi^-) S, S\to \mu^+\mu^-$, \newline lifetime of $c\tau=0.007\cm$.]{%
  \label{subfit:brazil:Kstar_mu_1:D}%
  \includegraphics[width=0.31\textwidth]{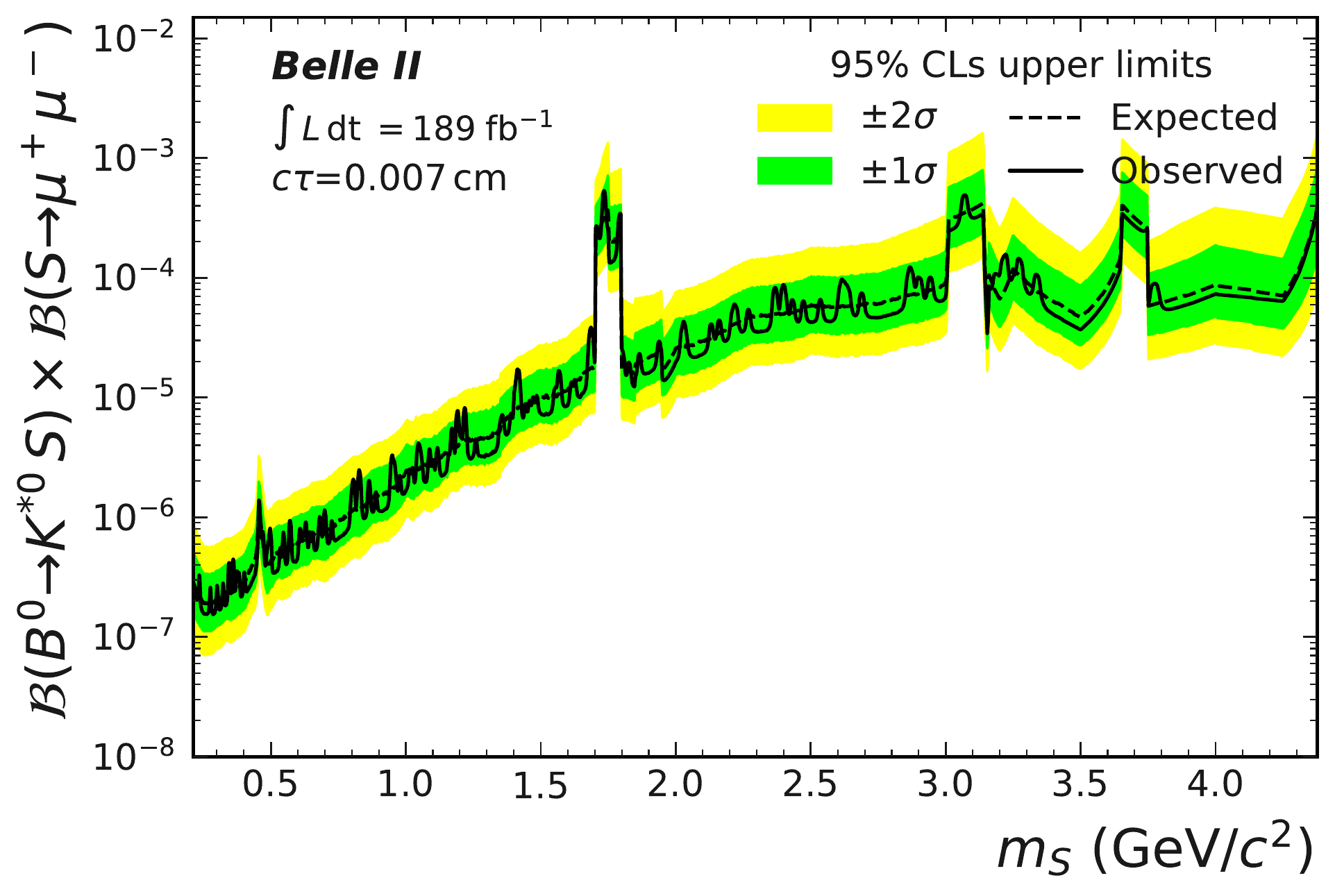}%
}%
\hspace*{\fill}
\subfigure[$\Bz\to \Kstarz(\to K^+\pi^-) S, S\to \mu^+\mu^-$, \newline lifetime of $c\tau=0.01\cm$.]{
  \label{subfit:brazil:Kstar_mu_1:E}%
  \includegraphics[width=0.31\textwidth]{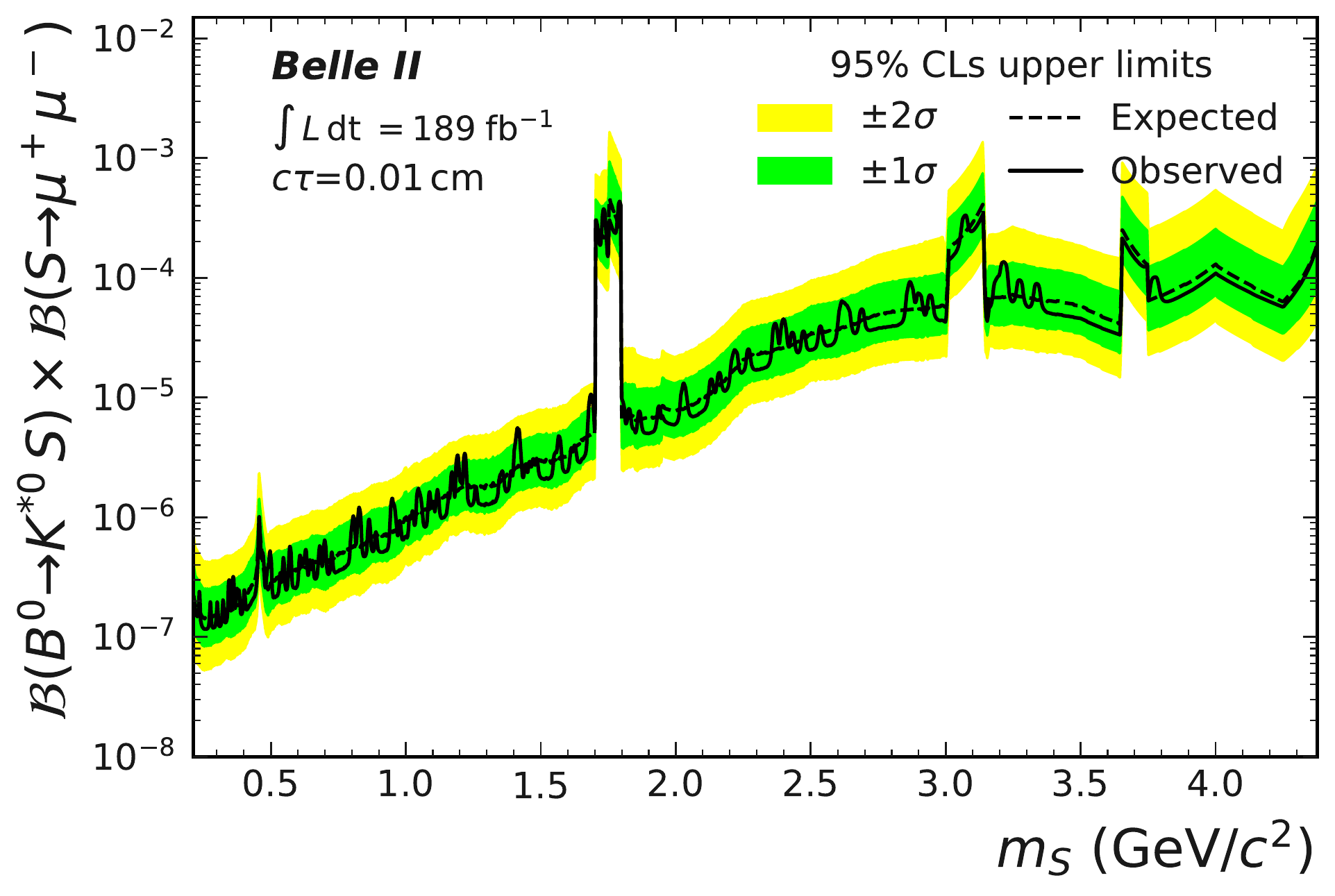}%
}%
\hspace*{\fill}
\subfigure[$\Bz\to \Kstarz(\to K^+\pi^-) S, S\to \mu^+\mu^-$, \newline lifetime of $c\tau=0.025\cm$.]{
  \label{subfit:brazil:Kstar_mu_1:F}%
  \includegraphics[width=0.31\textwidth]{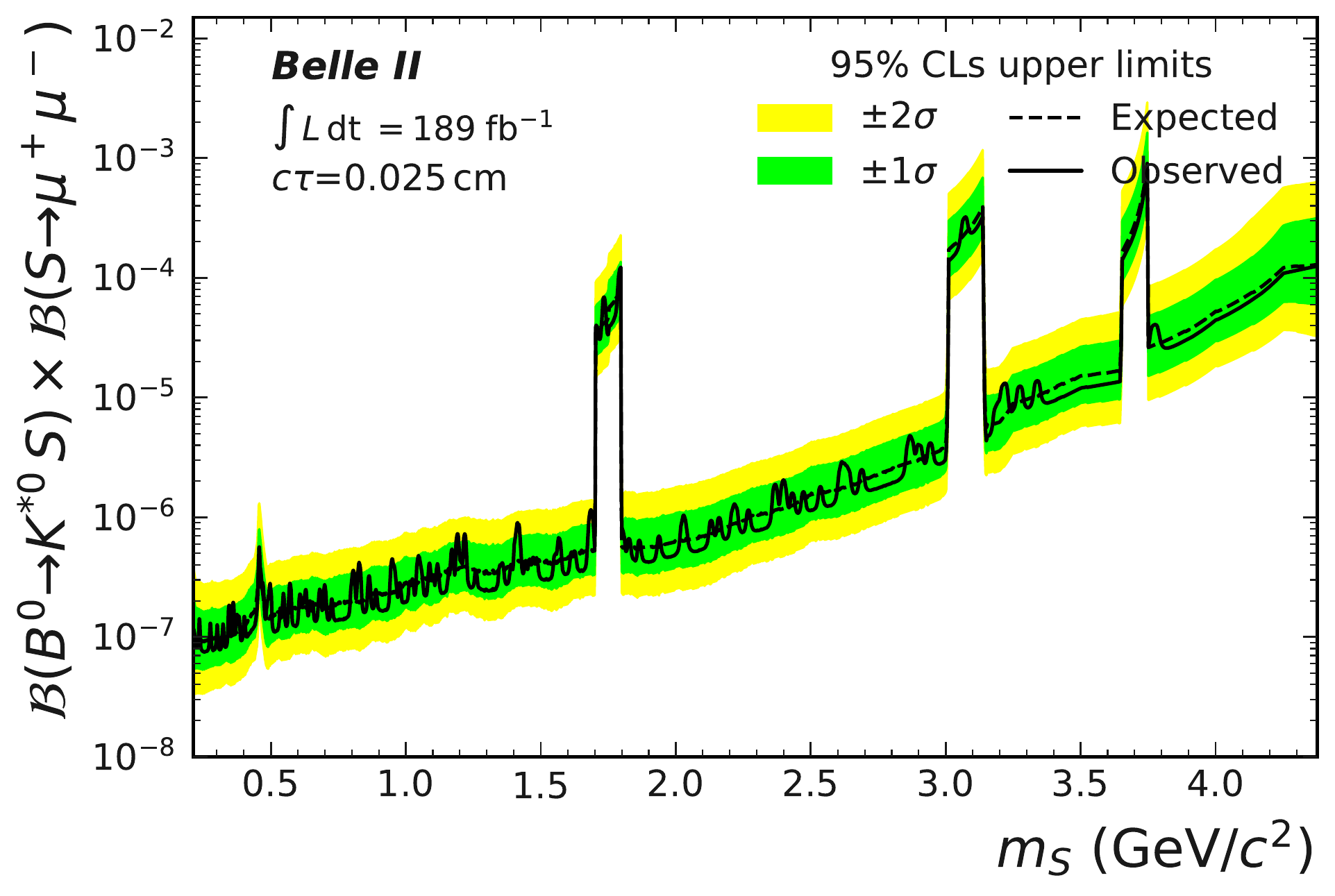}%
}
\subfigure[$\Bz\to \Kstarz(\to K^+\pi^-) S, S\to \mu^+\mu^-$, \newline lifetime of $c\tau=0.05\cm$.]{%
  \label{subfit:brazil:Kstar_mu_1:G}%
  \includegraphics[width=0.31\textwidth]{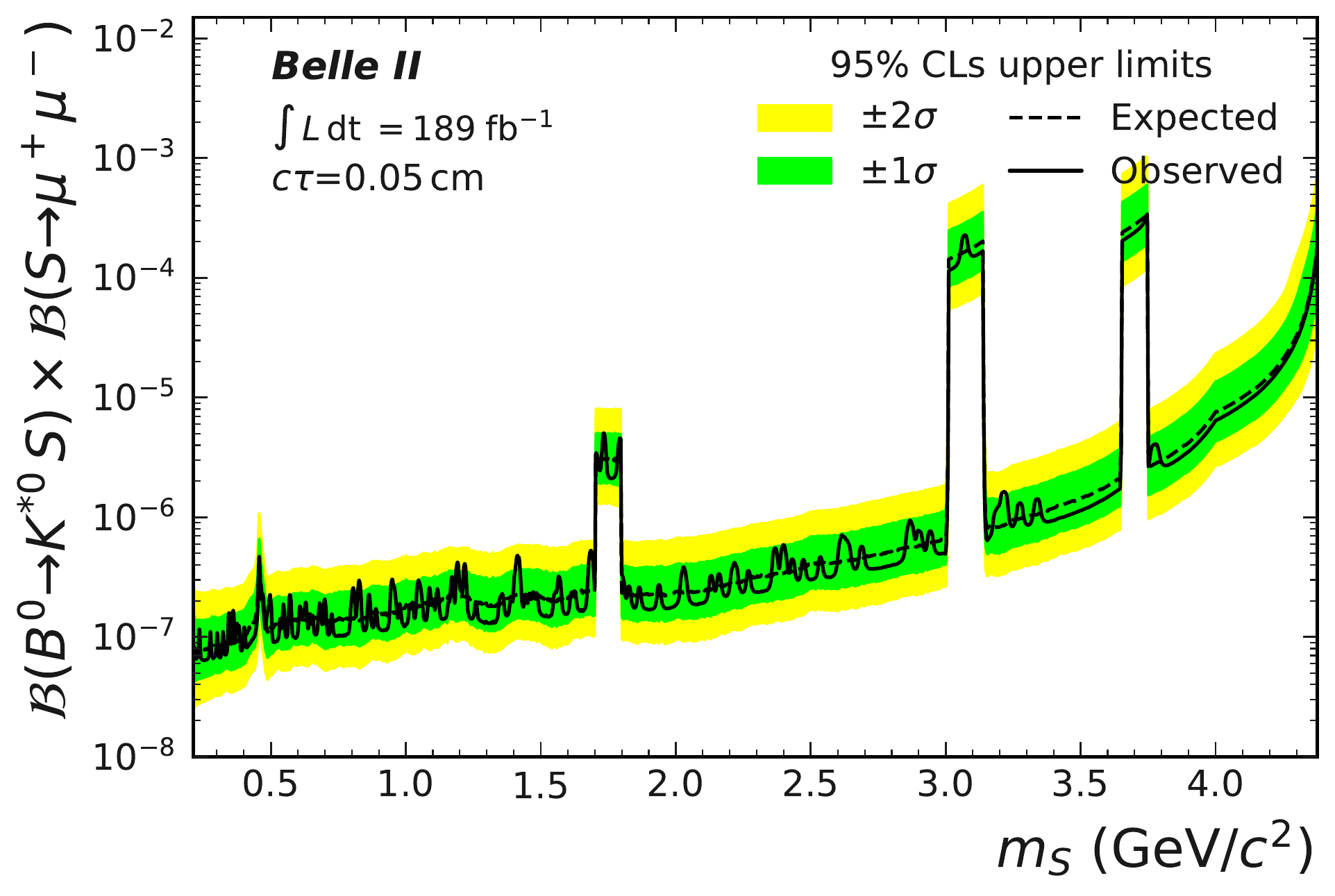}%
}%
\hspace*{\fill}
\subfigure[$\Bz\to \Kstarz(\to K^+\pi^-) S, S\to \mu^+\mu^-$, \newline lifetime of $c\tau=0.100\cm$.]{
  \label{subfit:brazil:Kstar_mu_1:H}%
  \includegraphics[width=0.31\textwidth]{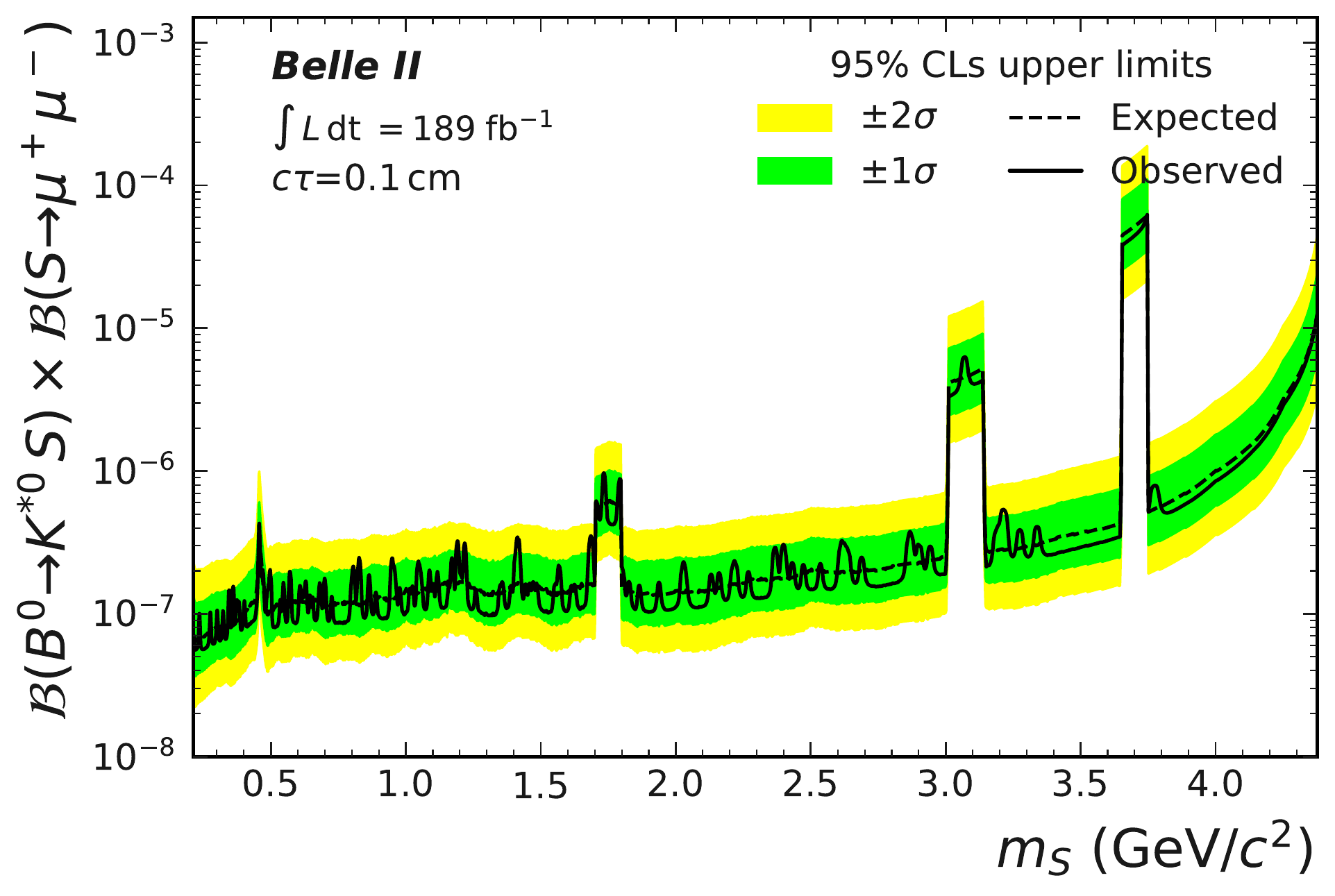}%
}%
\hspace*{\fill}
\subfigure[$\Bz\to \Kstarz(\to K^+\pi^-) S, S\to \mu^+\mu^-$, \newline lifetime of $c\tau=0.25\cm$.]{
  \label{subfit:brazil:Kstar_mu_1:I}%
  \includegraphics[width=0.31\textwidth]{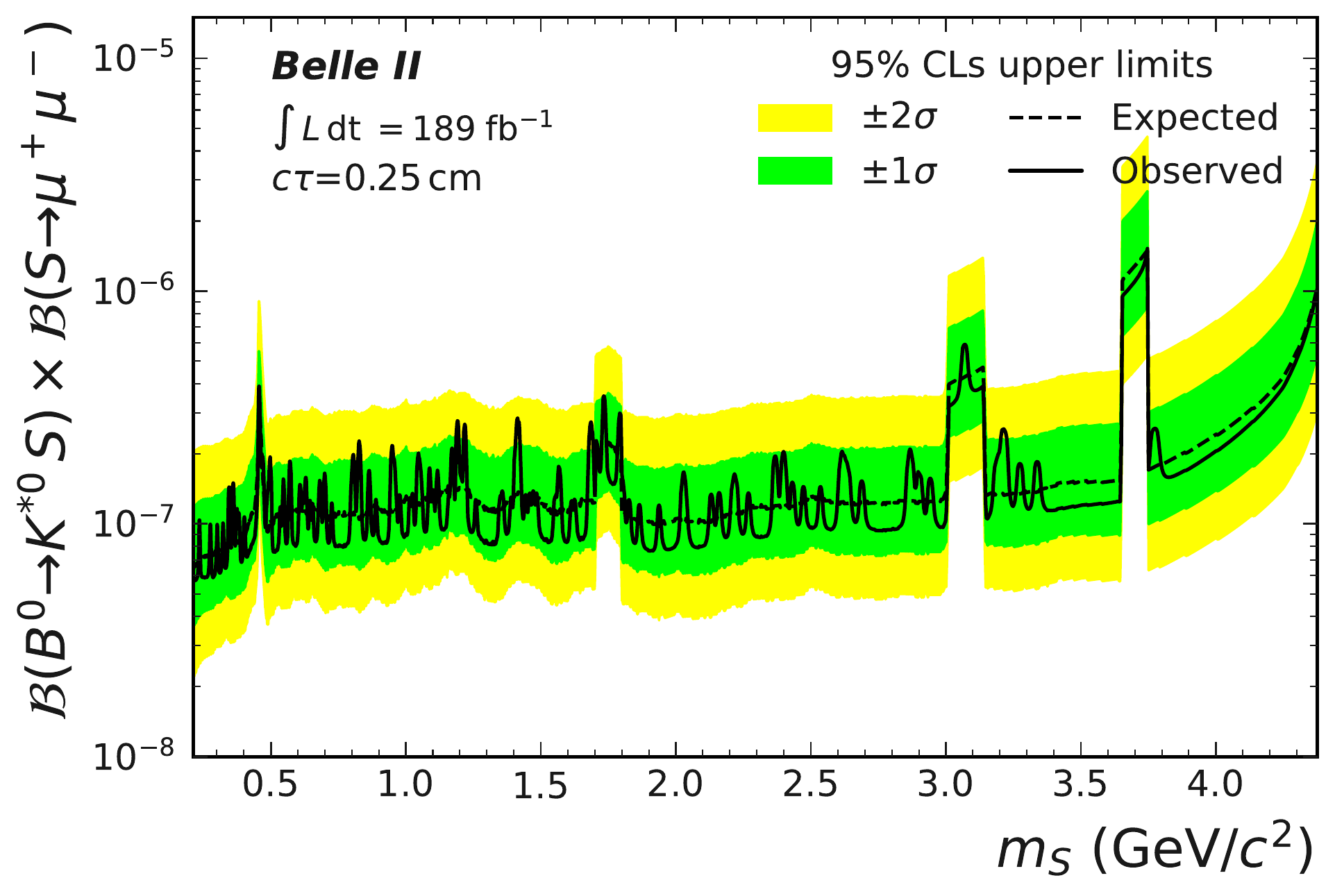}%
}
\subfigure[$\Bz\to \Kstarz(\to K^+\pi^-) S, S\to \mu^+\mu^-$, \newline lifetime of $c\tau=0.5\cm$.]{%
  \label{subfit:brazil:Kstar_mu_1:J}%
  \includegraphics[width=0.31\textwidth]{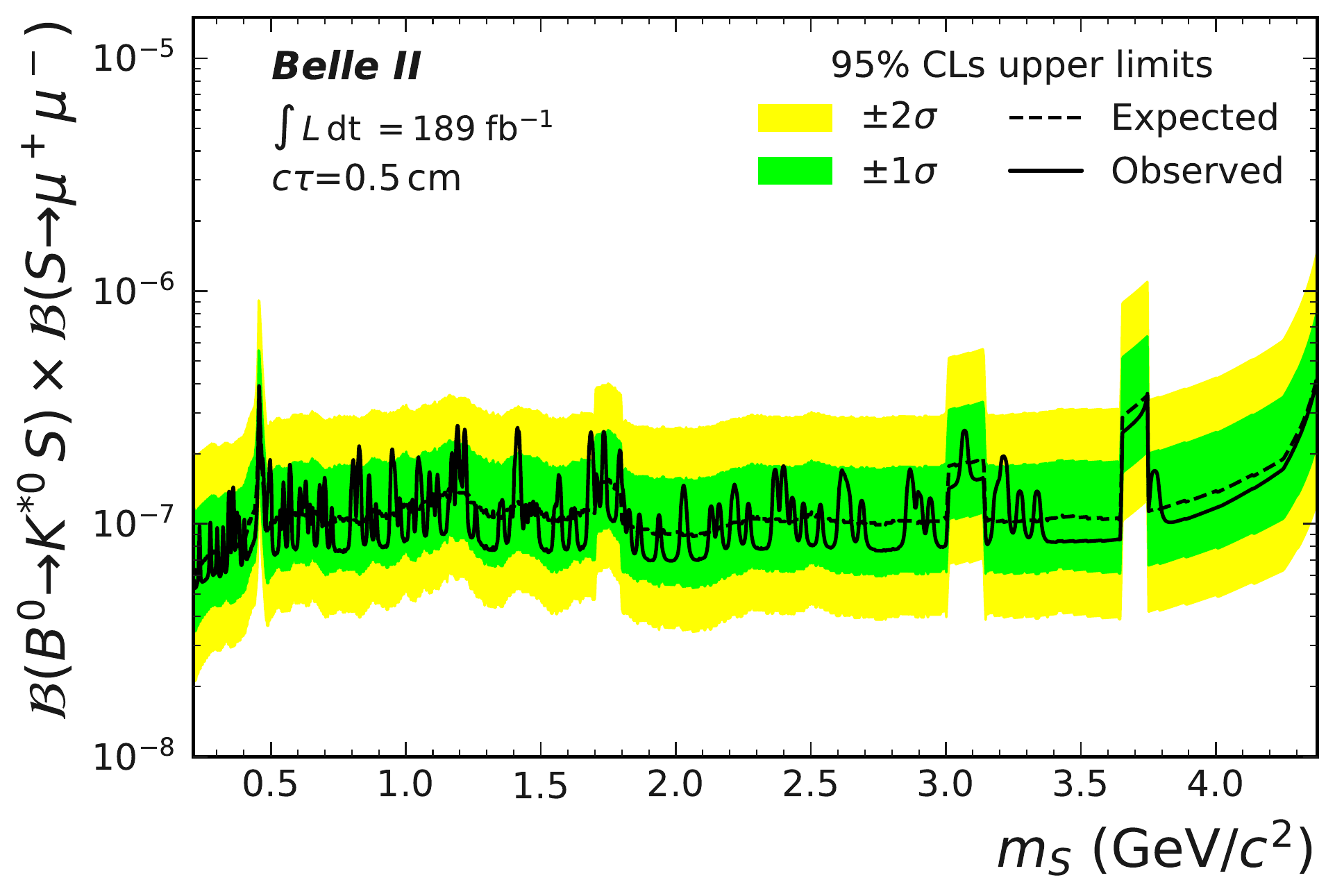}%
}%
\hspace*{\fill}
\subfigure[$\Bz\to \Kstarz(\to K^+\pi^-) S, S\to \mu^+\mu^-$, \newline lifetime of $c\tau=1\cm$.]{
  \label{subfit:brazil:Kstar_mu_1:K}%
  \includegraphics[width=0.31\textwidth]{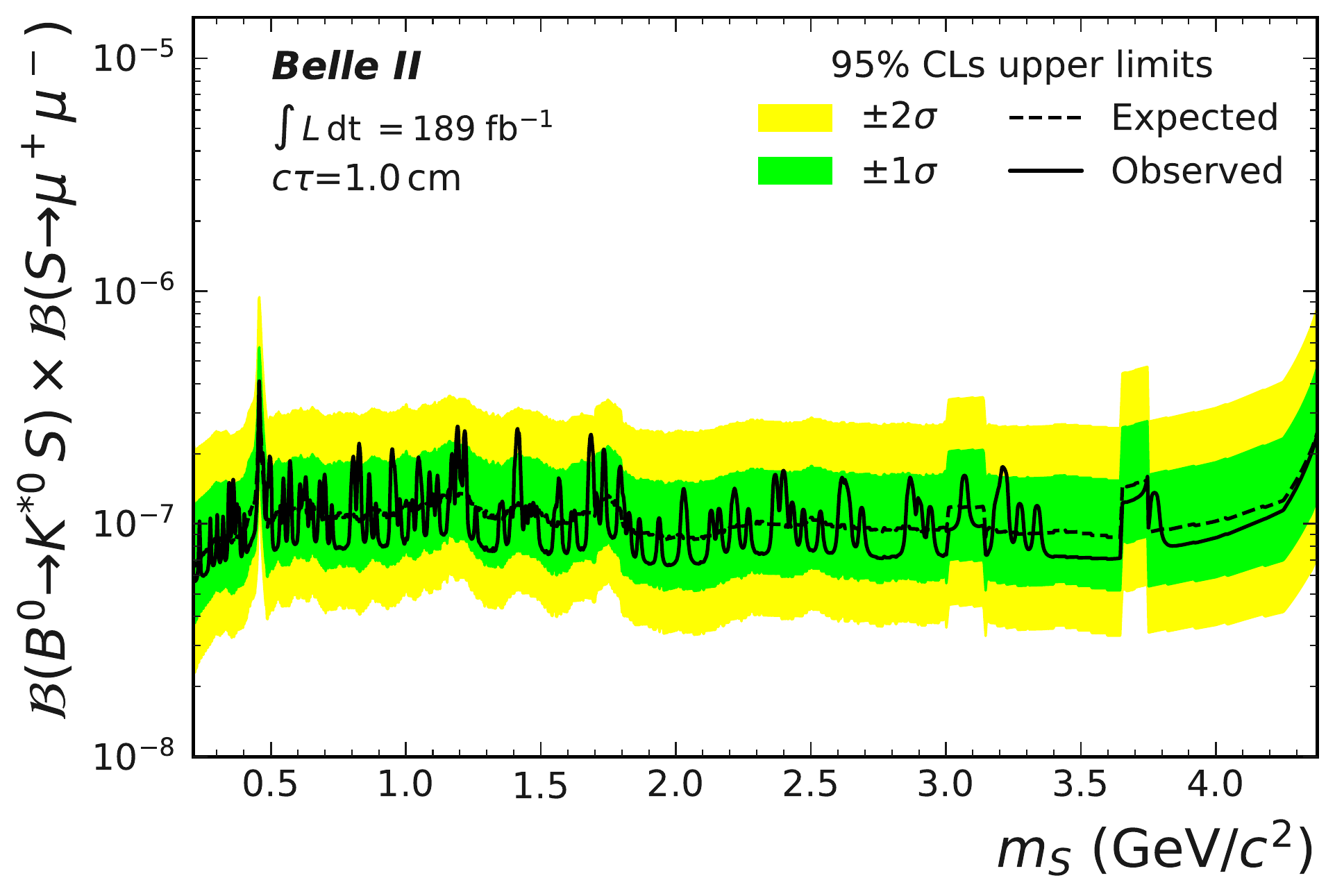}%
}%
\hspace*{\fill}
\subfigure[$\Bz\to \Kstarz(\to K^+\pi^-) S, S\to \mu^+\mu^-$, \newline lifetime of $c\tau=2.5\cm$.]{
  \label{subfit:brazil:Kstar_mu_1:L}%
  \includegraphics[width=0.31\textwidth]{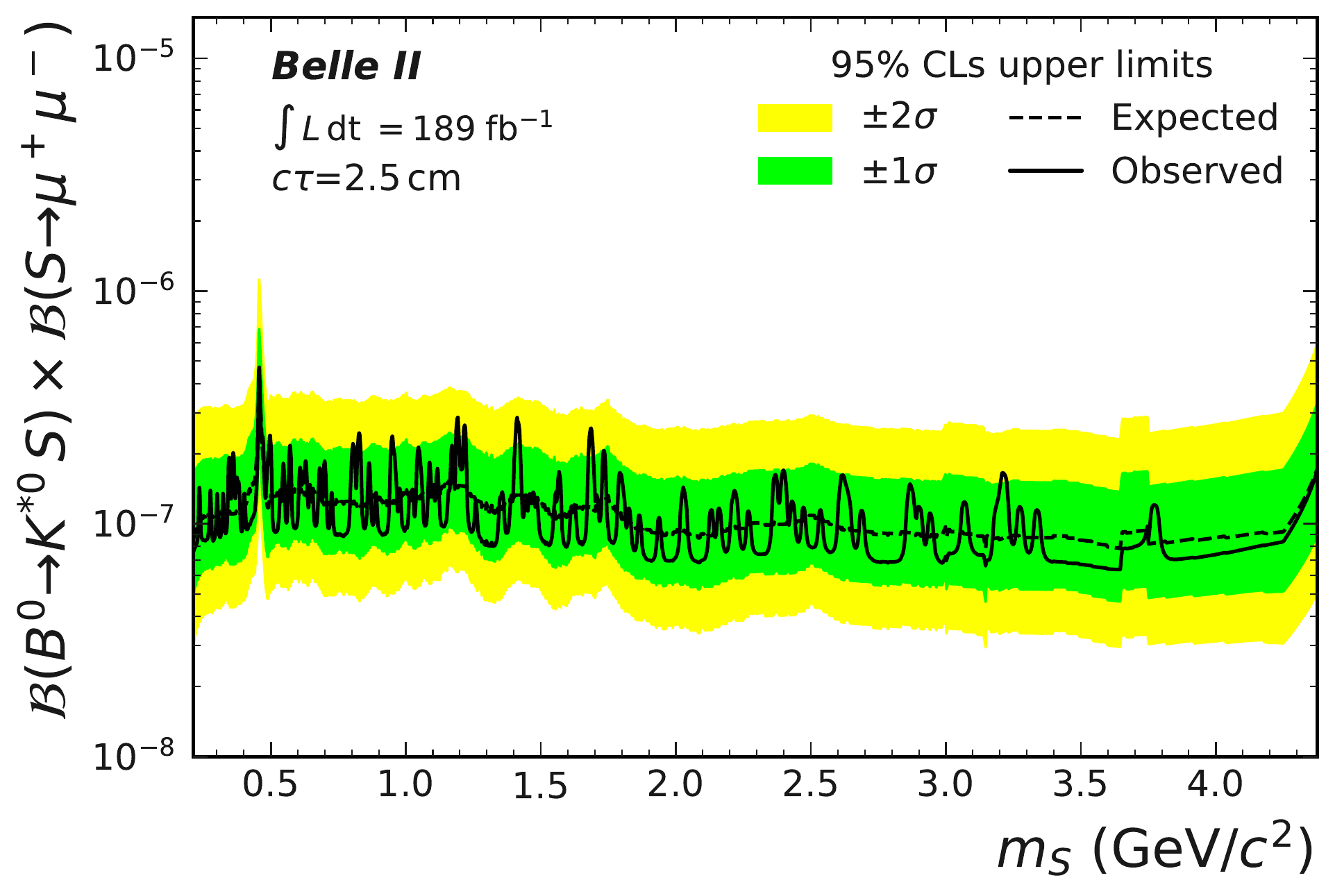}%
}
\caption{Expected and observed limits on the product of branching fractions $\mathcal{B}(B^0\to \Kstarz(\to K^+\pi^-) S) \times \mathcal{B}(S\to \mu^+\mu^-)$ for lifetimes \hbox{$0.001 < c\tau < 2.5\,\cm$}.}\label{subfit:brazil:Kstar_mu_1}
\end{figure*}

\begin{figure*}[ht]%
\subfigure[$\Bz\to \Kstarz(\to K^+\pi^-) S, S\to \mu^+\mu^-$, \newline lifetime of $c\tau=5\cm$.]{%
  \label{subfit:brazil:Kstar_mu_2:A}%
  \includegraphics[width=0.31\textwidth]{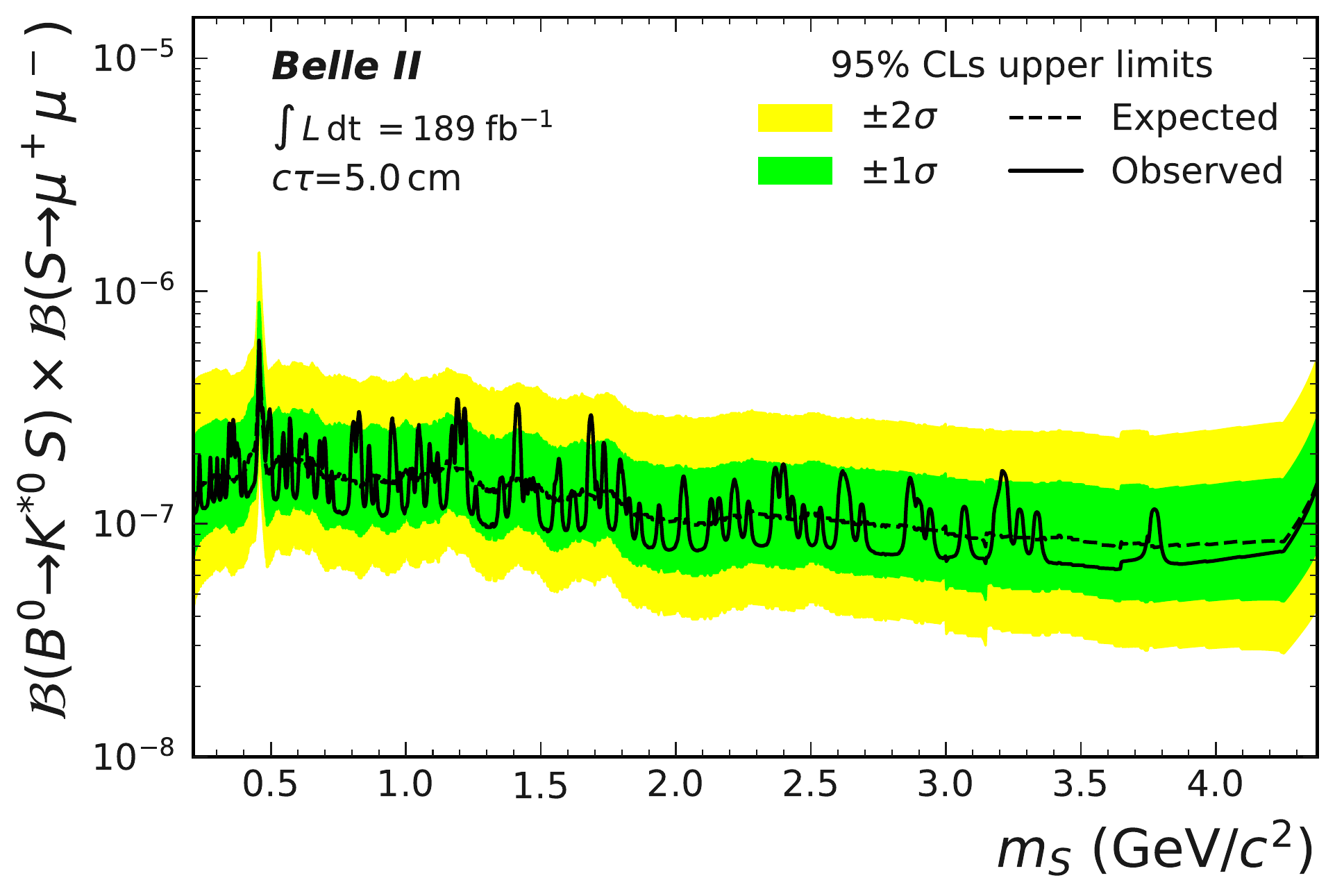}%
}%
\hspace*{\fill}
\subfigure[$\Bz\to \Kstarz(\to K^+\pi^-) S, S\to \mu^+\mu^-$, \newline lifetime of $c\tau=10\cm$.]{
  \label{subfit:brazil:Kstar_mu_2:B}%
  \includegraphics[width=0.31\textwidth]{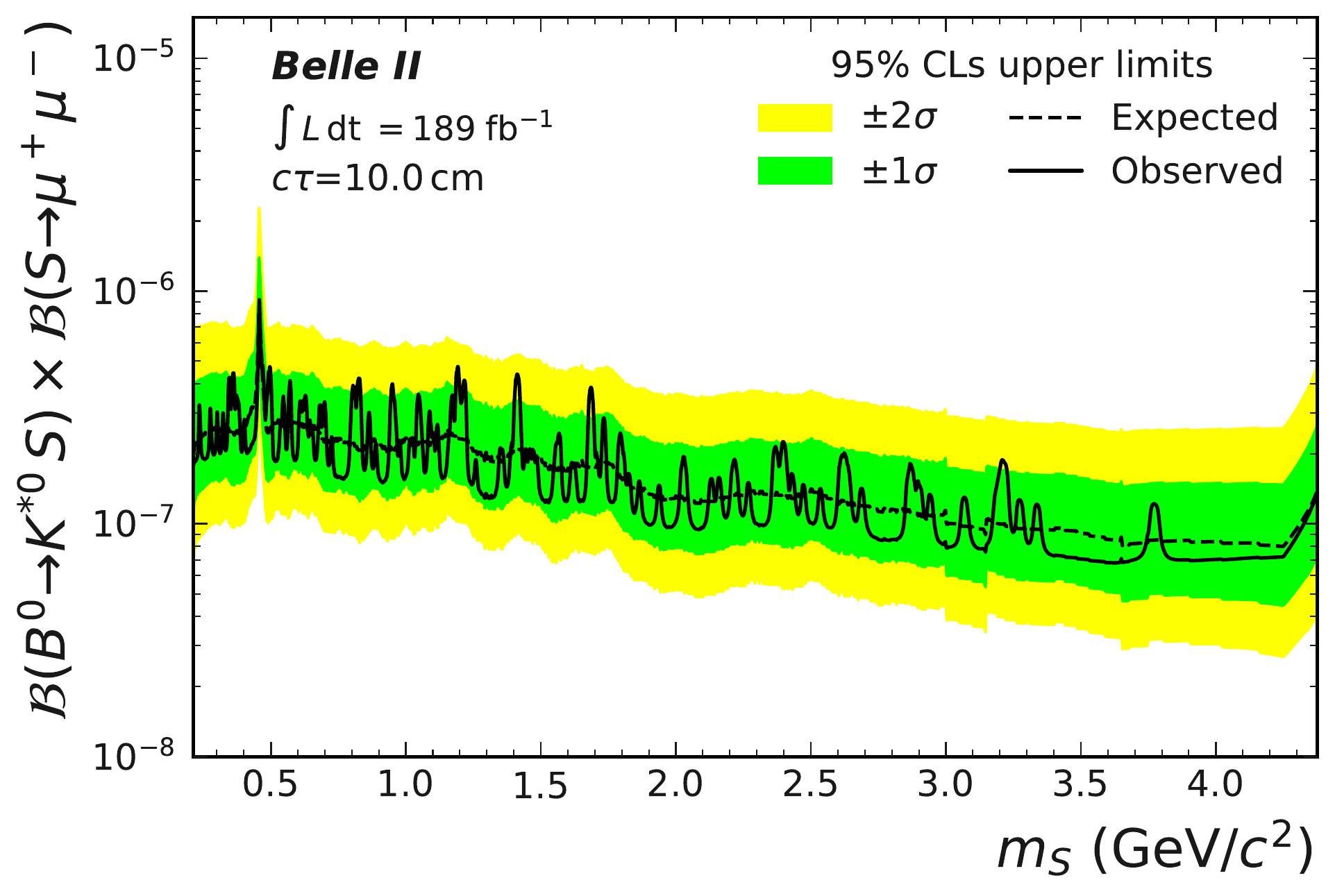}%
}%
\hspace*{\fill}
\subfigure[$\Bz\to \Kstarz(\to K^+\pi^-) S, S\to \mu^+\mu^-$, \newline lifetime of $c\tau=25\cm$.]{
  \label{subfit:brazil:Kstar_mu_2:C}%
  \includegraphics[width=0.31\textwidth]{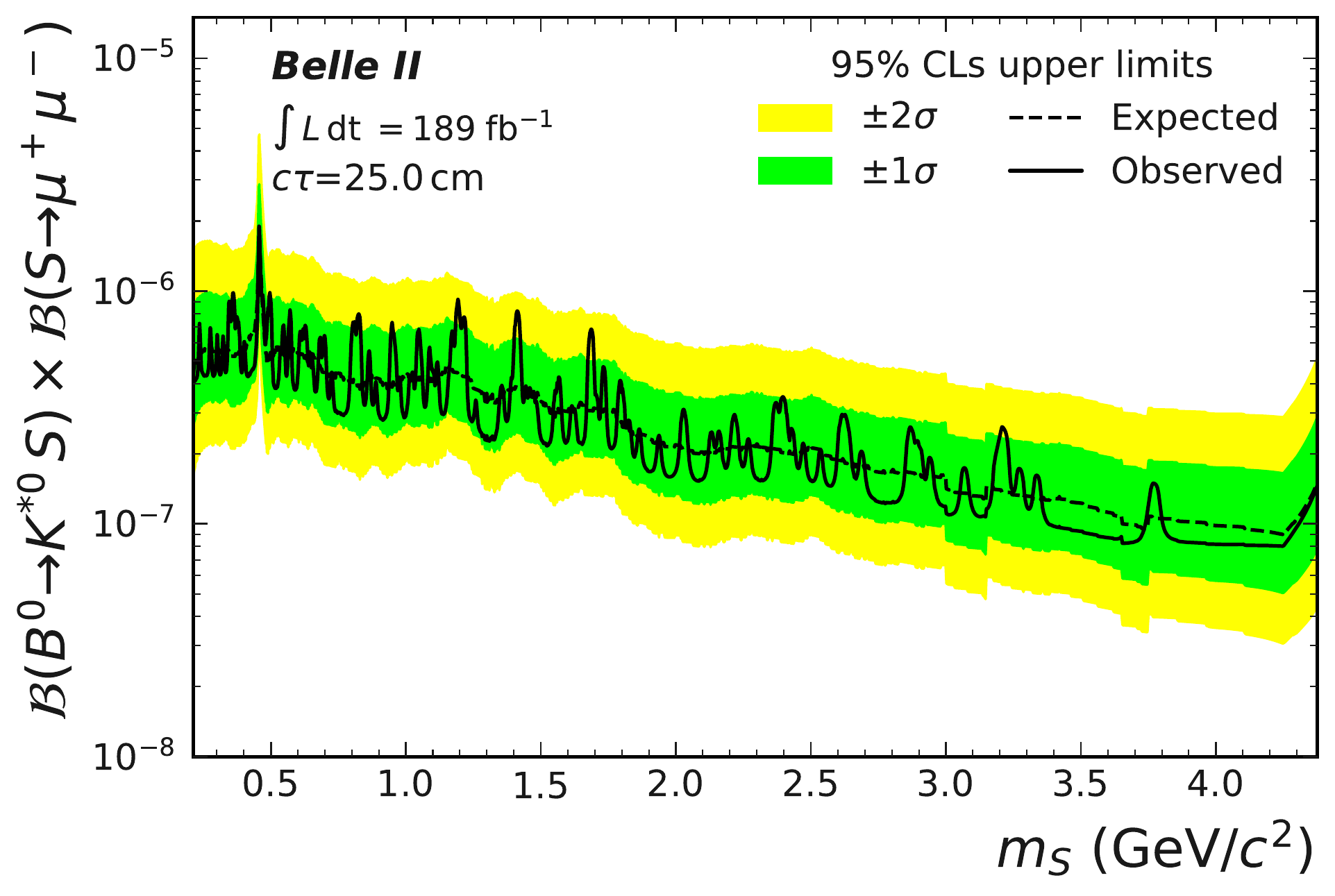}%
}
\subfigure[$\Bz\to \Kstarz(\to K^+\pi^-) S, S\to \mu^+\mu^-$, \newline lifetime of $c\tau=50\cm$.]{%
  \label{subfit:brazil:Kstar_mu_2:D}%
  \includegraphics[width=0.31\textwidth]{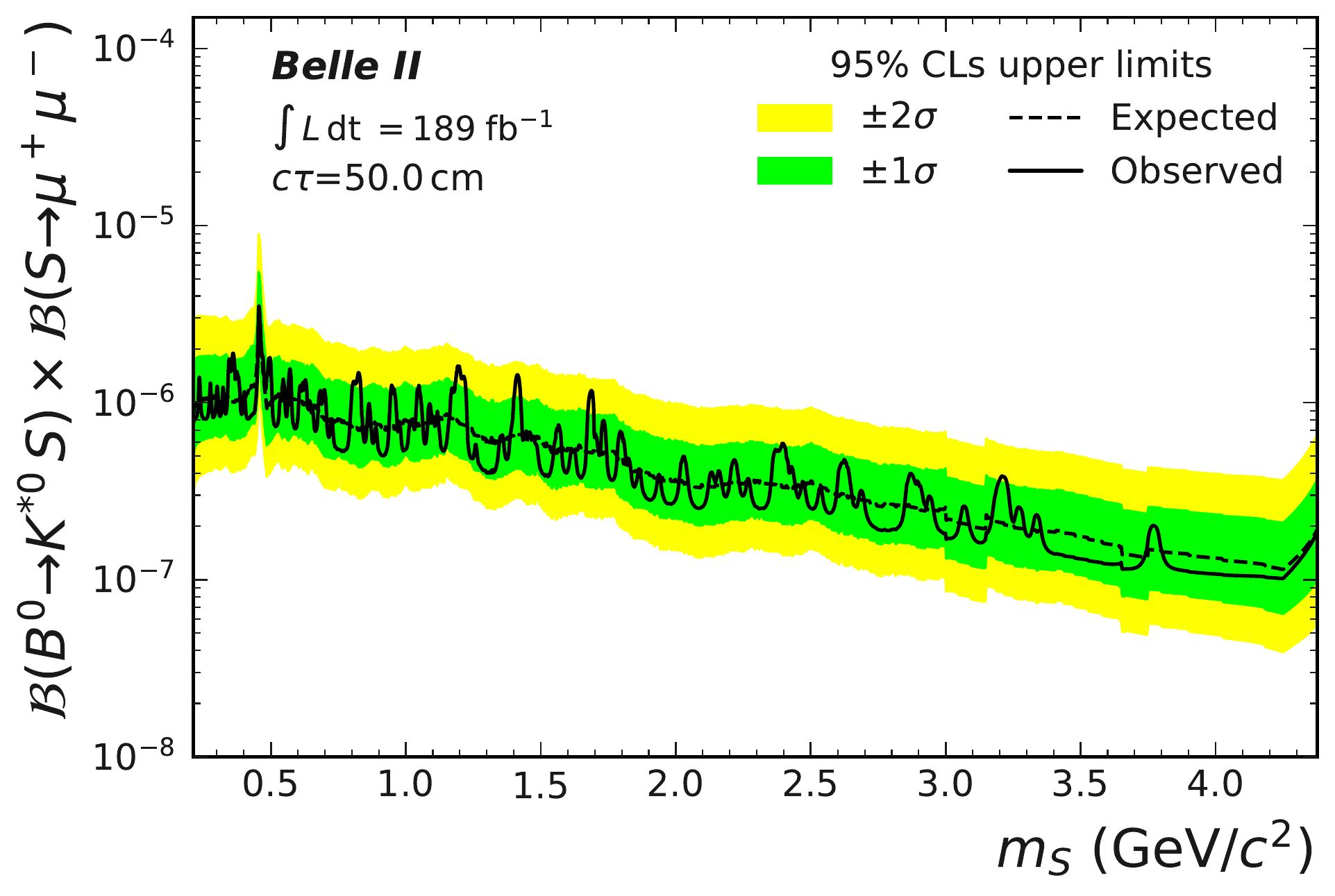}%
}%
\hspace*{\fill}
\subfigure[$\Bz\to \Kstarz(\to K^+\pi^-) S, S\to \mu^+\mu^-$, \newline lifetime of $c\tau=100\cm$.]{
  \label{subfit:brazil:Kstar_mu_2:E}%
  \includegraphics[width=0.31\textwidth]{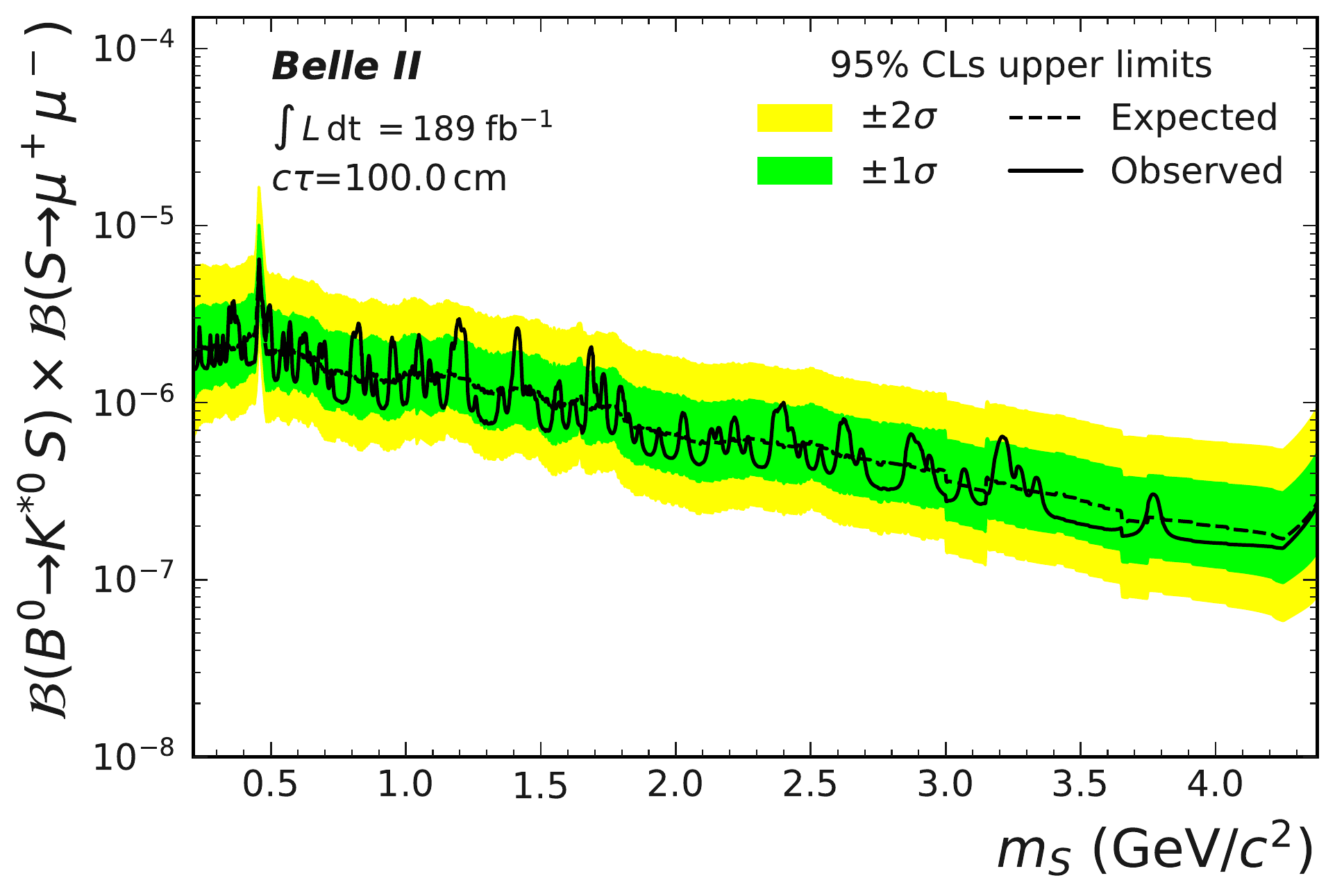}%
}%
\hspace*{\fill}
\subfigure[$\Bz\to \Kstarz(\to K^+\pi^-) S, S\to \mu^+\mu^-$, \newline lifetime of $c\tau=200\cm$.]{
  \label{subfit:brazil:Kstar_mu_2:F}%
  \includegraphics[width=0.31\textwidth]{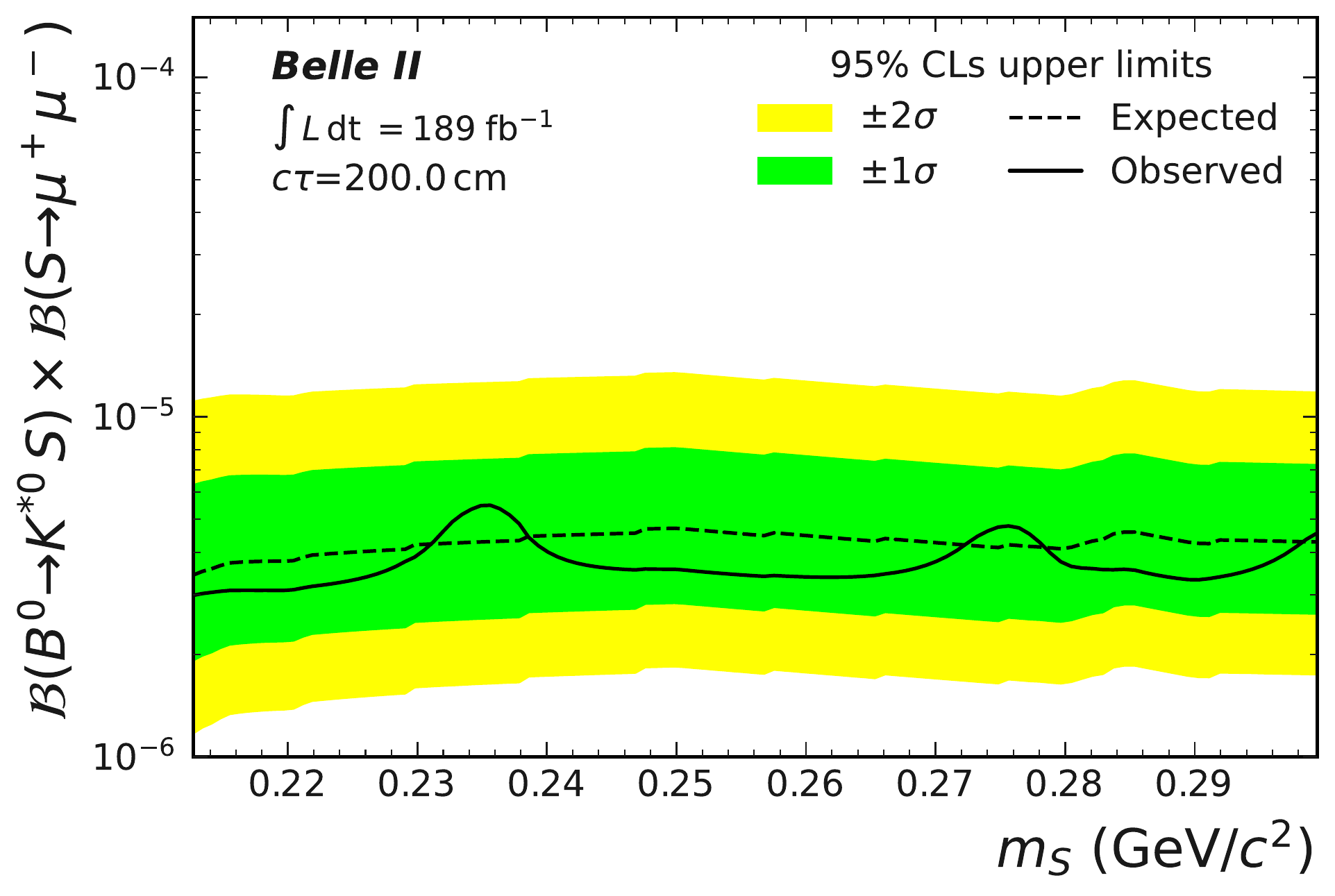}%
}
\subfigure[$\Bz\to \Kstarz(\to K^+\pi^-) S, S\to \mu^+\mu^-$, \newline lifetime of $c\tau=400\cm$.]{
  \label{subfit:brazil:Kstar_mu_2:G}%
  \includegraphics[width=0.31\textwidth]{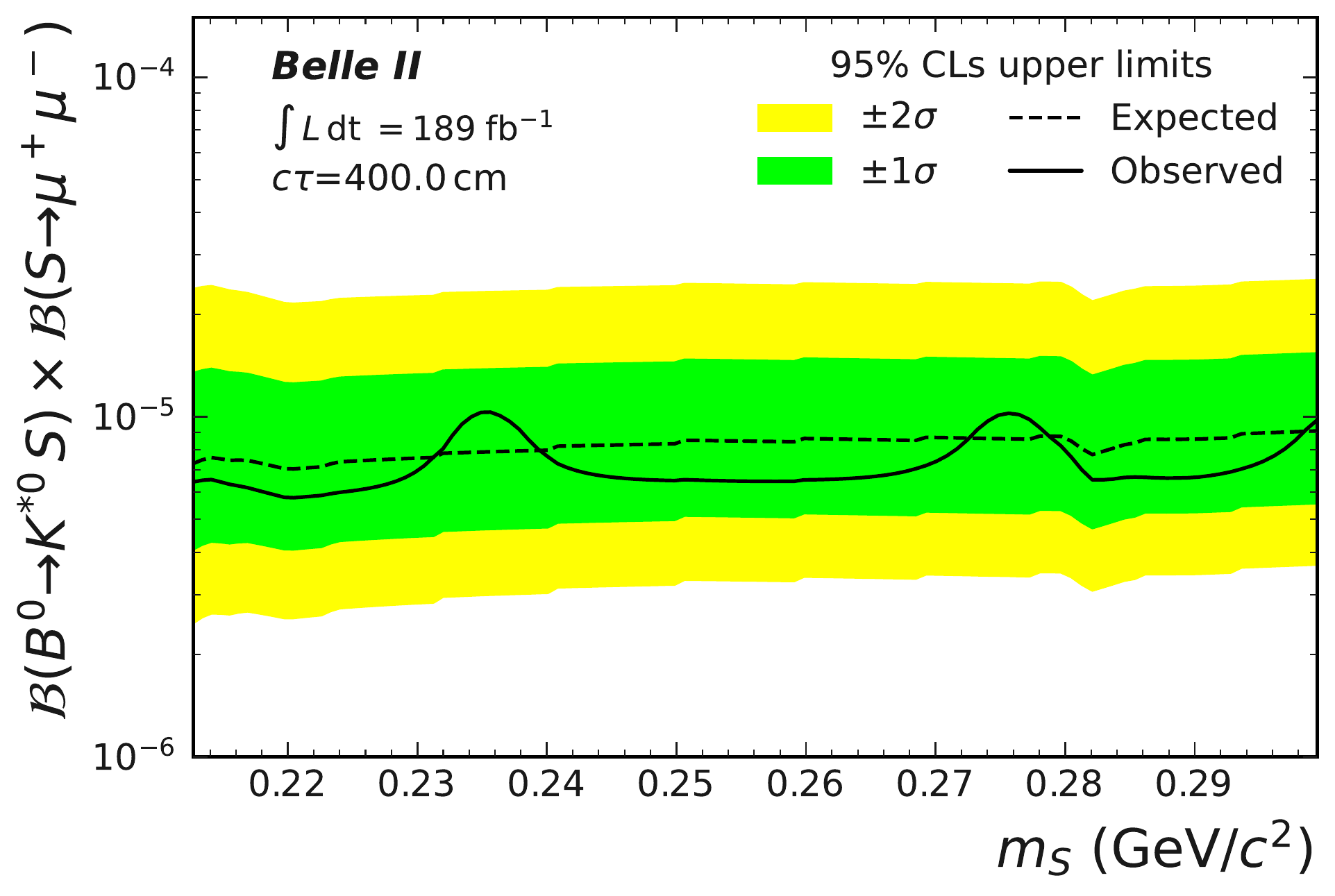}%
}
\caption{Expected and observed limits on the  product of branching fractions $\mathcal{B}(B^0\to \Kstarz(\to K^+\pi^-) S) \times \mathcal{B}(S\to \mu^+\mu^-)$ for lifetimes \hbox{$5 < c\tau < 400\,\cm$}.}\label{subfit:brazil:Kstar_mu_2}
\end{figure*}


\begin{figure*}[ht]%
\subfigure[$B^+\to K^+S, S\to \pi^+\pi^-$, \newline lifetime of $c\tau=0.001\cm$.]{%
  \label{subfit:brazil:Kp_pi_1:A}%
  \includegraphics[width=0.31\textwidth]{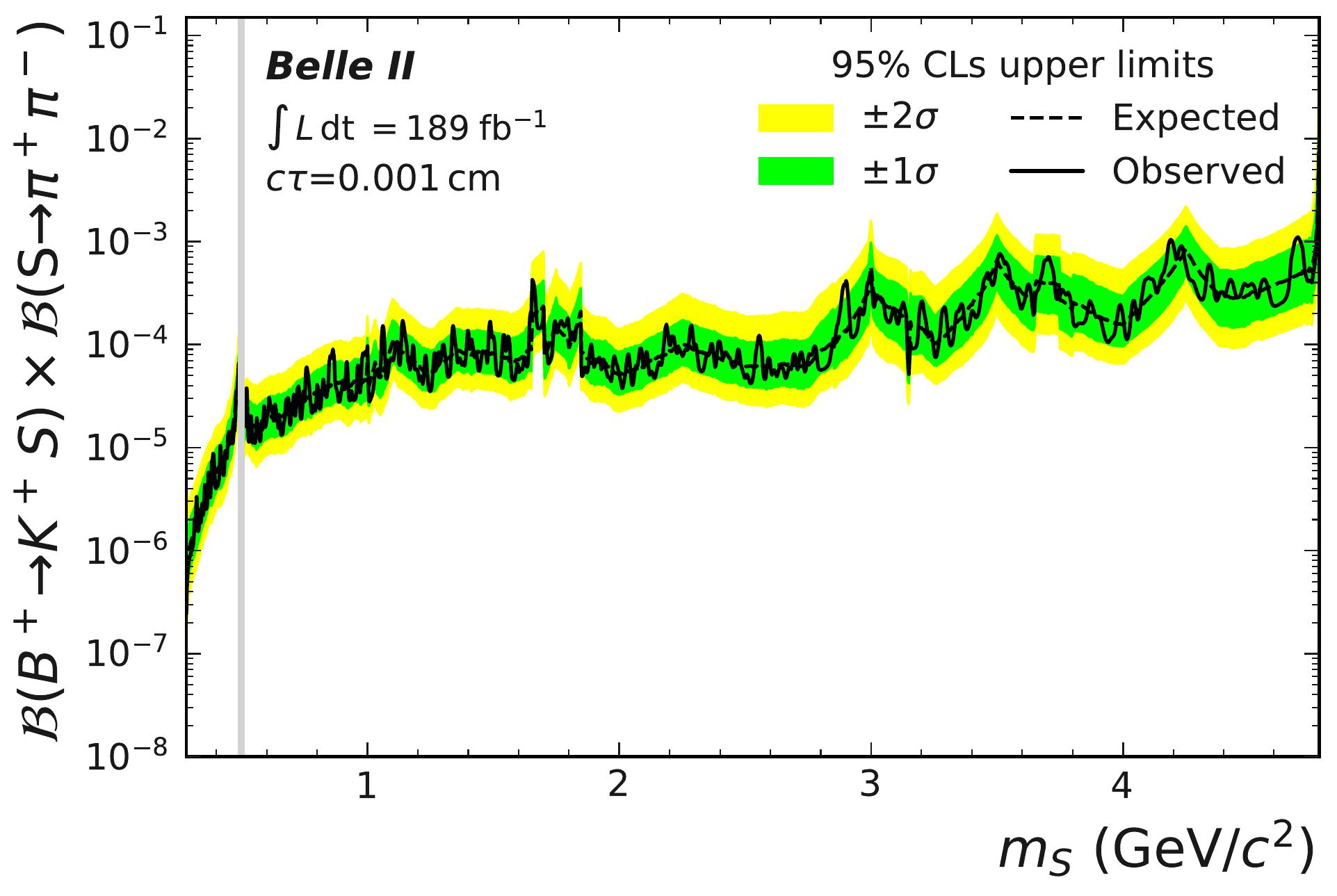}%
}%
\hspace*{\fill}
\subfigure[$B^+\to K^+S, S\to \pi^+\pi^-$, \newline lifetime of $c\tau=0.003\cm$.]{
  \label{subfit:brazil:Kp_pi_1:B}%
  \includegraphics[width=0.31\textwidth]{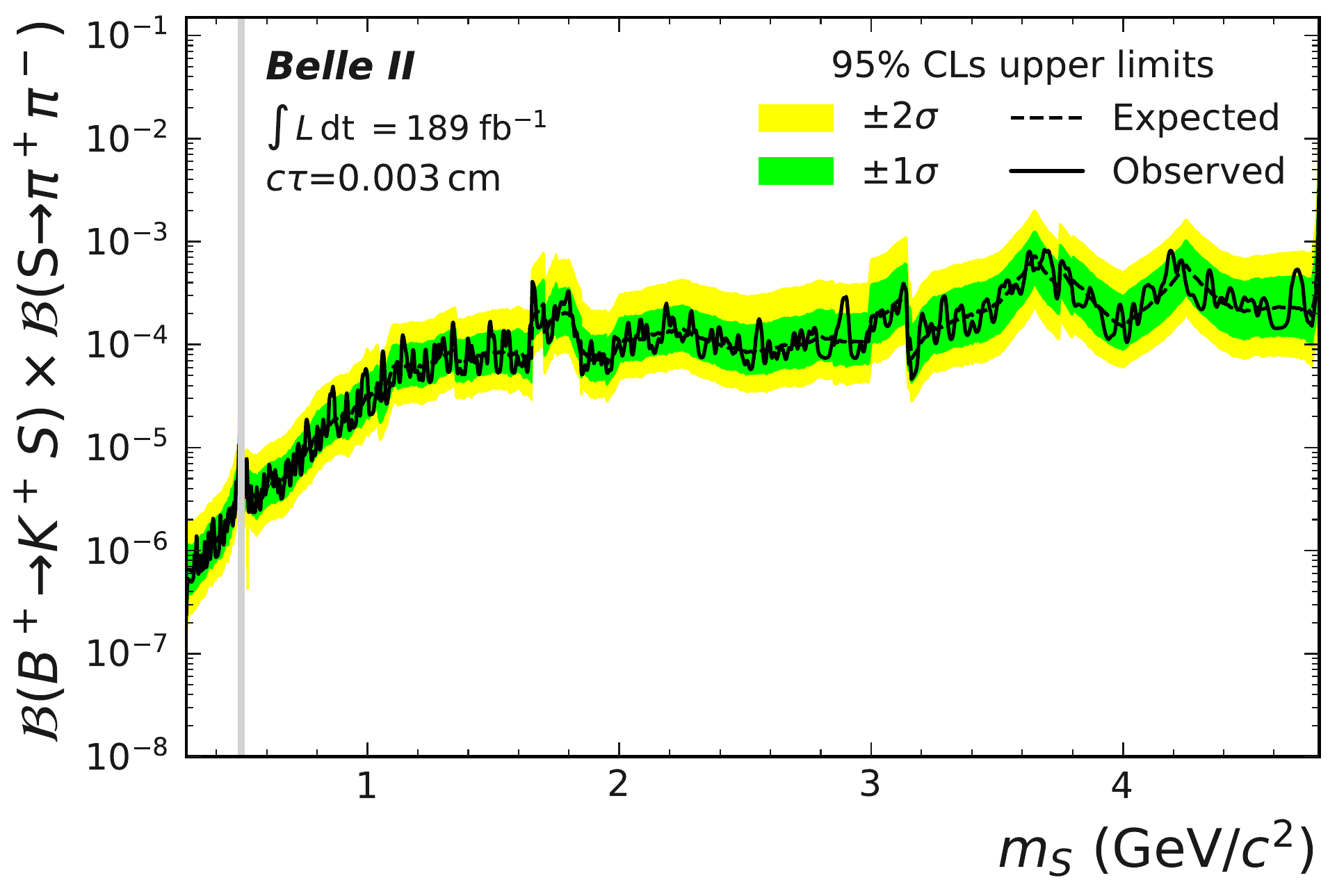}%
}%
\hspace*{\fill}
\subfigure[$B^+\to K^+S, S\to \pi^+\pi^-$, \newline lifetime of $c\tau=0.005\cm$.]{
  \label{subfit:brazil:Kp_pi_1:C}%
  \includegraphics[width=0.31\textwidth]{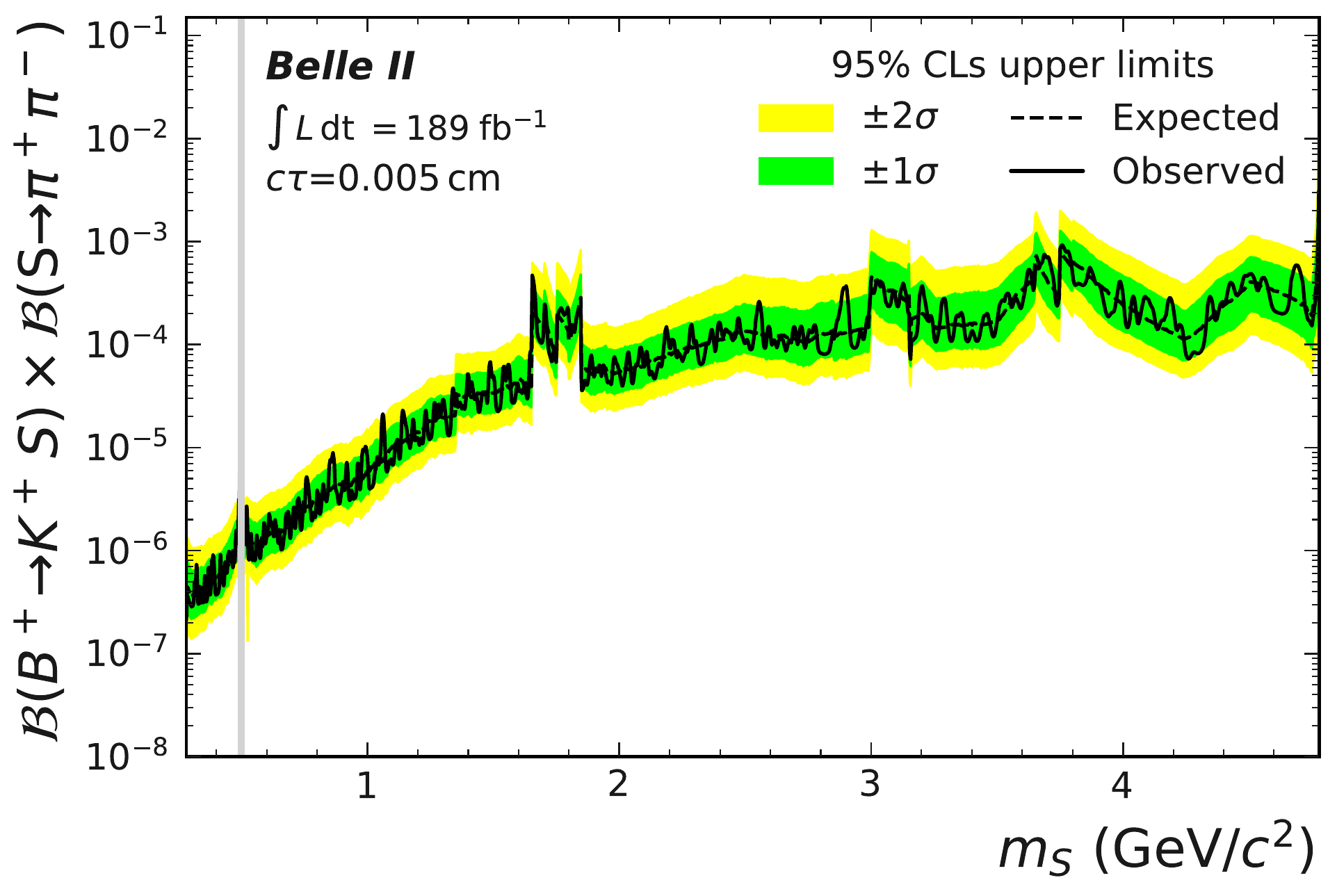}%
}
\subfigure[$B^+\to K^+S, S\to \pi^+\pi^-$, \newline lifetime of $c\tau=0.007\cm$.]{%
  \label{subfit:brazil:Kp_pi_1:D}%
  \includegraphics[width=0.31\textwidth]{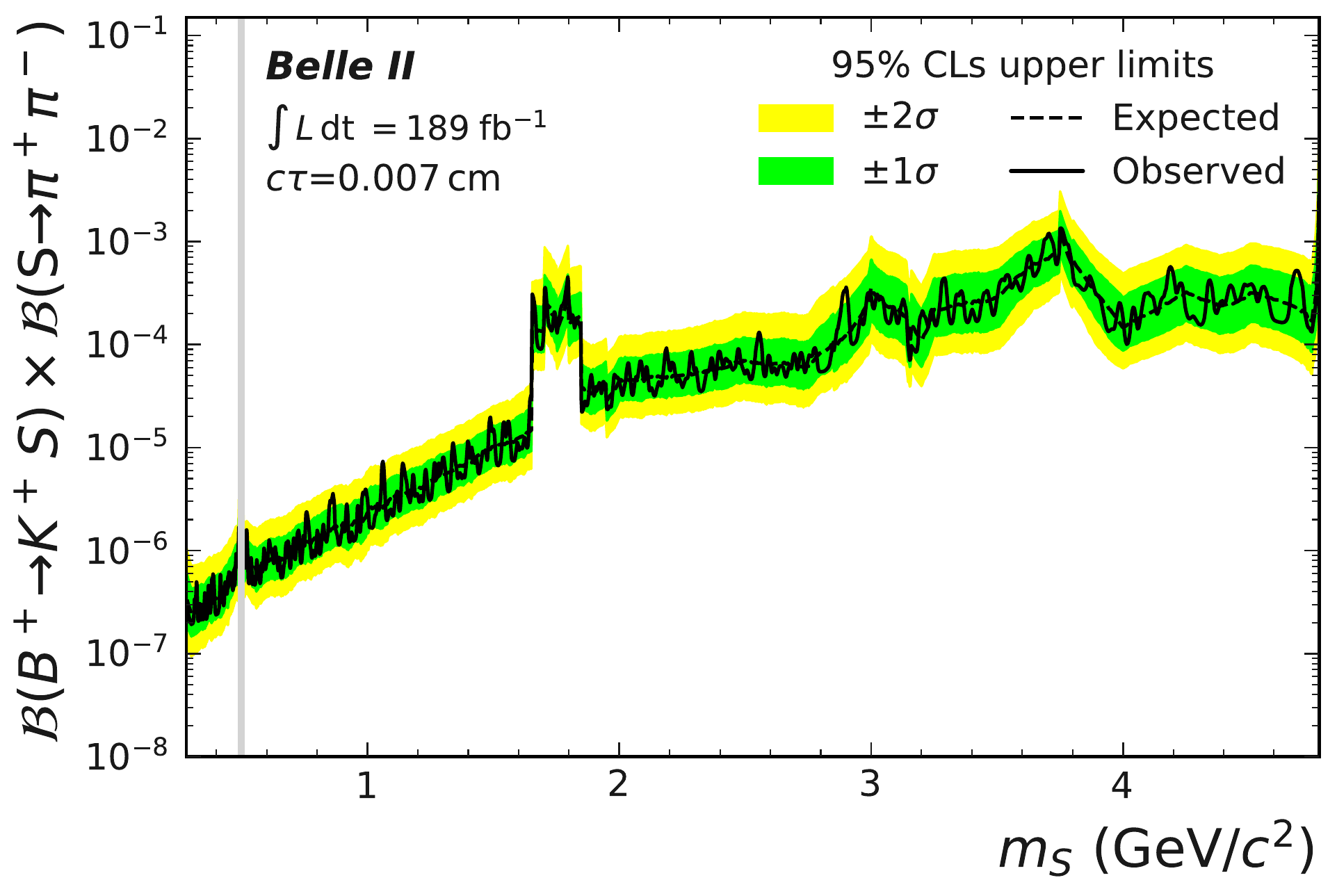}%
}%
\hspace*{\fill}
\subfigure[$B^+\to K^+S, S\to \pi^+\pi^-$, \newline lifetime of $c\tau=0.01\cm$.]{
  \label{subfit:brazil:Kp_pi_1:E}%
  \includegraphics[width=0.31\textwidth]{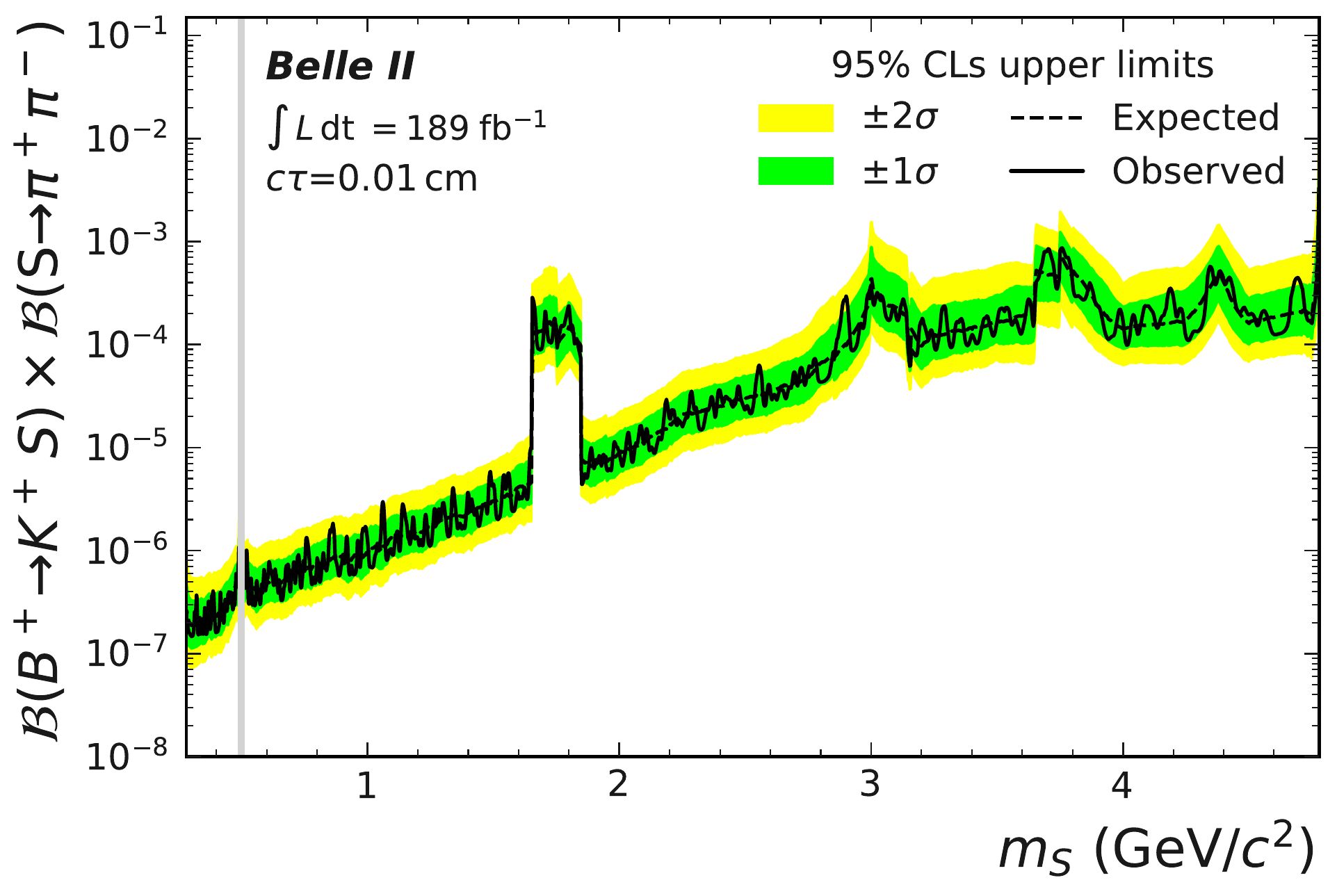}%
}%
\hspace*{\fill}
\subfigure[$B^+\to K^+S, S\to \pi^+\pi^-$, \newline lifetime of $c\tau=0.025\cm$.]{
  \label{subfit:brazil:Kp_pi_1:F}%
  \includegraphics[width=0.31\textwidth]{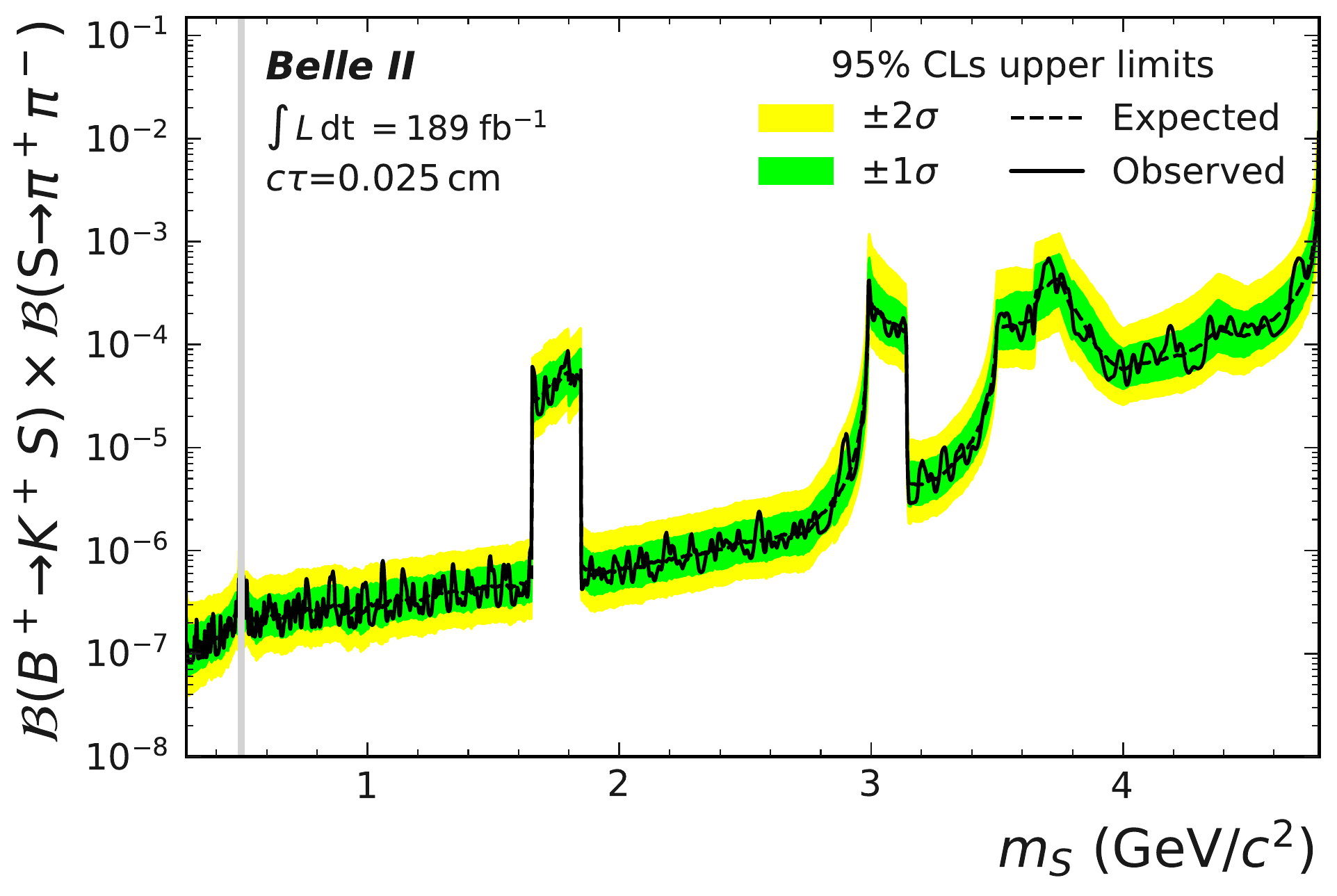}%
}
\subfigure[$B^+\to K^+S, S\to \pi^+\pi^-$, \newline lifetime of $c\tau=0.05\cm$.]{%
  \label{subfit:brazil:Kp_pi_1:G}%
  \includegraphics[width=0.31\textwidth]{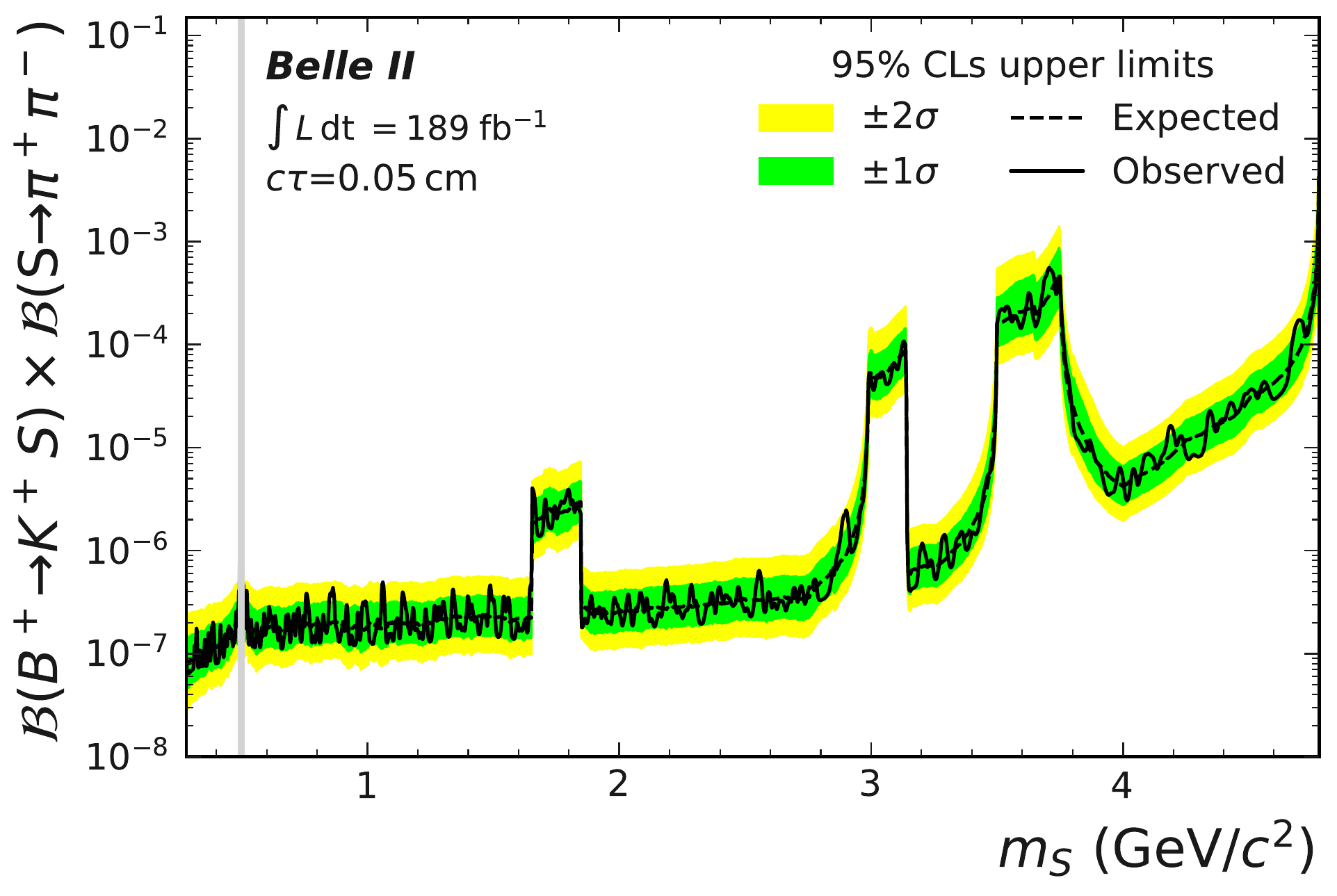}%
}%
\hspace*{\fill}
\subfigure[$B^+\to K^+S, S\to \pi^+\pi^-$, \newline lifetime of $c\tau=0.100\cm$.]{
  \label{subfit:brazil:Kp_pi_1:H}%
  \includegraphics[width=0.31\textwidth]{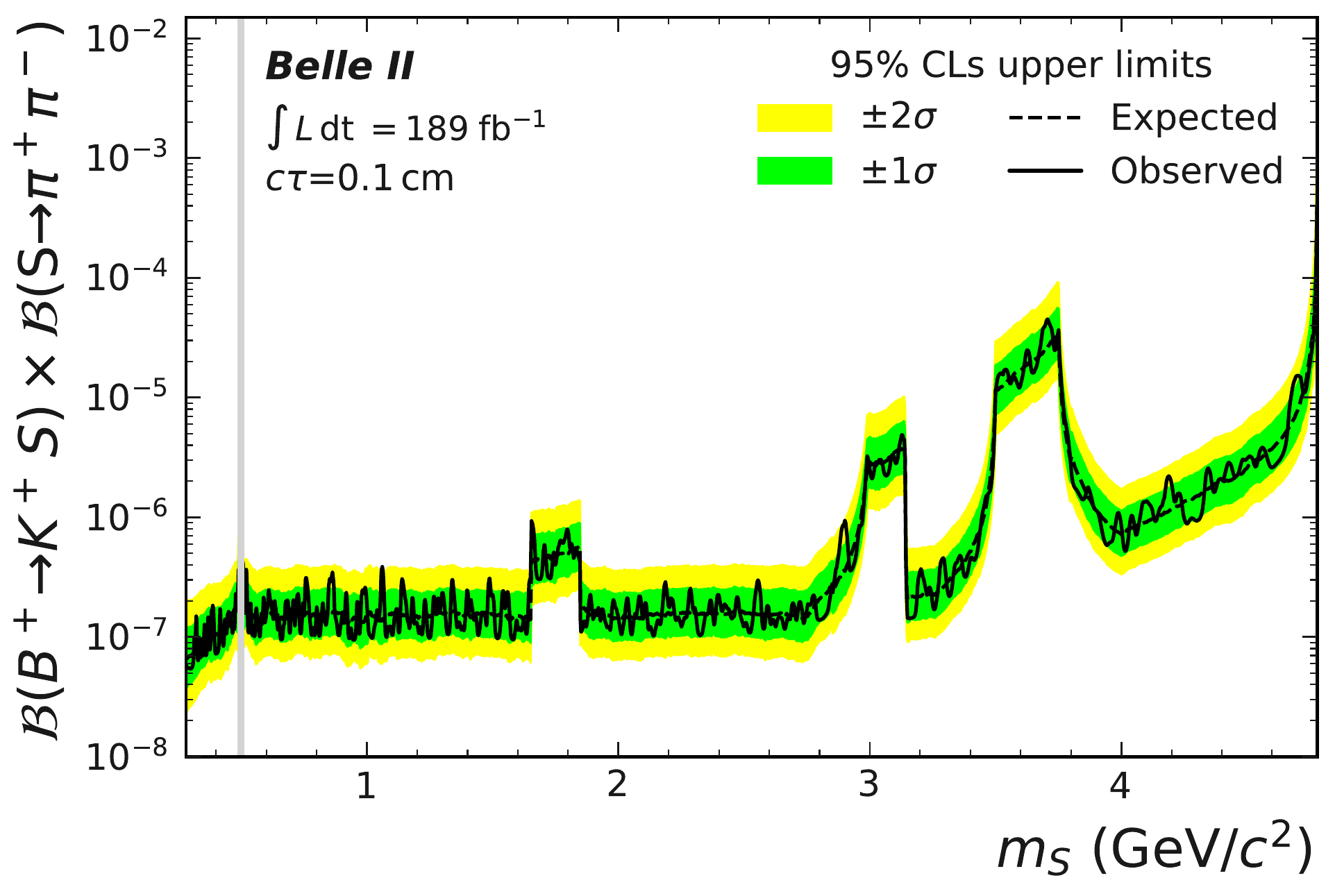}%
}%
\hspace*{\fill}
\subfigure[$B^+\to K^+S, S\to \pi^+\pi^-$, \newline lifetime of $c\tau=0.25\cm$.]{
  \label{subfit:brazil:Kp_pi_1:I}%
  \includegraphics[width=0.31\textwidth]{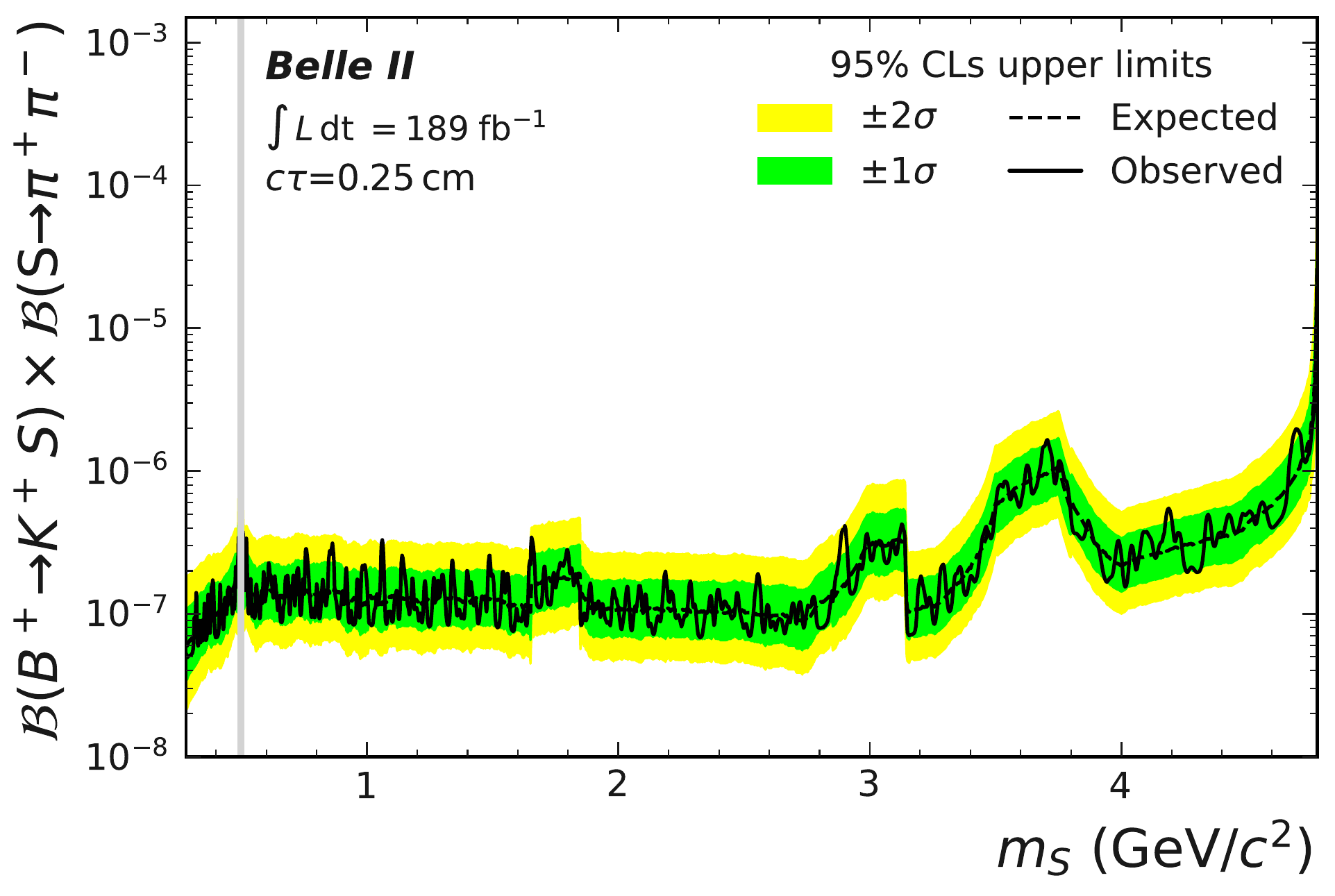}%
}
\subfigure[$B^+\to K^+S, S\to \pi^+\pi^-$, \newline lifetime of $c\tau=0.5\cm$.]{%
  \label{subfit:brazil:Kp_pi_1:J}%
  \includegraphics[width=0.31\textwidth]{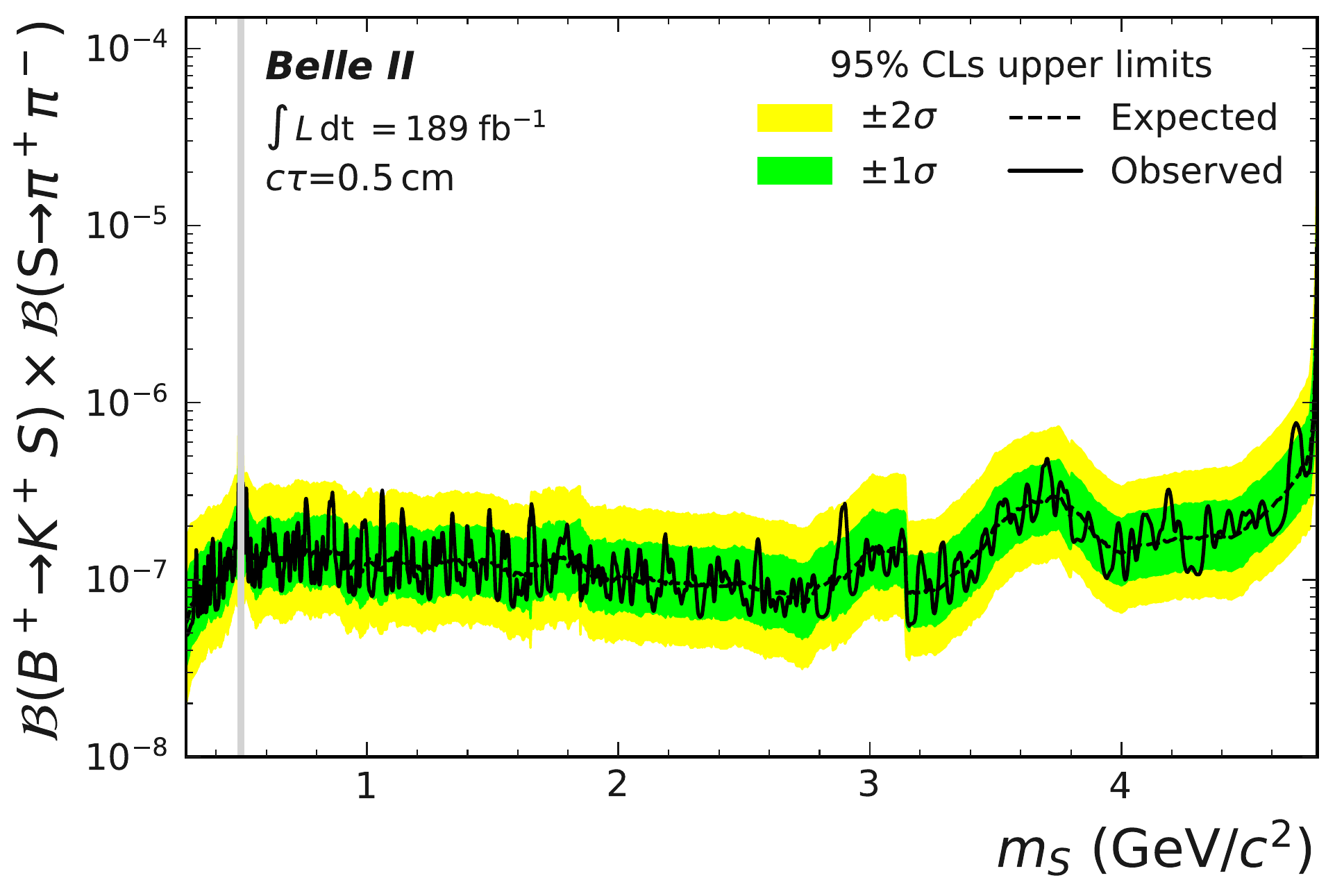}%
}%
\hspace*{\fill}
\subfigure[$B^+\to K^+S, S\to \pi^+\pi^-$, \newline lifetime of $c\tau=1\cm$.]{
  \label{subfit:brazil:Kp_pi_1:K}%
  \includegraphics[width=0.31\textwidth]{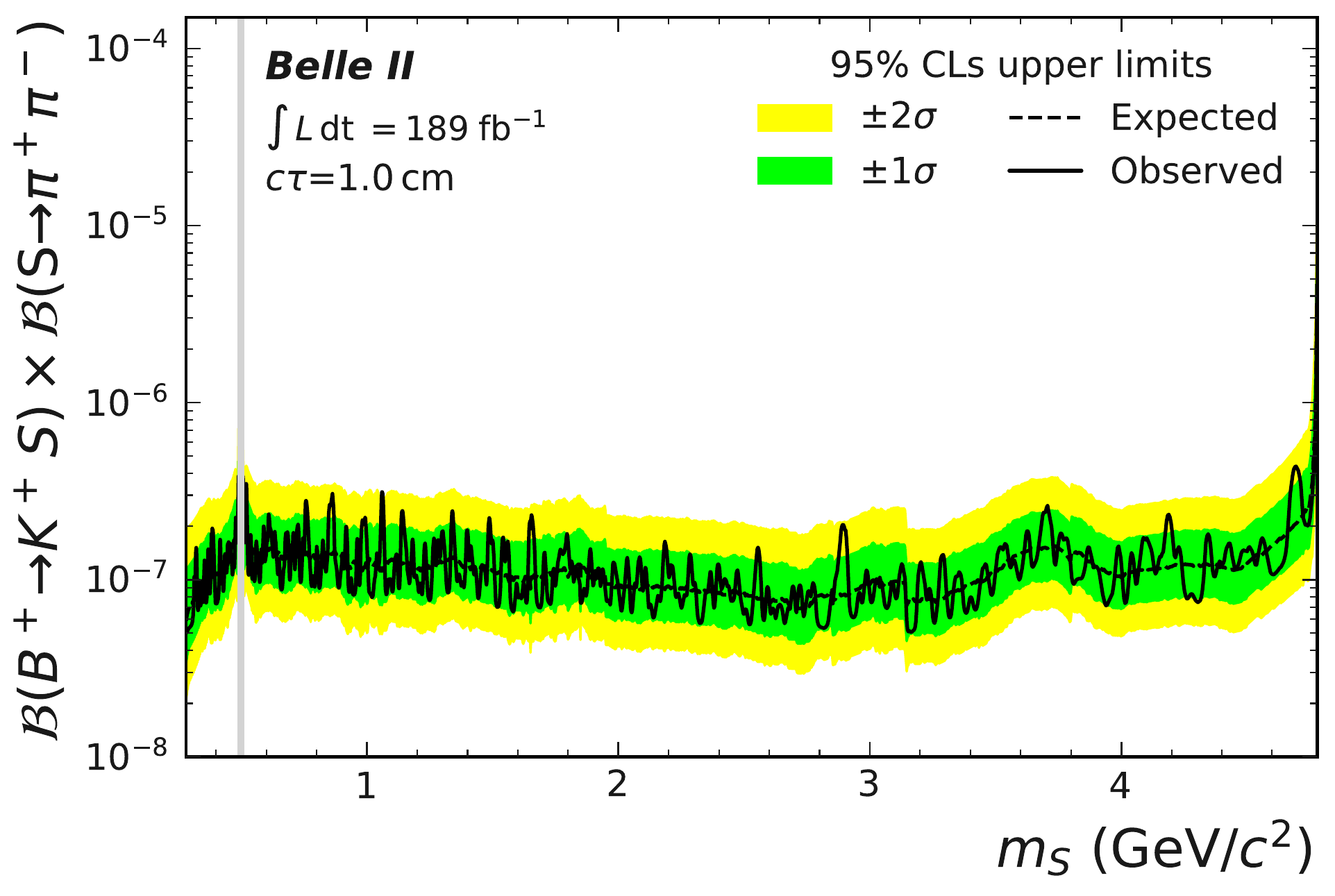}%
}%
\hspace*{\fill}
\subfigure[$B^+\to K^+S, S\to \pi^+\pi^-$, \newline lifetime of $c\tau=2.5\cm$.]{
  \label{subfit:brazil:Kp_pi_1:L}%
  \includegraphics[width=0.31\textwidth]{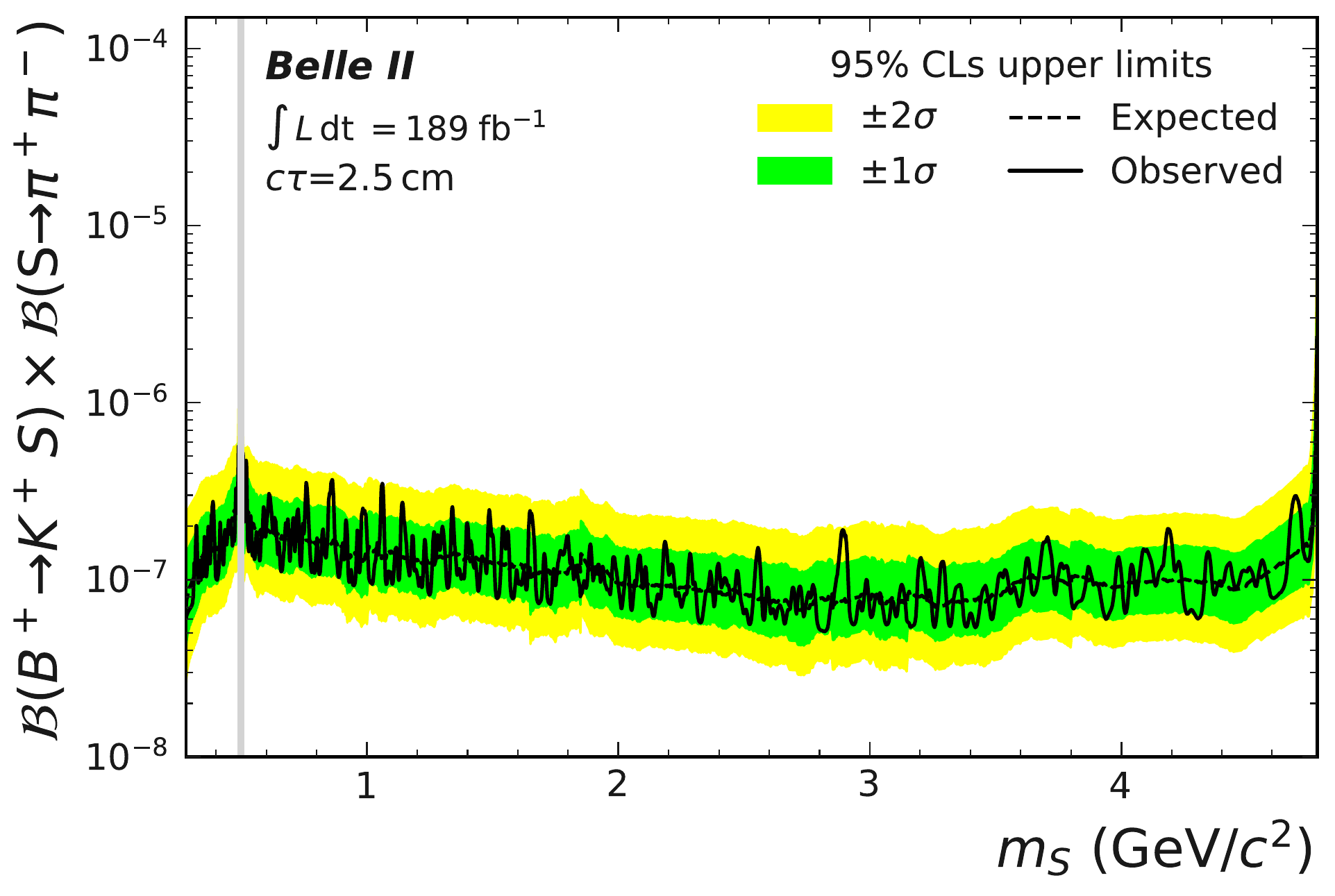}%
}
\caption{Expected and observed limits on the  product of branching fractions $\mathcal{B}(B^+\to K^+S) \times \mathcal{B}(S\to \pi^+\pi^-)$ for lifetimes \hbox{$0.001 < c\tau < 2.5\,\cm$}. The region corresponding to the fully-vetoed $\KS$ for $S\to\pi^+\pi^-$ is marked in gray.}\label{subfit:brazil:Kp_pi_1}
\end{figure*}

\begin{figure*}[ht]%
\subfigure[$B^+\to K^+S, S\to \pi^+\pi^-$, \newline lifetime of $c\tau=5\cm$.]{%
  \label{subfit:brazil:Kp_pi_2:A}%
  \includegraphics[width=0.31\textwidth]{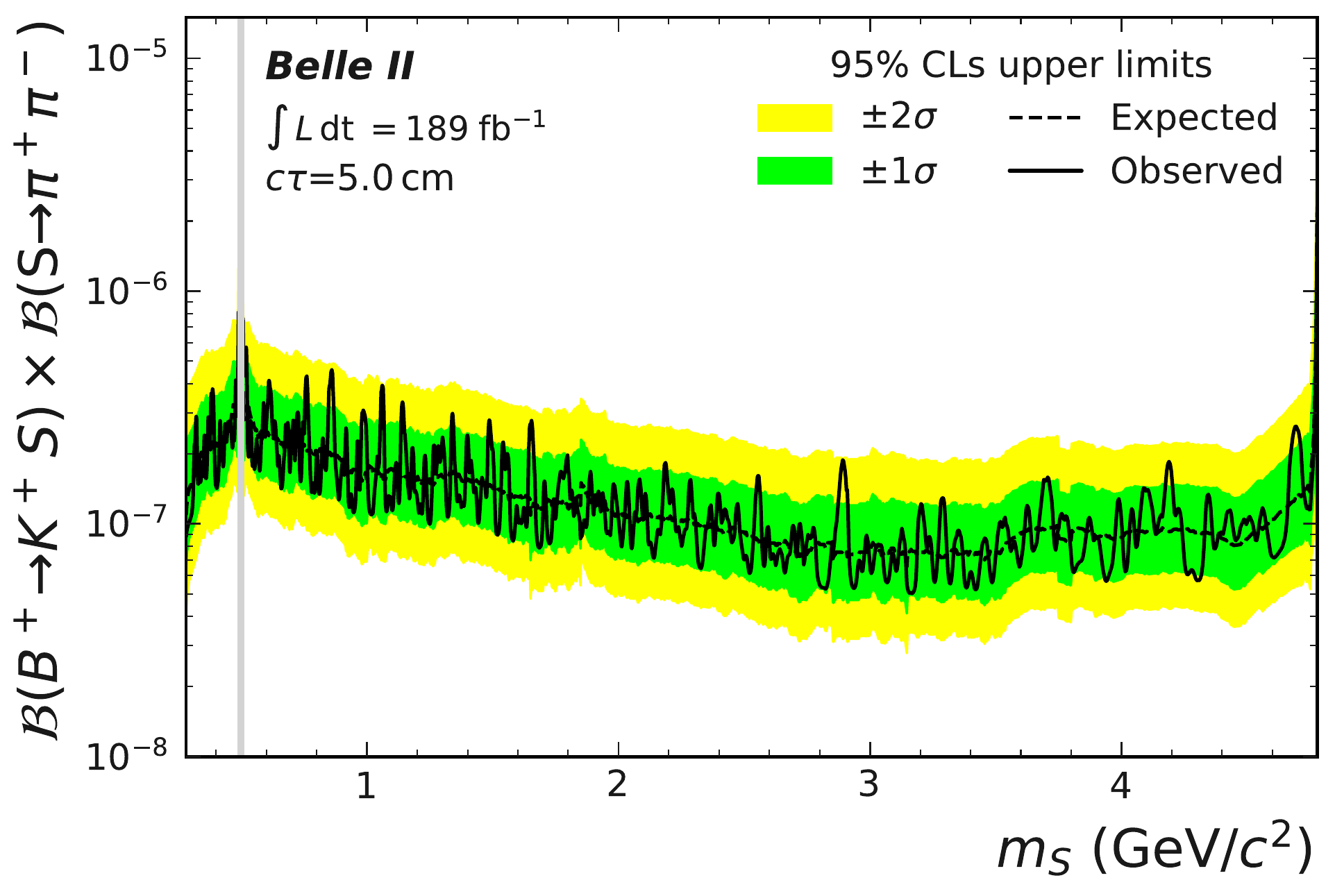}%
}%
\hspace*{\fill}
\subfigure[$B^+\to K^+S, S\to \pi^+\pi^-$, \newline lifetime of $c\tau=10\cm$.]{
  \label{subfit:brazil:Kp_pi_2:B}%
  \includegraphics[width=0.31\textwidth]{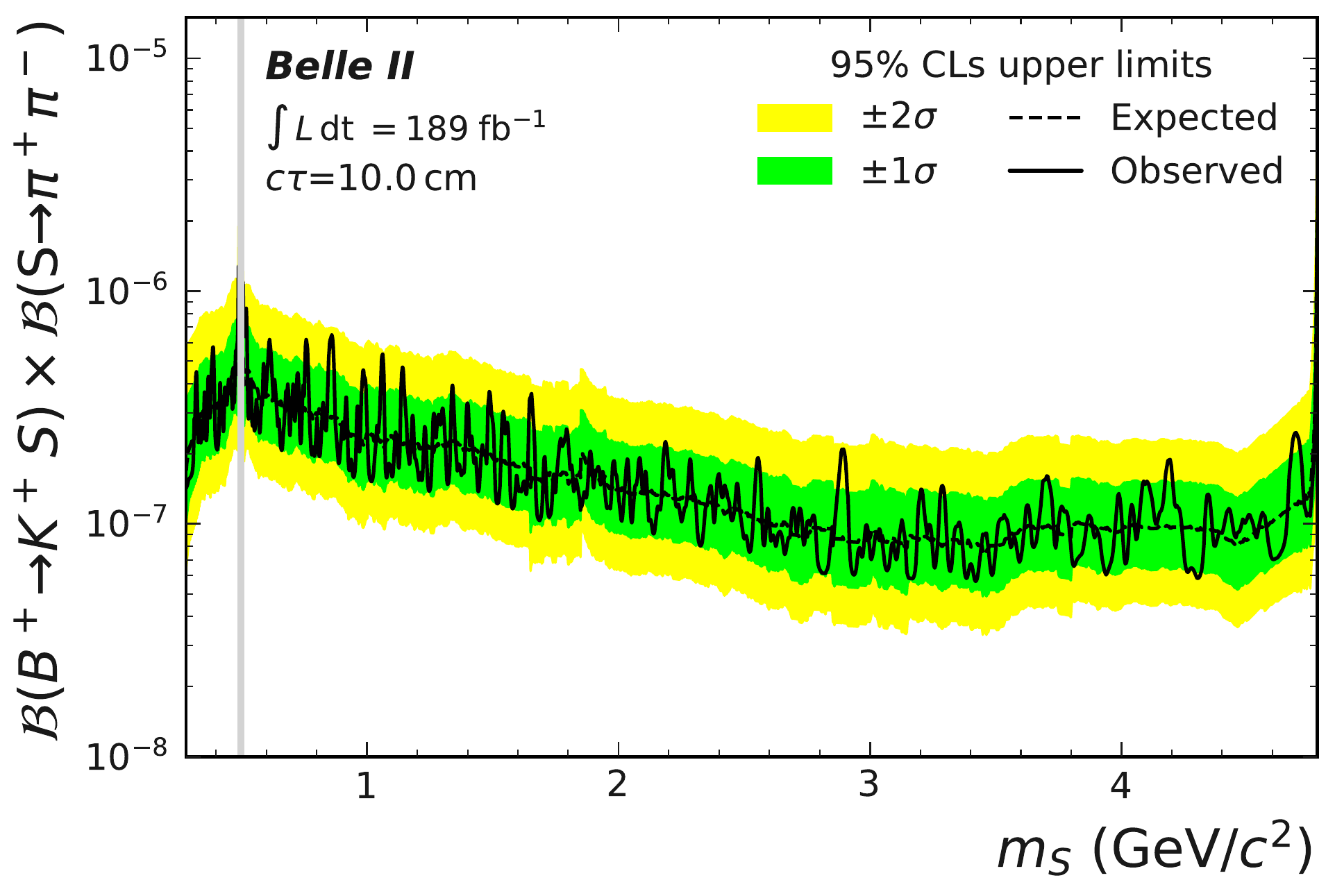}%
}%
\hspace*{\fill}
\subfigure[$B^+\to K^+S, S\to \pi^+\pi^-$, \newline lifetime of $c\tau=25\cm$.]{
  \label{subfit:brazil:Kp_pi_2:C}%
  \includegraphics[width=0.31\textwidth]{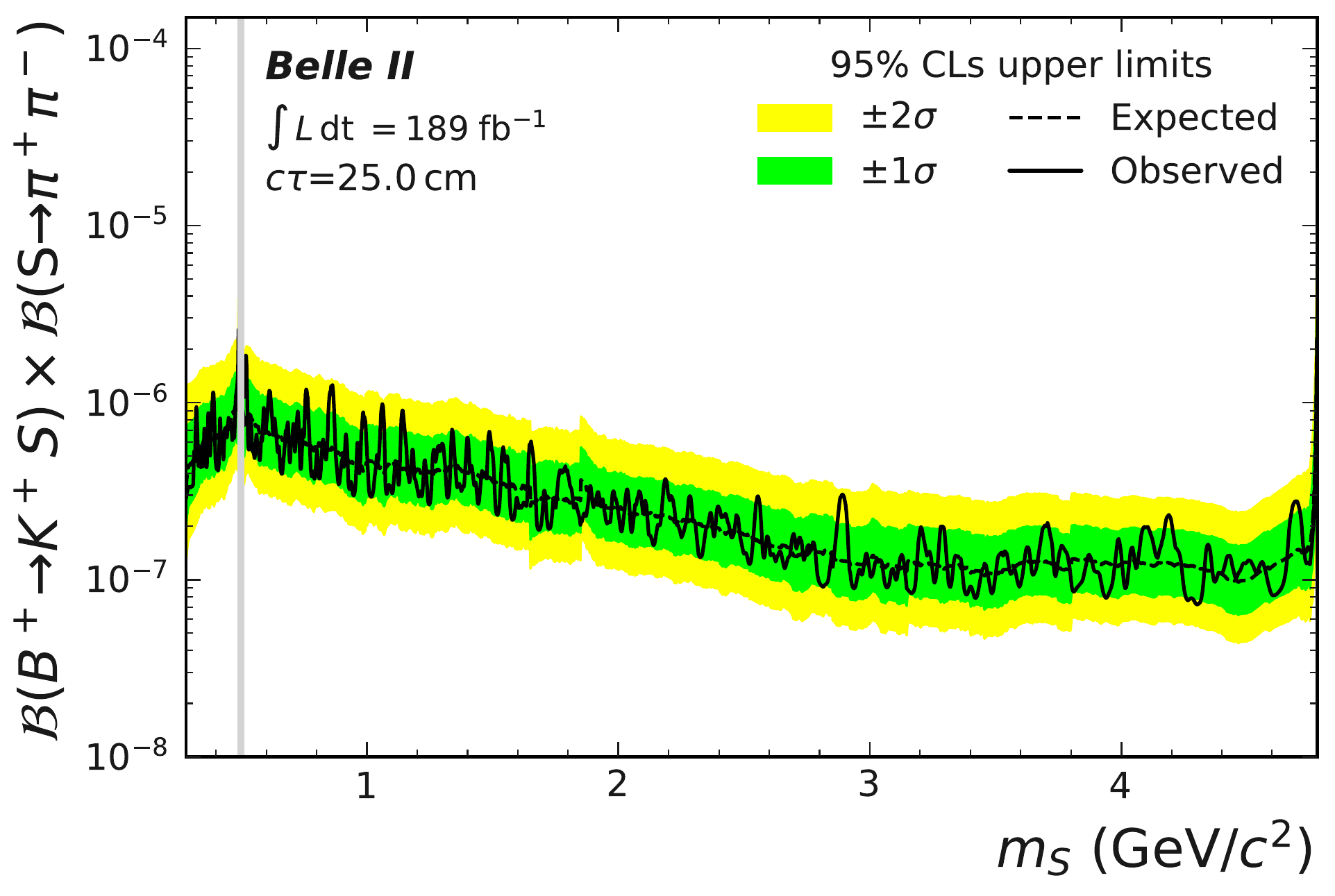}%
}
\subfigure[$B^+\to K^+S, S\to \pi^+\pi^-$, \newline lifetime of $c\tau=50\cm$.]{%
  \label{subfit:brazil:Kp_pi_2:D}%
  \includegraphics[width=0.31\textwidth]{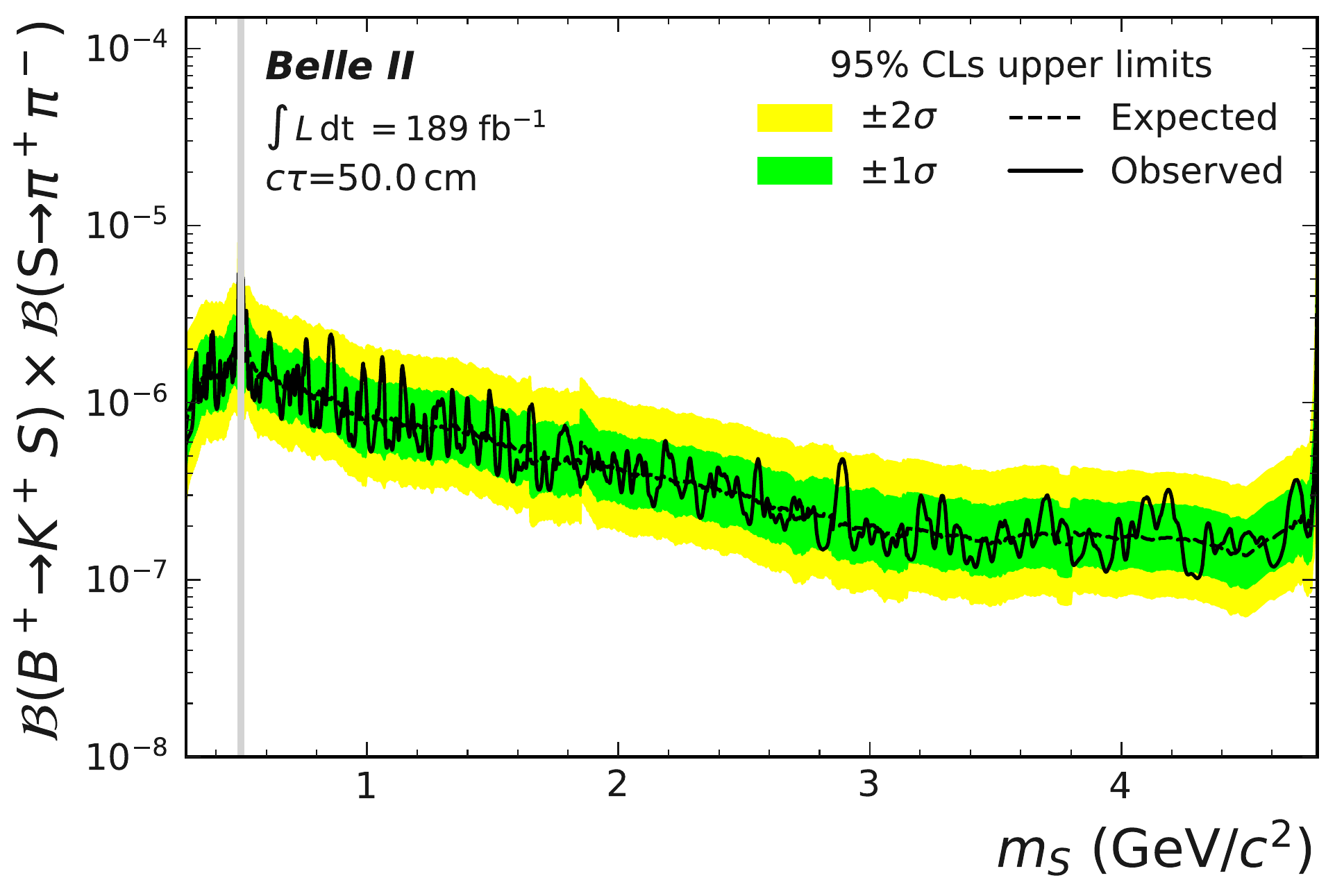}%
}%
\hspace*{\fill}
\subfigure[$B^+\to K^+S, S\to \pi^+\pi^-$, \newline lifetime of $c\tau=100\cm$.]{
  \label{subfit:brazil:Kp_pi_2:E}%
  \includegraphics[width=0.31\textwidth]{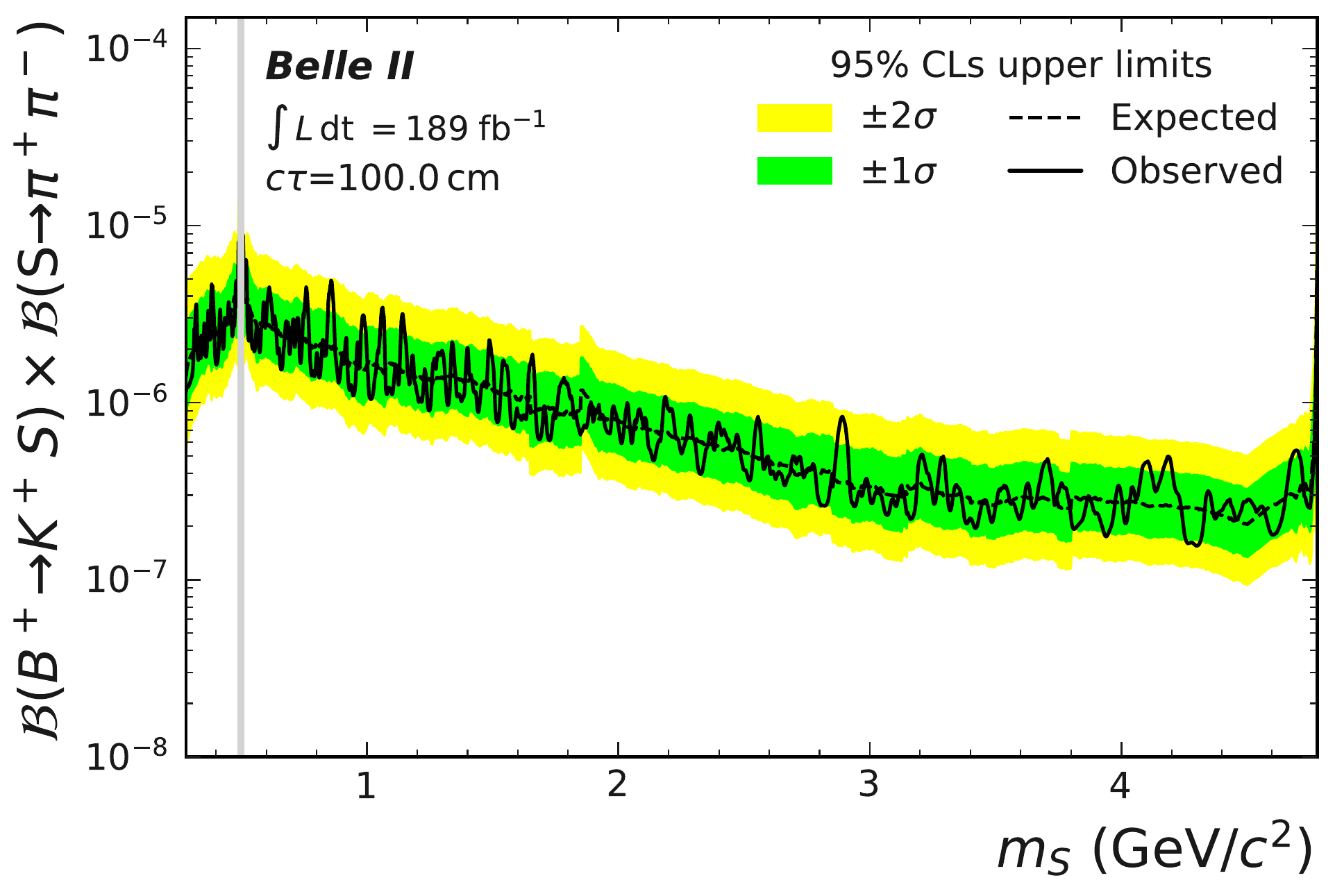}%
}%
\hspace*{\fill}
\subfigure[$B^+\to K^+S, S\to \pi^+\pi^-$, \newline lifetime of $c\tau=200\cm$.]{
  \label{subfit:brazil:Kp_pi_2:F}%
  \includegraphics[width=0.31\textwidth]{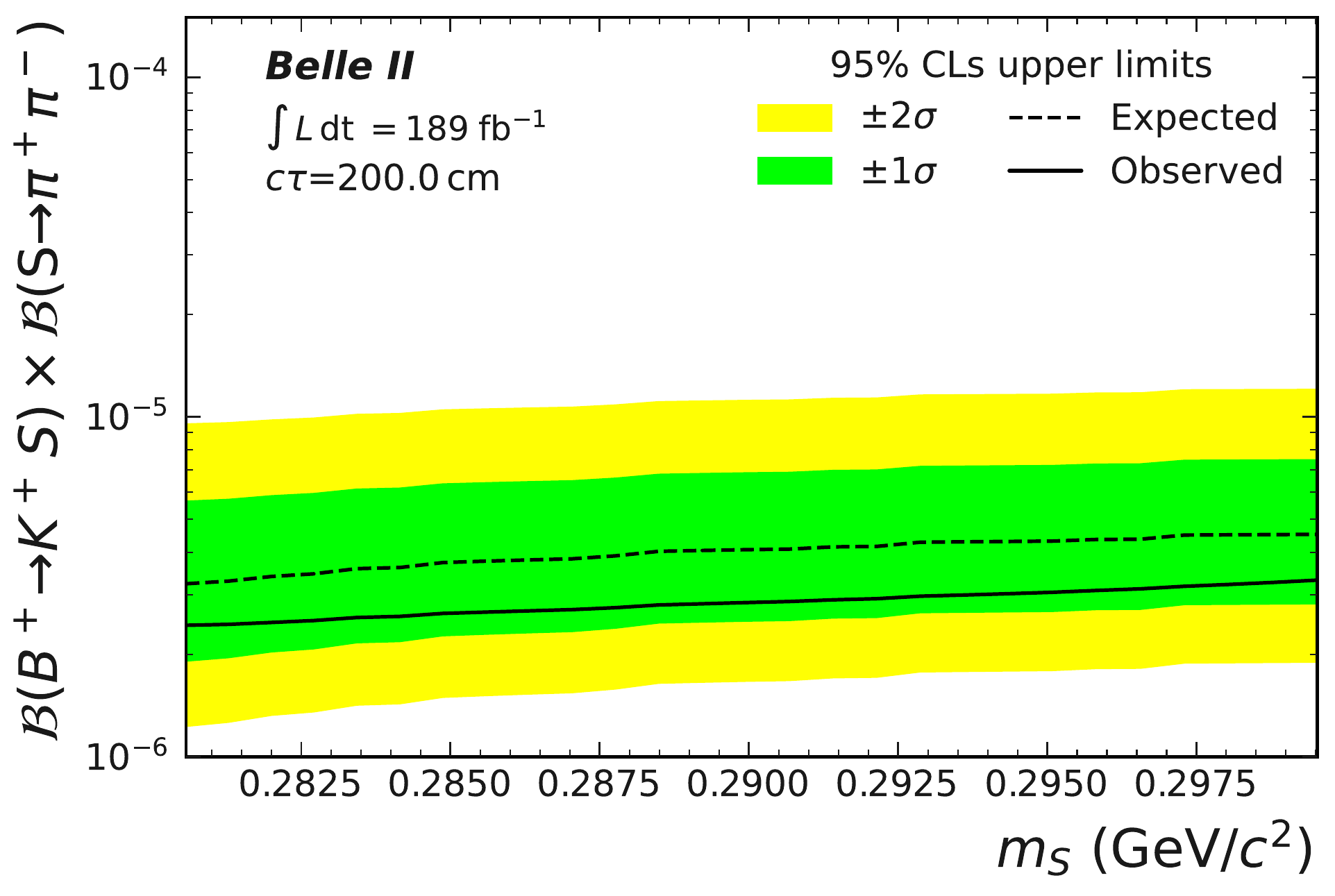}%
}
\subfigure[$B^+\to K^+S, S\to \pi^+\pi^-$, \newline lifetime of $c\tau=400\cm$.]{
  \label{subfit:brazil:Kp_pi_2:G}%
  \includegraphics[width=0.31\textwidth]{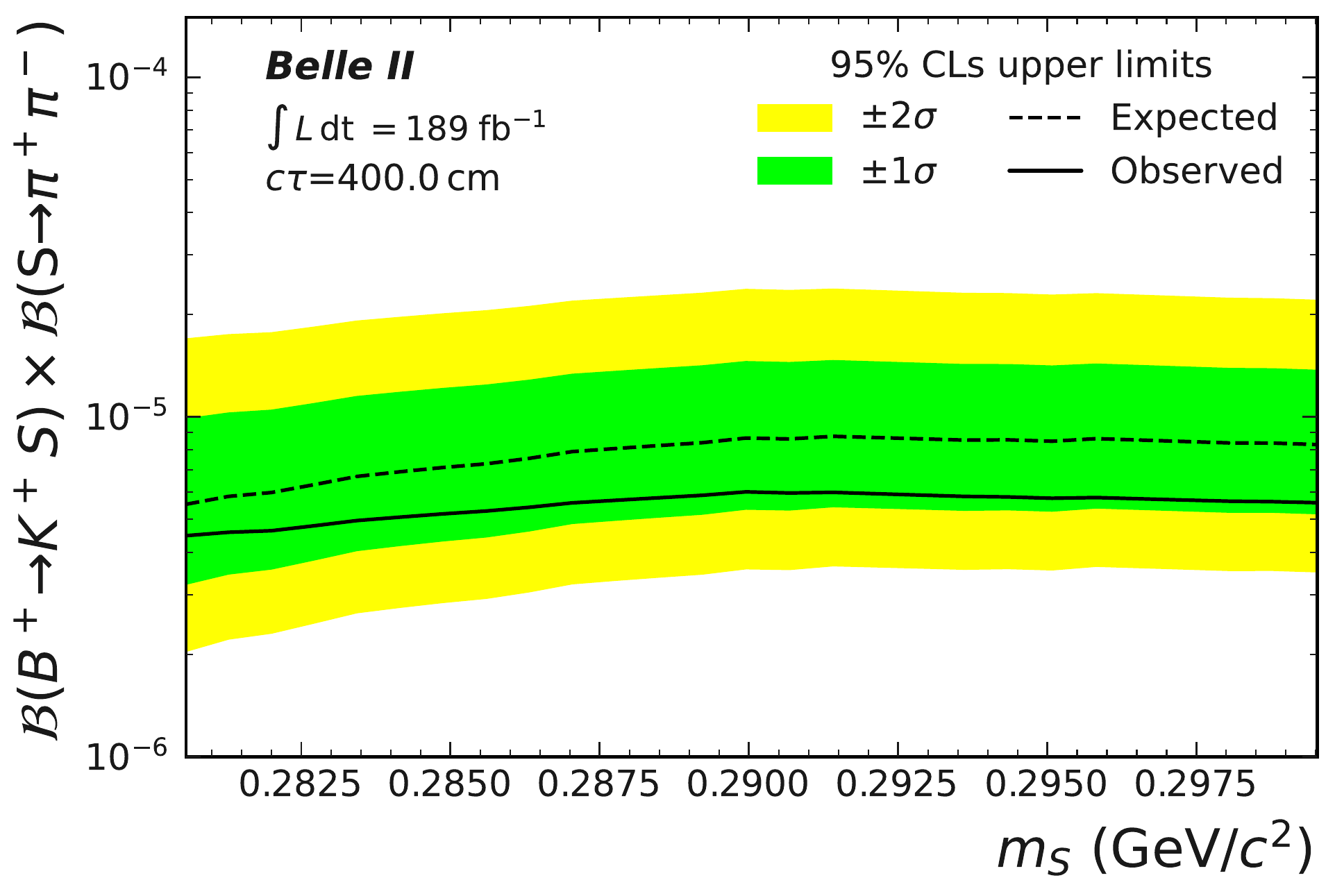}%
}
\caption{Expected and observed limits on the  product of branching fractions $\mathcal{B}(B^+\to K^+S) \times \mathcal{B}(S\to \pi^+\pi^-)$ for \\lifetimes \hbox{$5 < c\tau < 400\,\cm$}. The region corresponding to the fully-vetoed $\KS$ for $S\to\pi^+\pi^-$ is marked in gray.}\label{subfit:brazil:Kp_pi_2}
\end{figure*}

\begin{figure*}[ht]%
\subfigure[$\Bz\to \Kstarz(\to K^+\pi^-) S, S\to \pi^+\pi^-$, \newline lifetime of $c\tau=0.001\cm$.]{%
  \label{subfit:brazil:Kstar_pi_1:A}%
  \includegraphics[width=0.31\textwidth]{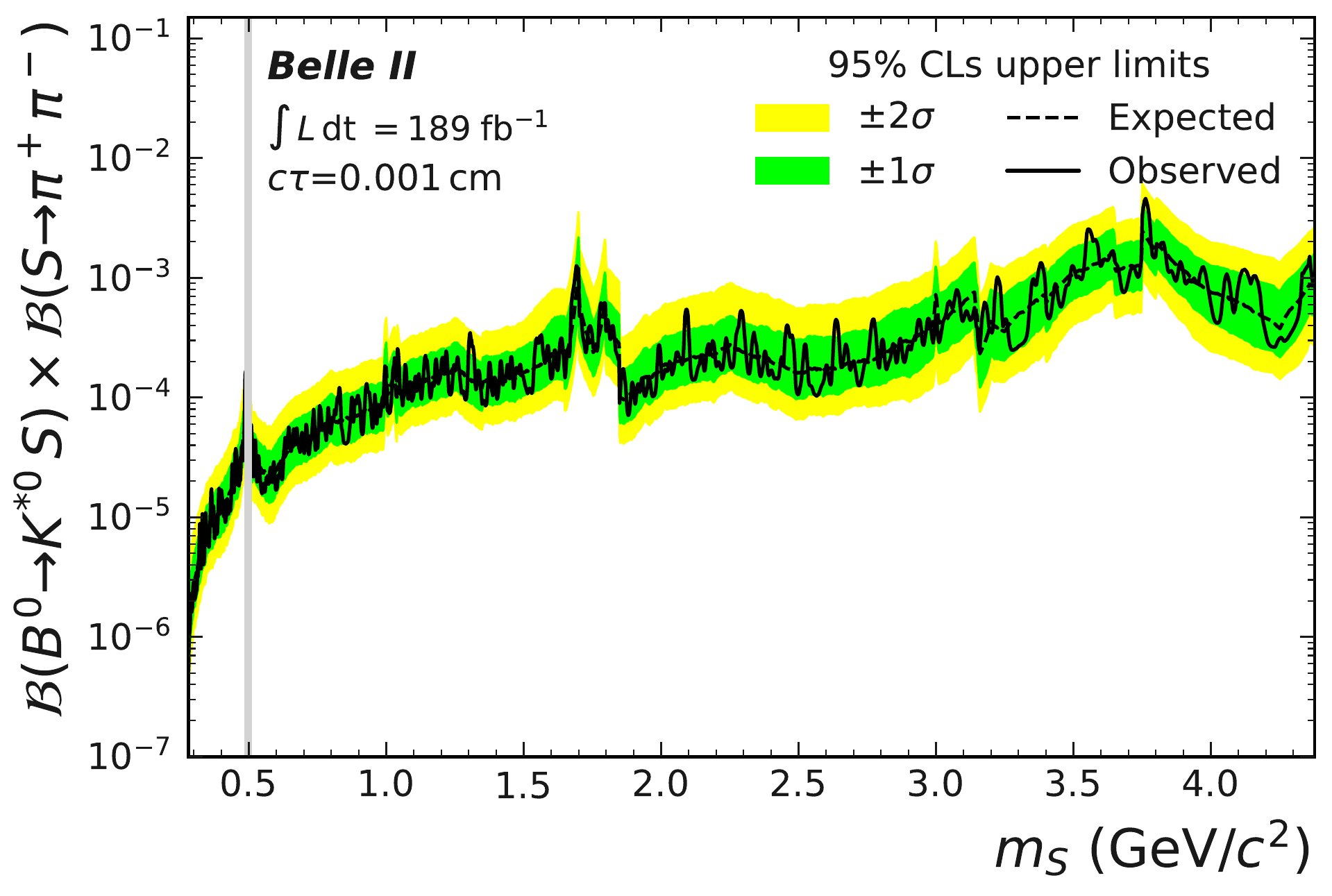}%
}%
\hspace*{\fill}
\subfigure[$\Bz\to \Kstarz(\to K^+\pi^-) S, S\to \pi^+\pi^-$, \newline lifetime of $c\tau=0.003\cm$.]{
  \label{subfit:brazil:Kstar_pi_1:B}%
  \includegraphics[width=0.31\textwidth]{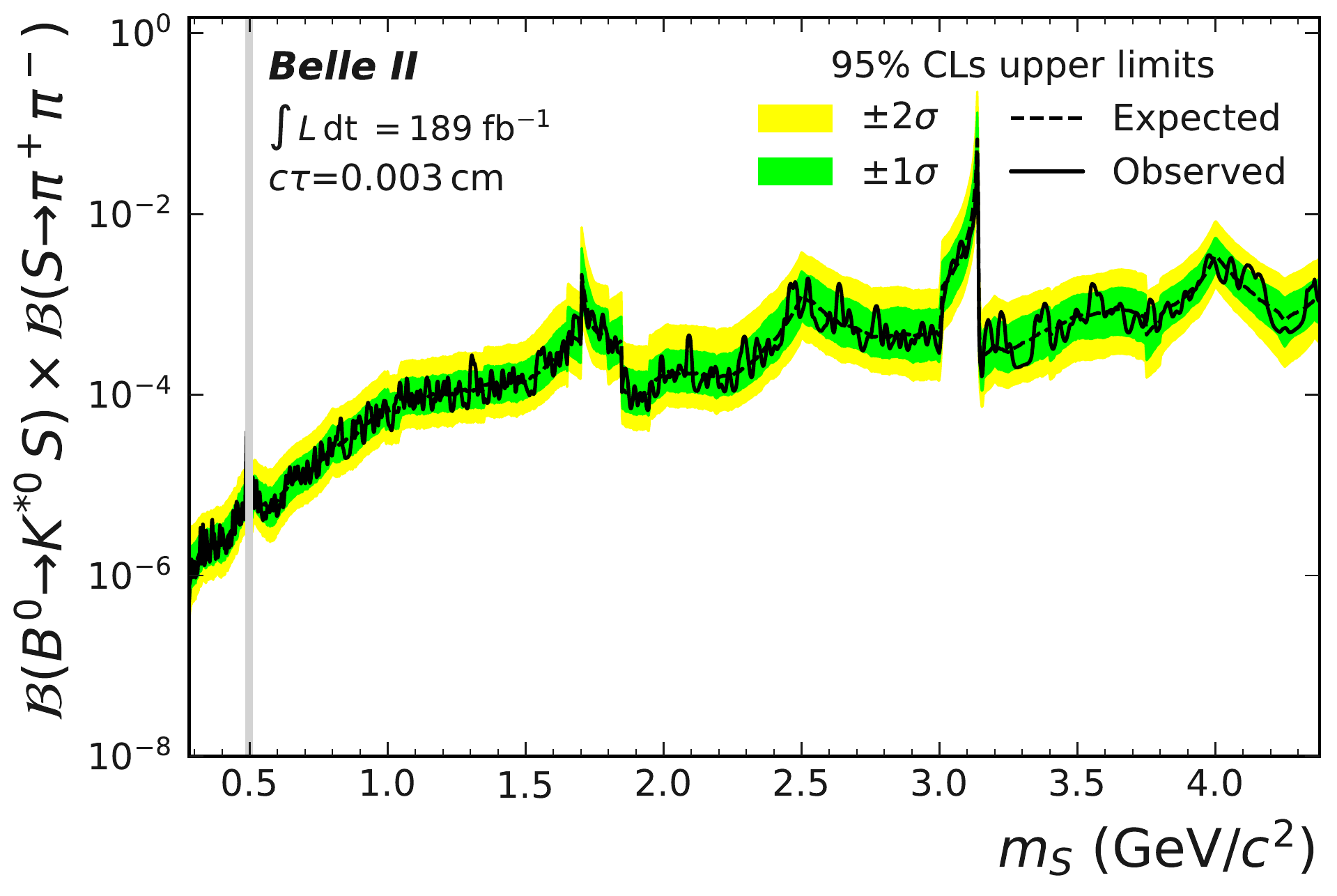}%
}%
\hspace*{\fill}
\subfigure[$\Bz\to \Kstarz(\to K^+\pi^-) S, S\to \pi^+\pi^-$, \newline lifetime of $c\tau=0.005\cm$.]{
  \label{subfit:brazil:Kstar_pi_1:C}%
  \includegraphics[width=0.31\textwidth]{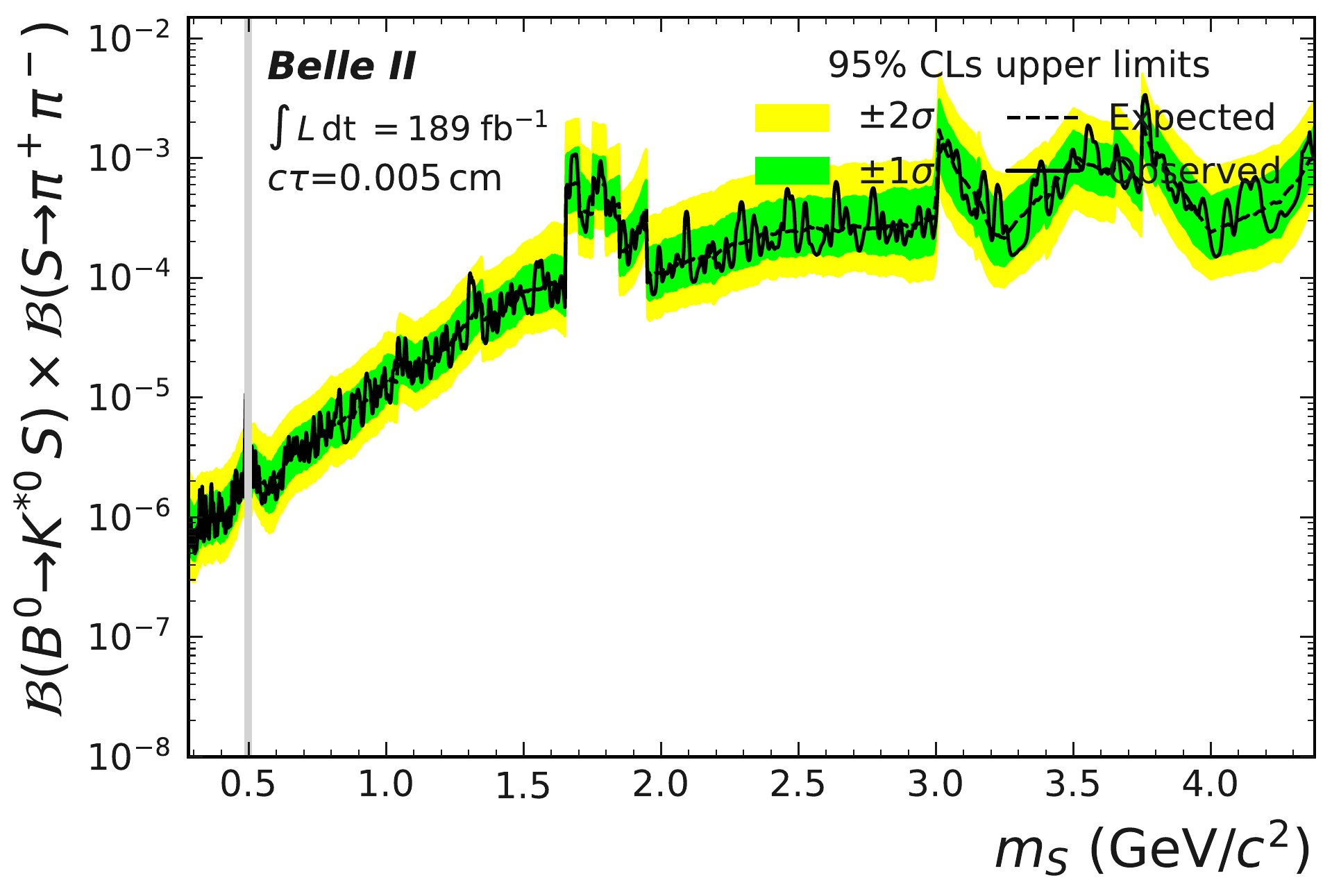}%
}
\subfigure[$\Bz\to \Kstarz(\to K^+\pi^-) S, S\to \pi^+\pi^-$, \newline lifetime of $c\tau=0.007\cm$.]{%
  \label{subfit:brazil:Kstar_pi_1:D}%
  \includegraphics[width=0.31\textwidth]{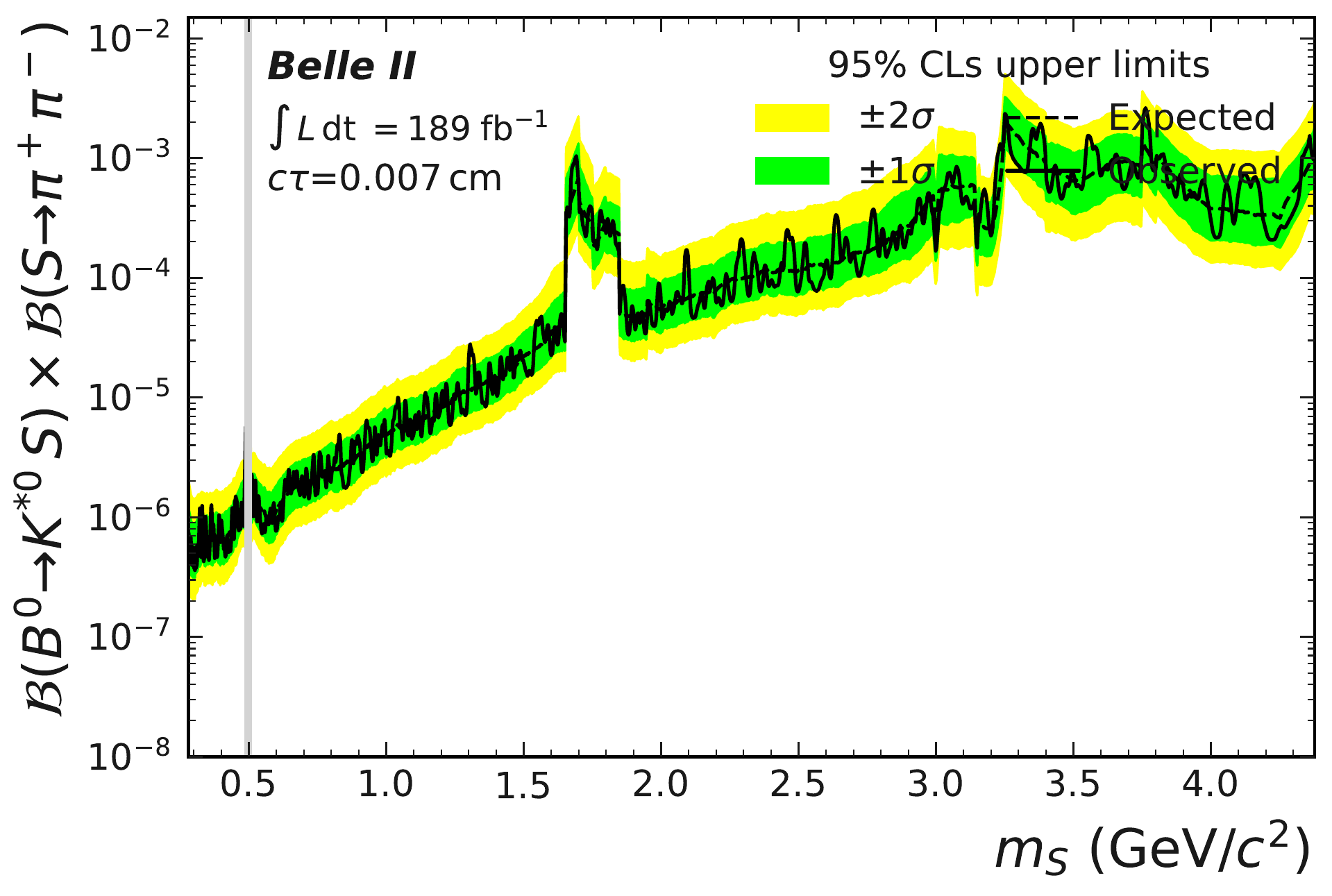}%
}%
\hspace*{\fill}
\subfigure[$\Bz\to \Kstarz(\to K^+\pi^-) S, S\to \pi^+\pi^-$, \newline lifetime of $c\tau=0.01\cm$.]{
  \label{subfit:brazil:Kstar_pi_1:E}%
  \includegraphics[width=0.31\textwidth]{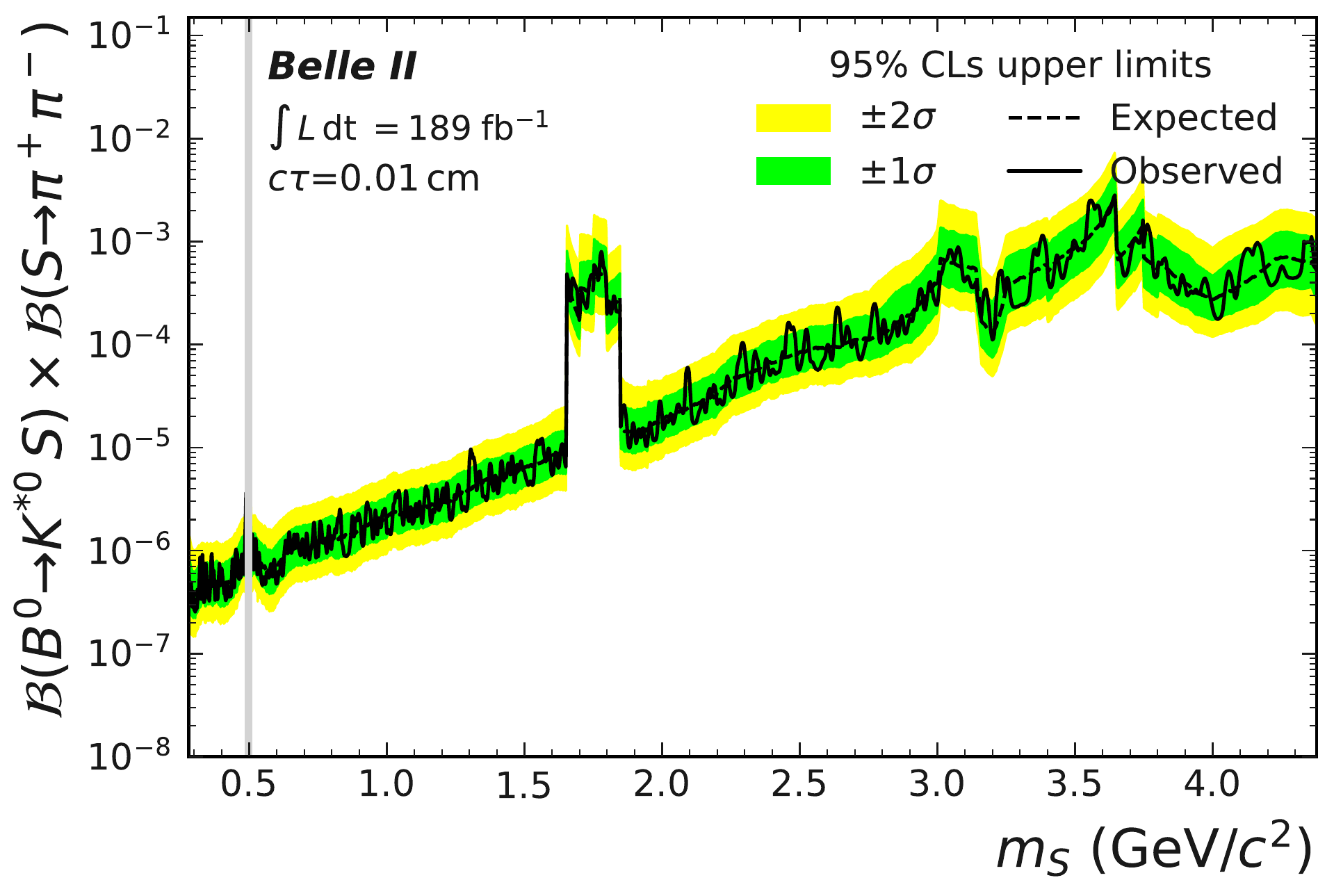}%
}%
\hspace*{\fill}
\subfigure[$\Bz\to \Kstarz(\to K^+\pi^-) S, S\to \pi^+\pi^-$, \newline lifetime of $c\tau=0.025\cm$.]{
  \label{subfit:brazil:Kstar_pi_1:F}%
  \includegraphics[width=0.31\textwidth]{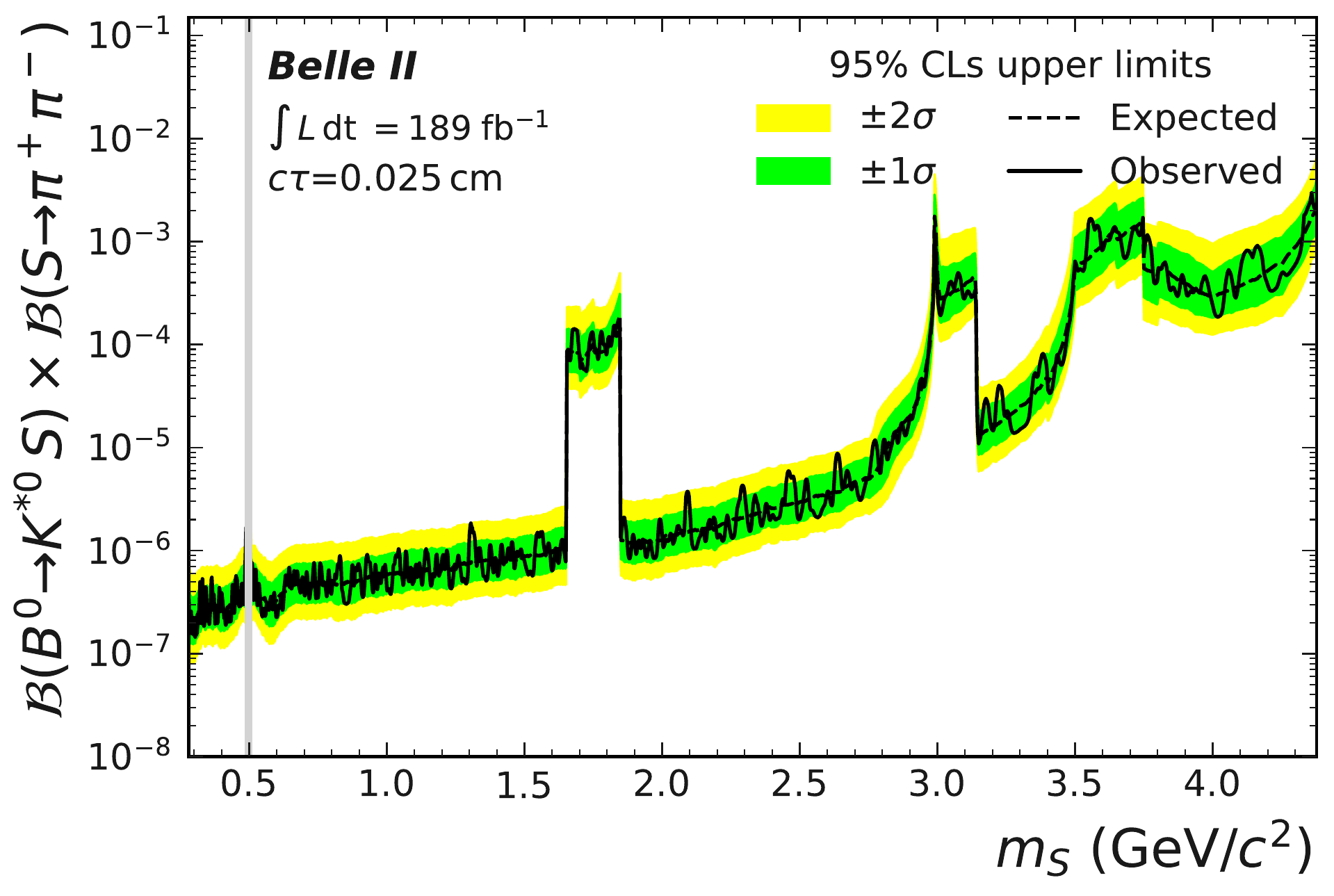}%
}
\subfigure[$\Bz\to \Kstarz(\to K^+\pi^-) S, S\to \pi^+\pi^-$, \newline lifetime of $c\tau=0.05\cm$.]{%
  \label{subfit:brazil:Kstar_pi_1:G}%
  \includegraphics[width=0.31\textwidth]{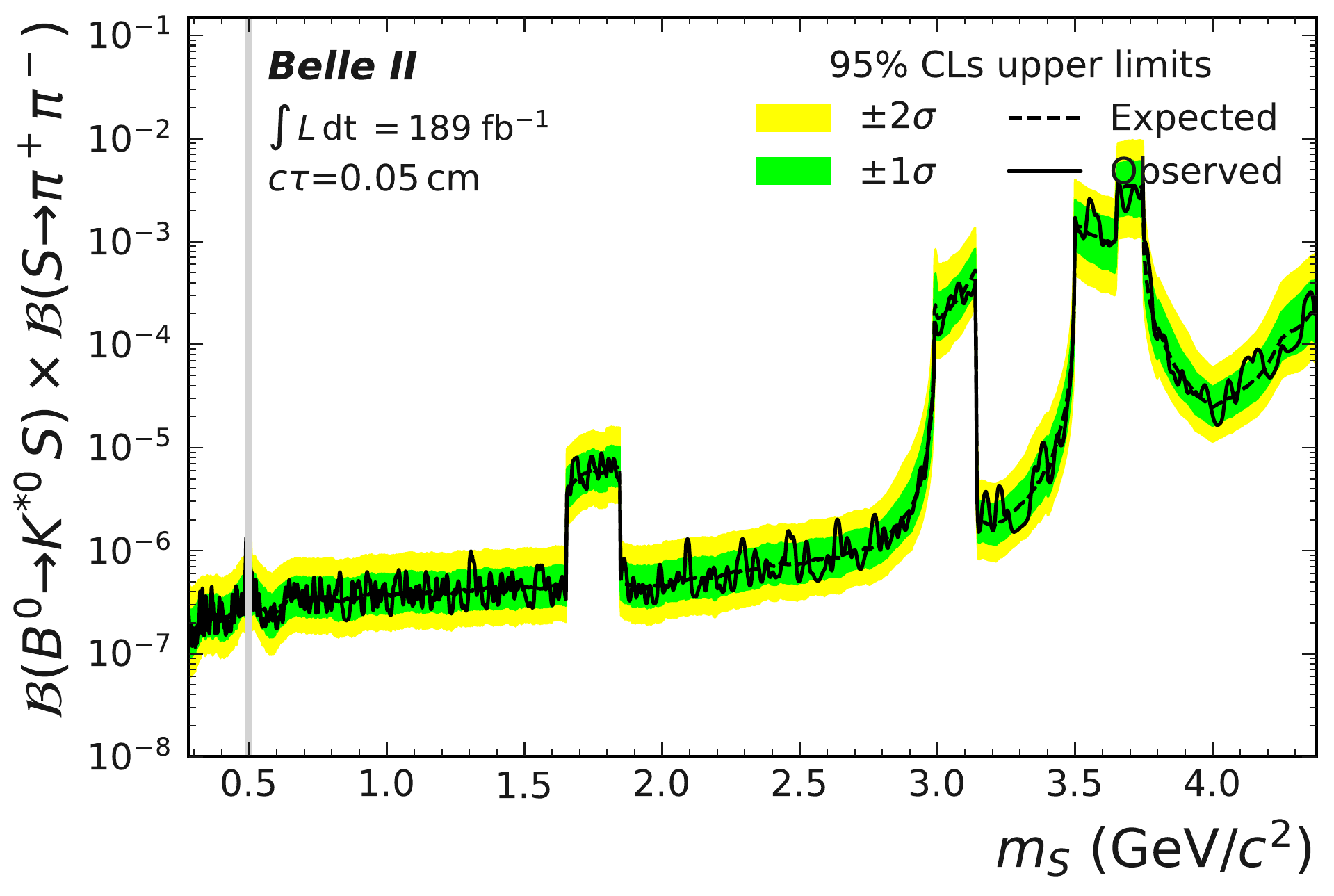}%
}%
\hspace*{\fill}
\subfigure[$\Bz\to \Kstarz(\to K^+\pi^-) S, S\to \pi^+\pi^-$, \newline lifetime of $c\tau=0.100\cm$.]{
  \label{subfit:brazil:Kstar_pi_1:H}%
  \includegraphics[width=0.31\textwidth]{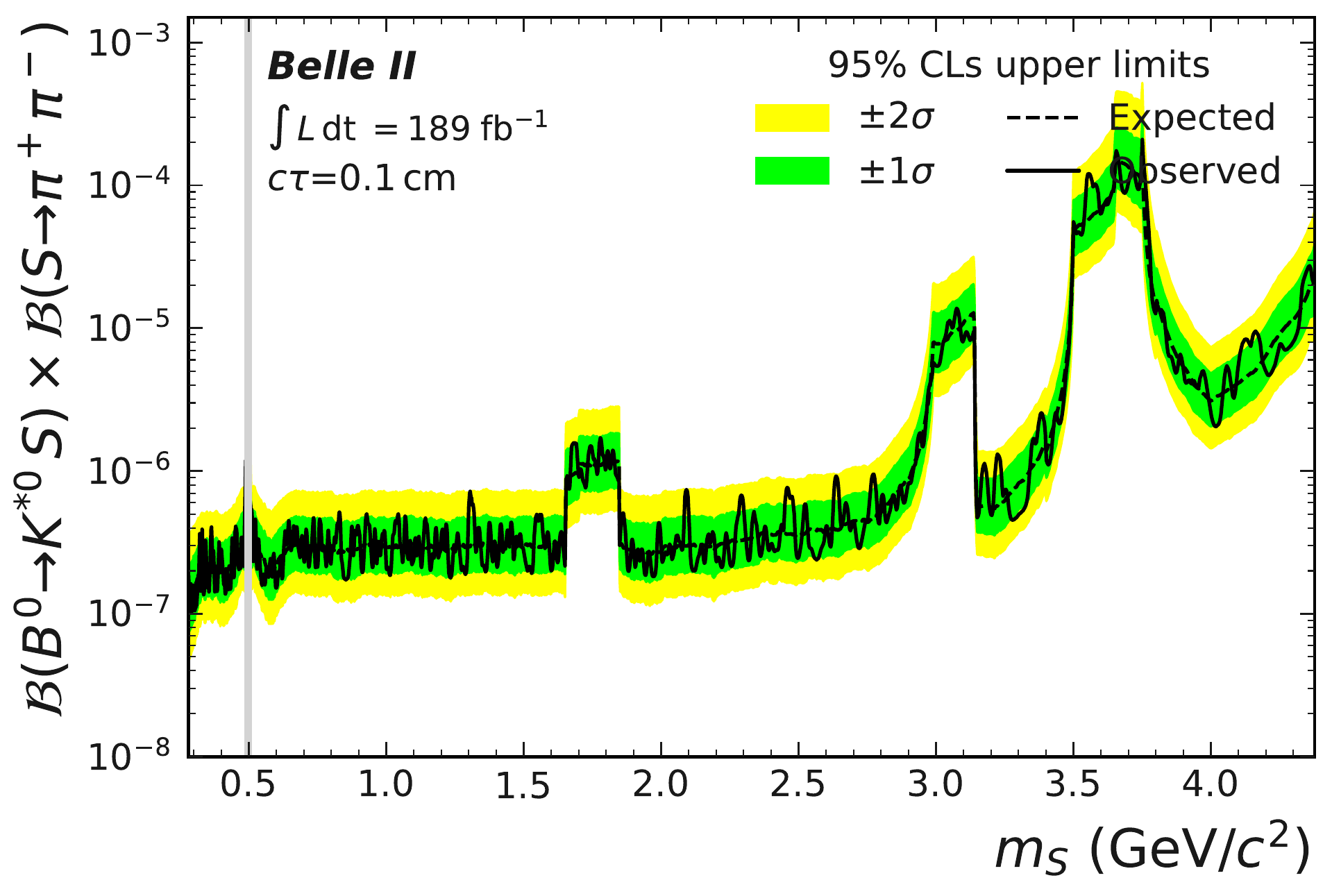}%
}%
\hspace*{\fill}
\subfigure[$\Bz\to \Kstarz(\to K^+\pi^-) S, S\to \pi^+\pi^-$, \newline lifetime of $c\tau=0.25\cm$.]{
  \label{subfit:brazil:Kstar_pi_1:I}%
  \includegraphics[width=0.31\textwidth]{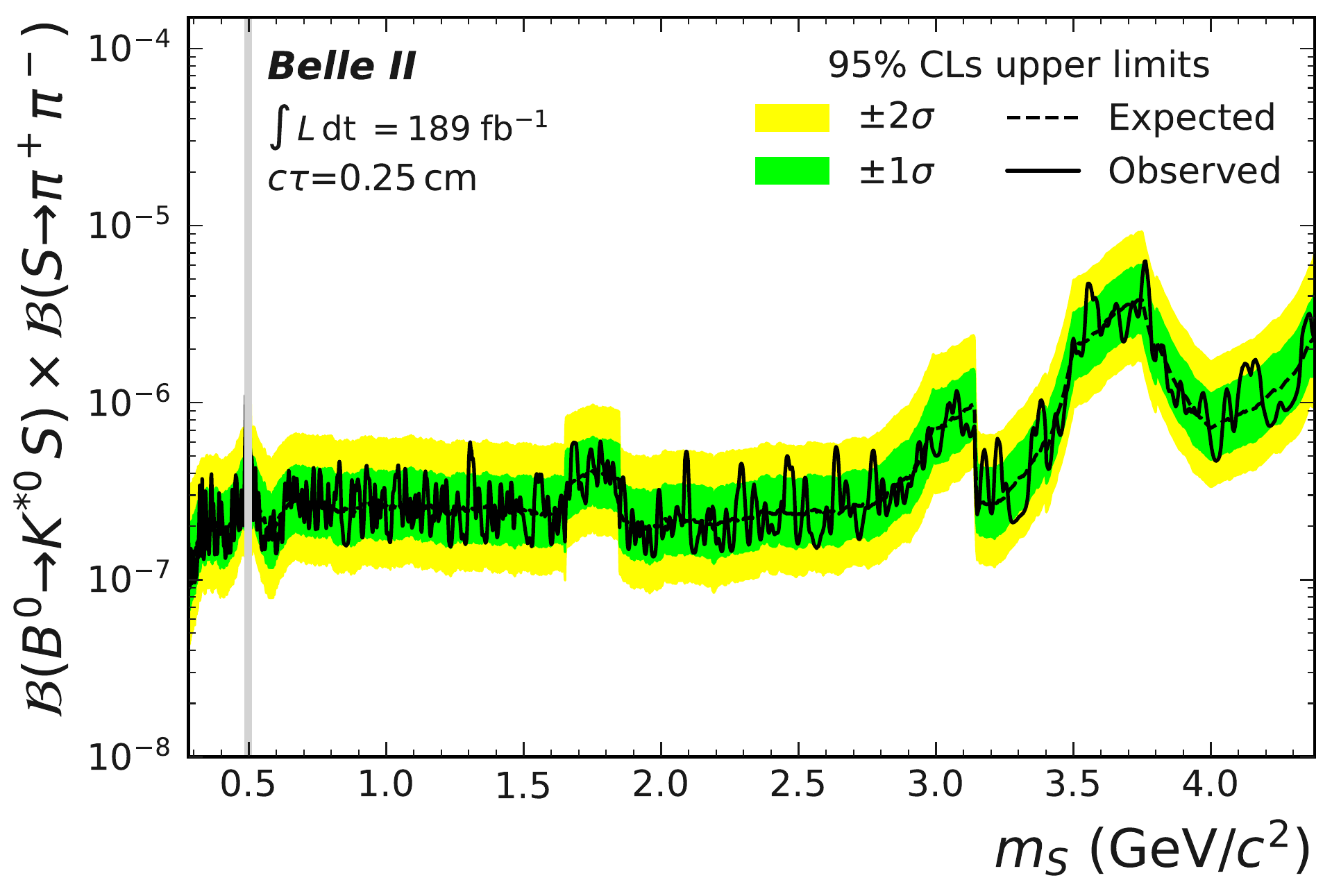}%
}
\subfigure[$\Bz\to \Kstarz(\to K^+\pi^-) S, S\to \pi^+\pi^-$, \newline lifetime of $c\tau=0.5\cm$.]{%
  \label{subfit:brazil:Kstar_pi_1:J}%
  \includegraphics[width=0.31\textwidth]{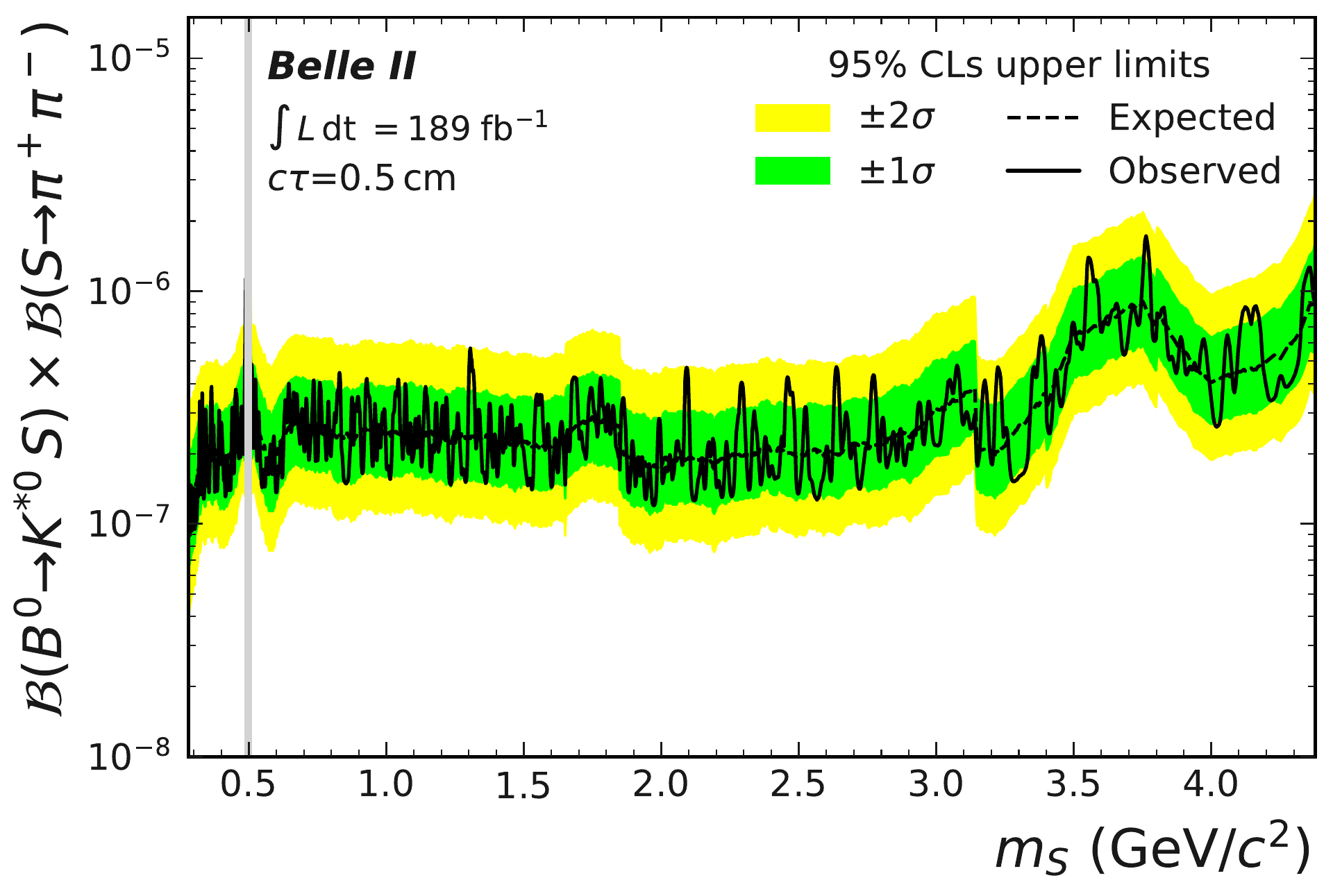}%
}%
\hspace*{\fill}
\subfigure[$\Bz\to \Kstarz(\to K^+\pi^-) S, S\to \pi^+\pi^-$, \newline lifetime of $c\tau=1\cm$.]{
  \label{subfit:brazil:Kstar_pi_1:K}%
  \includegraphics[width=0.31\textwidth]{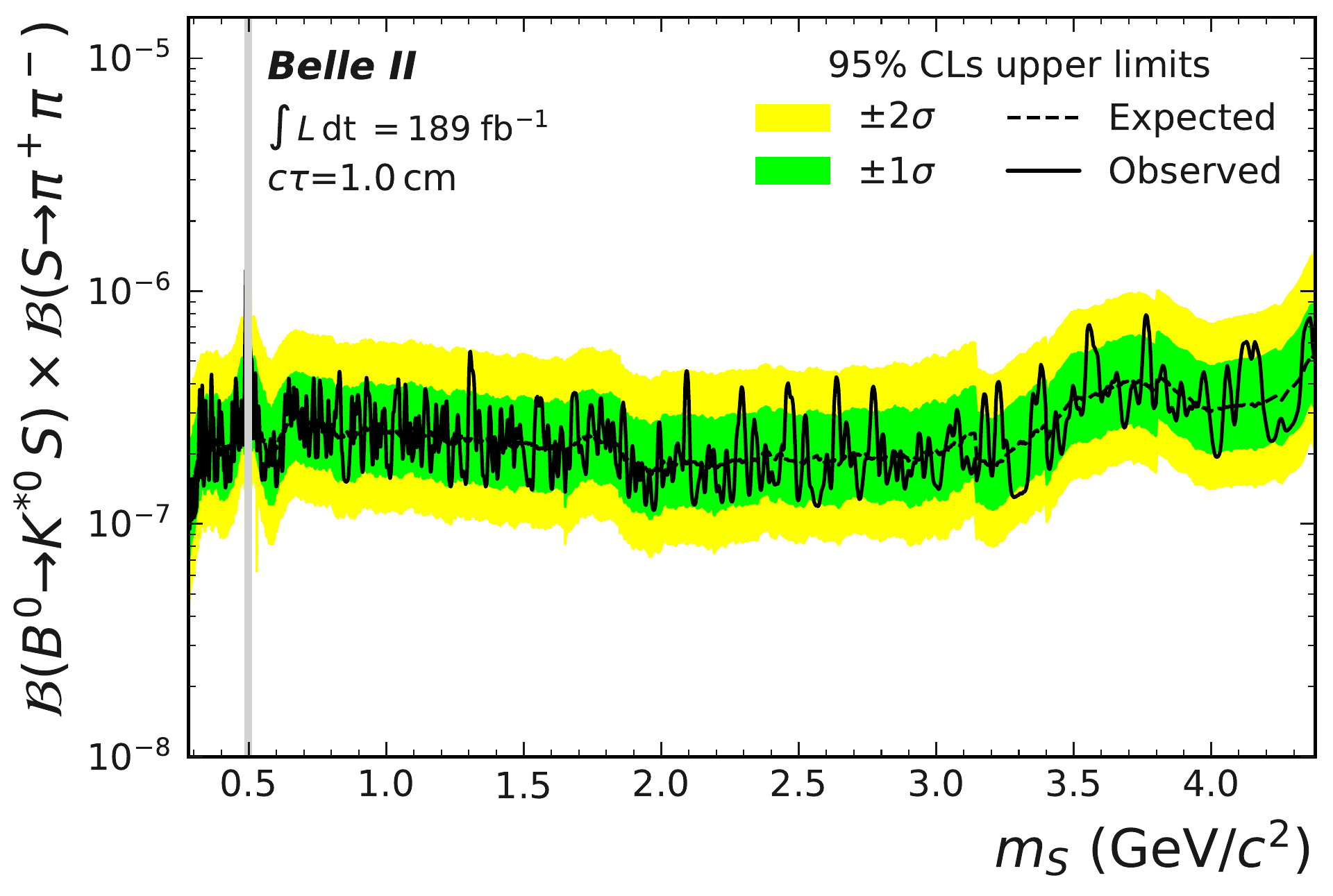}%
}%
\hspace*{\fill}
\subfigure[$\Bz\to \Kstarz(\to K^+\pi^-) S, S\to \pi^+\pi^-$, \newline lifetime of $c\tau=2.5\cm$.]{
  \label{subfit:brazil:Kstar_pi_1:L}%
  \includegraphics[width=0.31\textwidth]{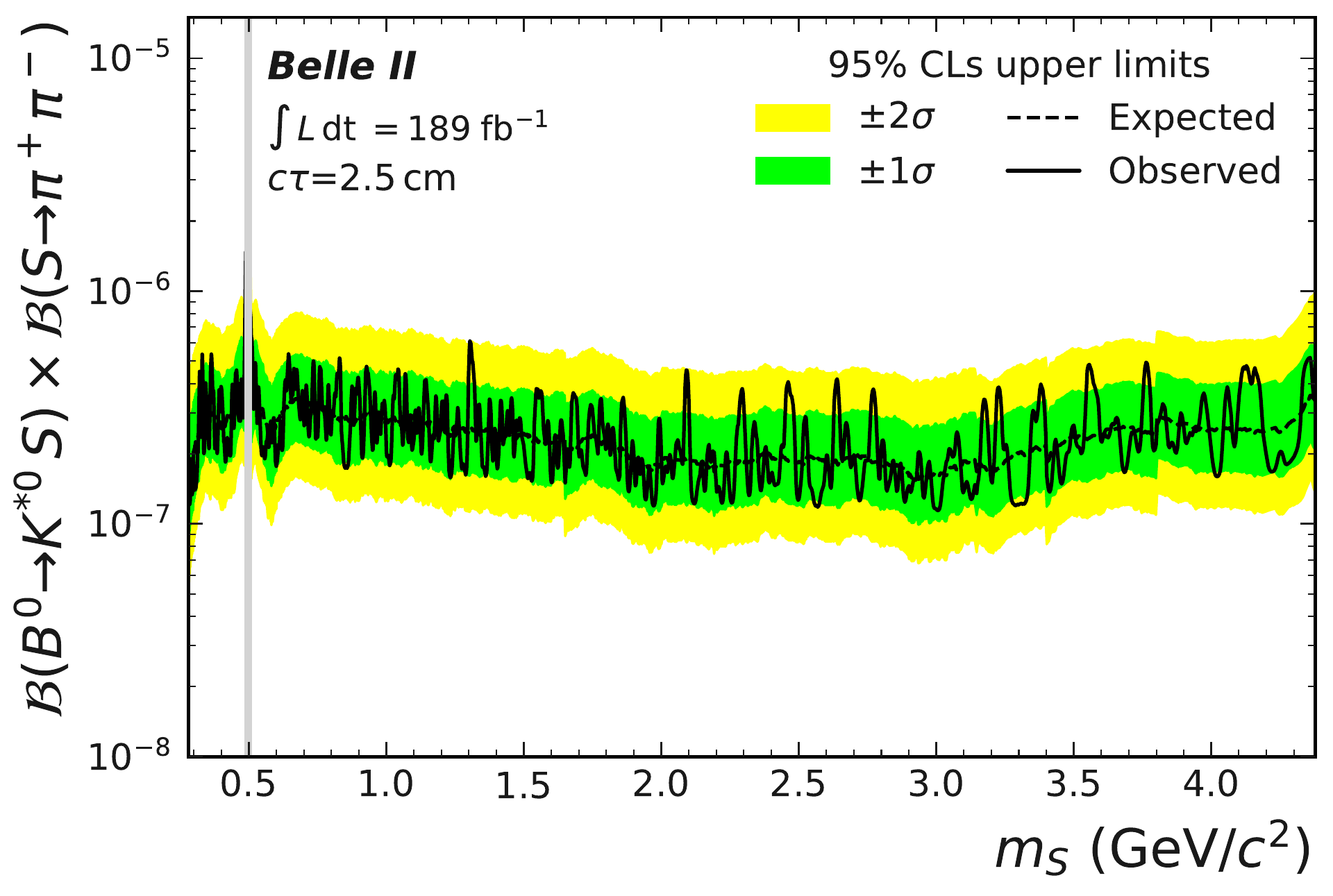}%
}
\caption{Expected and observed limits on the product of branching fractions $\mathcal{B}(B^0\to \Kstarz(\to K^+\pi^-) S) \times \mathcal{B}(S\to \pi^+\pi^-)$ for \\lifetimes \hbox{$0.001 < c\tau < 2.5\,\cm$}. The region corresponding to the fully-vetoed $\KS$ for $S\to\pi^+\pi^-$ is marked in gray.}\label{subfit:brazil:Kstar_pi_1}
\end{figure*}

\begin{figure*}[ht]%
\subfigure[$\Bz\to \Kstarz(\to K^+\pi^-) S, S\to \pi^+\pi^-$, \newline lifetime of $c\tau=5\cm$.]{%
  \label{subfit:brazil:Kstar_pi_2:A}%
  \includegraphics[width=0.31\textwidth]{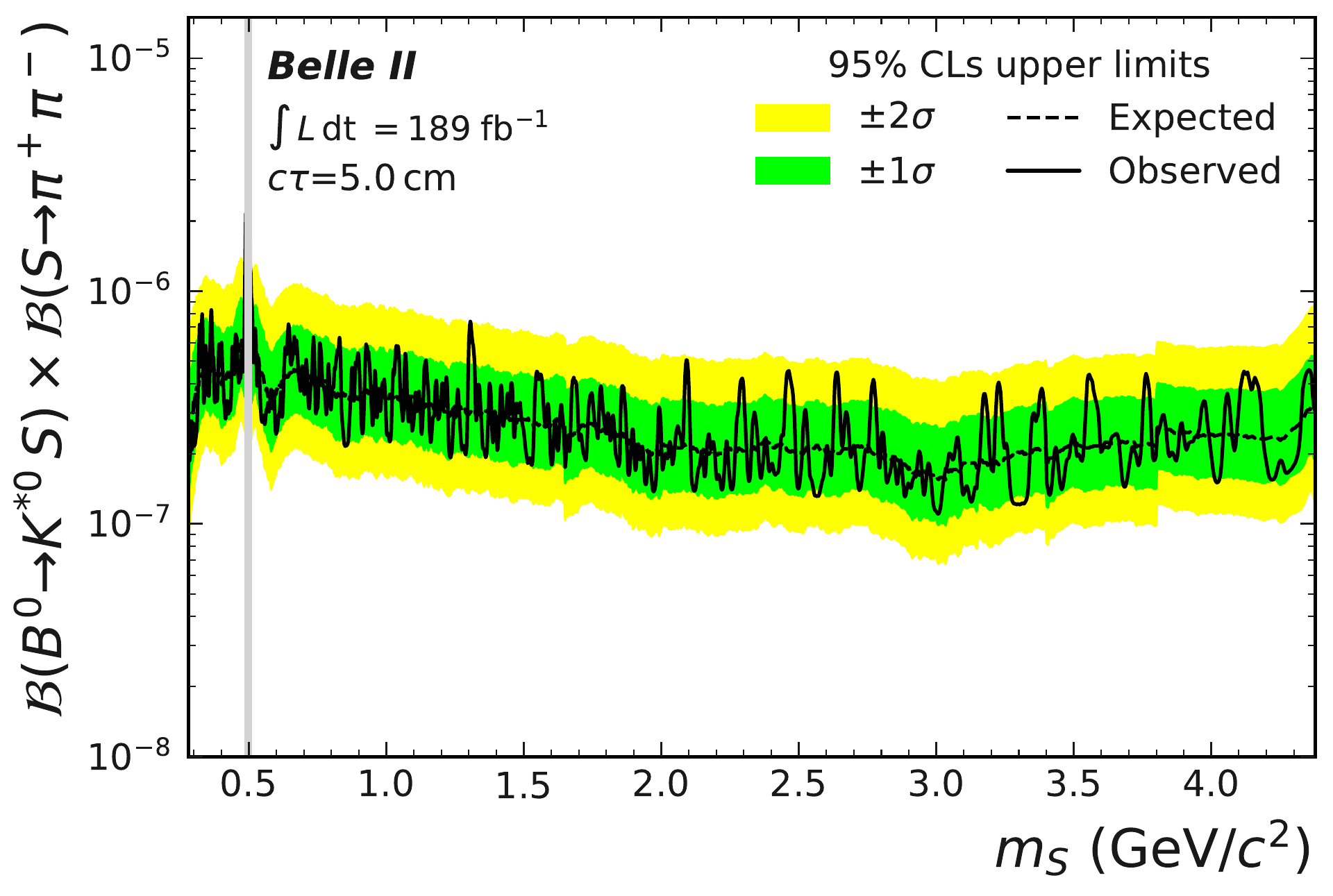}%
}%
\hspace*{\fill}
\subfigure[$\Bz\to \Kstarz(\to K^+\pi^-) S, S\to \pi^+\pi^-$, \newline lifetime of $c\tau=10\cm$.]{
  \label{subfit:brazil:Kstar_pi_2:B}%
  \includegraphics[width=0.31\textwidth]{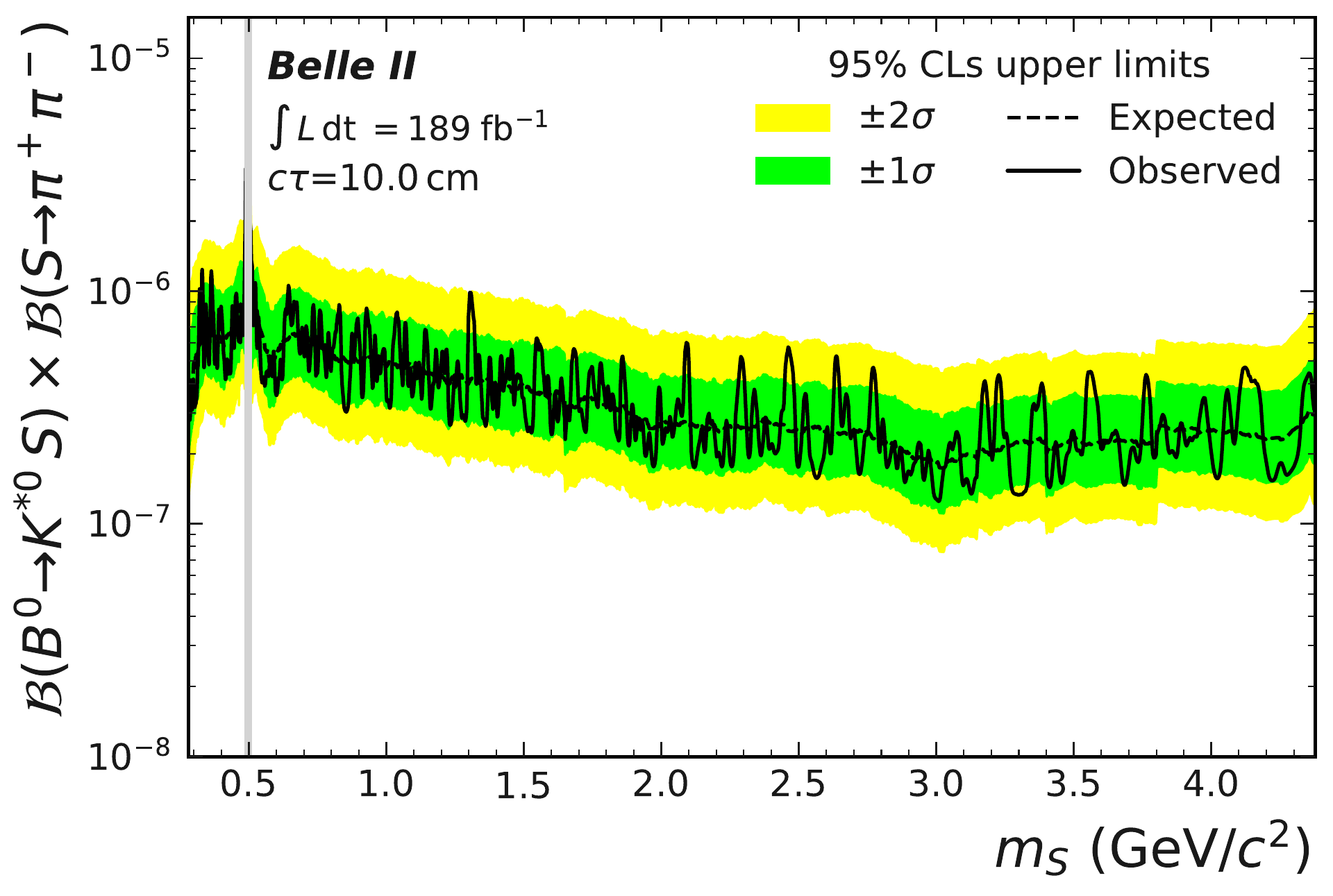}%
}%
\hspace*{\fill}
\subfigure[$\Bz\to \Kstarz(\to K^+\pi^-) S, S\to \pi^+\pi^-$, \newline lifetime of $c\tau=25\cm$.]{
  \label{subfit:brazil:Kstar_pi_2:C}%
  \includegraphics[width=0.31\textwidth]{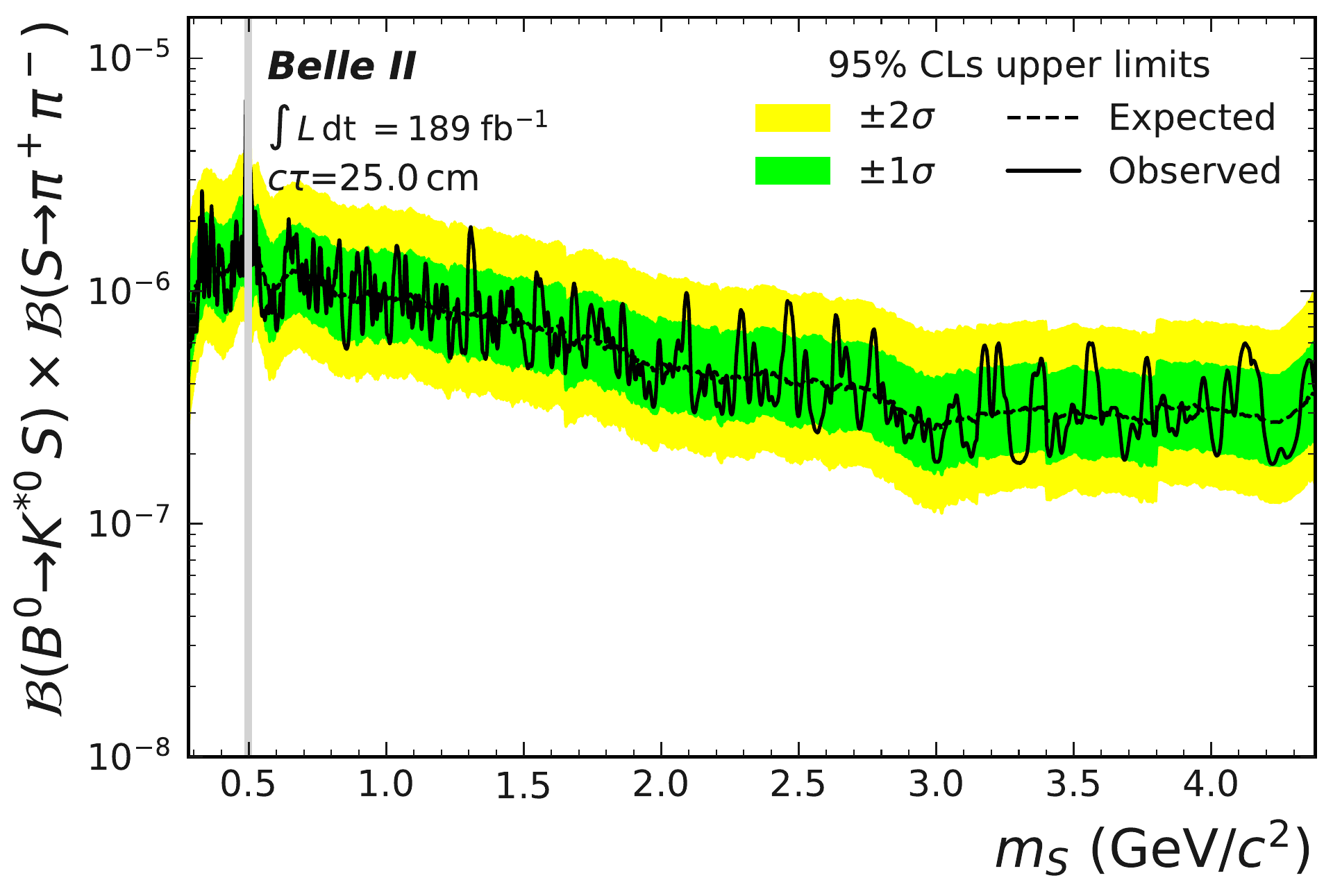}%
}
\subfigure[$\Bz\to \Kstarz(\to K^+\pi^-) S, S\to \pi^+\pi^-$, \newline lifetime of $c\tau=50\cm$.]{%
  \label{subfit:brazil:Kstar_pi_2:D}%
  \includegraphics[width=0.31\textwidth]{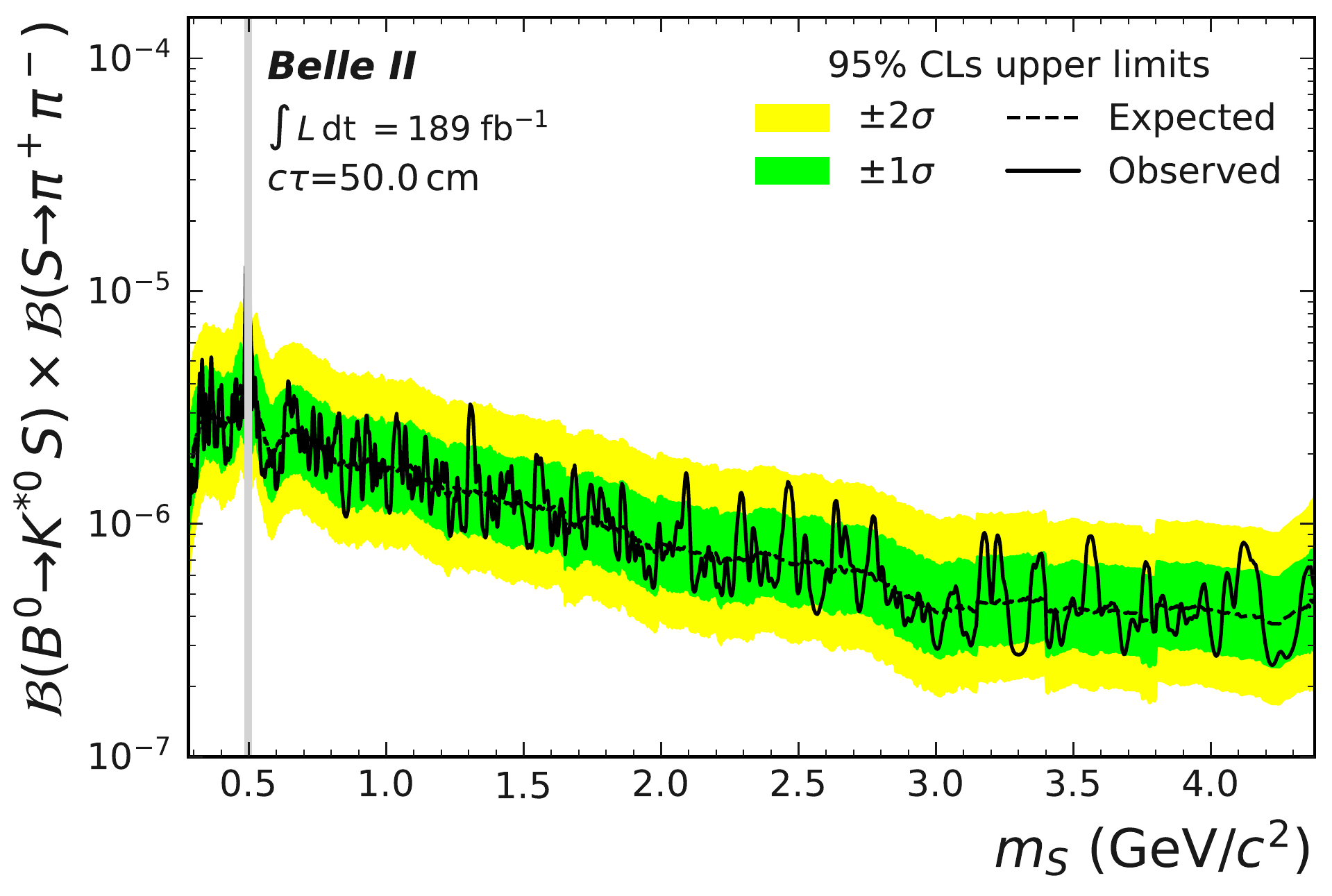}%
}%
\hspace*{\fill}
\subfigure[$\Bz\to \Kstarz(\to K^+\pi^-) S, S\to \pi^+\pi^-$, \newline lifetime of $c\tau=100\cm$.]{
  \label{subfit:brazil:Kstar_pi_2:E}%
  \includegraphics[width=0.31\textwidth]{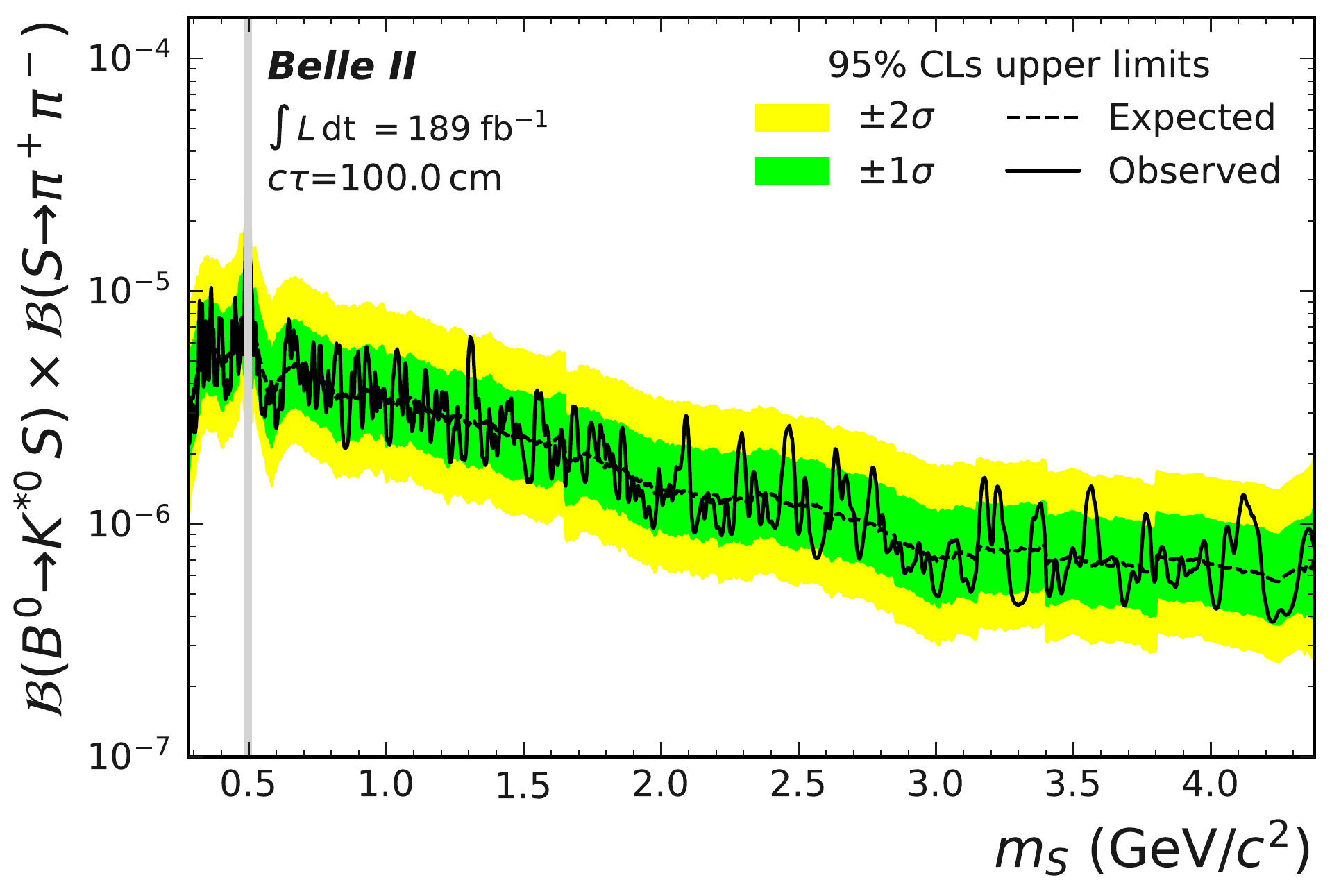}%
}%
\hspace*{\fill}
\subfigure[$\Bz\to \Kstarz(\to K^+\pi^-) S, S\to \pi^+\pi^-$, \newline lifetime of $c\tau=200\cm$.]{
  \label{subfit:brazil:Kstar_pi_2:F}%
  \includegraphics[width=0.31\textwidth]{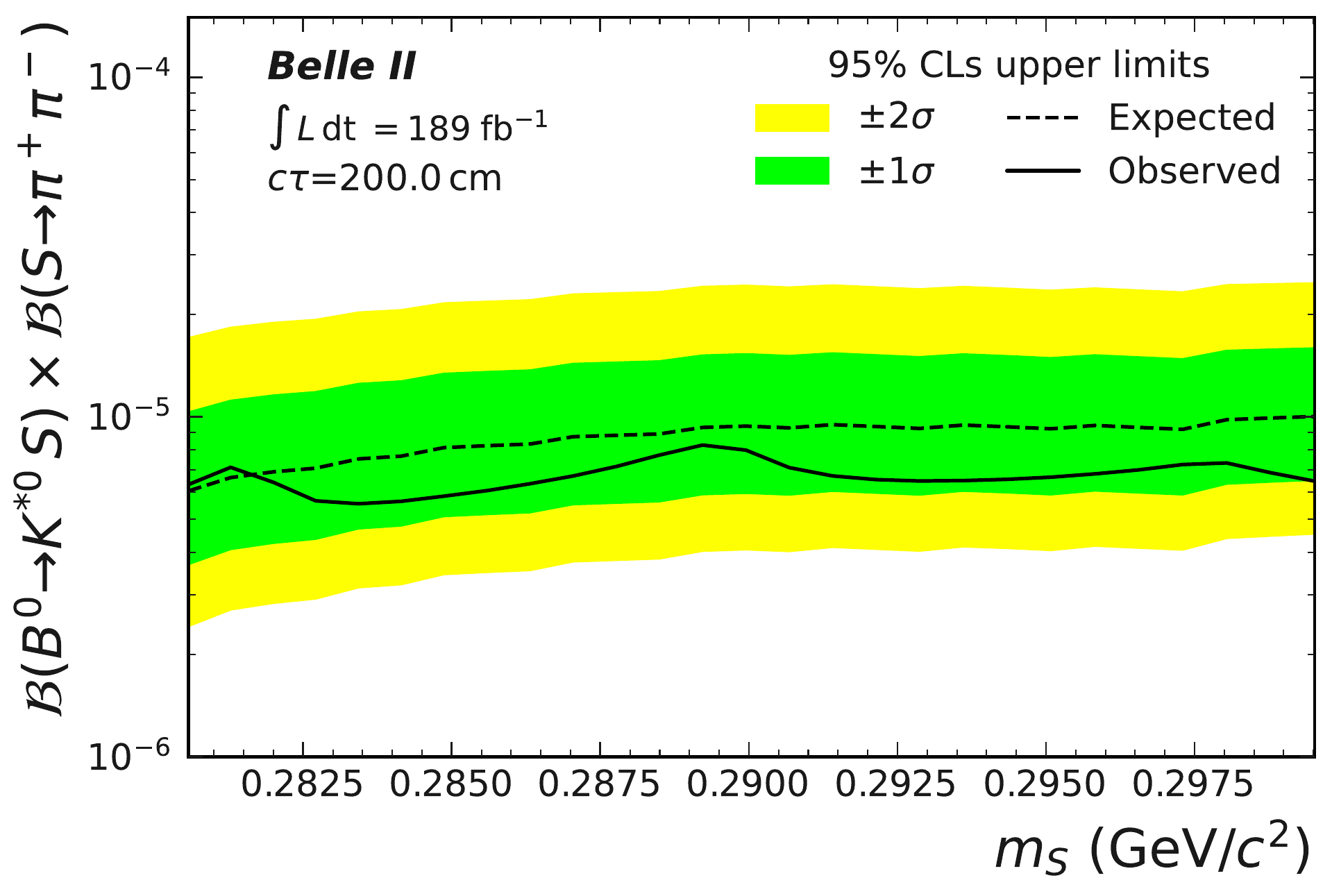}%
}
\subfigure[$\Bz\to \Kstarz(\to K^+\pi^-) S, S\to \pi^+\pi^-$, \newline lifetime of $c\tau=400\cm$.]{
  \label{subfit:brazil:Kstar_pi_2:G}%
  \includegraphics[width=0.31\textwidth]{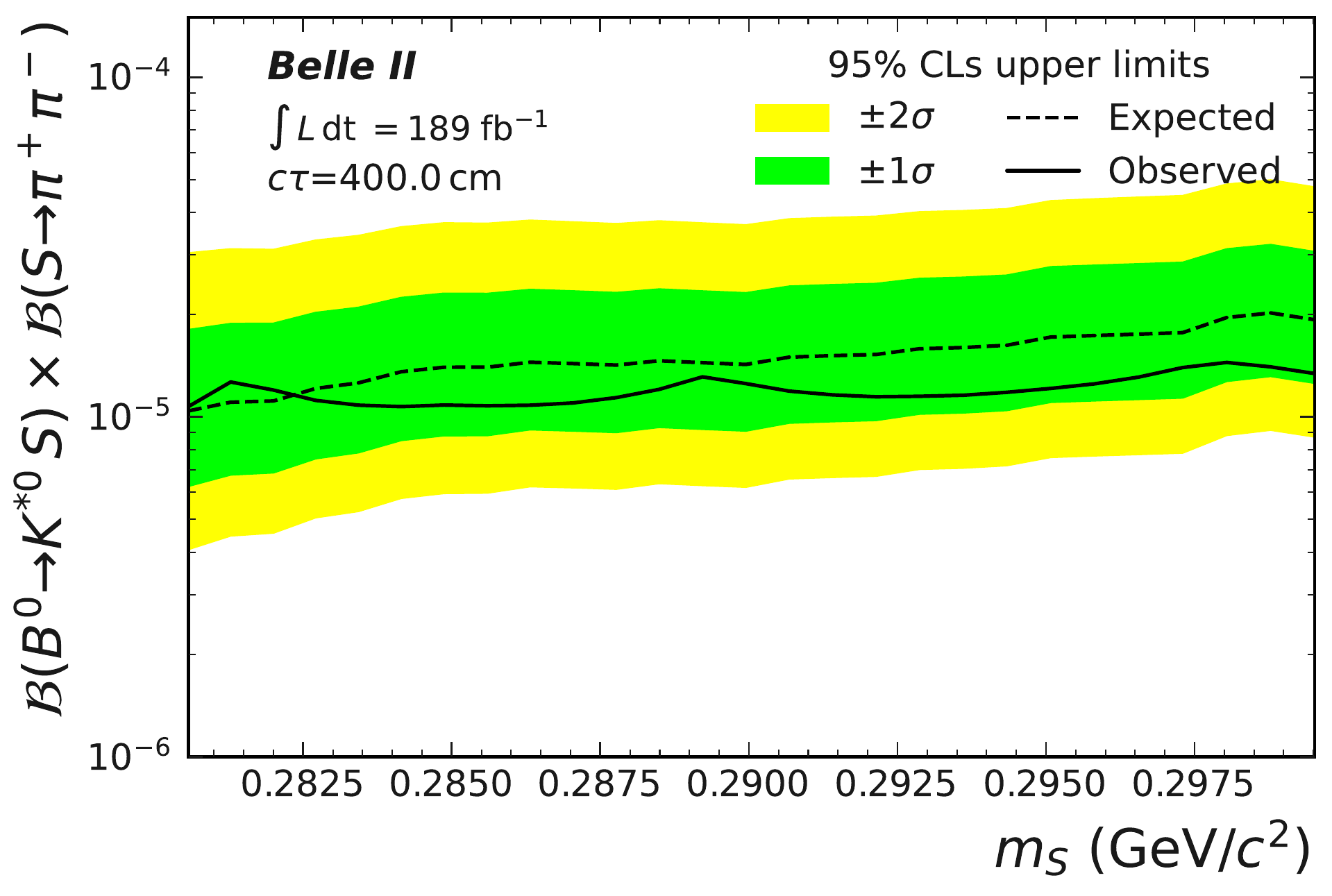}%
}
\caption{Expected and observed limits on the product of branching fractions $\mathcal{B}(B^0\to \Kstarz(\to K^+\pi^-) S) \times \mathcal{B}(S\to \pi^+\pi^-)$ for lifetimes \hbox{$5 < c\tau < 400\,\cm$}. The region corresponding to the fully-vetoed $\KS$ for $S\to\pi^+\pi^-$ is marked in gray.}\label{subfit:brazil:Kstar_pi_2}
\end{figure*}


\begin{figure*}[ht]%
\subfigure[$B^+\to K^+S, S\to K^+K^-$, \newline lifetime of $c\tau=0.001\cm$.]{%
  \label{subfit:brazil:Kp_K_1:A}%
  \includegraphics[width=0.31\textwidth]{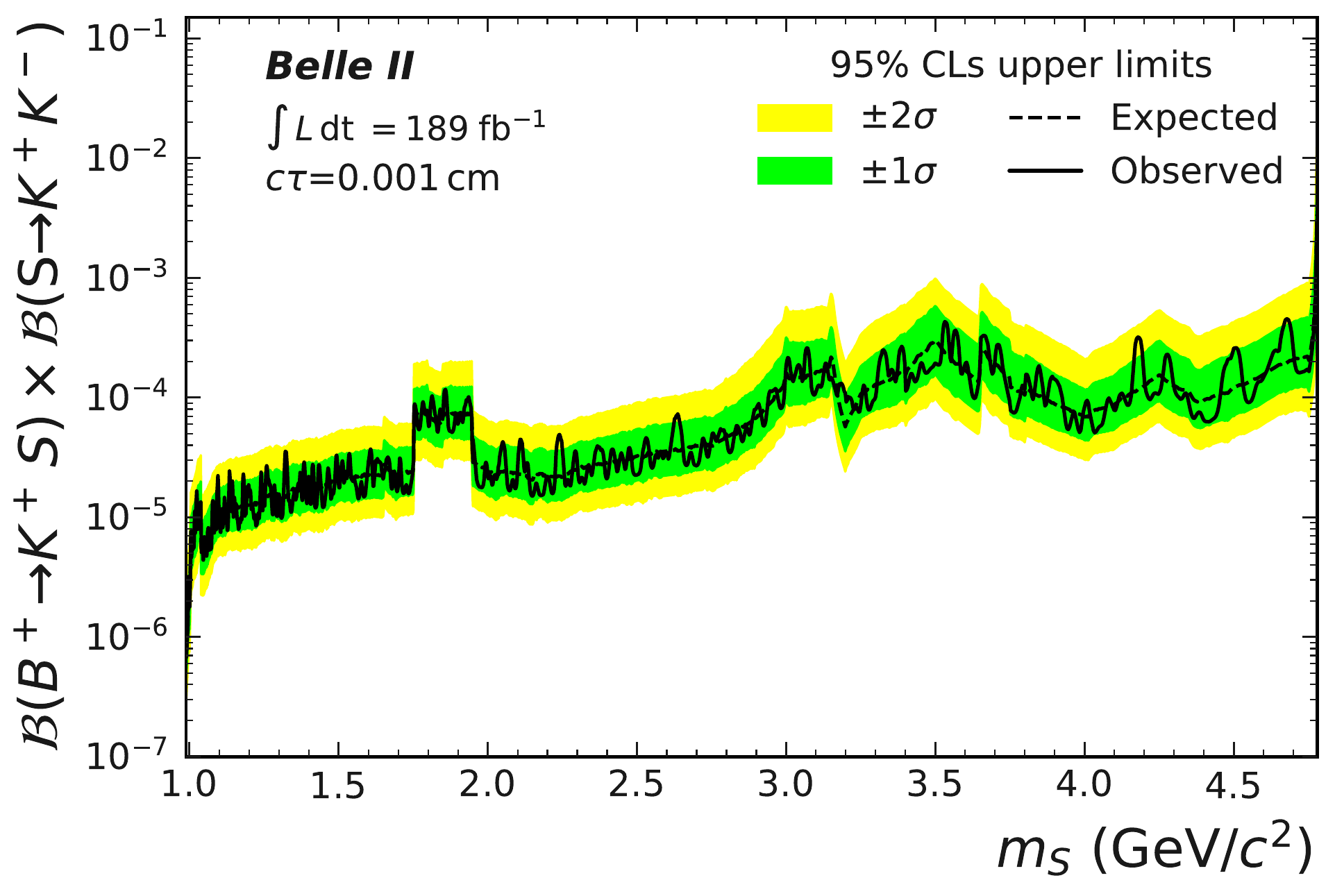}%
}%
\hspace*{\fill}
\subfigure[$B^+\to K^+S, S\to K^+K^-$, \newline lifetime of $c\tau=0.003\cm$.]{
  \label{subfit:brazil:Kp_K_1:B}%
  \includegraphics[width=0.31\textwidth]{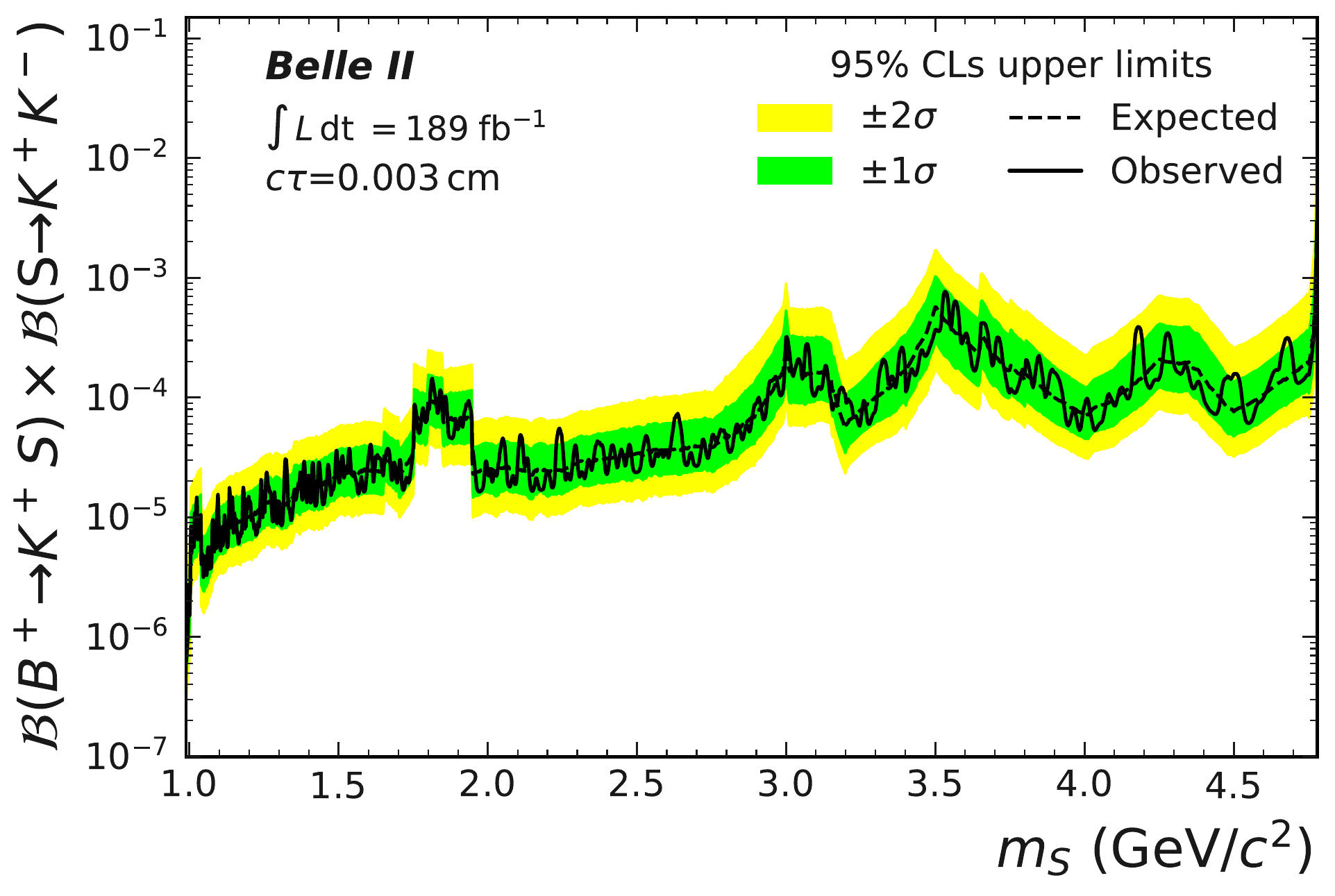}%
}%
\hspace*{\fill}
\subfigure[$B^+\to K^+S, S\to K^+K^-$, \newline lifetime of $c\tau=0.005\cm$.]{
  \label{subfit:brazil:Kp_K_1:C}%
  \includegraphics[width=0.31\textwidth]{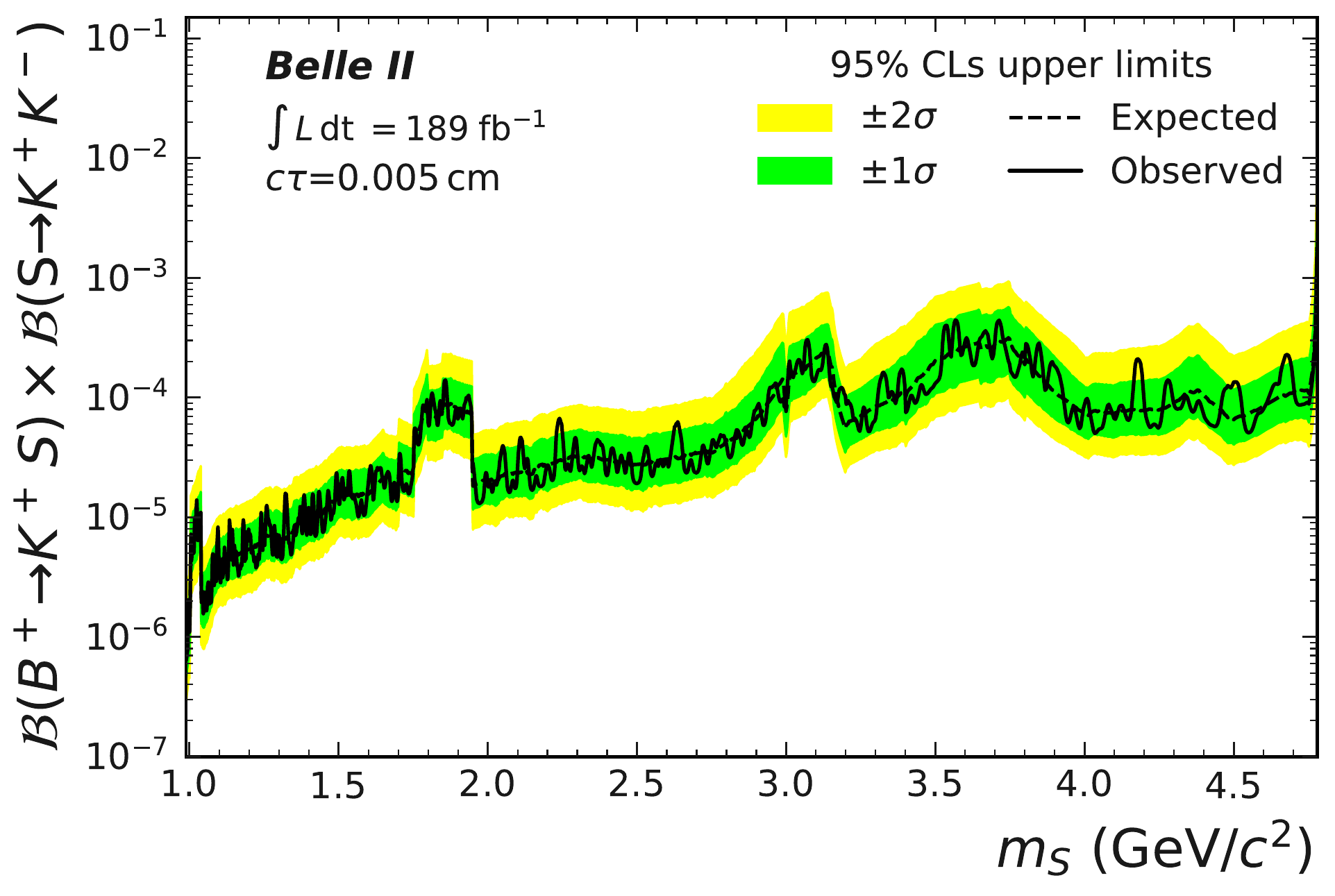}%
}
\subfigure[$B^+\to K^+S, S\to K^+K^-$, \newline lifetime of $c\tau=0.007\cm$.]{%
  \label{subfit:brazil:Kp_K_1:D}%
  \includegraphics[width=0.31\textwidth]{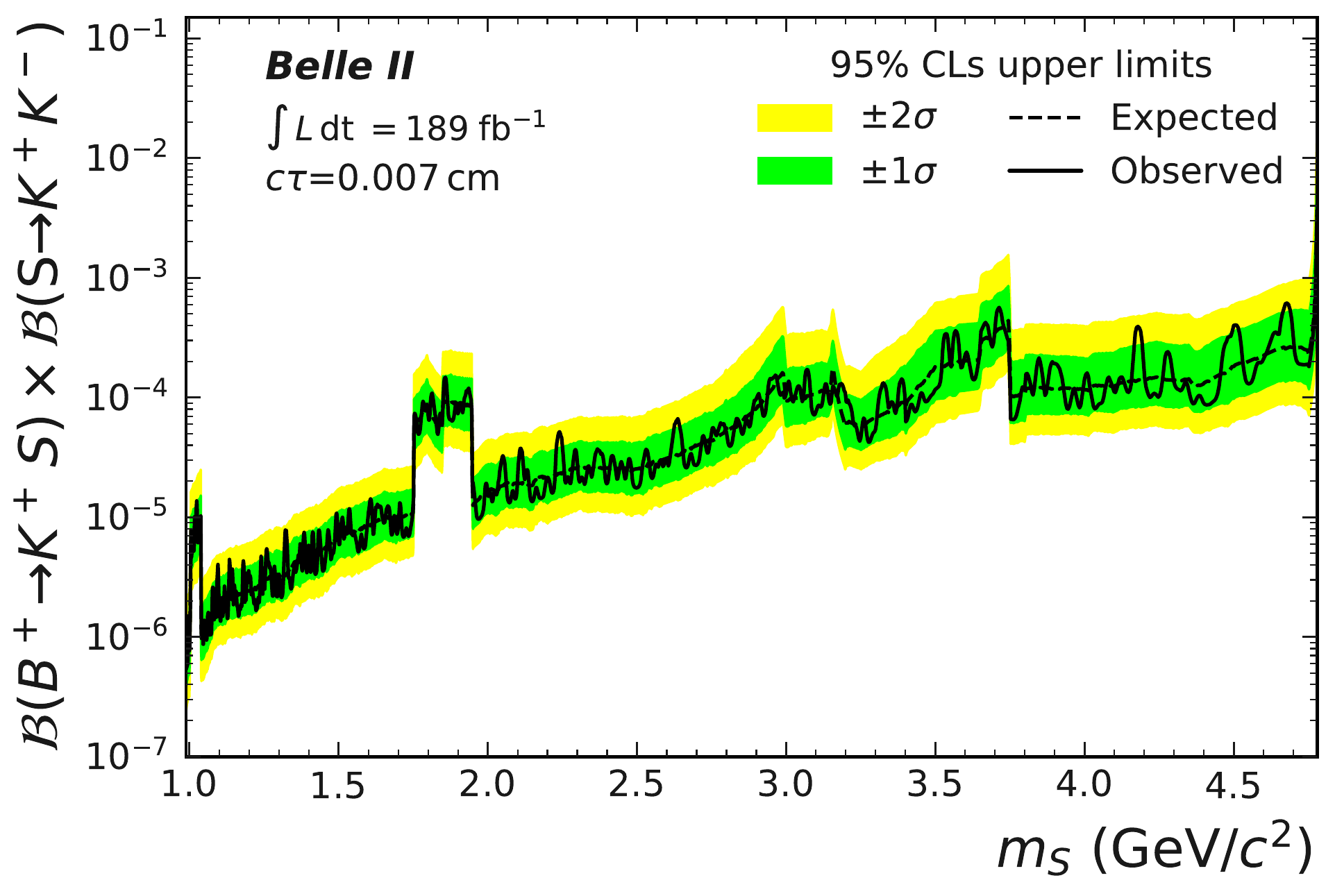}%
}%
\hspace*{\fill}
\subfigure[$B^+\to K^+S, S\to K^+K^-$, \newline lifetime of $c\tau=0.01\cm$.]{
  \label{subfit:brazil:Kp_K_1:E}%
  \includegraphics[width=0.31\textwidth]{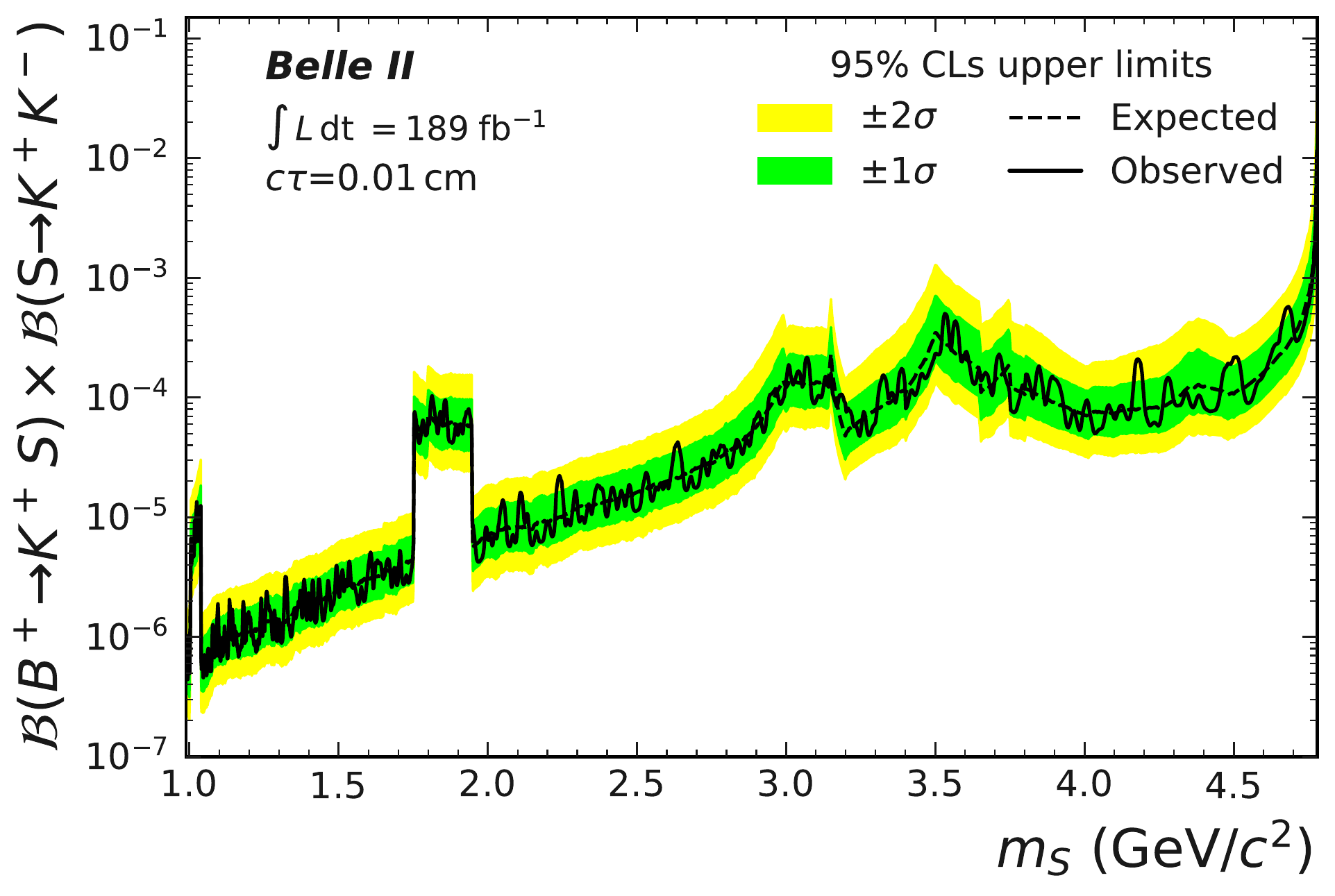}%
}%
\hspace*{\fill}
\subfigure[$B^+\to K^+S, S\to K^+K^-$, \newline lifetime of $c\tau=0.025\cm$.]{
  \label{subfit:brazil:Kp_K_1:F}%
  \includegraphics[width=0.31\textwidth]{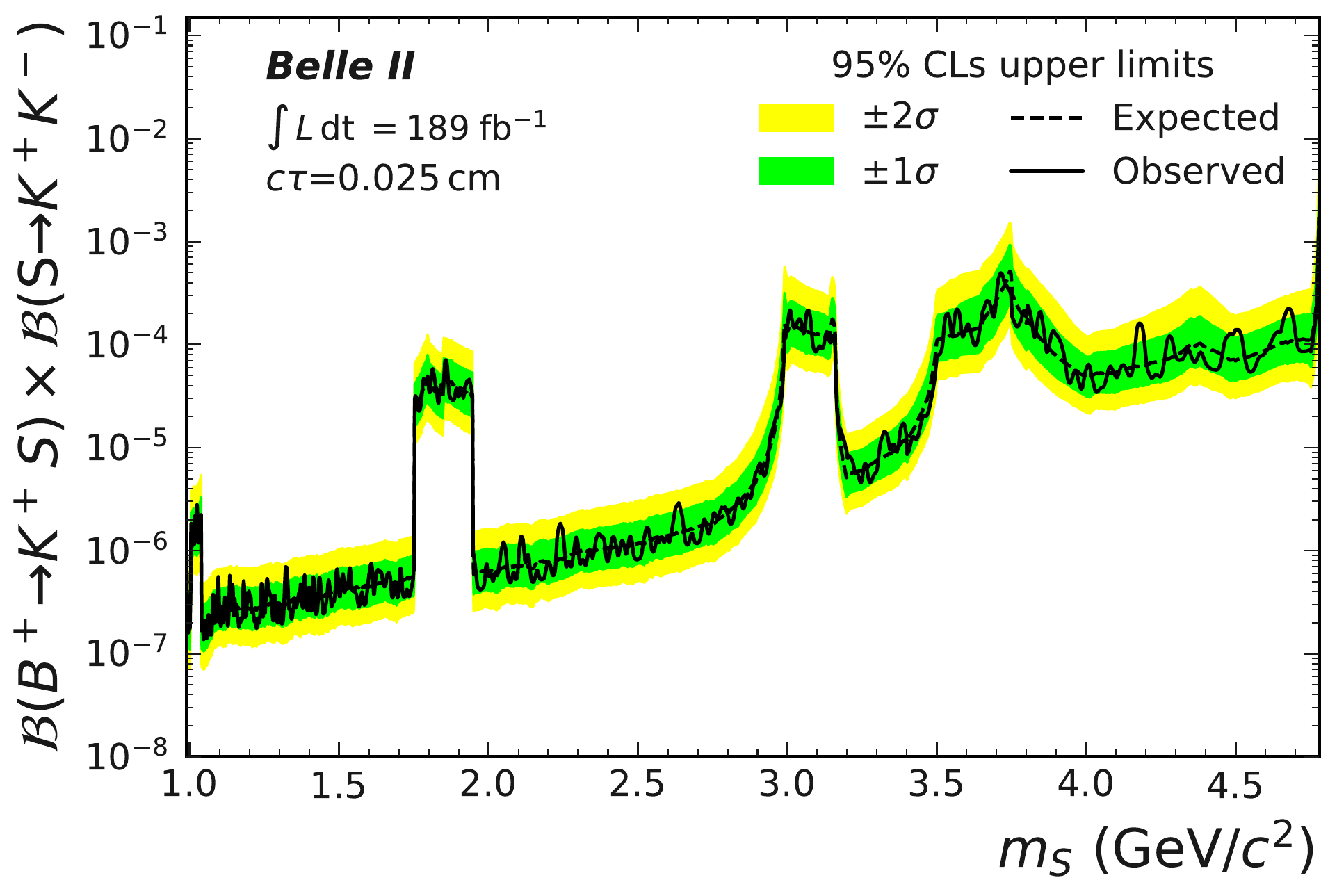}%
}
\subfigure[$B^+\to K^+S, S\to K^+K^-$, \newline lifetime of $c\tau=0.05\cm$.]{%
  \label{subfit:brazil:Kp_K_1:G}%
  \includegraphics[width=0.31\textwidth]{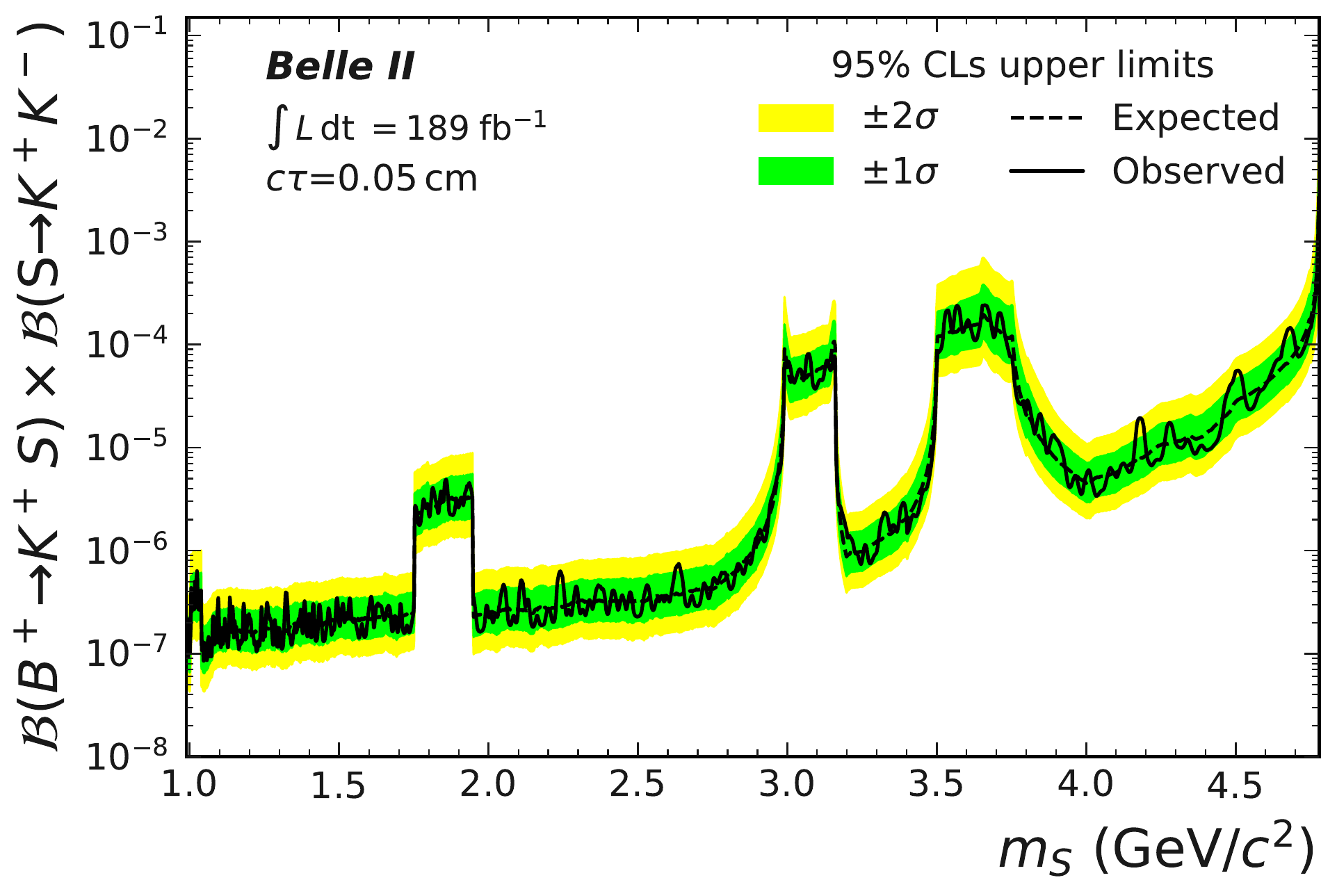}%
}%
\hspace*{\fill}
\subfigure[$B^+\to K^+S, S\to K^+K^-$, \newline lifetime of $c\tau=0.100\cm$.]{
  \label{subfit:brazil:Kp_K_1:H}%
  \includegraphics[width=0.31\textwidth]{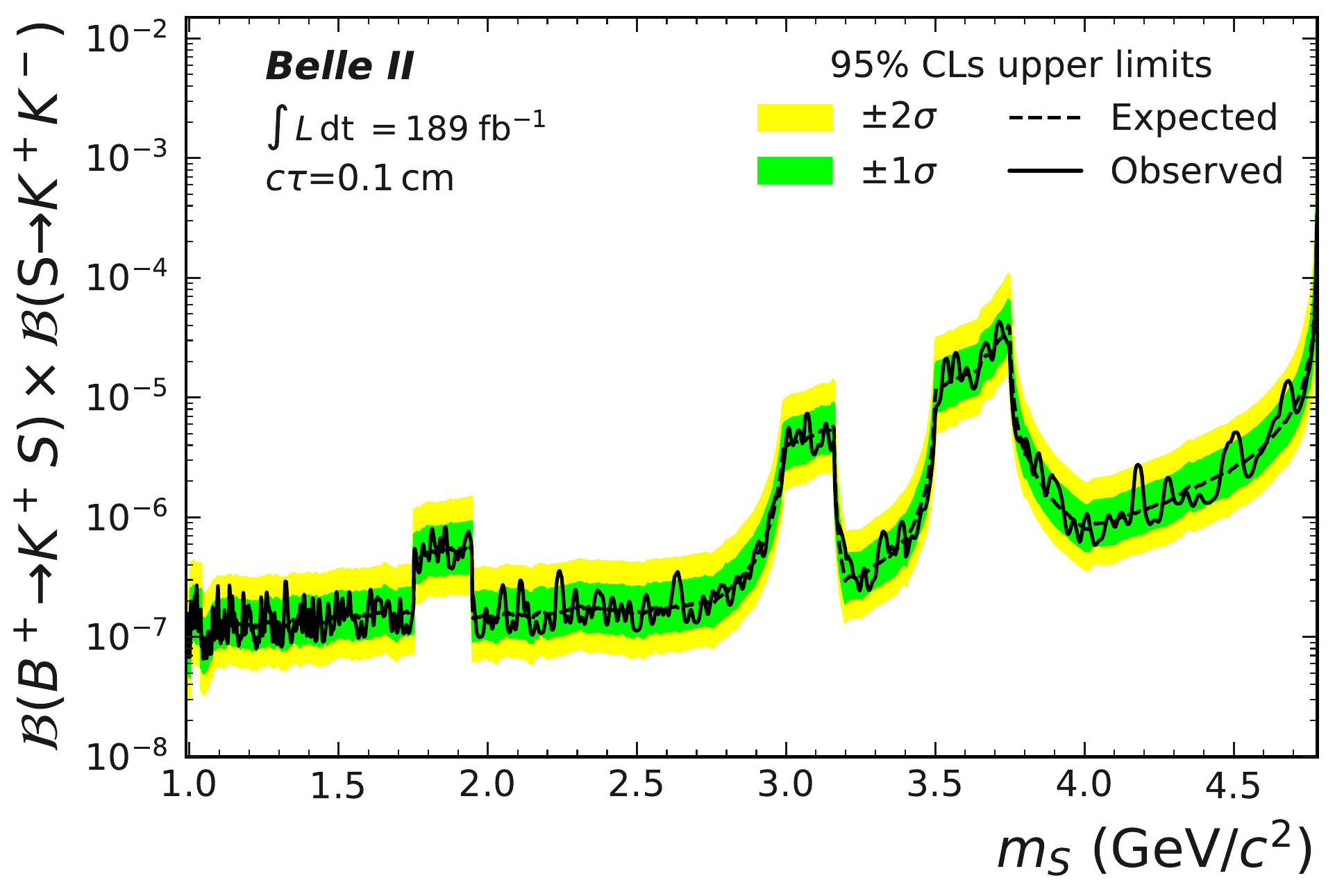}%
}%
\hspace*{\fill}
\subfigure[$B^+\to K^+S, S\to K^+K^-$, \newline lifetime of $c\tau=0.25\cm$.]{
  \label{subfit:brazil:Kp_K_1:I}%
  \includegraphics[width=0.31\textwidth]{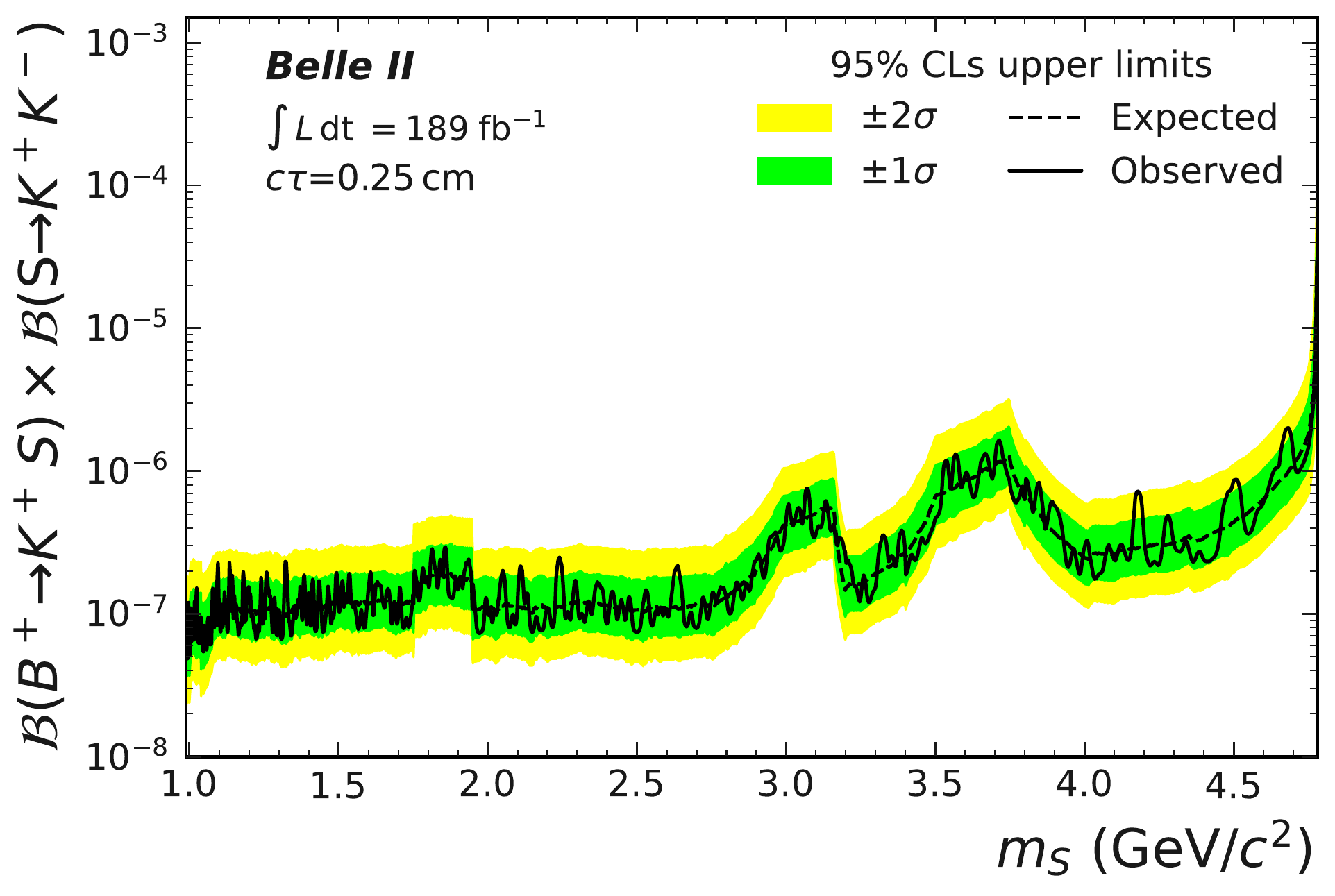}%
}
\subfigure[$B^+\to K^+S, S\to K^+K^-$, \newline lifetime of $c\tau=0.5\cm$.]{%
  \label{subfit:brazil:Kp_K_1:J}%
  \includegraphics[width=0.31\textwidth]{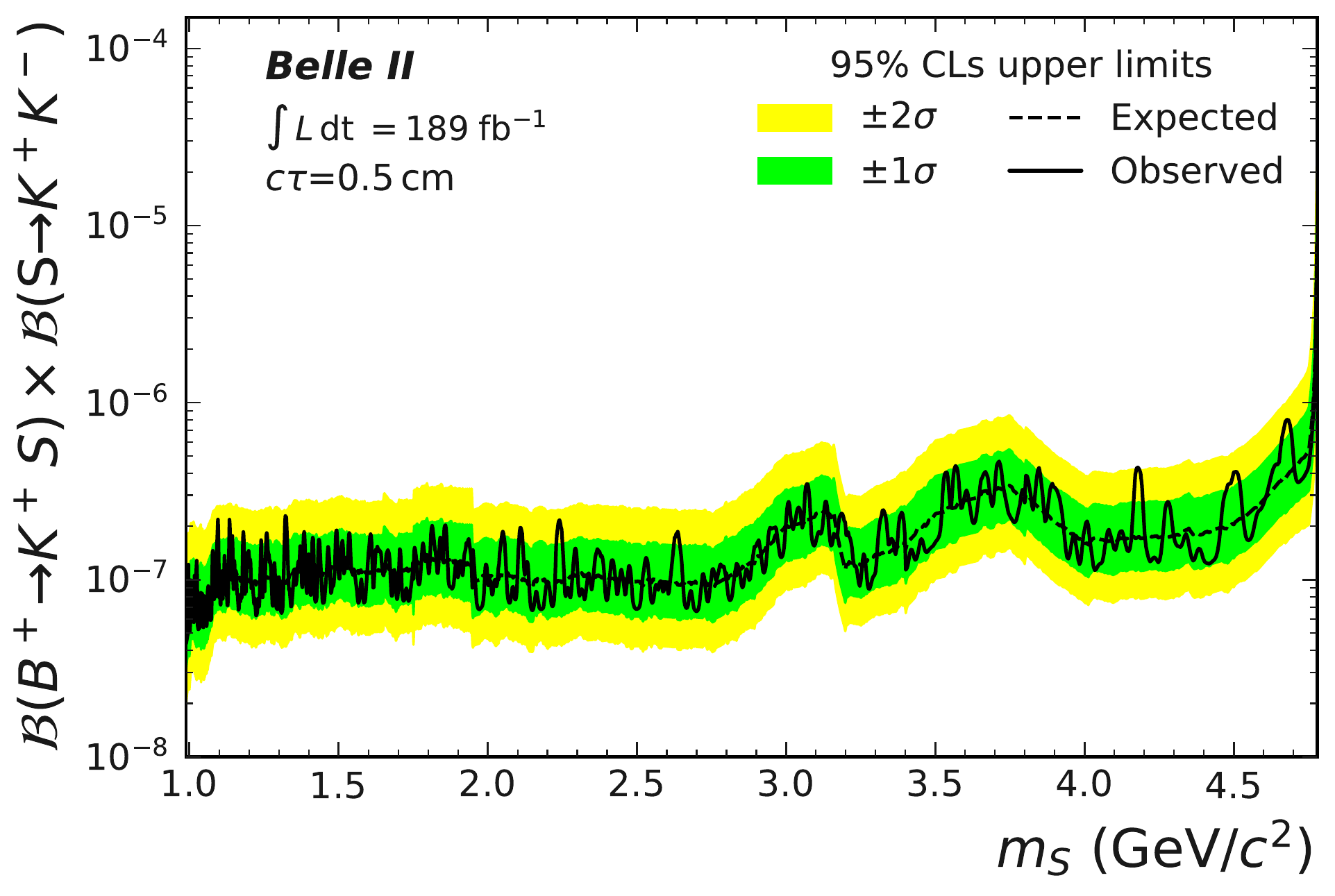}%
}%
\hspace*{\fill}
\subfigure[$B^+\to K^+S, S\to K^+K^-$, \newline lifetime of $c\tau=1\cm$.]{
  \label{subfit:brazil:Kp_K_1:K}%
  \includegraphics[width=0.31\textwidth]{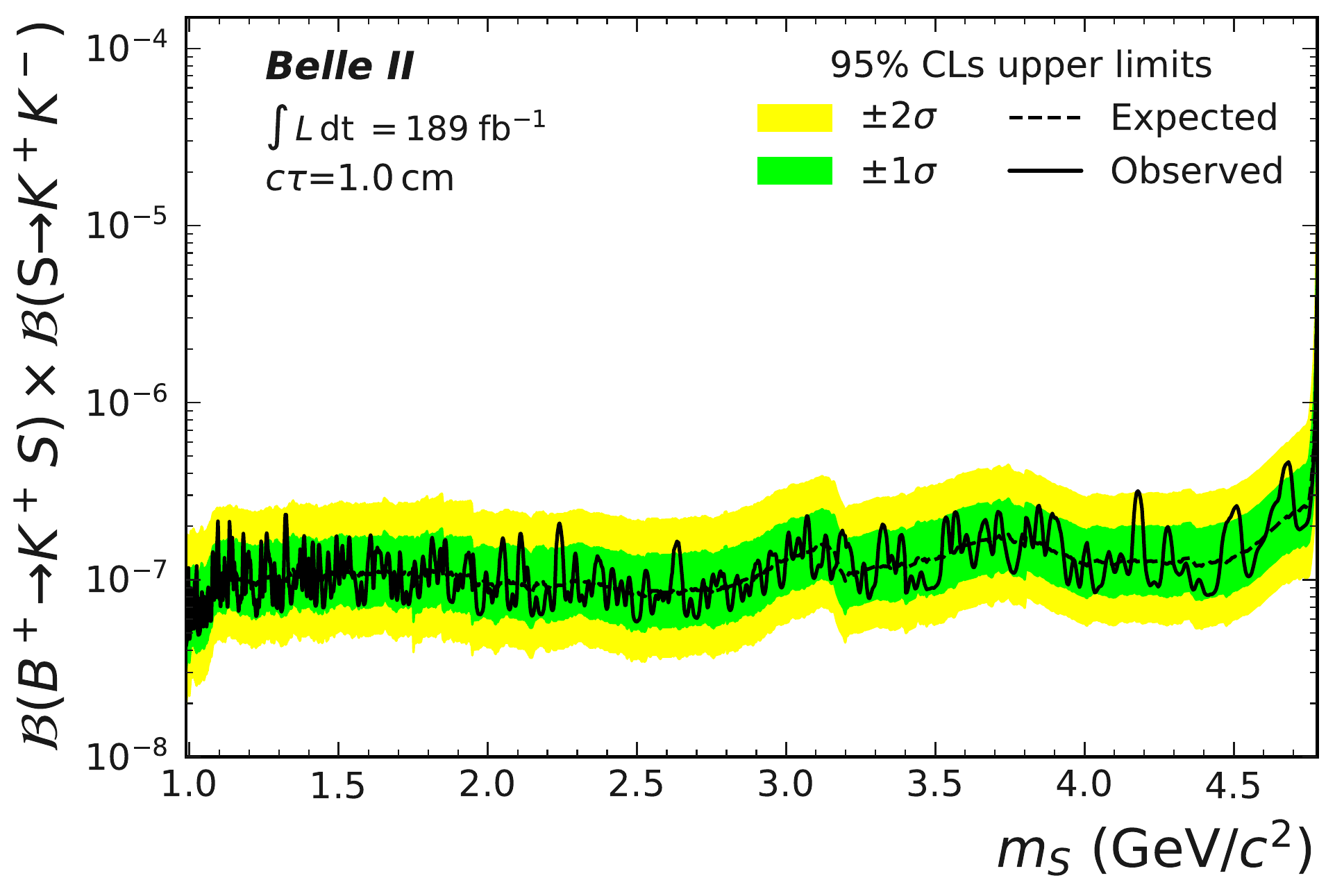}%
}%
\hspace*{\fill}
\subfigure[$B^+\to K^+S, S\to K^+K^-$, \newline lifetime of $c\tau=2.5\cm$.]{
  \label{subfit:brazil:Kp_K_1:L}%
  \includegraphics[width=0.31\textwidth]{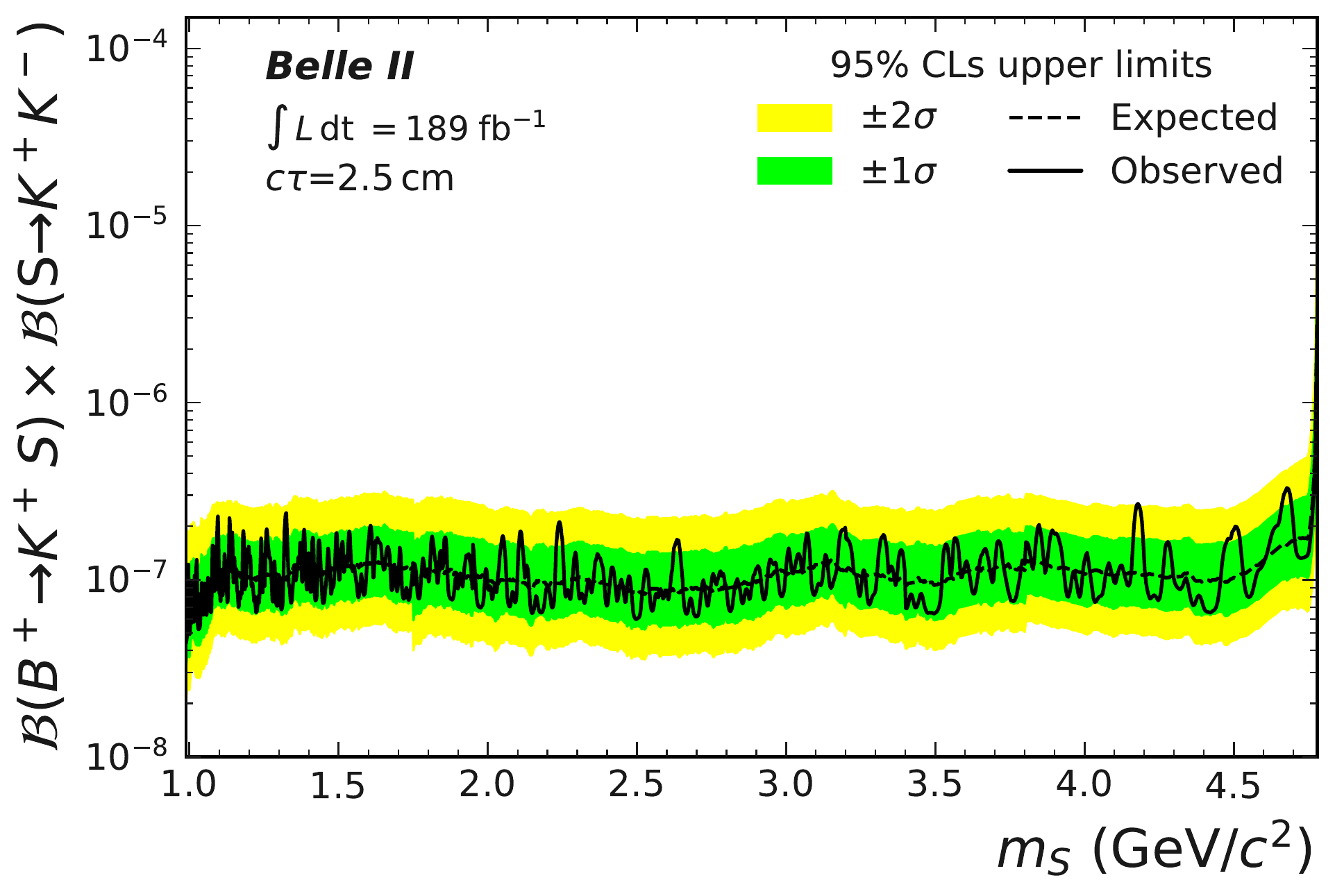}%
}
\caption{Expected and observed limits on the product of branching fractions $\mathcal{B}(B^+\to K S) \times \mathcal{B}(S\to K^+K^-)$ for lifetimes \hbox{$0.001 < c\tau < 2.5\,\cm$}.}\label{subfit:brazil:Kp_K_1}
\end{figure*}

\begin{figure*}[ht]%
\subfigure[$B^+\to K^+S, S\to K^+K^-$, \newline lifetime of $c\tau=5\cm$.]{%
  \label{subfit:brazil:Kp_K_2:A}%
  \includegraphics[width=0.31\textwidth]{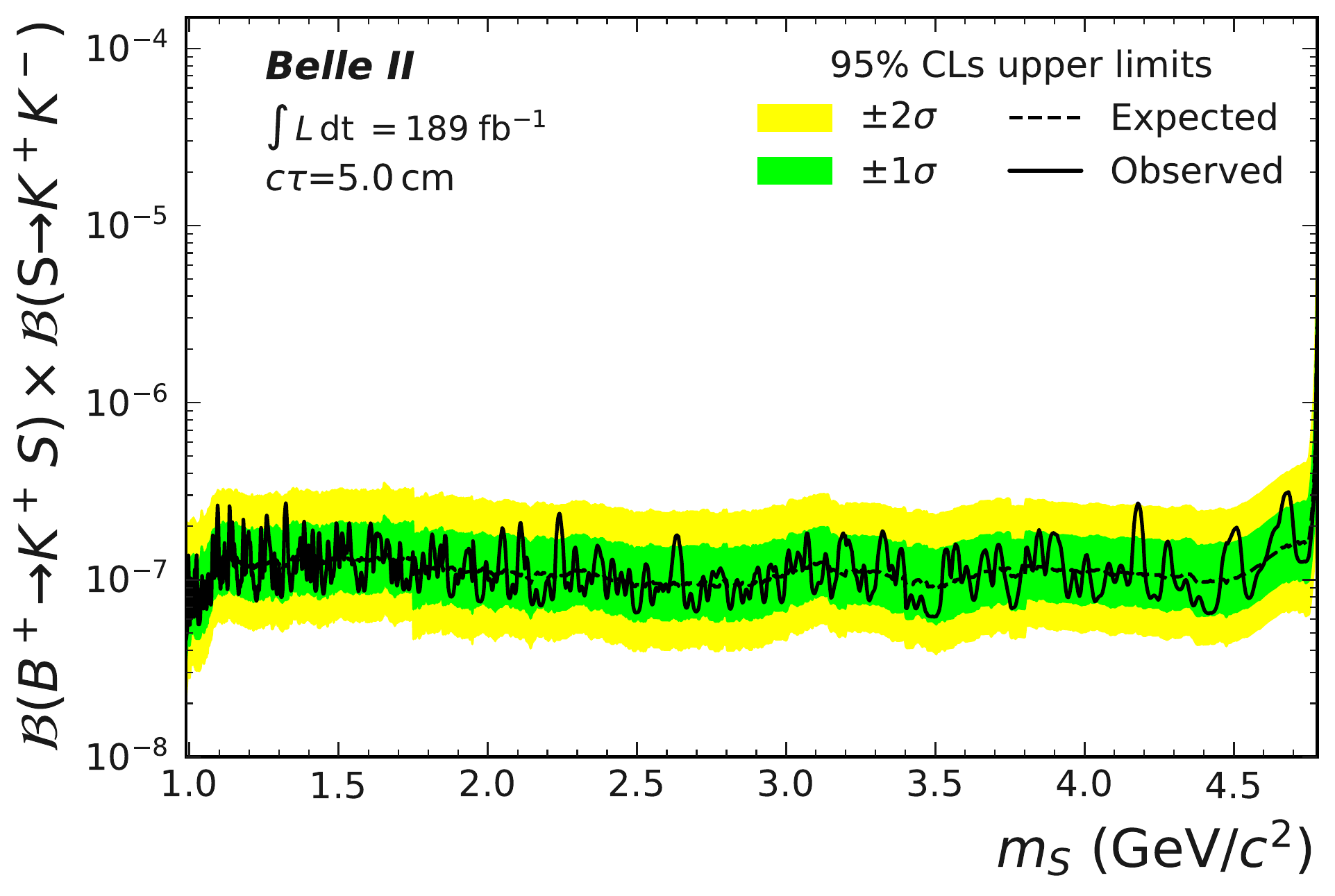}%
}%
\hspace*{\fill}
\subfigure[$B^+\to K^+S, S\to K^+K^-$, \newline lifetime of $c\tau=10\cm$.]{
  \label{subfit:brazil:Kp_K_2:B}%
  \includegraphics[width=0.31\textwidth]{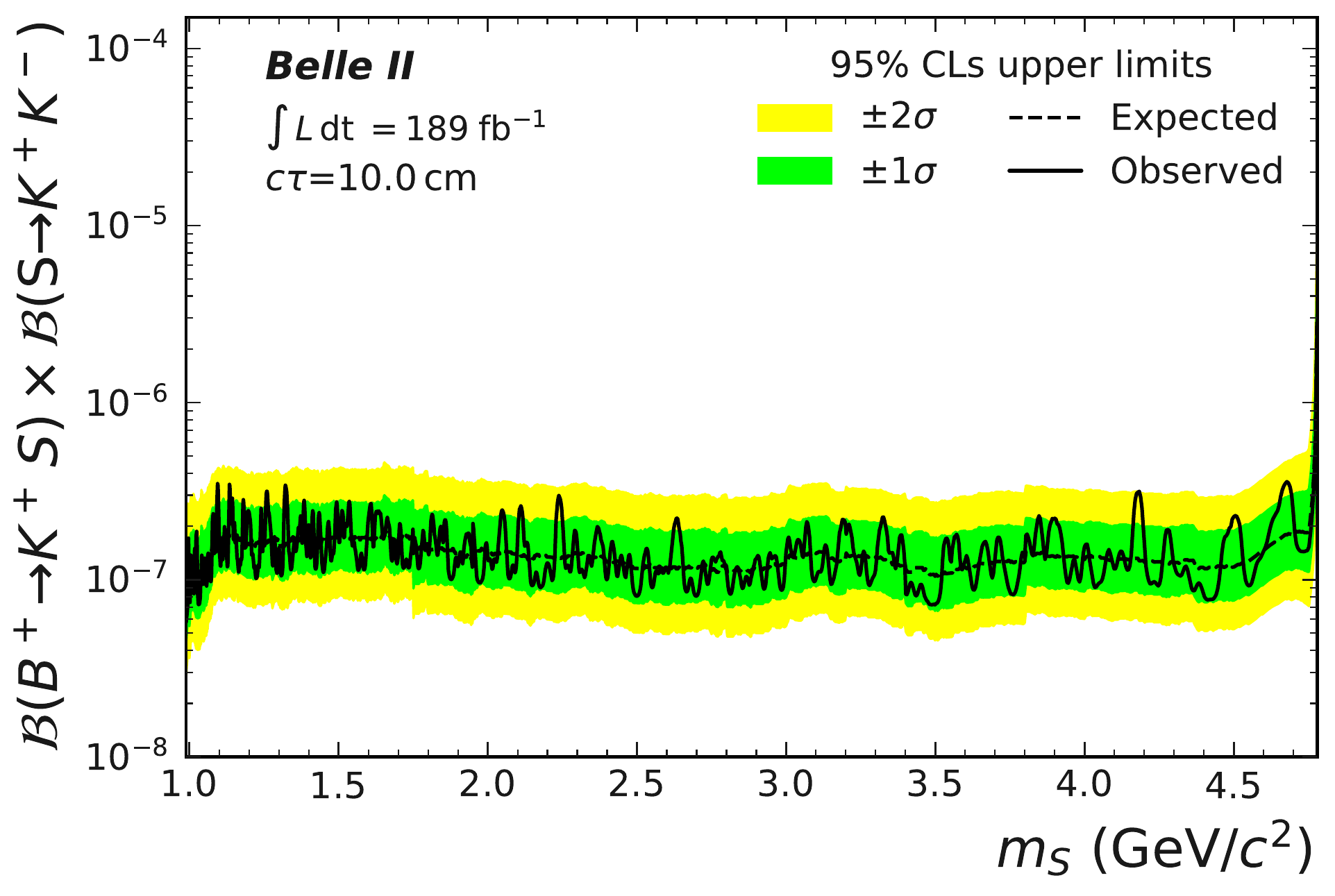}%
}%
\hspace*{\fill}
\subfigure[$B^+\to K^+S, S\to K^+K^-$, \newline lifetime of $c\tau=25\cm$.]{
  \label{subfit:brazil:Kp_K_2:C}%
  \includegraphics[width=0.31\textwidth]{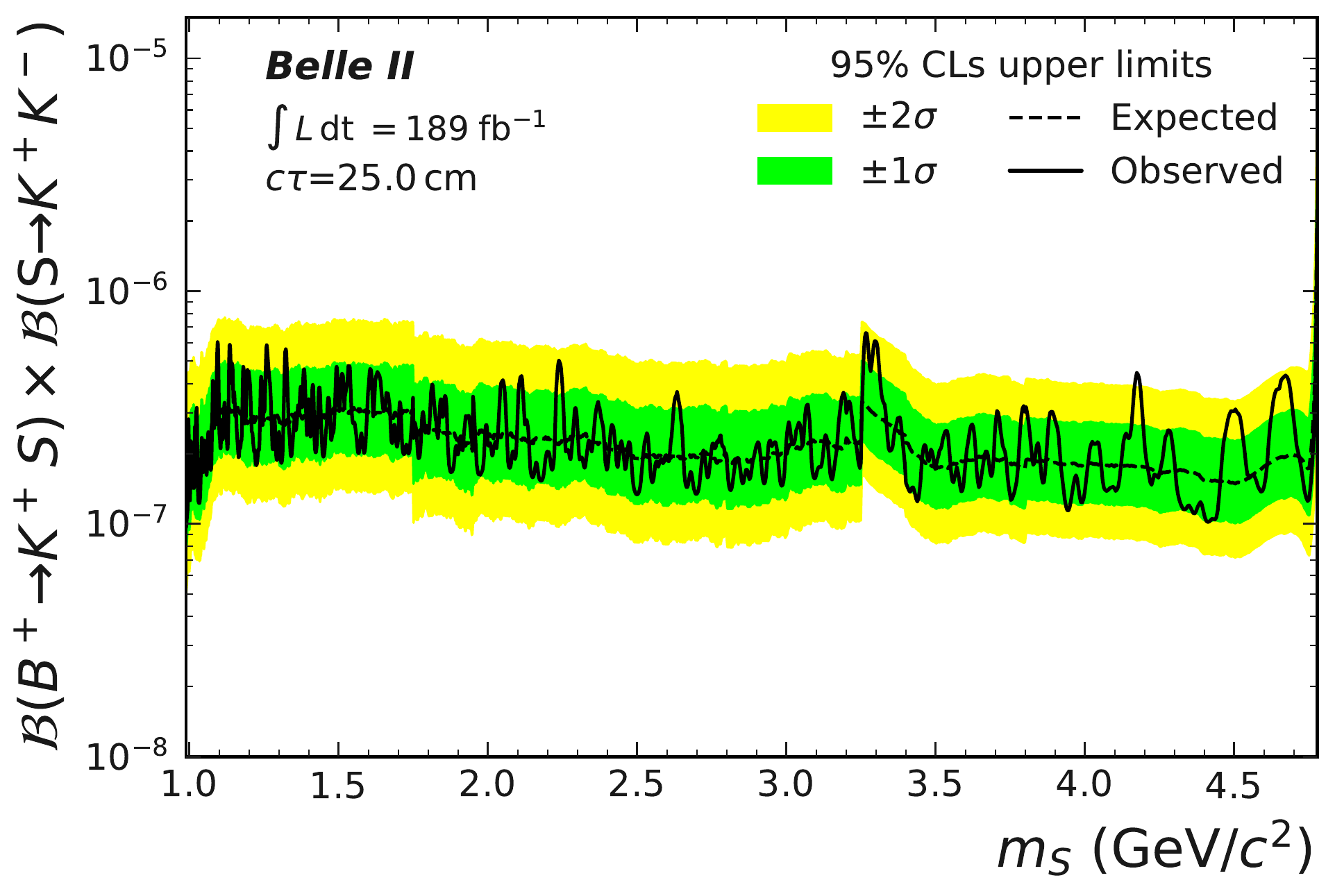}%
}
\subfigure[$B^+\to K^+S, S\to K^+K^-$, \newline lifetime of $c\tau=50\cm$.]{%
  \label{subfit:brazil:Kp_K_2:D}%
  \includegraphics[width=0.31\textwidth]{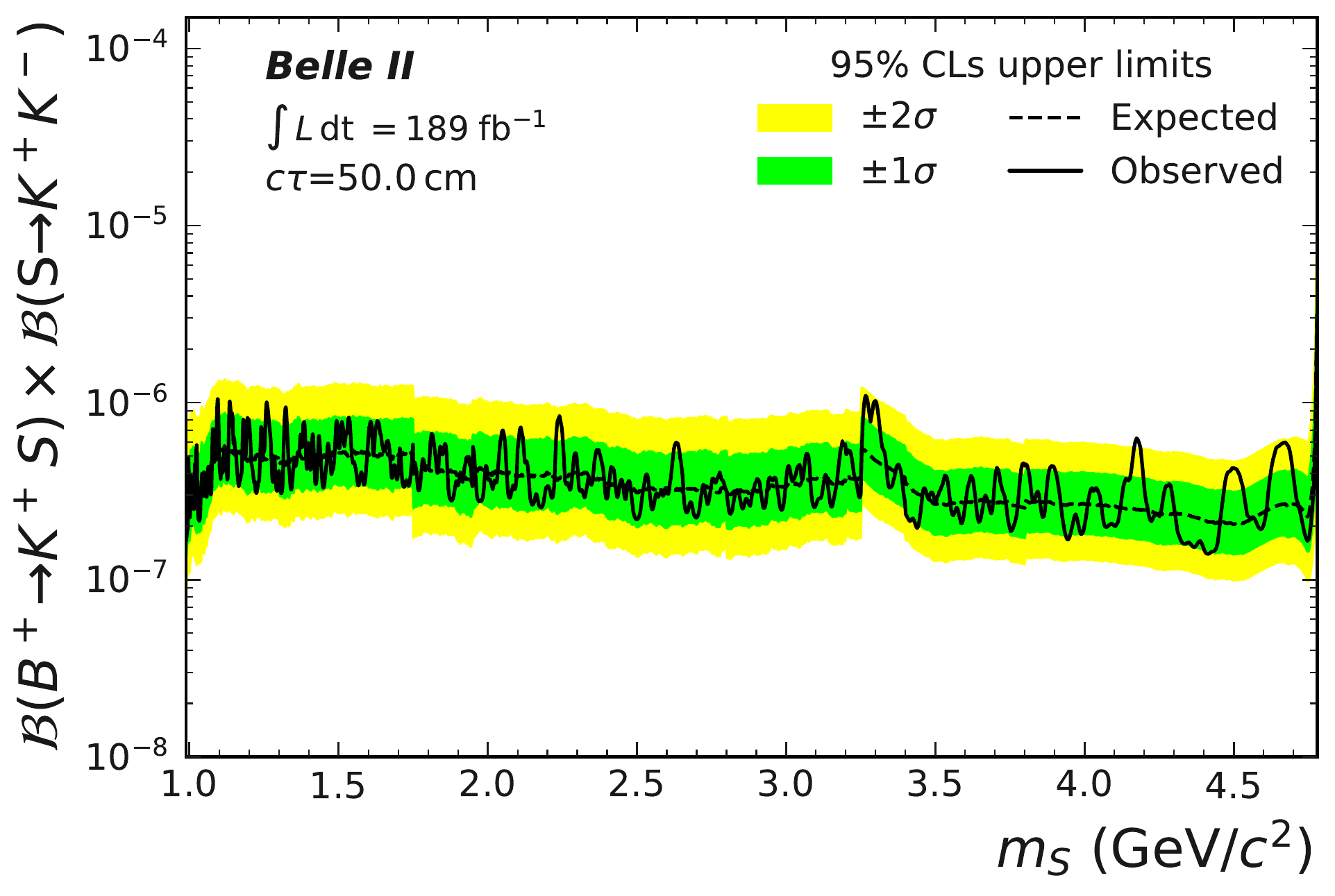}%
}%
\subfigure[$B^+\to K^+S, S\to K^+K^-$, \newline lifetime of $c\tau=100\cm$.]{
  \label{subfit:brazil:Kp_K_2:E}%
  \includegraphics[width=0.31\textwidth]{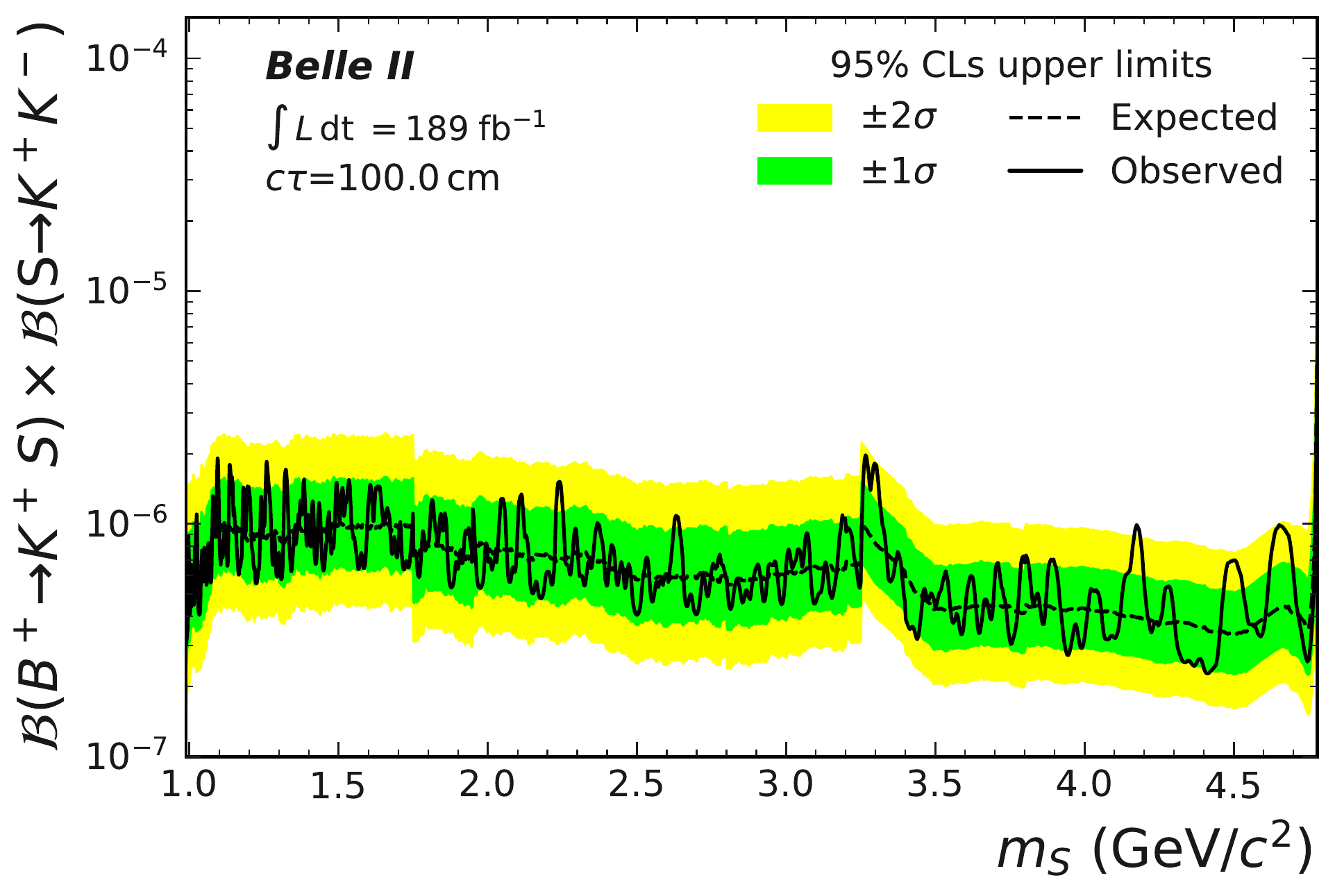}%
}%
\caption{Expected and observed limits on the product of branching fractions $\mathcal{B}(B^+\to K S) \times \mathcal{B}(S\to K^+K^-)$ for lifetimes \hbox{$5 < c\tau < 100\,\cm$}.}\label{subfit:brazil:Kp_K_2}
\end{figure*}

\begin{figure*}[ht]%
\subfigure[$\Bz\to \Kstarz(\to K^+\pi^-) S, S\to K^+K^-$, \newline lifetime of $c\tau=0.001\cm$.]{%
  \label{subfit:brazil:Kstar_K_1:A}%
  \includegraphics[width=0.31\textwidth]{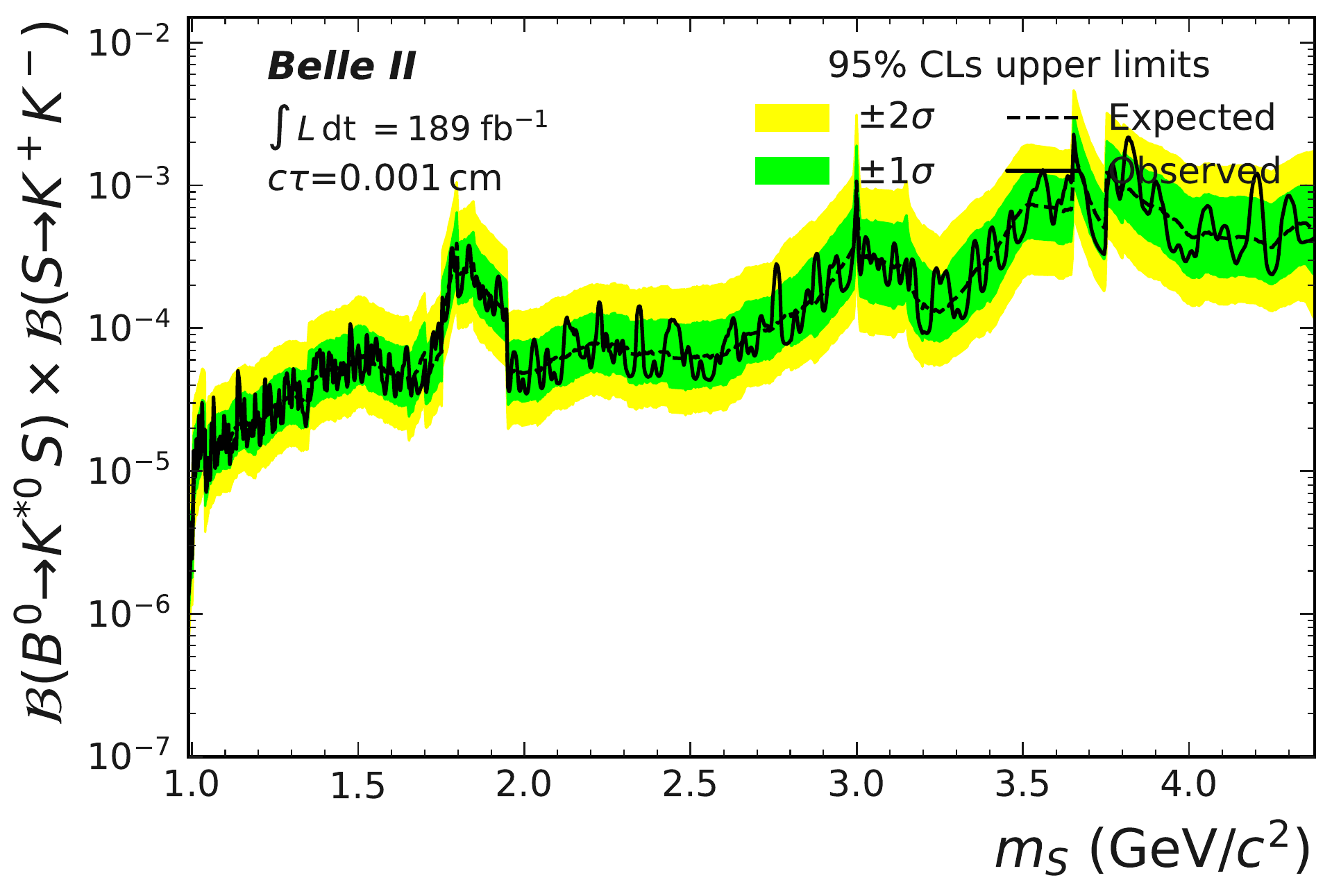}%
}%
\hspace*{\fill}
\subfigure[$\Bz\to \Kstarz(\to K^+\pi^-) S, S\to K^+K^-$, \newline lifetime of $c\tau=0.003\cm$.]{
  \label{subfit:brazil:Kstar_K_1:B}%
  \includegraphics[width=0.31\textwidth]{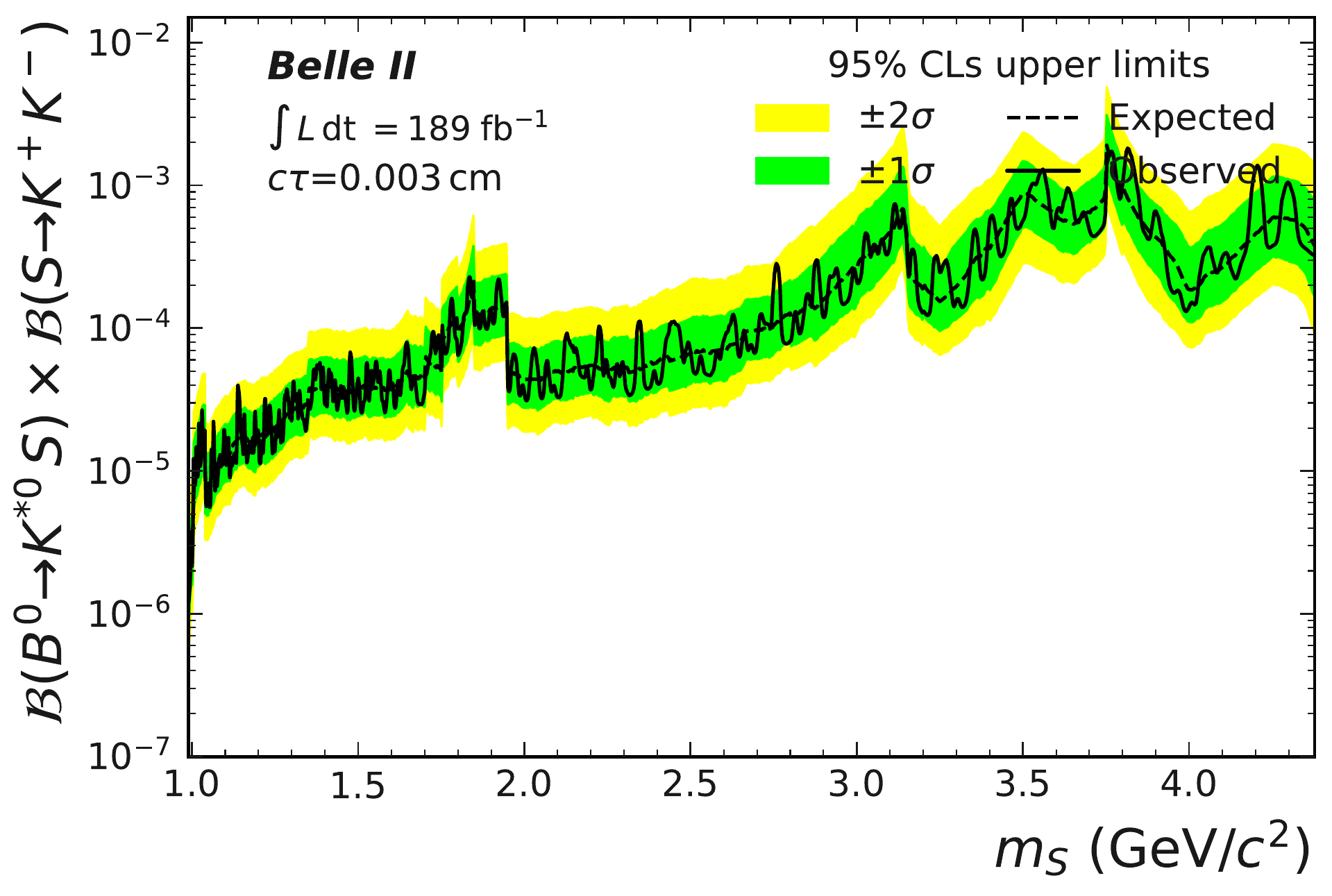}%
}%
\hspace*{\fill}
\subfigure[$\Bz\to \Kstarz(\to K^+\pi^-) S, S\to K^+K^-$, \newline lifetime of $c\tau=0.005\cm$.]{
  \label{subfit:brazil:Kstar_K_1:C}%
  \includegraphics[width=0.31\textwidth]{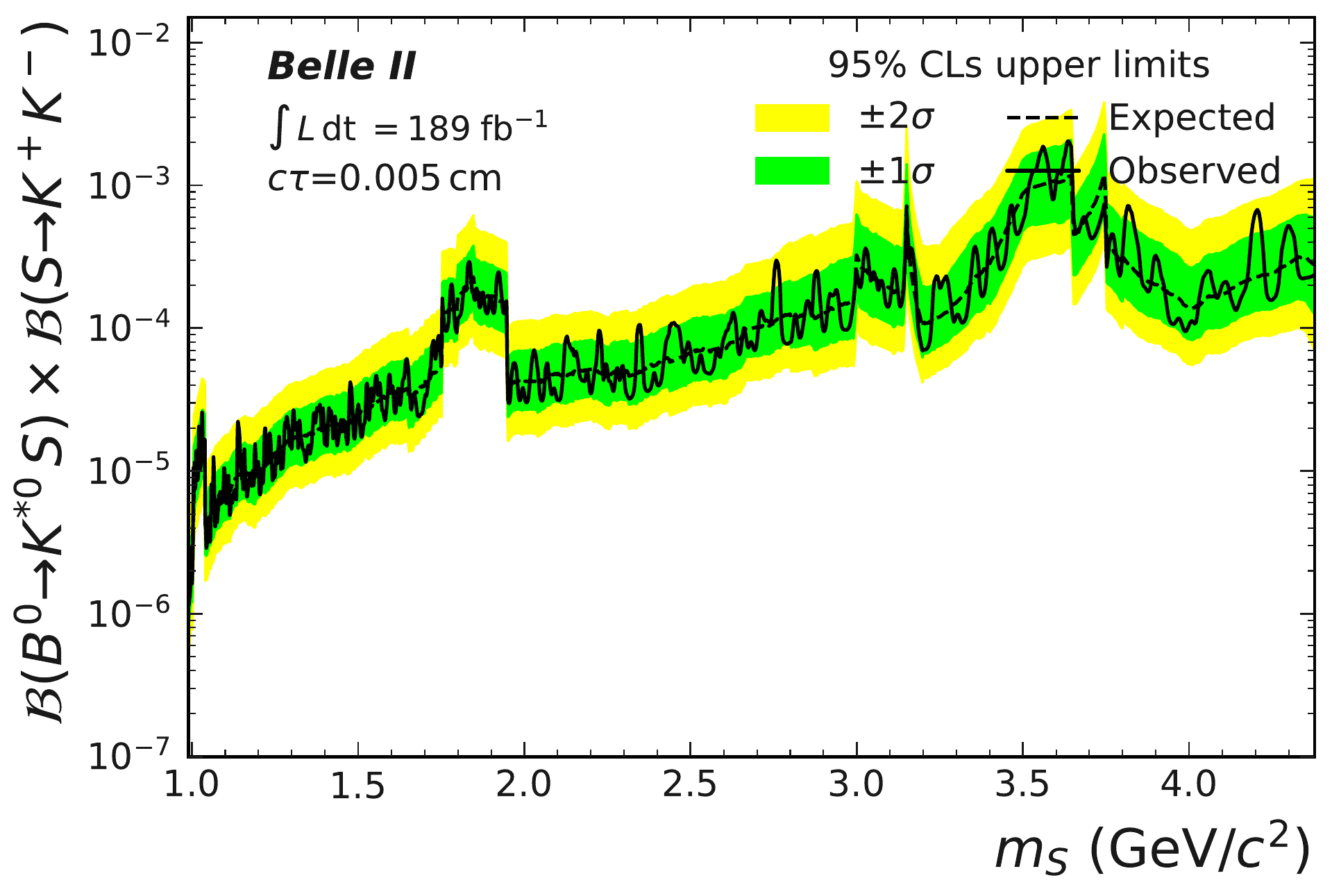}%
}
\subfigure[$\Bz\to \Kstarz(\to K^+\pi^-) S, S\to K^+K^-$, \newline lifetime of $c\tau=0.007\cm$.]{%
  \label{subfit:brazil:Kstar_K_1:D}%
  \includegraphics[width=0.31\textwidth]{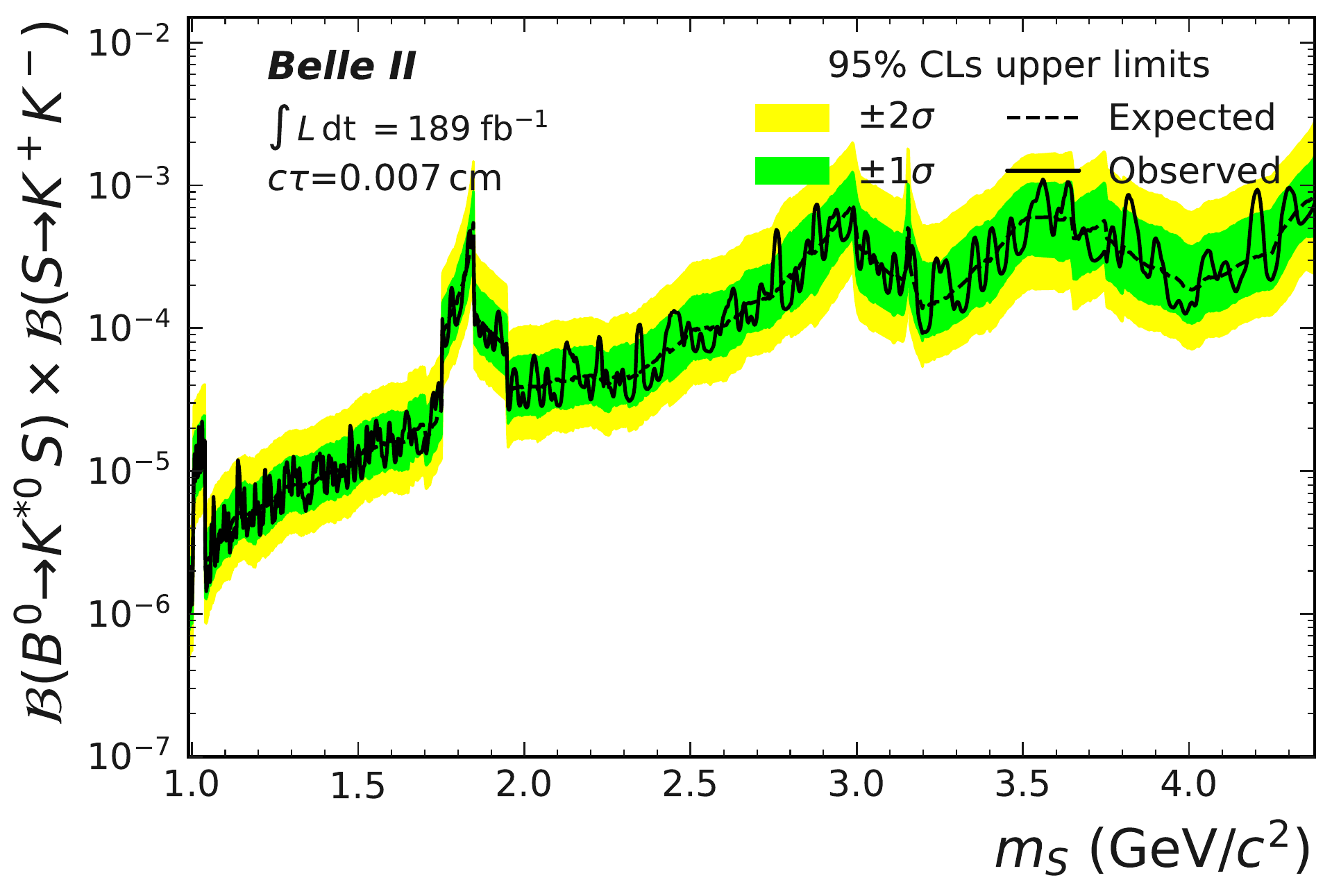}%
}%
\hspace*{\fill}
\subfigure[$\Bz\to \Kstarz(\to K^+\pi^-) S, S\to K^+K^-$, \newline lifetime of $c\tau=0.01\cm$.]{
  \label{subfit:brazil:Kstar_K_1:E}%
  \includegraphics[width=0.31\textwidth]{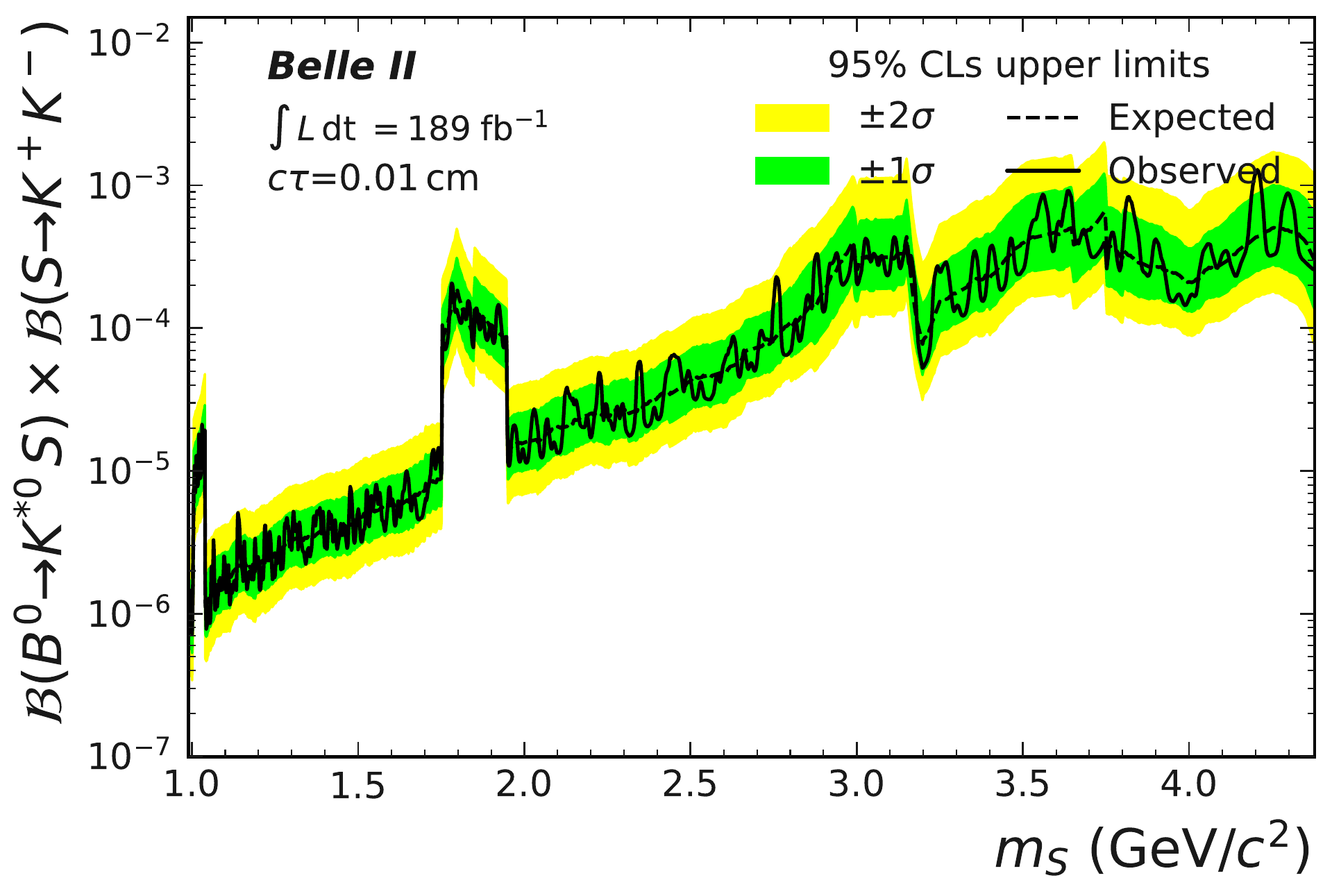}%
}%
\hspace*{\fill}
\subfigure[$\Bz\to \Kstarz(\to K^+\pi^-) S, S\to K^+K^-$, \newline lifetime of $c\tau=0.025\cm$.]{
  \label{subfit:brazil:Kstar_K_1:F}%
  \includegraphics[width=0.31\textwidth]{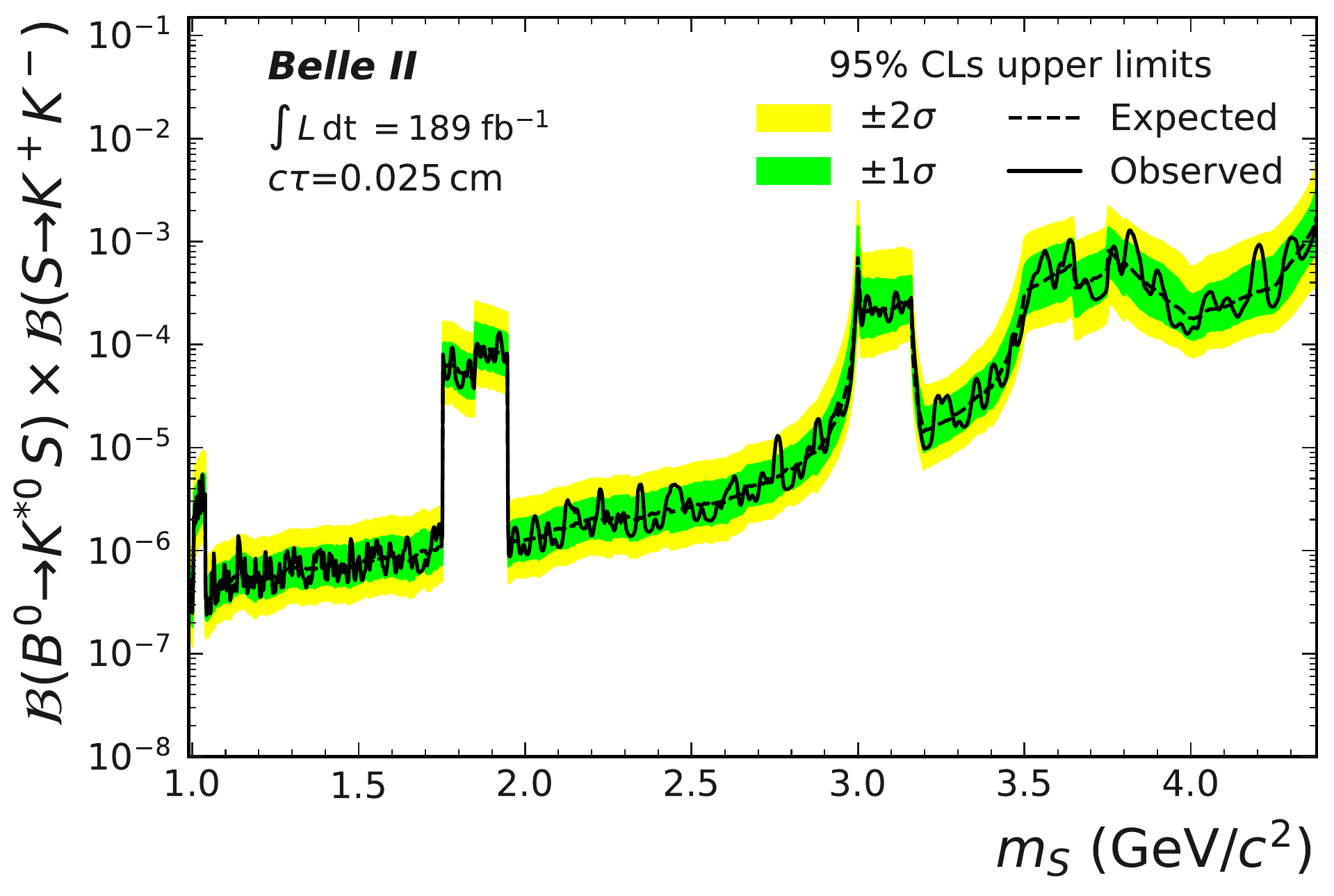}%
}
\subfigure[$\Bz\to \Kstarz(\to K^+\pi^-) S, S\to K^+K^-$, \newline lifetime of $c\tau=0.05\cm$.]{%
  \label{subfit:brazil:Kstar_K_1:G}%
  \includegraphics[width=0.31\textwidth]{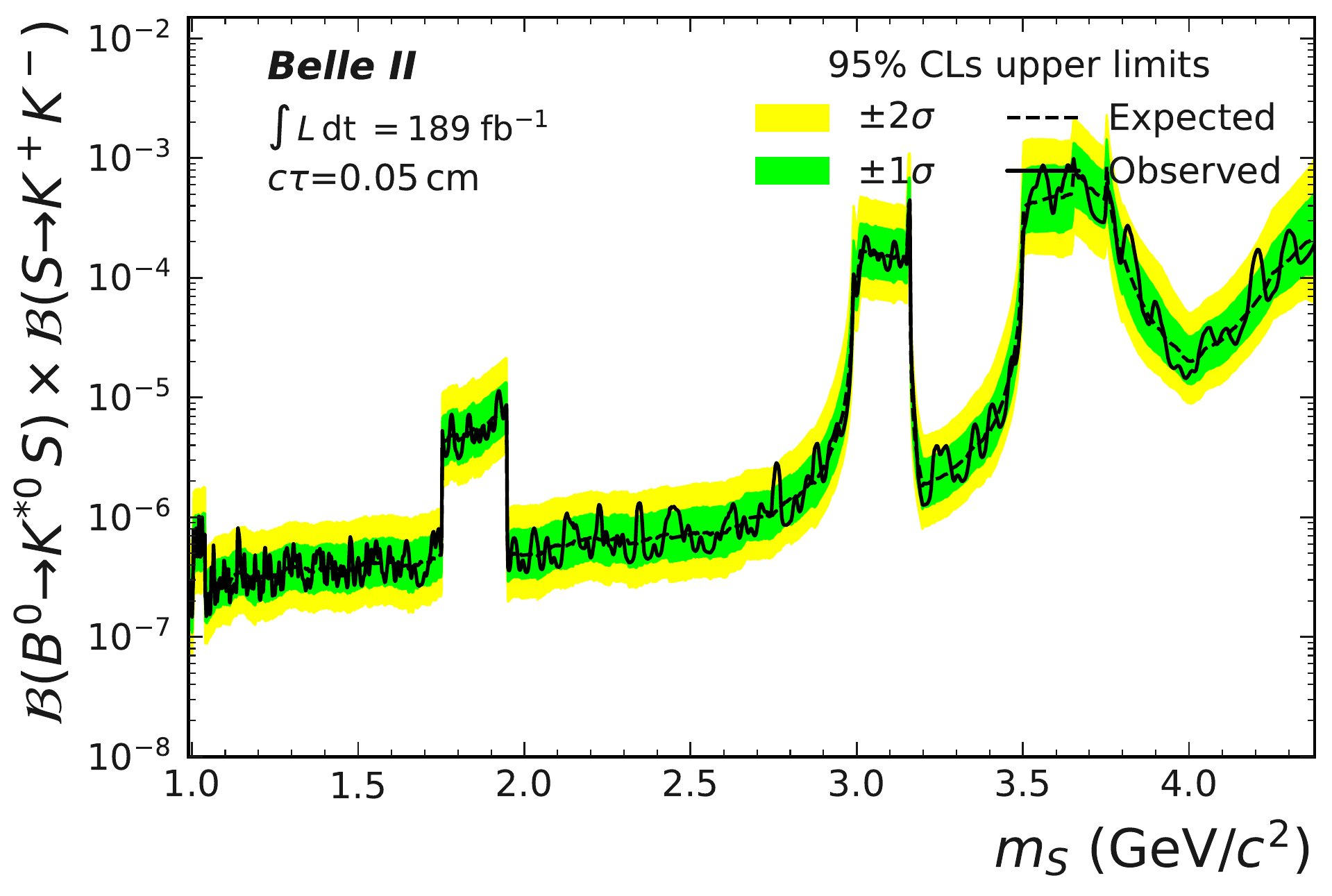}%
}%
\hspace*{\fill}
\subfigure[$\Bz\to \Kstarz(\to K^+\pi^-) S, S\to K^+K^-$, \newline lifetime of $c\tau=0.100\cm$.]{
  \label{subfit:brazil:Kstar_K_1:H}%
  \includegraphics[width=0.31\textwidth]{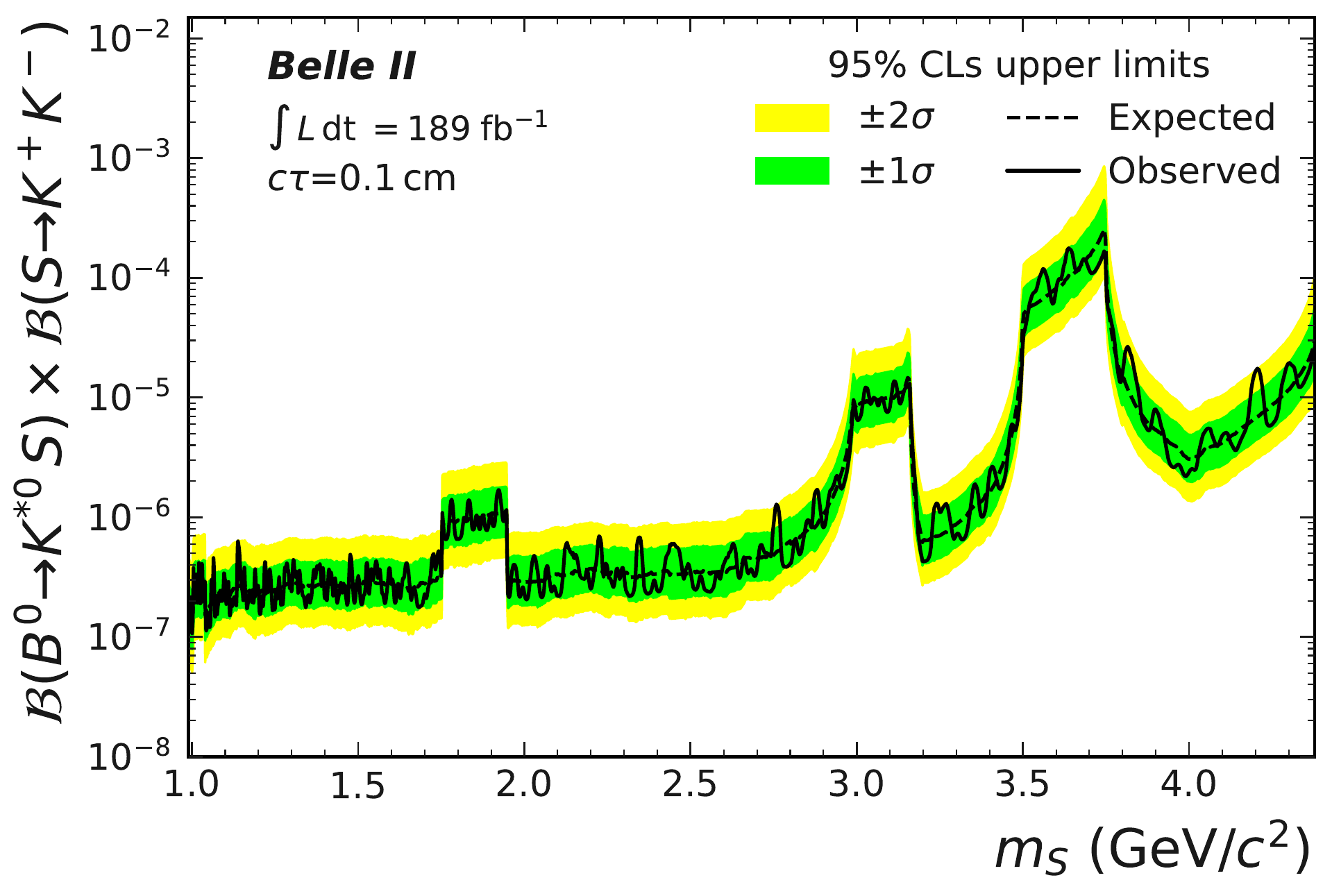}%
}%
\hspace*{\fill}
\subfigure[$\Bz\to \Kstarz(\to K^+\pi^-) S, S\to K^+K^-$, \newline lifetime of $c\tau=0.25\cm$.]{
  \label{subfit:brazil:Kstar_K_1:I}%
  \includegraphics[width=0.31\textwidth]{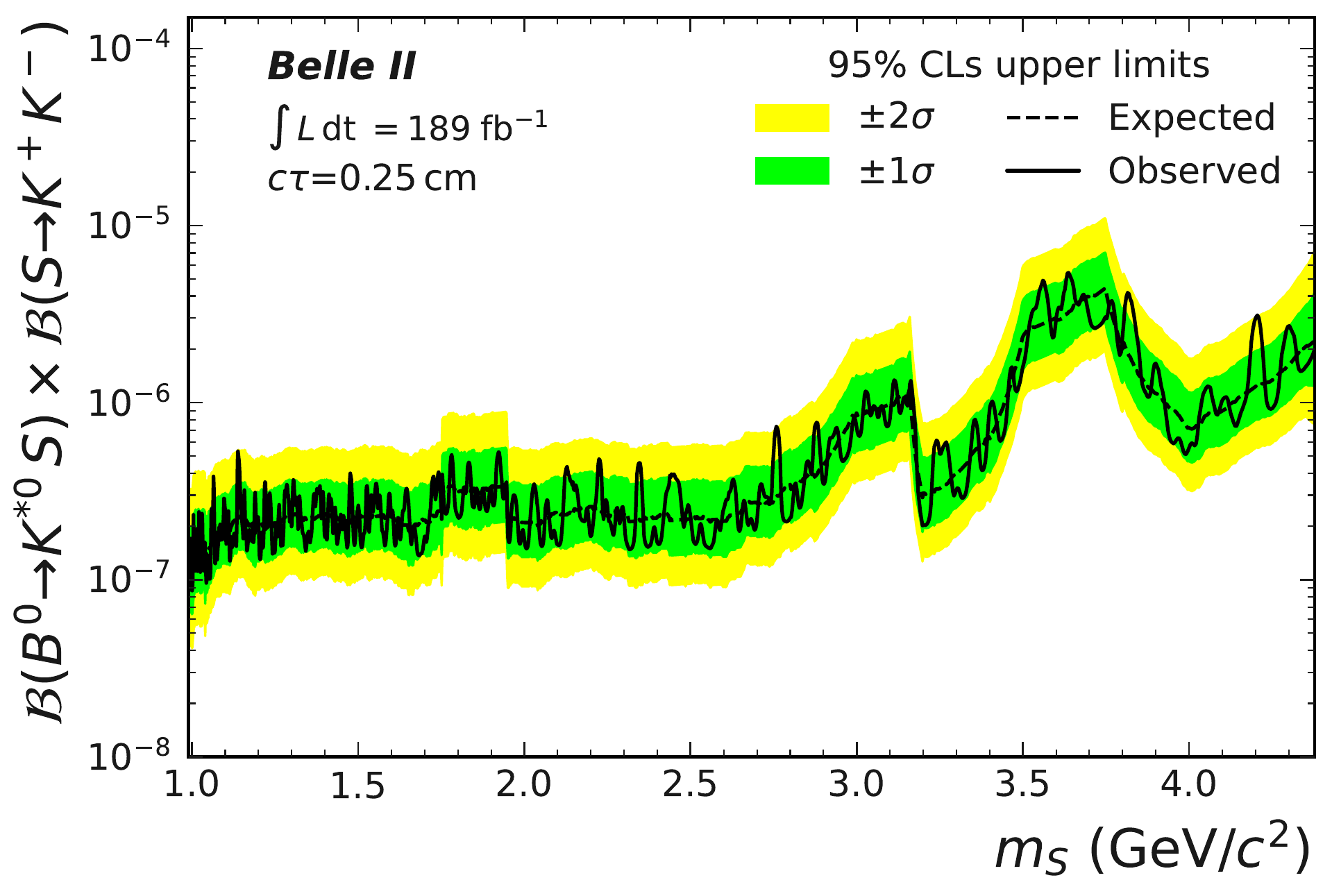}%
}
\subfigure[$\Bz\to \Kstarz(\to K^+\pi^-) S, S\to K^+K^-$, \newline lifetime of $c\tau=0.5\cm$.]{%
  \label{subfit:brazil:Kstar_K_1:J}%
  \includegraphics[width=0.31\textwidth]{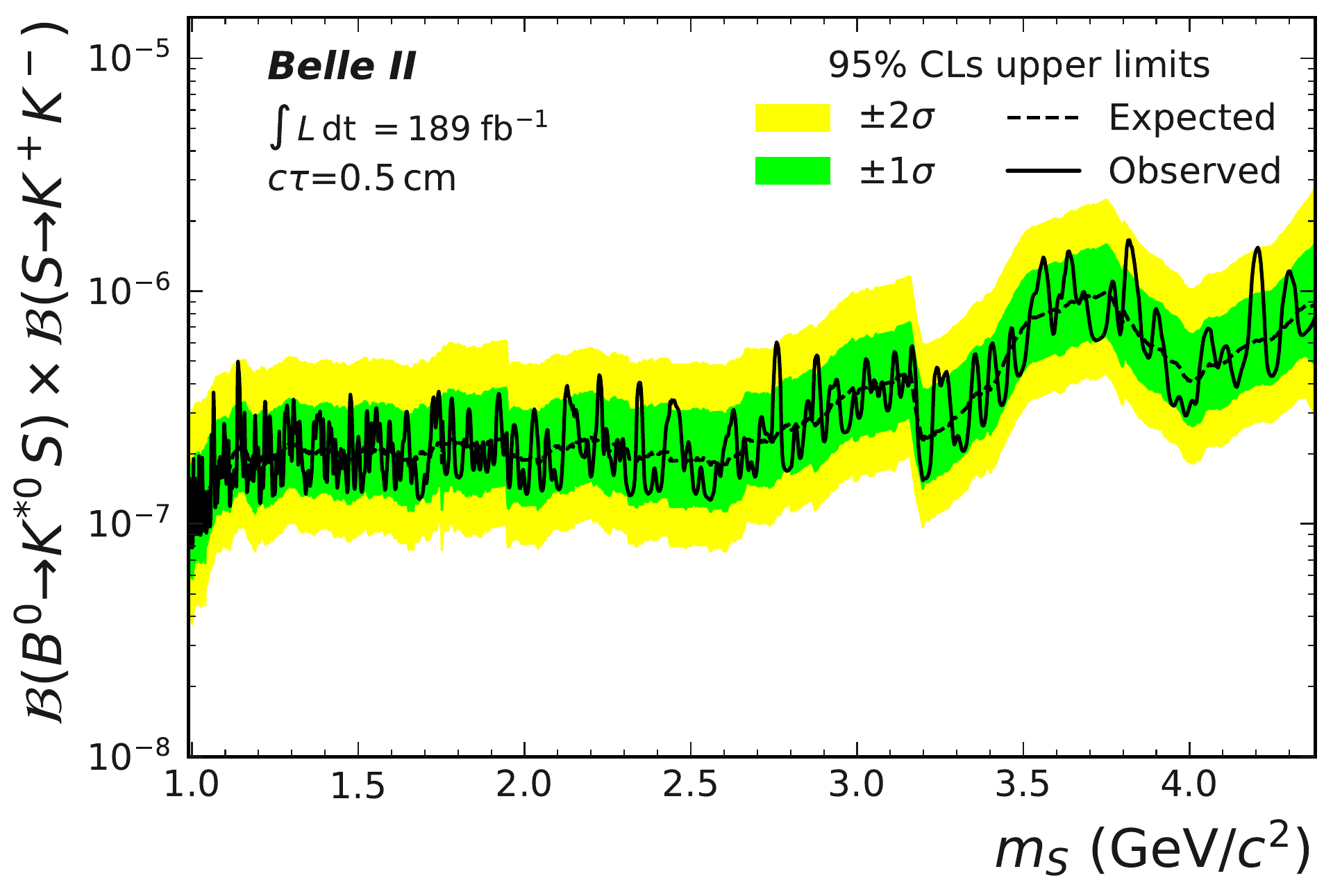}%
}%
\hspace*{\fill}
\subfigure[$\Bz\to \Kstarz(\to K^+\pi^-) S, S\to K^+K^-$, \newline lifetime of $c\tau=1\cm$.]{
  \label{subfit:brazil:Kstar_K_1:K}%
  \includegraphics[width=0.31\textwidth]{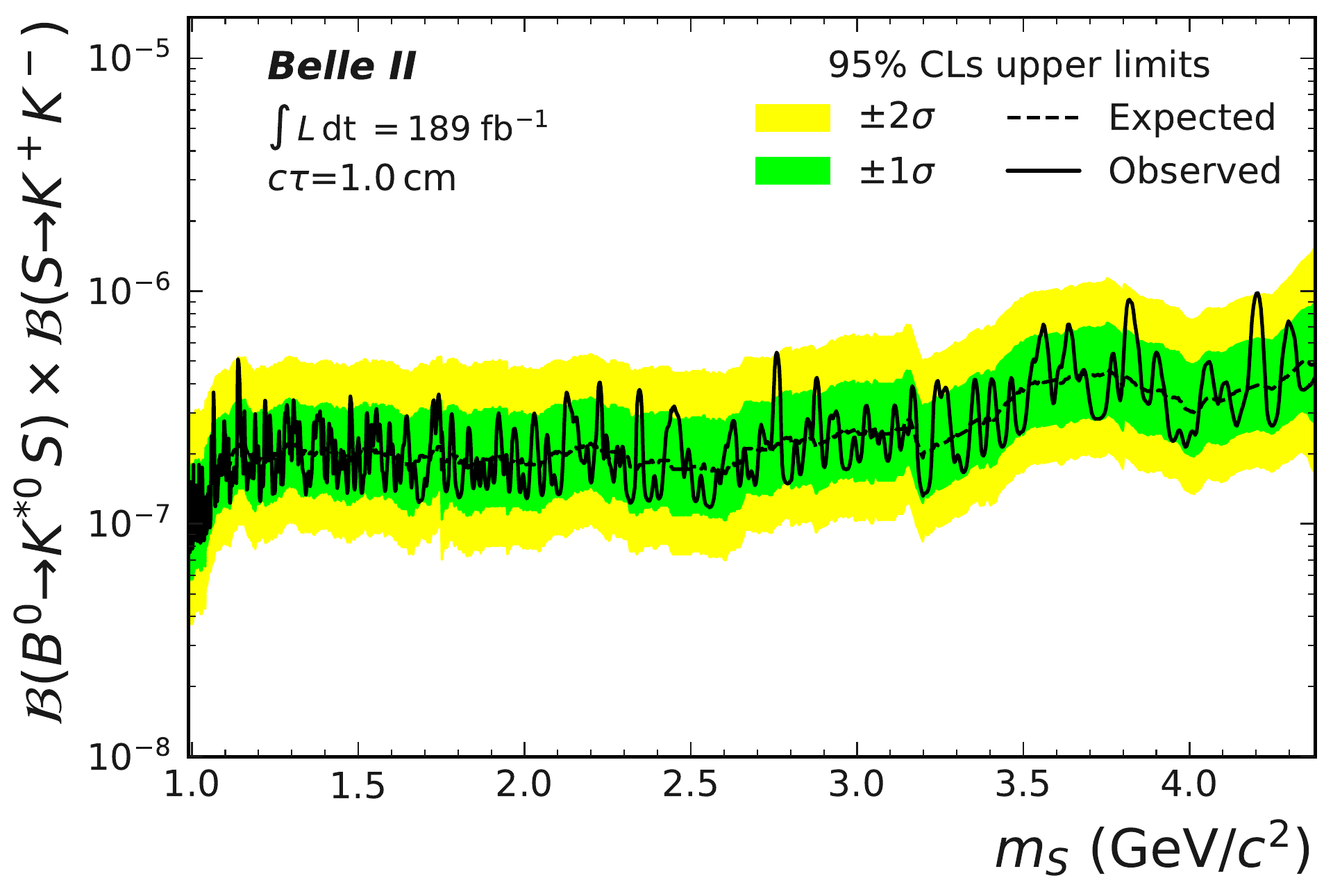}%
}%
\hspace*{\fill}
\subfigure[$\Bz\to \Kstarz(\to K^+\pi^-) S, S\to K^+K^-$, \newline lifetime of $c\tau=2.5\cm$.]{
  \label{subfit:brazil:Kstar_K_1:L}%
  \includegraphics[width=0.31\textwidth]{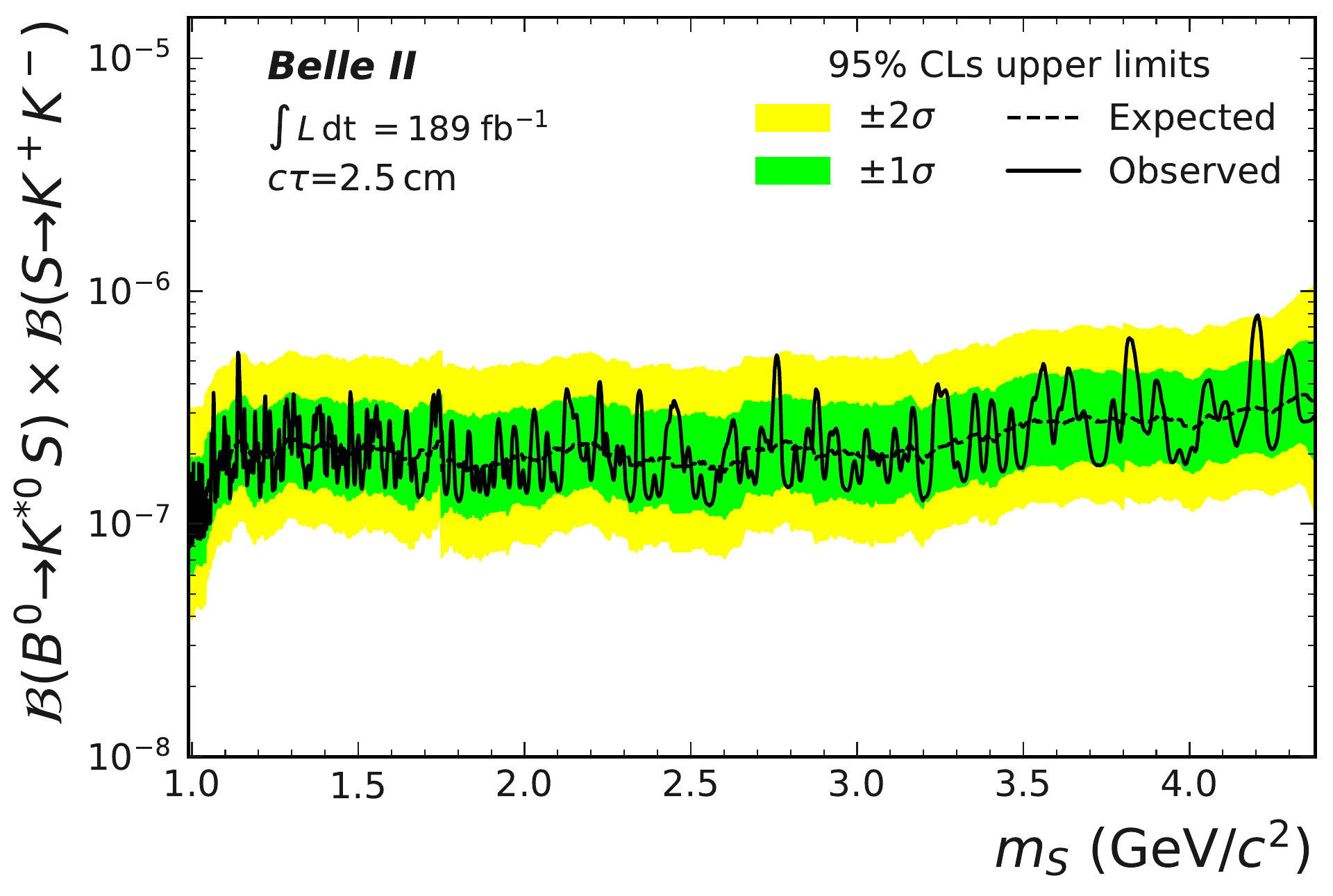}%
}
\caption{Expected and observed limits on the product of branching fractions $\mathcal{B}(B^0\to \Kstarz(\to K^+\pi^-) S) \times \mathcal{B}(S\to K^+K^-)$ for lifetimes \hbox{$0.001 < c\tau < 2.5\,\cm$}.}\label{subfit:brazil:Kstar_K_1}
\end{figure*}

\begin{figure*}[ht]%
\subfigure[$\Bz\to \Kstarz(\to K^+\pi^-) S, S\to K^+K^-$, \newline lifetime of $c\tau=5\cm$.]{%
  \label{subfit:brazil:Kstar_K_2:A}%
  \includegraphics[width=0.31\textwidth]{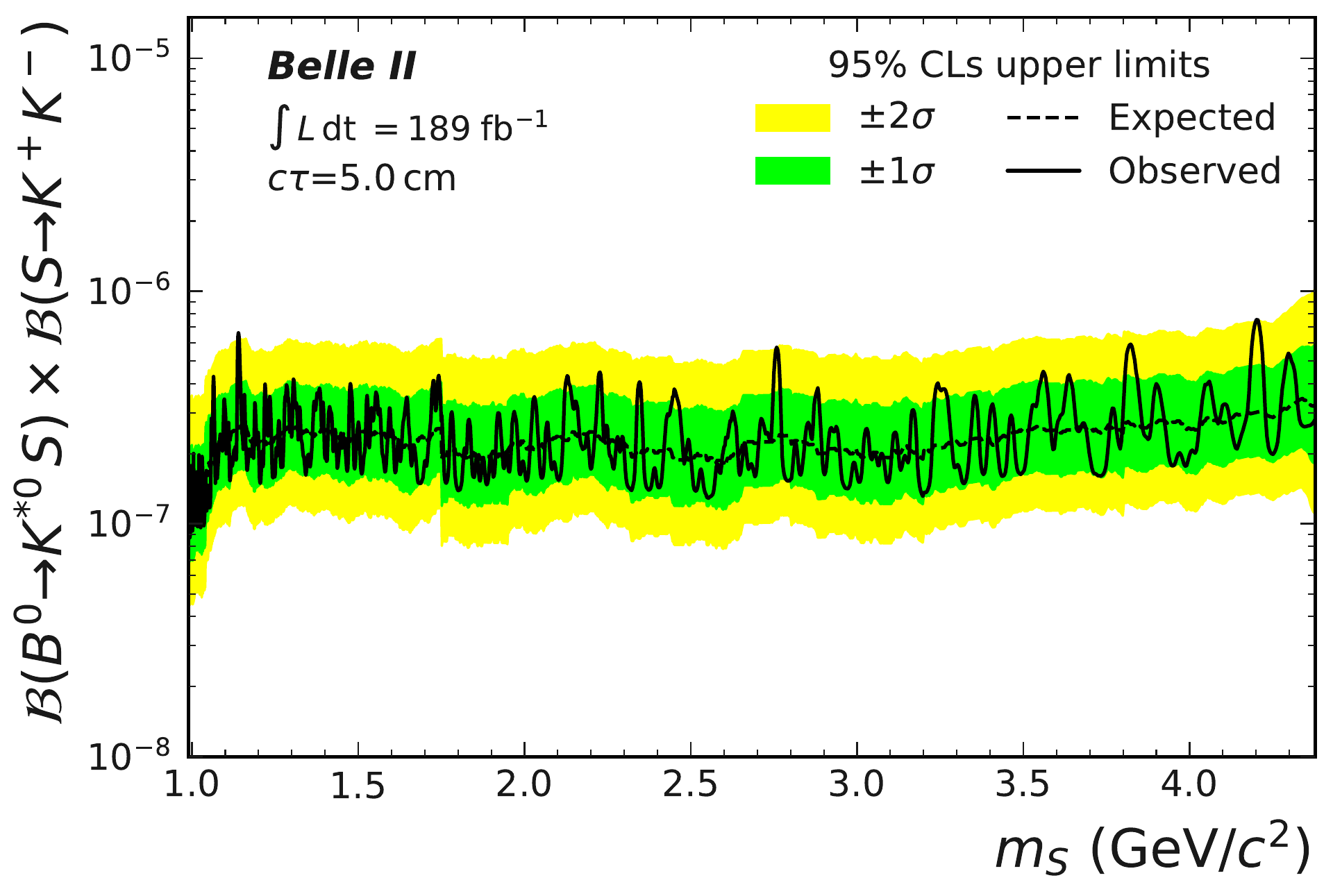}%
}%
\hspace*{\fill}
\subfigure[$\Bz\to \Kstarz(\to K^+\pi^-) S, S\to K^+K^-$, \newline lifetime of $c\tau=10\cm$.]{
  \label{subfit:brazil:Kstar_K_2:B}%
  \includegraphics[width=0.31\textwidth]{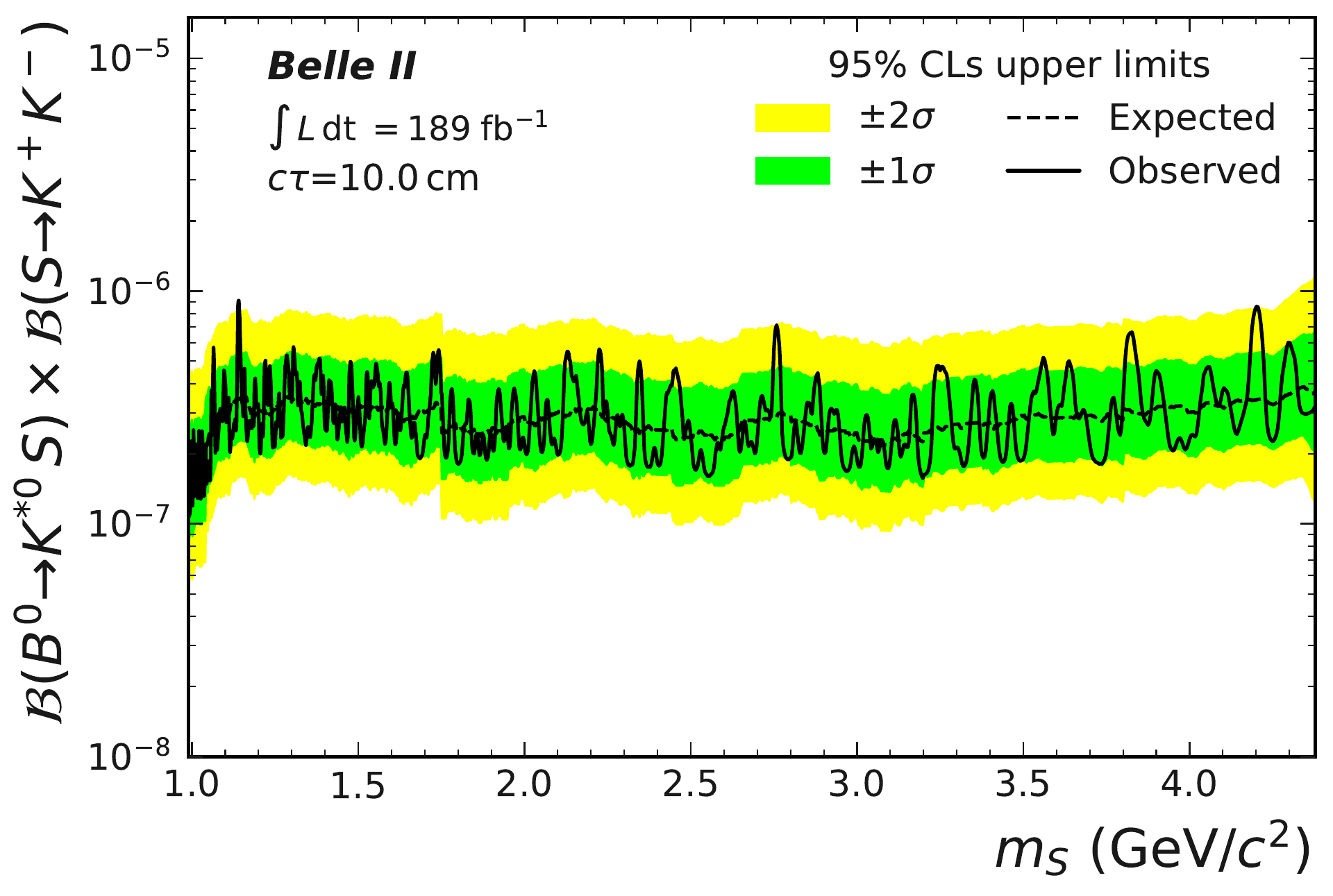}%
}%
\hspace*{\fill}
\subfigure[$\Bz\to \Kstarz(\to K^+\pi^-) S, S\to K^+K^-$, \newline lifetime of $c\tau=25\cm$.]{
  \label{subfit:brazil:Kstar_K_2:C}%
  \includegraphics[width=0.31\textwidth]{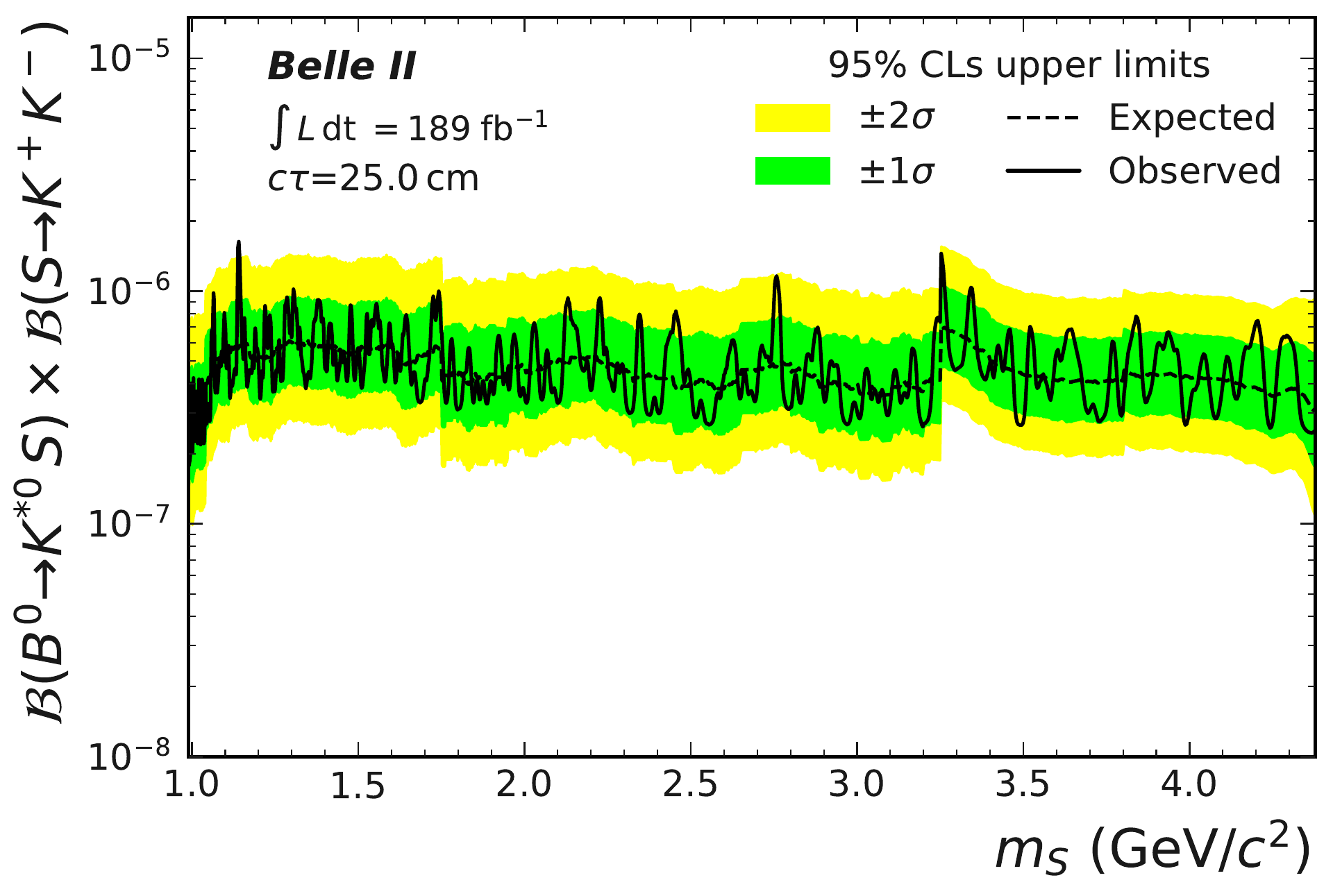}%
}
\subfigure[$\Bz\to \Kstarz(\to K^+\pi^-) S, S\to K^+K^-$, \newline lifetime of $c\tau=50\cm$.]{%
  \label{subfit:brazil:Kstar_K_2:D}%
  \includegraphics[width=0.31\textwidth]{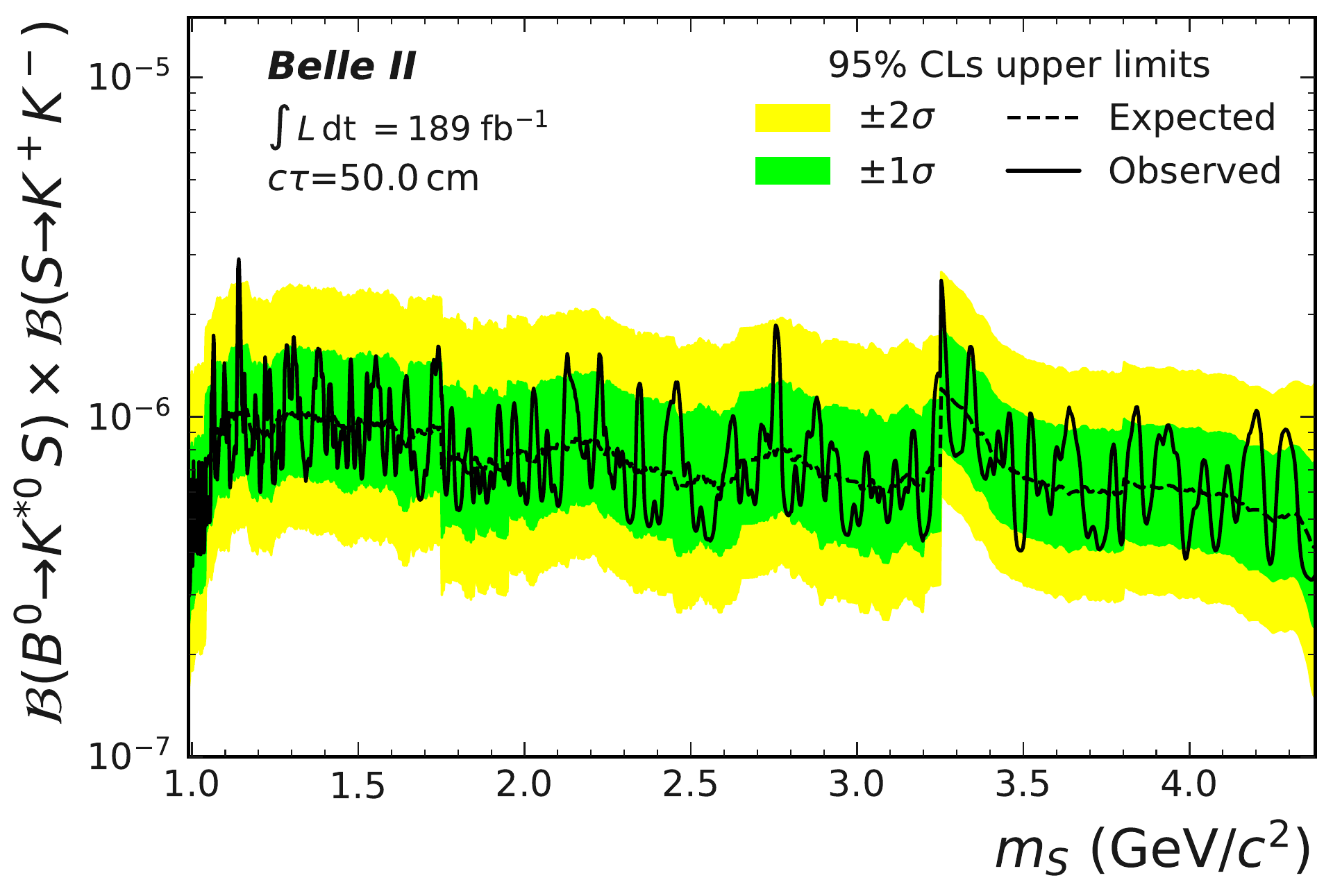}%
}%
\subfigure[$\Bz\to \Kstarz(\to K^+\pi^-) S, S\to K^+K^-$, \newline lifetime of $c\tau=100\cm$.]{
  \label{subfit:brazil:Kstar_K_2:E}%
  \includegraphics[width=0.31\textwidth]{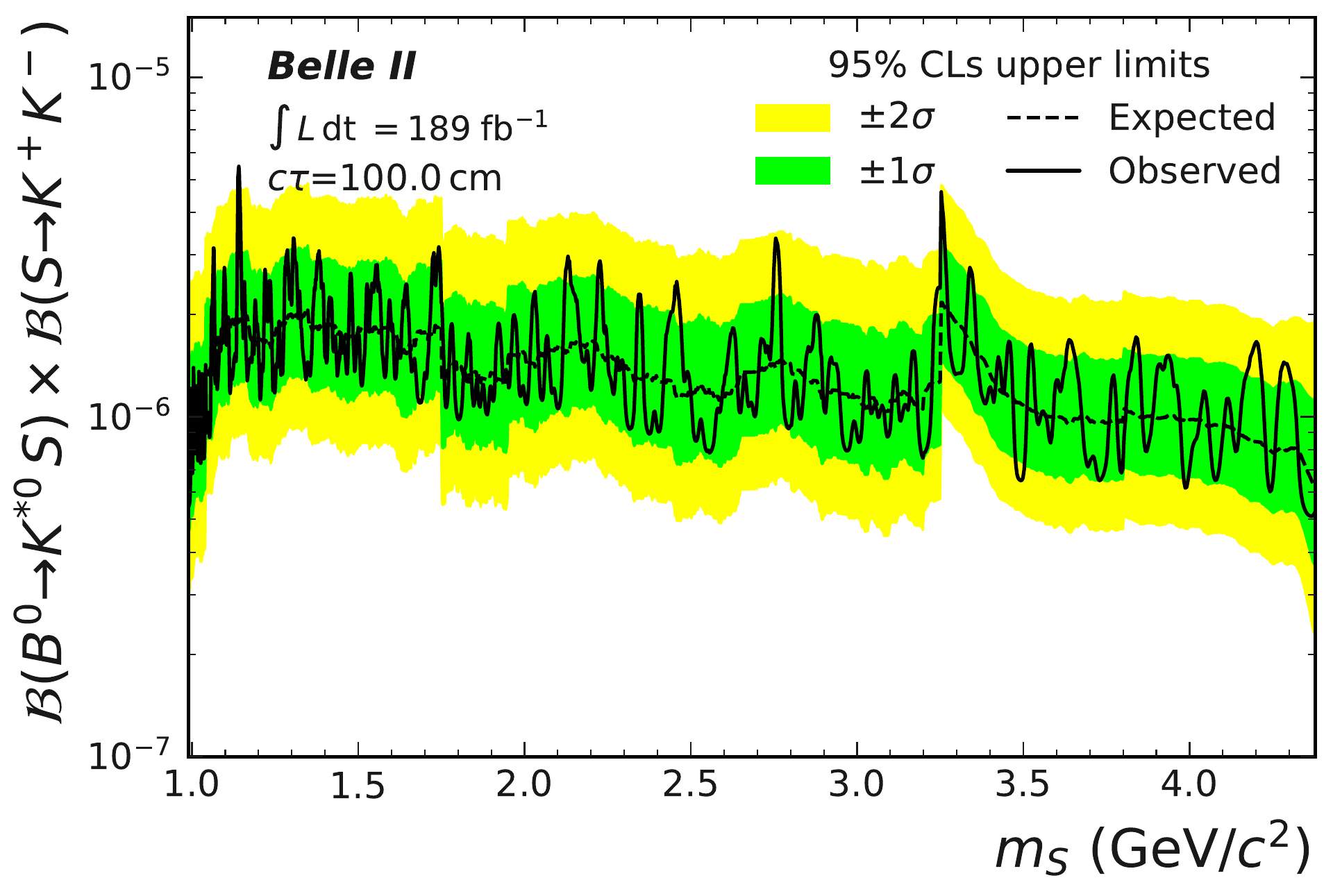}%
}%
\caption{Expected and observed limits on the product of branching fractions $\mathcal{B}(B^0\to \Kstarz(\to K^+\pi^-) S) \times \mathcal{B}(S\to K^+K^-)$ for lifetimes \hbox{$5 < c\tau < 100\,\cm$}.}\label{subfit:brazil:Kstar_K_2}
\end{figure*} 
}

\end{document}